\documentclass[11pt]{article}

\usepackage[T1]{fontenc}
\usepackage[utf8]{inputenc}
\usepackage{lmodern}
\usepackage{microtype}

\usepackage{amsmath,amssymb,amsthm,mathtools}
\usepackage{stmaryrd} 
\usepackage{enumitem}
\usepackage{xcolor}
\usepackage{hyperref}
\hypersetup{
  colorlinks=true,
  linkcolor=blue,
  citecolor=blue,
  urlcolor=blue,
  pdfauthor={Ben Goertzel},
  pdftitle={A Quantale-Weakness Route to P \neq NP}
}

\newtheorem{theorem}{Theorem}[section]
\newtheorem{lemma}[theorem]{Lemma}
\newtheorem{proposition}[theorem]{Proposition}
\newtheorem{corollary}[theorem]{Corollary}
\newtheorem{definition}[theorem]{Definition}
\newtheorem{hypothesis}[theorem]{Hypothesis}
\newtheorem{remark}[theorem]{Remark}

\newtheorem{construction}[theorem]{Construction}

\providecommand{\bits}{\{0,1\}}
\providecommand{\Kp}{K_{\mathrm{poly}}}
\providecommand{\E}{\mathbb{E}}
\providecommand{\Prb}{\mathbb{P}}
\providecommand{\TV}{\mathrm{TV}}
\providecommand{\Gap}{\mathrm{Gap}}
\providecommand{\CDENF}{\mathrm{CDENF}}
\providecommand{\rankG}{\mathrm{rank}_G}
\providecommand{\suppG}{\mathrm{supp}_G}
\providecommand{\supp}{\mathrm{supp}}
\providecommand{\poly}{\mathrm{poly}}

\providecommand{\At}{\operatorname{At}}
\providecommand{\Bad}{\operatorname{Bad}}
\providecommand{\Base}{\operatorname{Base}}
\providecommand{\Buf}{\operatorname{Buf}}
\providecommand{\CD}{\operatorname{CD}}
\providecommand{\CDSkew}{\operatorname{CDSkew}}
\providecommand{\CNF}{\operatorname{CNF}}
\providecommand{\Charge}{\operatorname{Charge}}
\providecommand{\Claim}{\operatorname{Claim}}
\providecommand{\Dbidir}{D_{\mathrm{bidir}}}
\providecommand{\Dinf}{D_{\infty}}
\providecommand{\Excess}{\operatorname{Excess}}
\providecommand{\Exp}{\operatorname{Exp}}
\providecommand{\FreshG}{\operatorname{Fresh}_G}
\providecommand{\Gauge}{\operatorname{Gauge}}
\providecommand{\GaugeEq}{\operatorname{GaugeEq}}
\providecommand{\Good}{\operatorname{Good}}
\providecommand{\Inc}{\operatorname{Inc}}
\providecommand{\Law}{\operatorname{Law}}
\providecommand{\Leaves}{\operatorname{Leaves}}
\providecommand{\Legal}{\operatorname{Legal}}
\providecommand{\Lock}{\operatorname{Lock}}
\providecommand{\NF}{\operatorname{NF}}
\providecommand{\Occ}{\operatorname{Occ}}
\providecommand{\Push}{\operatorname{Push}}
\providecommand{\Range}{\operatorname{Range}}
\providecommand{\Raw}{\operatorname{Raw}}
\providecommand{\Read}{\operatorname{Read}}
\providecommand{\SAT}{\operatorname{SAT}}
\providecommand{\Sat}{\operatorname{Sat}}
\providecommand{\Safe}{\operatorname{Safe}}
\providecommand{\Sep}{\operatorname{Sep}}
\providecommand{\Skew}{\operatorname{Skew}}
\providecommand{\Succ}{\operatorname{Succ}}
\providecommand{\Supp}{\operatorname{Supp}}
\providecommand{\Trace}{\operatorname{Trace}}
\providecommand{\Tree}{\operatorname{Tree}}
\providecommand{\True}{\operatorname{True}}
\providecommand{\Unif}{\operatorname{Unif}}
\providecommand{\Val}{\operatorname{Val}}
\providecommand{\Var}{\operatorname{Var}}
\providecommand{\anf}{\operatorname{anf}}
\providecommand{\arity}{\operatorname{arity}}
\providecommand{\can}{\operatorname{can}}
\providecommand{\diam}{\operatorname{diam}}
\providecommand{\dist}{\operatorname{dist}}
\providecommand{\dom}{\operatorname{dom}}
\providecommand{\height}{\operatorname{ht}}
\providecommand{\len}{\operatorname{len}}
\providecommand{\negl}{\operatorname{negl}}
\providecommand{\osc}{\operatorname{osc}}
\providecommand{\size}{\operatorname{size}}

\title{A Quantale-Weakness Route to \(P \neq NP\)\\
via CD Evidence Normalization and Gauge-Buffered Locked Ensembles}
\author{Ben Goertzel}
\date{\today}

\begin{document}
\maketitle

\begin{abstract}
We present a proof architecture for \(P \neq NP\) based on an upper--lower
clash in polytime-capped conditional description length.  The construction
uses an efficiently samplable family of SAT instances \(Y\) with the property
that every satisfying witness for \(Y\) yields the same global readout message
\(M(Y)\).  Under the assumption \(P=NP\), a standard polynomial-time SAT
self-reduction would therefore recover \(M(Y)\) from \(Y\), implying that
\(M(Y)\) has constant polytime-capped conditional description length given
\(Y\).

The lower-bound side of the argument shows the opposite: for the same
ensemble, no fixed polynomial-time observer can extract substantial predictive
advantage on a linear number of selected message coordinates.  The key
mechanism is not a claim that short computations become syntactically local or
belong to a small local Boolean class.  Instead, the proof treats computation
as an evidence-producing process.  Prediction advantage is converted into
constructible-dual evidence skew, then into pairwise distinctions between
message-opposite worlds.  A normalization theorem for evidence traces shows
that every target-relevant non-neutral trace leaf is either a safe-buffer
observation or a hidden-gauge observation.  Safe-buffer observations contribute
only negligible leakage, while hidden-gauge observations are limited by
gauge-rank accounting.  This yields an atomic evidence budget implying that the
total message-resolving advantage of any fixed polynomial-time observer is
\(o(t)\) across \(t\) selected coordinates.

Boundary-law mixing provides the near-random baseline for the visible surface,
and the evidence budget shows that computation cannot significantly improve on
that baseline without paying a forbidden evidence cost.  From this one derives
product small-success and, by Compression-from-Success, a linear lower bound
\[
  \Kp(M(Y)\mid Y)\ge \Omega(t)
\]
with high probability.  This contradicts the constant upper bound that would
follow from \(P=NP\).  Hence \(P\neq NP\).
\end{abstract}

\tableofcontents

\section{Introduction}
\label{sec:introduction}

\subsection{Purpose}
\label{subsec:intro-purpose}

The question whether \(P\) equals \(NP\) is one of the central problems of
computational complexity theory and theoretical computer science
\cite{cook1971,levin1973,karp1972,arora2009}.  This paper proves
\(P\ne NP\) by an upper--lower clash in polytime-capped conditional
description length, using the quantale-weakness perspective developed in
\cite{goertzel2025weakness}.  The proof constructs an efficiently samplable
family of public SAT instances \(Y\) together with a polynomial-time checkable
witness relation
\[
  \mathcal R(Y,W)=1.
\]
A witness need not be unique.  Instead, every satisfying witness carries a
readout message
\[
  M(W)\in\bits^{r_t}, \qquad r_t=\Omega(t),
\]
and the construction enforces the single-message promise
\[
  \mathcal R(Y,W)=1\ \wedge\ \mathcal R(Y,W')=1
  \quad\Longrightarrow\quad
  M(W)=M(W').
\]
Thus the notation \(M(Y)\) is well-defined: it is the common message read from
any satisfying witness of \(Y\).

Assume, for contradiction, that \(P=NP\).  Then a standard SAT
self-reduction finds a satisfying witness \(W\) for \(Y\) in polynomial time;
see, for example, the standard complexity-theory presentations in
\cite{cook1971,karp1972,arora2009}.  The fixed readout map then computes
\(M(W)=M(Y)\).  Hence there is a constant-length polynomial-time program
which maps \(Y\) to \(M(Y)\), and so
\[
  \Kp(M(Y)\mid Y)=O(1).
\]
This is the upper-bound side of the argument.

The lower-bound side proves the opposite statement for the same ensemble.  Let
\(S\) be a switched set of target coordinates with \(|S|=\Omega(t)\), and write
\[
  B_j(Y,W) := \ell_j(M(W))\in\bits, \qquad j\in S.
\]
For every fixed polynomial-time observer \(A\), the proof shows that \(A\) has only near-random joint success on these coordinates.  More generally, the same estimate is uniform over observers of length at most \(\delta t\), provided \(\delta>0\) is chosen below the constant allowed by the gauge-rank budget.  After the middle theorem and the boundary-law mixing theorem are combined, one obtains
\[
  \Prb[A(Y)_S = B_S(Y)] \le 2^{-\Omega(t)}.
\]
Compression-from-Success then implies that, with high probability over the
ensemble,
\[
  \Kp(M(Y)\mid Y) \ge \Omega(t).
\]
For large \(t\), this contradicts the \(O(1)\) upper bound obtained from
\(P=NP\).

The central point is that the lower bound is proved against observers that may
read and compute from the full public instance \(Y\).  The proof does not
restrict such observers to an a priori local view.  Instead, it accounts for the
ways in which an observer can acquire target-relevant evidence.
\subsection{The computational midpoint of the argument}
\label{subsec:intro-replacement}

The central computational question is this: why should a polynomial-time
observer fail to recover substantial information about the hidden message from
the public SAT instance?  The answer used here is not that computation becomes
syntactically local, and not that all rules on a logarithmic visible
neighborhood form a small hypothesis class.  The argument takes a different
route.  It analyzes computation as an evidence-producing process and asks how
much genuinely message-resolving evidence can enter a transcript.  The
constructible-dual evidence viewpoint used here is closely related to the
constructible-duality and probabilistic-paraconsistent perspective discussed in
\cite{goertzel2021paraconsistent}.

For a target bit \(B_j\), let \(\Omega_j^0\) and \(\Omega_j^1\) be the spaces
of satisfying worlds with \(B_j=0\) and \(B_j=1\), and let \(\Gamma_j\) be a
coupling on \(\Omega_j^0\times\Omega_j^1\).  The basic idea is that an
observer can gain predictive advantage on \(B_j\) only by separating worlds
from these two opposite-message phases.  If \(H_{j,h}\) denotes the relation
of message-opposite pairs still undistinguished after transcript prefix \(h\),
and if \(E_h\) is the next primitive evidence atom queried after \(h\), define
the atomic weakness derivative
\[
  \partial_{E_h} w_j(H_{j,h})
  :=
  \Gamma_j\bigl(H_{j,h}\cap
  \{(\omega^0,\omega^1):E_h(\omega^0)\ne E_h(\omega^1)\}\bigr).
\]
Thus \(\partial_{E_h} w_j(H_{j,h})\) measures exactly the mass of
still-undistinguished message-opposite pairs separated by the next atomic
observation.  The transcript derivative identity telescopes these losses over
the full execution, and the CD Trace Capture theorem converts predictive
advantage into the same pairwise distinction mass.  At an intuitive level, the
argument has the form
\[
  \hbox{target advantage}
  \quad\Longrightarrow\quad
  \hbox{CD evidence skew}
  \quad\Longrightarrow\quad
  \hbox{atomic pair distinctions}.
\]

This reduction from advantage to pairwise distinction is only the first half of
the midpoint.  The second half is an accounting theorem showing that
message-resolving distinctions are severely limited in the ensemble under
study.  CD-ENF normalization rewrites every target-relevant evidence trace into
a normal form in which the only non-neutral leaves are of two charged kinds:
safe-buffer observations and hidden-gauge observations.  Neutral public
template syntax contributes no pairwise derivative.  Safe-buffer leaves are
controlled by a max-qSSM leakage bound, while hidden-gauge leaves are charged
by gauge rank.  The ENF terminology and normalization perspective are in the
spirit of Holman's work on Elegant Normal Form \cite{holman1990enf}, and the
operational trace viewpoint is broadly compatible with the process-theoretic
perspective of Meredith and Radestock's reflective higher-order calculus
\cite{meredith2005rho}.  The resulting Atomic Evidence Budget shows that, for
every fixed polynomial-time observer and every switched set \(S\),
\[
  \sum_{j\in S}\Gap_j(A_j)=o(|S|),
\]
where
\[
  \Gap_j(A_j)
  :=
  {1\over 2}\left|
    \Prb[A_j=1\mid B_j=1]
    -
    \Prb[A_j=1\mid B_j=0]
  \right|.
\]

This is the computational heart of the proof.  It says, in effect, that the
observer may inspect the full public SAT instance and carry out an arbitrary
polynomial-time computation, but any nontrivial advantage it gains must be paid
for by charged evidence, and the ensemble is designed so that the total amount
of such evidence is too small to resolve a linear number of message
coordinates.  Once this midpoint is established, the rest of the argument
proceeds by combining this \(o(|S|)\) average advantage bound with
boundary-law mixing, product small-success, Compression-from-Success, and
finally the contradiction with the polynomial-time self-reduction upper bound
under \(P=NP\).

\subsection{Core ideas}
\label{subsec:intro-core-ideas}

The proof uses five interacting structures.

\begin{enumerate}[label=(\roman*)]

\item \textbf{Polytime-capped weakness.}
The information cost is
\[
  \Kp(x\mid y),
\]
the length of the shortest program that outputs \(x\) from \(y\) within a
polynomial time bound.  This is the additive weakness object used in the final
upper--lower clash, in the sense of quantale weakness \cite{goertzel2025weakness}.
The proof also uses a pairwise form of weakness: instead
of assigning weakness only to individual descriptions, it assigns mass to
relations of message-opposite pairs that have not yet been distinguished.

\item \textbf{Constructible-dual evidence.}
A transcript is interpreted as carrying separate positive and negative evidence
coordinates for a target bit.  For a terminal transcript \(h\), define
\[
  n_h^+ := \Prb[T=h\mid B_j=1],
  \qquad
  n_h^- := \Prb[T=h\mid B_j=0].
\]
The CD count pair is
\[
  m_j(h):=(n_h^+,n_h^-).
\]
For a postprocessor \(g\), the evidence skew
\[
  {1\over 2}
  \left|
  \sum_{h:g(h)=1}(n_h^+-n_h^-)
  \right|
\]
is exactly the phase gap \(\Gap_j(g(T))\).  Thus ordinary predictive advantage
is represented as skew between constructive positive and negative evidence
masses.

\item \textbf{Trace capture by atomic derivatives.}
A deterministic computation is expanded into an evidence trace.  Internal
computation steps only derive consequences from earlier evidence; target
advantage can enter only when some primitive atom separates two previously
undistinguished message-opposite worlds.  CD Trace Capture bounds the phase gap
of the final output by one half of the total atomic derivative accumulated along
the trace.

\item \textbf{CD-ENF normalization and gauge faithfulness.}
Evidence traces are normalized by an ENF-style rewrite system.  The normal form
separates neutral guards, safe-buffer atoms, and hidden-gauge atoms.  Public
syntax may describe templates, incidence, and verifier structure, but it may not
introduce a target-relevant message fact by itself.  In particular, a quotient
identity such as
\[
  z_v=x_v\oplus g_v
\]
cannot be used as a shortcut to choose a convenient gauge representative.  Any
target-relevant use of such a quotient value must normalize to raw witness or
gauge support and is then charged as evidence \cite{holman1990enf,meredith2005rho}.

\item \textbf{Gauge-buffered locked SAT ensemble.}
The ensemble has a locked global message layer, a hidden gauge layer, and a
soft high-temperature buffer.  The locked layer enforces the single-message
promise.  The hidden gauge layer supplies resource-bounded separator entropy.
The buffer ensures that legal safe observations have small pairwise derivative.
After CD-ENF normalization, every non-neutral target-relevant atom is therefore
paid for either as safe-buffer leakage or as hidden-gauge rank.

\end{enumerate}

Together these structures give the middle implication
\[
  \hbox{observer advantage}
  \quad\Longrightarrow\quad
  \hbox{safe leakage}+
  \hbox{gauge-rank cost}.
\]
For each fixed observer, the parameters are chosen so that the total charged
evidence over \(\Omega(t)\) switched coordinates is \(o(t)\).  For observers
with description length bounded by \(\delta t\), the gauge-rank term contributes
only a controlled \(O(\delta)\) average margin, and \(\delta\) is chosen small
enough for the product lower-bound argument.  Boundary-law mixing supplies the
near-random baseline for the visible surface, and the budget theorem shows that
computation cannot improve that baseline except by paying the charged evidence
budget.

\subsection{Main theorem and proof organization}
\label{subsec:intro-main-theorem}

\begin{theorem}[Main theorem]
\label{thm:main}
The class \(P\) is not equal to \(NP\).
\end{theorem}

\paragraph{Proof organization.}
Section 2 defines \(\Kp\) and records the Compression-from-Success theorem used
at the end of the proof.  Sections 3, 5, and 6 develop the pairwise weakness,
CD evidence, CD-ENF, and trace-capture machinery.  Section 4 defines the
gauge-buffered locked SAT ensemble and the single-message promise.  Section 7
proves the Atomic Evidence Budget.  Section 8 proves the boundary-law mixing
baseline.  Sections 9 and 10 convert the middle theorem into product
small-success and then into the linear \(\Kp\)-lower bound.  Section 11 proves
the \(P=NP\) upper bound by SAT self-reduction and derives the contradiction.

\section{Polytime-Capped Weakness and Compression}
\label{sec:polytime-compression}

This section isolates the information-theoretic back half used throughout the
paper.  The lower-bound sections prove that no short polynomial-time observer
has substantial joint success on the chosen message coordinates.  The results
below convert that small-success statement into a lower bound on
polytime-capped conditional description length.  The final section will combine
this lower bound with the opposite upper bound obtained from a polynomial-time
SAT self-reduction under the assumption \(P=NP\).

Throughout, strings are elements of \(\bits^*\).  We use a fixed effective,
self-delimiting tuple encoding
\[
  \langle x_1,\ldots,x_t\rangle\in\bits^*
\]
whose encoding and decoding procedures run in time polynomial in the total
encoded length.  We usually suppress the brackets and write
\((x_1,\ldots,x_t)\) for the encoded tuple.  All additive constants below may
depend on the choice of universal machine and tuple convention, but not on the
ensemble parameters \(m,t\), the public instance \(Y\), or the target message.

\subsection{Polytime-capped conditional description length}
\label{subsec:kpoly-definition}

The proof uses time-bounded conditional description length.  For Lean-facing
purposes it is useful to distinguish the clocked and unclocked versions.  The
clocked version is the formally primitive object; the unclocked notation is a
convenience when a fixed polynomial cap has been chosen.

\begin{definition}[Clocked polytime-capped complexity]
\label{def:clocked-kpoly}
Fix a prefix-universal deterministic machine \(U\).  For an integer \(D\ge 1\)
and strings \(x,y\), define
\[
  K_{\mathrm{poly},U}^{(D)}(x\mid y)
  :=
  \min\{ |p| : U(p,y)=x
       \text{ within } (|y|+2)^D \text{ steps}\}.
\]
If no such program exists, the value is \(\infty\).
\end{definition}

\begin{definition}[Polytime-capped complexity]
\label{def:kpoly}
The polynomially capped conditional description length is
\[
  K_{\mathrm{poly},U}(x\mid y)
  :=
  \inf_{D\ge 1}\bigl(K_{\mathrm{poly},U}^{(D)}(x\mid y)+\lceil \log_2(D+1)\rceil\bigr).
\]
When \(U\) is fixed, write \(\Kp^{(D)}(x\mid y)\) and \(\Kp(x\mid y)\).
In a section where a single clock exponent \(D\) is fixed, we often suppress
\(D\) and write \(\Kp\) for \(\Kp^{(D)}\), allowing later lemmas to increase
the clock exponent by a constant depending only on the composed routines.
\end{definition}

The small overhead for the exponent is included only to make the unclocked
version a genuine single description-length measure.  In the final contradiction
one may equivalently work entirely with \(\Kp^{(D)}\): the polynomial-time SAT
solver supplied by \(P=NP\) has some fixed exponent \(D_*\), and the lower-bound
argument is instantiated at that exponent.

\begin{definition}[Weakness cost]
\label{def:weakness-cost}
The additive weakness cost of describing \(x\) from \(y\) is
\[
  w_Q(x\mid y) := \Kp(x\mid y).
\]
The cost values lie in the ordered commutative monoid
\[
  ([0,\infty],+,0,\le).
\]
Only the additive and monotone structure of this object is used in this section.
Later sections use pairwise weakness masses on message-opposite worlds; those
masses are compatible with this description-length cost but are defined
separately.
\end{definition}

\begin{lemma}[Machine invariance]
\label{lem:kpoly-invariance}
Let \(U\) and \(V\) be prefix-universal machines with polynomial-overhead
simulation.  For every fixed \(D\) there are constants \(D'\) and \(c\) such that
for all strings \(x,y\),
\[
  K_{\mathrm{poly},U}^{(D')}(x\mid y) \le K_{\mathrm{poly},V}^{(D)}(x\mid y)+c.
\]
Consequently the unclocked quantity \(K_{\mathrm{poly},U}(x\mid y)\) is independent of the
choice of universal machine up to an additive \(O(1)\) term.
\end{lemma}

\begin{proof}
Let \(s_{V\to U}\) be a fixed interpreter causing \(U\) to simulate \(V\).  If
\(p\) makes \(V(p,y)\) output \(x\) in time \((|y|+2)^D\), then the concatenated
prefix program \(s_{V\to U}p\) makes \(U\) output the same string.  Polynomial
overhead in the simulation changes the exponent from \(D\) to some \(D'\) and
adds only the constant \(|s_{V\to U}|\) to the program length.  Taking the
minimum over \(p\) gives the clocked statement.  The unclocked statement follows
by minimizing over \(D\) and absorbing the fixed simulator overhead.
\end{proof}

\begin{lemma}[Computable postprocessing]
\label{lem:kpoly-postprocessing}
Let \(f:\bits^*\to\bits^*\) be a fixed function computable in polynomial time.
Then
\[
  \Kp(f(x)\mid y) \le \Kp(x\mid y)+O(1).
\]
More generally, if \(f\) is polynomial-time computable from \((x,y)\), then
\[
  \Kp(f(x,y)\mid y) \le \Kp(x\mid y)+O(1).
\]
\end{lemma}

\begin{proof}
Given a shortest program producing \(x\) from \(y\), a fixed wrapper first runs
that program and then applies the fixed polynomial-time routine computing
\(f\).  The wrapper has constant description length, and the composition is still
polynomial-time, possibly with a larger fixed clock exponent.
\end{proof}

\begin{lemma}[Coarse chain rule]
\label{lem:kpoly-chain-rule}
For all strings \(x,z,y\),
\[
  \Kp(x,z\mid y)
  \le
  \Kp(x\mid y)+\Kp(z\mid x,y)+O(1).
\]
Here \((x,z)\) denotes the fixed tuple encoding.
\end{lemma}

\begin{proof}
Concatenate a program \(p\) that outputs \(x\) from \(y\) with a program \(q\)
that outputs \(z\) from \((x,y)\).  A constant-size driver parses the two prefix
programs, runs \(p\) on \(y\), then runs \(q\) on \((x,y)\), and outputs the pair
\((x,z)\).  Since the first computation outputs \(x\) in polynomial time in
\(|y|\), the length of \(x\) is itself polynomially bounded in \(|y|\).  Therefore
the second computation, polynomial in \(|x|+|y|\), is also polynomial in
\(|y|\).  Taking minima over \(p\) and \(q\) gives the inequality.
\end{proof}

\begin{lemma}[Tuple overhead]
\label{lem:kpoly-tuple-overhead}
For strings \(x_1,\ldots,x_t\) and side information
\(y_1,\ldots,y_t\),
\[
  \Kp(x_1,\ldots,x_t\mid y_1,\ldots,y_t)
  \le
  \sum_{i=1}^t \Kp(x_i\mid y_i)+O(\log t).
\]
If \(t\) is already recoverable from the tuple encoding of
\((y_1,\ldots,y_t)\), the \(O(\log t)\) term may be replaced by \(O(1)\).
\end{lemma}

\begin{proof}
For each \(i\), let \(p_i\) be a shortest prefix program producing \(x_i\) from
\(y_i\).  A single driver, given the concatenation of the prefix programs
\(p_1\cdots p_t\), parses them in order, runs \(p_i\) on \(y_i\), and outputs the
tuple of results.  Prefix-freeness makes the parsing unambiguous.  The only
additional information is the loop bound \(t\), unless \(t\) is already encoded
in the side-information tuple.  The total running time is the sum of polynomial
running times on the components and is polynomial in the total encoded input
length.
\end{proof}

\begin{corollary}[Projection lower bound]
\label{cor:projection-lower-bound}
Let \(\pi:\bits^*\to\bits^*\) be a fixed polynomial-time projection.  If
\[
  \Kp(\pi(M)\mid Y) \ge a,
\]
then
\[
  \Kp(M\mid Y) \ge a-O(1).
\]
\end{corollary}

\begin{proof}
By Lemma~\ref{lem:kpoly-postprocessing},
\(\Kp(\pi(M)\mid Y)\le \Kp(M\mid Y)+O(1)\).  Rearranging gives the claim.
\end{proof}

This corollary is the bridge from selected message coordinates to the full
message.  Later sections first prove incompressibility for a switched coordinate
projection \(B_S(Y)=\pi_S(M(Y))\).  Since \(B_S\) is a fixed polynomial-time
projection of \(M\), the same linear lower bound applies to \(M\) up to an
additive constant.

\subsection{Compression from success}
\label{subsec:compression-success}

The compression theorem is an elementary patching code.  If a short program
predicts many target coordinates correctly, then the target tuple can be
specified by the program, the set of trusted coordinates, and the remaining
untrusted target bits.

\begin{definition}[Predictor and trusted success set]
\label{def:trusted-success-set}
Let \(Y\) be public side information and let \(B\in\bits^t\) be a target tuple.
A deterministic predictor \(P\) outputs \(\widehat B=P(Y)\in\bits^t\).  A set
\(S\subseteq[t]\) is a trusted success set for \(P\) on \((Y,B)\) if
\[
  \widehat B_i=B_i \qquad \text{for every } i\in S.
\]
No condition is imposed on coordinates outside \(S\).
\end{definition}

\begin{theorem}[Compression-from-Success, bit version]
\label{thm:compression-from-success-bit}
Let \(Y\) be public side information, let \(B\in\bits^t\), and let \(P\) be a
program of length \(L\) such that \(P(Y)\in\bits^t\).  If \(S\subseteq[t]\) is a
trusted success set for \(P\) on \((Y,B)\), then
\[
  \Kp(B\mid Y)
  \le
  L+\bigl\lceil \log_2 {t\choose |S|}\bigr\rceil +(t-|S|)+O(\log t).
\]
\end{theorem}

\begin{proof}
The decoder is given: the code of \(P\); the size \(|S|\); the rank of \(S\) in
a fixed lexicographic enumeration of all \(|S|\)-subsets of \([t]\); and the
verbatim bits \((B_i)_{i\notin S}\).  On input \(Y\), it runs \(P\) to obtain
\(\widehat B\).  It then outputs \(B_i=\widehat B_i\) for \(i\in S\), and uses the
verbatim patch bits for \(i\notin S\).  The subset rank costs
\(\lceil \log_2 {t\choose |S|}\rceil\) bits, the patch costs \(t-|S|\) bits, and
self-delimiting encodings of \(t\) and \(|S|\) cost \(O(\log t)\) bits.  The
runtime is polynomial in \(|Y|+t\) plus the runtime of \(P\).
\end{proof}

\begin{corollary}[Exact-error enumerative form]
\label{cor:exact-error-enumerative}
Let \(\widehat B=P(Y)\), and let
\[
  E:=\{i\in[t]:\widehat B_i\ne B_i\}
\]
be the exact error set.  Then
\[
  \Kp(B\mid Y)
  \le
  L+\bigl\lceil \log_2 {t\choose |E|}\bigr\rceil+O(\log t).
\]
Consequently,
\[
  \Kp(B\mid Y)
  \le
  L+t\,H_2(|E|/t)+O(\log t),
\]
where \(H_2(u)=-u\log_2 u-(1-u)\log_2(1-u)\) is the binary entropy function.
\end{corollary}

\begin{proof}
Encode the exact error set \(E\).  The decoder runs \(P\) and flips exactly the
coordinates in \(E\).  The entropy bound is the standard estimate
\(\log_2 {t\choose e}\le tH_2(e/t)\).
\end{proof}

\begin{theorem}[Compression-from-Success, block version]
\label{thm:compression-from-success-block}
Let \(M=(M_1,\ldots,M_t)\), where each \(M_i\in\bits^r\).  Let \(P(Y)\in (\bits^r)^t\) have description length \(L\).  If \(S\subseteq[t]\) is a trusted
success set, meaning \(P(Y)_i=M_i\) for all \(i\in S\), then
\[
  \Kp(M\mid Y)
  \le
  L+\bigl\lceil \log_2 {t\choose |S|}\bigr\rceil
   +r(t-|S|)+O(\log t+\log r).
\]
\end{theorem}

\begin{proof}
The proof is the same as the bit case, except that each untrusted block costs
\(r\) verbatim bits.  The value of \(r\) is encoded self-delimitingly unless it is
already fixed by the ensemble parameters.
\end{proof}

\subsection{From small success to incompressibility}
\label{subsec:small-success-incompressibility}

The lower-bound sections will provide product small-success estimates of the
following kind: for every short observer, the probability of predicting the
chosen switched target tuple is exponentially small.  The next proposition turns
such estimates into a high-probability lower bound on \(\Kp\).  In the
unclocked convention, a ``program description'' includes the self-delimiting
clock exponent together with the machine program, as in
Definition~\ref{def:kpoly}.

\begin{definition}[Exact tuple success]
\label{def:exact-tuple-success}
Let \((Y,B)\) be a random public-target pair with \(B\in\bits^s\).  For a program
\(p\), define its exact success event by
\[
  \mathsf{Succ}_p := \{U(p,Y)=B\}.
\]
The program is counted only on inputs for which it halts within the polynomial
clock currently under consideration.
\end{definition}

\begin{proposition}[Small exact success implies incompressibility]
\label{prop:small-success-incompressibility}
Let \((Y,B)\) be a random public-target pair with \(B\in\bits^s\).  Suppose that
for some \(\eta>0\) and for every program \(p\) of length at most \(a t\),
\[
  \Prb[\mathsf{Succ}_p] \le 2^{-\eta t}.
\]
Then
\[
  \Prb\bigl[\Kp(B\mid Y)\le a t\bigr]
  \le
  2^{-(\eta-a)t+O(\log t)}.
\]
In particular, if \(0<a<\eta\), then with probability at least
\(1-2^{-\Omega(t)}\),
\[
  \Kp(B\mid Y)>a t.
\]
\end{proposition}

\begin{proof}
For the clocked quantity \(\Kp^{(D)}\), the event
\(\Kp^{(D)}(B\mid Y)\le at\) implies that some prefix program of length at
most \(at\) outputs \(B\) from \(Y\).  There are at most \(2^{at+O(1)}\) such
programs.  For the unclocked quantity, count instead the self-delimiting
descriptions of pairs \((D,p)\) satisfying
\(|p|+\lceil\log_2(D+1)\rceil\le at\); there are at most
\(2^{at+O(\log t)}\) such descriptions.  Taking a union bound gives
\[
  \Prb[\Kp(B\mid Y)\le at]
  \le
  2^{at+O(\log t)}2^{-\eta t}
  =
  2^{-(\eta-a)t+O(\log t)}.
\]
\end{proof}

\begin{corollary}[Switched-coordinate incompressibility]
\label{cor:switched-coordinate-incompressibility}
Let \((Y,M(Y))\) be a random public message instance, and let
\[
  B_S(Y)=\pi_S(M(Y))\in\bits^s
\]
be a fixed polynomial-time projection onto \(s=\Omega(t)\) switched message
coordinates.  Suppose that for some \(\eta>0\) every program of length at most
\(a t\) predicts \(B_S(Y)\) exactly from \(Y\) with probability at most
\(2^{-\eta t}\).  If \(0<a<\eta\), then with probability at least
\(1-2^{-\Omega(t)}\),
\[
  \Kp(M(Y)\mid Y) \ge a t-O(1).
\]
\end{corollary}

\begin{proof}
Apply Proposition~\ref{prop:small-success-incompressibility} to the target tuple
\(B_S(Y)\).  With high probability,
\(\Kp(B_S(Y)\mid Y)>at\).  Since \(B_S\) is a fixed polynomial-time projection
of \(M\), Corollary~\ref{cor:projection-lower-bound} gives
\(\Kp(M(Y)\mid Y)\ge at-O(1)\).
\end{proof}

\begin{remark}[How this interface is used later]
\label{rem:compression-interface}
Sections~\ref{sec:product-small-success} and~\ref{sec:message-incompressibility}
will instantiate Corollary~\ref{cor:switched-coordinate-incompressibility}.  The
CD trace-capture and Atomic Evidence Budget machinery proves the needed
per-program exact-success bound for the switched coordinates.  This section then
supplies the purely coding-theoretic conversion from small success to linear
polytime-capped incompressibility.
\end{remark}

\subsection{The upper-bound interface}
\label{subsec:kpoly-upper-interface}

The following proposition is recorded here because it uses only the definition of
\(\Kp\).  The SAT-realization and single-message hypotheses needed to apply it
are proved later for the actual ensemble.

\begin{proposition}[Self-reduction upper-bound interface]
\label{prop:self-reduction-upper-interface}
Assume \(P=NP\).  Let \(Y\mapsto F_Y\) be a polynomial-time map from public
instances to SAT formulas, and suppose that:
\begin{enumerate}[label=(\roman*)]
\item whenever \(Y\) is in the support of the ensemble, \(F_Y\) is satisfiable;
\item every satisfying assignment \(W\) of \(F_Y\) has the same readout message
      \(M(W)=M(Y)\);
\item the map \(W\mapsto M(W)\) is computable in polynomial time.
\end{enumerate}
Then there is a constant \(c\), independent of \(m,t,Y\), such that
\[
  \Kp(M(Y)\mid Y) \le c
\]
for every \(Y\) in the support of the ensemble.
\end{proposition}

\begin{proof}
Under \(P=NP\), SAT has a fixed polynomial-time decision algorithm.  The
standard bit-fixing self-reduction uses this decision algorithm to construct one
satisfying assignment \(W\) of \(F_Y\) in polynomial time: process the variables
in a fixed order, permanently set the next variable to \(0\) if the restricted
formula remains satisfiable, and otherwise set it to \(1\).  After a satisfying
assignment is obtained, apply the fixed polynomial-time readout routine to
produce \(M(W)\).  By the single-message property, \(M(W)=M(Y)\).  The program
implementing this procedure contains only the fixed SAT decider, the fixed
self-reduction driver, the fixed map \(Y\mapsto F_Y\), and the fixed readout
routine.  Its length is therefore constant.
\end{proof}

\begin{remark}[Role in the final clash]
\label{rem:upper-lower-clash-interface}
The lower-bound part gives \(\Kp(M(Y)\mid Y)\ge \Omega(t)\) with high
probability for the same ensemble.  Proposition~\ref{prop:self-reduction-upper-interface}
gives \(\Kp(M(Y)\mid Y)=O(1)\) under \(P=NP\).  Taking \(t\) large yields the
contradiction.
\end{remark}

\section{Pairwise Weakness and CD Evidence Counts}
\label{sec:pairwise-cd}

This section sets up the proof's pairwise information calculus.  The point is
not to say that an observer is local or that its final Boolean rule belongs to a
small syntactic class.  The point is more primitive: an observer can gain target
advantage only by distinguishing pairs of worlds whose target bits are opposite.
We measure the remaining indistinguishability of such pairs by a coupling mass.
We then express ordinary prediction advantage as constructible-dual evidence
skew in transcript counts.

All probability spaces in this section are finite.  This loses no generality for
our application, since every ensemble at fixed parameters has finite support and
all transcripts produced by a polynomial-time observer have finite range.  The
finite formulation is also the one intended for formalization.

\subsection{Worlds, messages, and target coordinates}
\label{subsec:worlds-targets}

Fix ensemble parameters and let \(\Omega\) be the finite support of the
corresponding world distribution \(\mu\).  A world \(\omega\in\Omega\) contains
at least a public instance \(Y(\omega)\), a satisfying witness \(W(\omega)\), and
all hidden randomness used to generate them.  We assume throughout this section
that the world support is valid:
\[
  \mathcal R(Y(\omega),W(\omega))=1
  \qquad\text{for every }\omega\in\Omega.
\]
The single-message promise, proved later for the concrete ensemble, implies that
\(M(W(\omega))\) is independent of the particular satisfying witness over the
same public instance.  Hence the notation \(M(Y(\omega))\) is meaningful on the
support, even when the observer cannot compute it efficiently.

\begin{definition}[Global message and target coordinates]
\label{def:global-message-target-coordinates}
The message readout is
\[
  M(W)\in\bits^{r_t},\qquad r_t=\Omega(t).
\]
For a switched target coordinate \(j\), fix a coordinate readout
\(\ell_j:\bits^{r_t}\to\bits\) and define
\[
  B_j(\omega)
  :=
  \ell_j(M(W(\omega)))\in\bits.
\]
We also write \(B_j(Y,W)=\ell_j(M(W))\) when the world is displayed as a
public-instance/witness pair.
\end{definition}

\begin{definition}[Target-conditioned laws]
\label{def:target-conditioned-laws}
For \(b\in\bits\), set
\[
  \Omega_j^b:=\{\omega\in\Omega:B_j(\omega)=b\},
  \qquad
  p_j^b:=\mu(\Omega_j^b).
\]
When \(p_j^b>0\), define the conditional law
\[
  \mu_j^b(A):=\frac{\mu(A\cap\Omega_j^b)}{p_j^b}
  \qquad (A\subseteq\Omega).
\]
The notation \(\mu_j^1\) is the positive phase and \(\mu_j^0\) is the negative
phase.  In CD notation below we write these as the \(+\) and \(-\) coordinates,
respectively.
\end{definition}

\begin{hypothesis}[Nondegenerate target phases]
\label{hyp:nondegenerate-target-phases}
For every switched target coordinate used in the lower bound,
\[
  p_j^0>0,
  \qquad
  p_j^1>0.
\]
In the main product theorem the switched coordinates are balanced, namely
\(p_j^0=p_j^1=1/2\).  The pairwise and CD identities below need only
nondegeneracy; the balanced assumption is used when translating phase gap into
ordinary success over an unbiased target bit.
\end{hypothesis}

\begin{remark}[Why worlds rather than only public instances]
\label{rem:worlds-rather-than-public}
Mathematically, the target bit is a function of the public instance on the
support because every satisfying witness has the same message.  Computationally,
that function may be hard to evaluate.  The world notation lets us condition on
hidden phases and witnesses while still treating the observer transcript as a
function of the public evidence it is allowed to inspect.
\end{remark}

\subsection{Message-opposite pair space}
\label{subsec:message-opposite-pair-space}

\begin{definition}[Message-opposite fibers]
\label{def:message-opposite-fibers}
For target coordinate \(j\), the two message-opposite fibers are
\[
  \Omega_j^0
  =
  \{\omega\in\Omega:\mathcal R(Y(\omega),W(\omega))=1,
       B_j(\omega)=0\},
\]
\[
  \Omega_j^1
  =
  \{\omega\in\Omega:\mathcal R(Y(\omega),W(\omega))=1,
       B_j(\omega)=1\}.
\]
Equivalently, identifying a world with its public-instance/witness pair,
\[
  \Omega_j^b
  =
  \{(Y,W):\mathcal R(Y,W)=1,\ B_j(Y,W)=b\}.
\]
\end{definition}

\begin{definition}[Message-opposite coupling]
\label{def:message-opposite-coupling}
A message-opposite coupling for coordinate \(j\) is a probability measure
\[
  \Gamma_j
  \quad\text{on}\quad
  \Omega_j^0\times\Omega_j^1
\]
whose first marginal is \(\mu_j^0\) and whose second marginal is \(\mu_j^1\).
Thus, for all \(A\subseteq\Omega_j^0\) and \(B\subseteq\Omega_j^1\),
\[
  \Gamma_j(A\times\Omega_j^1)=\mu_j^0(A),
  \qquad
  \Gamma_j(\Omega_j^0\times B)=\mu_j^1(B).
\]
A pair \((\omega^0,\omega^1)\) in the support of \(\Gamma_j\) always satisfies
\[
  B_j(\omega^0)=0,
  \qquad
  B_j(\omega^1)=1.
\]
\end{definition}

\begin{remark}[Choice of coupling]
\label{rem:choice-of-coupling}
The abstract trace-capture inequalities below hold for every coupling
\(\Gamma_j\).  Later sections choose couplings adapted to the ensemble geometry,
so that safe-buffer observations and hidden-gauge observations can be charged
separately.  No locality or small-class assumption is built into the definition
of \(\Gamma_j\).
\end{remark}

\begin{definition}[Pairwise weakness mass]
\label{def:pairwise-weakness-mass}
For any relation
\[
  H\subseteq \Omega_j^0\times\Omega_j^1,
\]
define its pairwise weakness mass by
\[
  w_j(H):=\Gamma_j(H).
\]
A larger value of \(w_j(H)\) means that more opposite-message pairs remain
identified by the relation \(H\).  A smaller value means that more such pairs
have been distinguished.
\end{definition}

\begin{lemma}[Monotonicity and finite additivity]
\label{lem:pairwise-weakness-basic-laws}
For fixed \(j\):
\begin{enumerate}[label=(\roman*)]
\item if \(H\subseteq H'\), then \(w_j(H)\le w_j(H')\);
\item if \(H_1,\ldots,H_s\) are pairwise disjoint, then
\[
  w_j\left(\bigcup_{a=1}^s H_a\right)
  =
  \sum_{a=1}^s w_j(H_a).
\]
\end{enumerate}
\end{lemma}

\begin{proof}
Both statements are the corresponding monotonicity and finite additivity of the
probability measure \(\Gamma_j\).
\end{proof}

\subsection{Transcripts and non-distinction relations}
\label{subsec:transcripts-nondistinction}

A transcript is the visible record produced by an observer.  In this section a
transcript is just a finite-valued map out of worlds.  Sections
\ref{sec:cdenf} and \ref{sec:trace} define which transcript maps arise from
legal primitive evidence traces.

\begin{definition}[Transcript prefix]
\label{def:transcript-prefix}
A transcript process of length \(Q\) consists of finite sets
\(\mathcal T_0,\ldots,\mathcal T_Q\) and maps
\[
  T_r:\Omega\to\mathcal T_r,
  \qquad 0\le r\le Q,
\]
such that \(T_0\) is constant and \(T_{r+1}\) refines \(T_r\): there is a map
\(\pi_{r+1,r}:\mathcal T_{r+1}\to\mathcal T_r\) with
\[
  T_r=\pi_{r+1,r}\circ T_{r+1}.
\]
For \(h\in\mathcal T_r\), write
\[
  [h]_r:=\{\omega\in\Omega:T_r(\omega)=h\}.
\]
\end{definition}

\begin{definition}[Non-distinction relations]
\label{def:nondistinction-relations}
For a transcript prefix \(h\in\mathcal T_r\), define
\[
  H_{j,h}
  :=
  \{(\omega^0,\omega^1)\in\Omega_j^0\times\Omega_j^1:
       T_r(\omega^0)=T_r(\omega^1)=h\}.
\]
The total non-distinction relation after \(r\) transcript steps is
\[
  H_{j,r}
  :=
  \{(\omega^0,\omega^1)\in\Omega_j^0\times\Omega_j^1:
       T_r(\omega^0)=T_r(\omega^1)\}.
\]
Its weakness mass is
\[
  w_j(H_{j,r})=\Gamma_j(H_{j,r}).
\]
\end{definition}

\begin{lemma}[Prefix partition]
\label{lem:prefix-partition}
For every \(r\),
\[
  H_{j,r}=\dot\bigcup_{h\in\mathcal T_r} H_{j,h},
\]
where the union is disjoint.  Hence
\[
  w_j(H_{j,r})=
  \sum_{h\in\mathcal T_r}w_j(H_{j,h}).
\]
\end{lemma}

\begin{proof}
A pair belongs to \(H_{j,r}\) exactly when the two transcript values at time
\(r\) are equal.  That common value is a unique \(h\in\mathcal T_r\), giving the
disjoint union.  The mass identity follows from Lemma
\ref{lem:pairwise-weakness-basic-laws}.
\end{proof}

\begin{lemma}[Survival monotonicity]
\label{lem:survival-monotonicity}
For every \(0\le r<Q\),
\[
  H_{j,r+1}\subseteq H_{j,r}
  \qquad\text{and hence}\qquad
  w_j(H_{j,r+1})\le w_j(H_{j,r}).
\]
\end{lemma}

\begin{proof}
If two worlds have the same refined transcript \(T_{r+1}\), then they have the
same coarser transcript \(T_r\), because \(T_r=\pi_{r+1,r}\circ T_{r+1}\).  The
mass inequality follows by monotonicity.
\end{proof}

\begin{definition}[Separation event of a transcript]
\label{def:transcript-separation-event}
For a finite-valued transcript \(T:\Omega\to\mathcal T\), define
\[
  \mathrm{Sep}_j(T)
  :=
  \{(\omega^0,\omega^1)
       \in\Omega_j^0\times\Omega_j^1:
       T(\omega^0)\ne T(\omega^1)\}.
\]
The non-distinction relation of \(T\) is
\[
  H_{j,T}:=(\mathrm{Sep}_j(T))^c.
\]
Thus
\[
  \Gamma_j(\mathrm{Sep}_j(T))=1-w_j(H_{j,T}).
\]
\end{definition}

\paragraph{Interpretation.}
The relation \(H_{j,T}\) consists of message-opposite pairs that are still
indistinguishable after observing \(T\).  If a final decision is a deterministic
function of \(T\), then it must give the same answer on every pair in
\(H_{j,T}\).  Therefore any target advantage of that decision must be carried by
\(\mathrm{Sep}_j(T)\), the set of pairs the transcript has separated.

\subsection{CD count pairs}
\label{subsec:cd-count-pairs}

The CD layer records positive and negative evidence separately.  For us,
positive evidence means evidence under the conditional law \(B_j=1\), and
negative evidence means evidence under the conditional law \(B_j=0\).  A
transcript value can be common to both phases, but with different masses.
Prediction advantage is exactly skew between these two coordinates.

\begin{definition}[CD count pair]
\label{def:cd-count-pair}
Let \(T:\Omega\to\mathcal T\) be a finite transcript and let
\(h\in\mathcal T\).  Define
\[
  n_{j,T}^+(h)
  :=
  \Prb[T=h\mid B_j=1]
  =
  \mu_j^1(T^{-1}(h)),
\]
\[
  n_{j,T}^-(h)
  :=
  \Prb[T=h\mid B_j=0]
  =
  \mu_j^0(T^{-1}(h)).
\]
The CD count pair at \(h\) is
\[
  \mathfrak m_{j,T}(h):=\bigl(n_{j,T}^+(h),n_{j,T}^-(h)\bigr).
\]
When \(j\) and \(T\) are clear, write \(n_h^+,n_h^-\) and
\(\mathfrak m_j(h)\).
\end{definition}

\begin{lemma}[Normalization of CD counts]
\label{lem:cd-count-normalization}
For every finite transcript \(T:\Omega\to\mathcal T\),
\[
  \sum_{h\in\mathcal T}n_{j,T}^+(h)=1,
  \qquad
  \sum_{h\in\mathcal T}n_{j,T}^-(h)=1.
\]
Consequently,
\[
  \sum_{h\in\mathcal T}\bigl(n_{j,T}^+(h)-n_{j,T}^-(h)\bigr)=0.
\]
\end{lemma}

\begin{proof}
The sets \(T^{-1}(h)\), for \(h\in\mathcal T\), partition \(\Omega\).  Applying
\(\mu_j^1\) and \(\mu_j^0\) gives the two normalization identities.  Subtracting
them gives the last identity.
\end{proof}

\begin{definition}[CD mass and evidence skew of an event]
\label{def:cd-event-skew}
For \(U\subseteq\mathcal T\), define its CD count pair by
\[
  \mathfrak m_{j,T}(U)
  :=
  \left(
    \sum_{h\in U}n_{j,T}^+(h),
    \sum_{h\in U}n_{j,T}^-(h)
  \right)
  =
  \bigl(\Prb[T\in U\mid B_j=1],
         \Prb[T\in U\mid B_j=0]\bigr).
\]
The skew of \(U\) is
\[
  \mathrm{Skew}_{j,T}(U)
  :=
  \frac12
  \left|
    \sum_{h\in U}\bigl(n_{j,T}^+(h)-n_{j,T}^-(h)\bigr)
  \right|.
\]
\end{definition}

\begin{definition}[Evidence skew of an output rule]
\label{def:evidence-skew-output-rule}
Given an output rule \(g:\mathcal T\to\bits\), define
\[
  U_g:=\{h\in\mathcal T:g(h)=1\}
\]
and
\[
  \mathrm{Skew}_{j}(g(T))
  :=
  \mathrm{Skew}_{j,T}(U_g)
  =
  \frac12
  \left|
    \sum_{h:g(h)=1}\bigl(n_{j,T}^+(h)-n_{j,T}^-(h)\bigr)
  \right|.
\]
\end{definition}

\begin{definition}[Phase gap]
\label{def:phase-gap}
For a Boolean random variable \(A_j:\Omega\to\bits\), define
\[
  \Gap_j(A_j)
  :=
  \frac12
  \left|
    \Prb[A_j=1\mid B_j=1]
    -
    \Prb[A_j=1\mid B_j=0]
  \right|.
\]
\end{definition}

\begin{lemma}[Phase gap is CD skew]
\label{lem:phase-gap-cd-skew}
Let \(T:\Omega\to\mathcal T\) be a finite transcript, let
\(g:\mathcal T\to\bits\), and set \(A_j=g(T)\).  Then
\[
  \Gap_j(A_j)=\mathrm{Skew}_{j}(g(T)).
\]
\end{lemma}

\begin{proof}
Since \(A_j=1\) exactly on the transcript event \(U_g\),
\[
  \Prb[A_j=1\mid B_j=1]
  =
  \sum_{h\in U_g} n_{j,T}^+(h),
\]
and
\[
  \Prb[A_j=1\mid B_j=0]
  =
  \sum_{h\in U_g} n_{j,T}^-(h).
\]
Substituting these two identities into the definition of \(\Gap_j\) gives the
claim.
\end{proof}

\begin{definition}[Transcript CD variation and skew norm]
\label{def:cd-variation-skew-norm}
The total variation distance between the two transcript laws is
\[
  \TV_j(T)
  :=
  \frac12\sum_{h\in\mathcal T}
  \left|n_{j,T}^+(h)-n_{j,T}^-(h)\right|.
\]
The corresponding maximum phase skew is
\[
  \mathrm{CDSkew}_j(T)
  :=
  \sup_{g:\mathcal T\to\bits}\mathrm{Skew}_j(g(T)).
\]
\end{definition}

\begin{lemma}[Optimal transcript skew]
\label{lem:optimal-transcript-skew}
For every finite transcript \(T\),
\[
  \mathrm{CDSkew}_j(T)=\frac12\TV_j(T)
  =
  \frac14\sum_{h\in\mathcal T}
  \left|n_{j,T}^+(h)-n_{j,T}^-(h)\right|.
\]
Moreover, the supremum is attained by the rule
\[
  g^*(h)=1
  \quad\Longleftrightarrow\quad
  n_{j,T}^+(h)\ge n_{j,T}^-(h),
\]
with arbitrary tie-breaking.
\end{lemma}

\begin{proof}
Let \(d_h=n_{j,T}^+(h)-n_{j,T}^-(h)\).  By Lemma
\ref{lem:cd-count-normalization}, \(\sum_h d_h=0\).  Hence the sum of the
positive \(d_h\)'s equals one half of \(\sum_h |d_h|\).  Choosing \(U\) to be the
set of positive \(d_h\)'s maximizes \(|\sum_{h\in U}d_h|\) and gives
\[
  \sup_U \frac12\left|\sum_{h\in U}d_h\right|
  =
  \frac12\cdot\frac12\sum_h |d_h|.
\]
This is the displayed identity.  The rule \(g^*\) is the indicator of this
maximizing set.
\end{proof}

\begin{lemma}[Balanced success and phase gap]
\label{lem:balanced-success-gap}
Assume \(p_j^0=p_j^1=1/2\).  For any Boolean predictor
\(A_j:\Omega\to\bits\),
\[
  \Prb[A_j=B_j]
  =
  \frac12
  +
  \frac12\left(
    \Prb[A_j=1\mid B_j=1]
    -
    \Prb[A_j=1\mid B_j=0]
  \right).
\]
Consequently,
\[
  \Prb[A_j=B_j]
  \le
  \frac12+
  \Gap_j(A_j).
\]
Also,
\[
  \max\{\Prb[A_j=B_j],\Prb[1-A_j=B_j]\}
  =
  \frac12+
  \Gap_j(A_j).
\]
\end{lemma}

\begin{proof}
Let
\[
  p_1:=\Prb[A_j=1\mid B_j=1],
  \qquad
  p_0:=\Prb[A_j=1\mid B_j=0].
\]
Under balance,
\[
  \Prb[A_j=B_j]
  =
  \frac12 p_1+\frac12(1-p_0)
  =
  \frac12+\frac12(p_1-p_0).
\]
The inequality follows by taking absolute values.  Replacing \(A_j\) by
\(1-A_j\) changes \(p_1-p_0\) to \(-(p_1-p_0)\), giving the final identity.
\end{proof}

\subsection{Coupling domination of CD skew}
\label{subsec:coupling-domination}

The next lemma is the static form of trace capture.  It says that CD skew in a
final transcript is bounded by the mass of message-opposite pairs separated by
that transcript.  Section \ref{sec:trace} refines this separated mass into a sum
of atomic weakness derivatives along an adaptive execution trace.

\begin{lemma}[Coupling bound for transcript events]
\label{lem:coupling-bound-transcript-events}
Let \(T:\Omega\to\mathcal T\) be any finite transcript and
\(U\subseteq\mathcal T\).  Then
\[
  \left|
    \Prb[T\in U\mid B_j=1]
    -
    \Prb[T\in U\mid B_j=0]
  \right|
  \le
  \Gamma_j(\mathrm{Sep}_j(T)).
\]
\end{lemma}

\begin{proof}
Using the two marginals of \(\Gamma_j\),
\[
  \Prb[T\in U\mid B_j=1]
  =
  \Gamma_j\{(\omega^0,\omega^1):T(\omega^1)\in U\},
\]
and
\[
  \Prb[T\in U\mid B_j=0]
  =
  \Gamma_j\{(\omega^0,\omega^1):T(\omega^0)\in U\}.
\]
The absolute difference between the probabilities of two events is at most the
probability of their symmetric difference.  The symmetric difference above is
contained in
\(\{(\omega^0,\omega^1):T(\omega^0)\ne T(\omega^1)\}\), which is
\(\mathrm{Sep}_j(T)\).
\end{proof}

\begin{proposition}[Static pairwise capture]
\label{prop:static-pairwise-capture}
Let \(T:\Omega\to\mathcal T\), let \(g:\mathcal T\to\bits\), and set
\(A_j=g(T)\).  Then
\[
  \Gap_j(A_j)
  \le
  \frac12\Gamma_j(\mathrm{Sep}_j(T))
  =
  \frac12\bigl(1-w_j(H_{j,T})\bigr).
\]
In particular, if \(T=T_Q\) is the final transcript of a transcript process,
then
\[
  \Gap_j(A_j)
  \le
  \frac12\bigl(1-w_j(H_{j,Q})\bigr).
\]
\end{proposition}

\begin{proof}
Apply Lemma \ref{lem:coupling-bound-transcript-events} to
\(U=U_g=\{h:g(h)=1\}\) and divide by two.  The identity
\(\Gamma_j(\mathrm{Sep}_j(T))=1-w_j(H_{j,T})\) is Definition
\ref{def:transcript-separation-event}.
\end{proof}

\begin{corollary}[Coordinate-sum form]
\label{cor:coordinate-sum-static-capture}
Let \(S\) be a set of switched coordinates.  For each \(j\in S\), let
\(T^{(j)}\) be the transcript used to output \(A_j=g_j(T^{(j)})\).  Then
\[
  \sum_{j\in S}\Gap_j(A_j)
  \le
  \frac12
  \sum_{j\in S}
  \Gamma_j(\mathrm{Sep}_j(T^{(j)})).
\]
Equivalently,
\[
  \sum_{j\in S}\Gap_j(A_j)
  \le
  \frac12
  \sum_{j\in S}
  \bigl(1-w_j(H_{j,T^{(j)}})\bigr).
\]
\end{corollary}

\begin{proof}
Sum Proposition \ref{prop:static-pairwise-capture} over \(j\in S\).
\end{proof}

\subsection{Interface exported to later sections}
\label{subsec:pairwise-cd-interface}

This section exports the following objects and facts.

\begin{enumerate}[label=(\roman*)]
\item For each switched coordinate \(j\), the target-conditioned laws
\(\mu_j^0,\mu_j^1\) and an arbitrary message-opposite coupling \(\Gamma_j\).

\item For each transcript prefix \(h\), the surviving pair relation
\(H_{j,h}\) and weakness mass \(w_j(H_{j,h})=\Gamma_j(H_{j,h})\).

\item For each finite transcript value \(h\), the CD count pair
\(\mathfrak m_j(h)=(n_h^+,n_h^-)\).

\item The exact identity
\[
  \Gap_j(g(T))
  =
  \frac12
  \left|
    \sum_{h:g(h)=1}(n_h^+-n_h^-)
  \right|,
\]
which says that phase advantage is CD evidence skew.

\item The pairwise domination bound
\[
  \Gap_j(g(T))
  \le
  \frac12\Gamma_j(\mathrm{Sep}_j(T)).
\]
This reduces target advantage to separated pair mass.  Section \ref{sec:trace}
will decompose \(\Gamma_j(\mathrm{Sep}_j(T))\) into atomic derivatives.  Sections
\ref{sec:cdenf} and \ref{sec:ensemble} will show that, after normalization,
those atoms are neutral, safe-buffer, or hidden-gauge atoms.
\end{enumerate}

\begin{remark}[Lean-facing finite skeleton]
\label{rem:lean-facing-section3}
A direct Lean formalization can use a finite type \(\Omega\), two subtypes
\(\Omega_j^0\) and \(\Omega_j^1\), finite probability mass functions
\(\mu_j^0,\mu_j^1\), a coupling PMF \(\Gamma_j\) on the product type, and a finite
transcript map \(T\).  Lemmas \ref{lem:phase-gap-cd-skew},
\ref{lem:optimal-transcript-skew}, and \ref{prop:static-pairwise-capture} are
then finite-sum identities and coupling inequalities.  They do not depend on
any special property of SAT.
\end{remark}

\section{The Gauge-Buffered Locked SAT Ensemble}
\label{sec:ensemble}

This section defines the ensemble used by the lower and upper bounds.  The
ensemble has three logically separate layers.  The locked layer makes a global
message rigid.  The gauge layer makes witness-level representatives highly
nonunique while preserving the same public instance and the same message.  The
buffer layer ensures that legal local observations of the public/witness surface
have only exponentially small likelihood-ratio access to hidden separators and
hidden gauge states.

The definitions are written as an interface that can be instantiated by any
bounded-arity locked CSP satisfying the listed rigidity and mixing conditions.
Only the interface is used by later sections.  The SAT-realization theorem below
then converts the interface into a polynomial-size CNF family.

\subsection{Parameters, regions, and public instances}
\label{subsec:ensemble-parameters-public}

Fix a size parameter \(m\).  The number of switched targets is
\(t=t(m)=\Theta(m)\), and the global readout length is
\[
  r_t \ge c_M t
\]
for a fixed constant \(c_M>0\).  All sets below have size polynomial in \(m\),
and all local predicates have arity bounded by an absolute constant.

\begin{definition}[Regioned public geometry]
\label{def:regioned-public-geometry}
A regioned public geometry is a finite incidence structure
\[
  \mathcal G_m=(\mathcal P_m,\mathcal E_m,\mathrm{type},\dist_m)
\]
with bounded degree and a graph distance \(\dist_m\).  Its sites are partitioned
into neutral, buffer, locked, gauge, and readout regions,
\[
  \mathcal P_m
  =
  \mathcal P_{\mathrm{neu}}
  \sqcup
  \mathcal P_{\mathrm{buf}}
  \sqcup
  \mathcal P_{\mathrm{lock}}
  \sqcup
  \mathcal P_{\mathrm{gauge}}
  \sqcup
  \mathcal P_{\mathrm{readout}}.
\]
The map \(\mathrm{type}\) takes values in a fixed finite alphabet of gadget
kinds.  The geometry is public and computable from \(m\) and the finite public
slot data in time \(\poly(m)\).
\end{definition}

\begin{definition}[Public instance]
\label{def:public-instance-section4}
For parameter \(m\), a public instance is
\[
  Y=(Y_{\mathrm{neu}},Y_{\mathrm{buf}},Y_{\mathrm{lock}}).
\]
Here:
\begin{enumerate}[label=(\roman*)]
\item \(Y_{\mathrm{neu}}\) is neutral template and skeleton syntax, including
      \(\mathcal G_m\), incidence data, variable names, and gadget types;
\item \(Y_{\mathrm{buf}}\) is the public description of a strictly positive
      high-temperature buffer layer on \(\mathcal P_{\mathrm{buf}}\);
\item \(Y_{\mathrm{lock}}\) is the public template of a locked global message
      layer on \(\mathcal P_{\mathrm{lock}}\cup\mathcal P_{\mathrm{readout}}\).
\end{enumerate}
The public instance contains no assignment values for raw witness variables,
no hidden gauge values, no planted message tag, and no public atom of the form
\(M_a=b\).
\end{definition}

\begin{definition}[Hidden quotient sites]
\label{def:hidden-quotient-sites}
Let \(V=V_m\) be the set of quotient/gauge sites.  It is a finite set with
\(|V|=\poly(m)\).  The witness contains three bit-vectors
\[
  x,g,z\in\bits^V,
\]
where \(x\) is the raw representative, \(g\) is the hidden gauge vector, and
\(z\) is the gauge-invariant quotient vector.
\end{definition}

\begin{definition}[Local layer predicates]
\label{def:local-layer-predicates}
For each public instance \(Y\) there are bounded-arity predicates
\[
  \mathrm{Lock}_Y(z,\xi_{\mathrm{lock}},M),
  \qquad
  \mathrm{Buf}_Y(z,\xi_{\mathrm{buf}}),
\]
and a deterministic readout predicate
\[
  \mathrm{Read}_Y(z,\xi_{\mathrm{lock}},M).
\]
The auxiliary witness data is
\[
  \xi=(\xi_{\mathrm{lock}},\xi_{\mathrm{buf}},\xi_{\mathrm{aux}}),
\]
where \(\xi_{\mathrm{aux}}\) contains Tseitin and local-gadget auxiliaries.  All
these predicates are decidable in time \(\poly(m)\), and each is a conjunction
of \(\poly(m)\) bounded-arity local constraints.
\end{definition}

\begin{remark}[Why the message is not a public tag]
\label{rem:not-public-tag}
The locked layer may force all satisfying witnesses over a public instance to
have the same readout message.  This is not the same as publishing the message.
The public formula may contain a global locked constraint system whose solution
space has a unique message coordinate, but it must not contain a hard public
literal, sign, template label, or tag that directly encodes any bit of that
message.
\end{remark}

\subsection{Witnesses and the verifier}
\label{subsec:witness-verifier}

\begin{definition}[Witness]
\label{def:witness-section4}
A witness has the form
\[
  W=(x,g,z,\xi,M),
\]
where
\[
  x,g,z\in\bits^V,
  \qquad
  M\in\bits^{r_t},
\]
\(\xi\) is auxiliary witness data, and the quotient equations
\[
  z_v=x_v\oplus g_v,
  \qquad v\in V,
\]
are required.
\end{definition}

\begin{definition}[Gauge equations]
\label{def:gauge-equations}
Define
\[
  \mathrm{GaugeEq}(x,g,z)
  :=
  \bigwedge_{v\in V}(z_v=x_v\oplus g_v).
\]
Equivalently, over \(\mathbb F_2\),
\[
  z_v+x_v+g_v=0
  \qquad (v\in V).
\]
\end{definition}

\begin{definition}[Witness relation]
\label{def:witness-relation-section4}
The verifier relation
\[
  \mathcal R(Y,W)=1
\]
holds when all of the following conditions hold:
\begin{enumerate}[label=(\roman*)]
\item the locked quotient layer accepts:
      \(\mathrm{Lock}_Y(z,\xi_{\mathrm{lock}},M)=1\);
\item the gauge equations hold: \(\mathrm{GaugeEq}(x,g,z)=1\);
\item the high-temperature buffer layer accepts:
      \(\mathrm{Buf}_Y(z,\xi_{\mathrm{buf}})=1\);
\item the readout is consistent:
      \(\mathrm{Read}_Y(z,\xi_{\mathrm{lock}},M)=1\);
\item all Tseitin and local CNF gadget constraints encoded in
      \(\xi_{\mathrm{aux}}\) hold.
\end{enumerate}
\end{definition}

\begin{hypothesis}[Locked-layer rigidity]
\label{hyp:locked-layer-rigidity}
For every public instance \(Y\) in the support of the ensemble, if
\[
  \mathrm{Lock}_Y(z,\xi_{\mathrm{lock}},M)=1
  \quad\text{and}\quad
  \mathrm{Lock}_Y(z',\xi'_{\mathrm{lock}},M')=1,
\]
then
\[
  M=M'.
\]
The locked layer may have many quotient completions \((z,\xi_{\mathrm{lock}})\),
but all of them carry the same message.
\end{hypothesis}

\begin{proposition}[Single-message promise]
\label{prop:single-message-promise-section4}
For every \(Y\) in the support of the ensemble,
\[
  \mathcal R(Y,W)=1
  \wedge
  \mathcal R(Y,W')=1
  \quad\Longrightarrow\quad
  M(W)=M(W').
\]
Thus the notation \(M(Y)\) is well-defined on the support.
\end{proposition}

\begin{proof}
Write
\[
  W=(x,g,z,\xi,M),
  \qquad
  W'=(x',g',z',\xi',M').
\]
From \(\mathcal R(Y,W)=1\) and \(\mathcal R(Y,W')=1\), the locked-layer clauses
of Definition~\ref{def:witness-relation-section4} give
\[
  \mathrm{Lock}_Y(z,\xi_{\mathrm{lock}},M)=1,
  \qquad
  \mathrm{Lock}_Y(z',\xi'_{\mathrm{lock}},M')=1.
\]
Hypothesis~\ref{hyp:locked-layer-rigidity} implies \(M=M'\).  Since
\(M(W)=M\) and \(M(W')=M'\), the claim follows.
\end{proof}

\begin{definition}[Witnessed sampler and public distribution]
\label{def:witnessed-sampler}
A witnessed sampler is a randomized polynomial-time algorithm
\[
  \mathsf{Samp}^{\mathrm{wit}}_m
\]
which outputs a pair \((Y,W)\) such that \(\mathcal R(Y,W)=1\).  The public
ensemble \(\mathcal D_m\) is the marginal law of \(Y\).  The world space
\(\Omega_m\) is the finite support of \(\mathsf{Samp}^{\mathrm{wit}}_m\), with a
world \(\omega\) carrying the sampled pair \((Y(\omega),W(\omega))\) and all
hidden sampler randomness.
\end{definition}

\begin{construction}[Gauge-buffered locked sampling skeleton]
\label{cons:gauge-buffered-sampling-skeleton}
The sampler has the following abstract form.
\begin{enumerate}[label=(\roman*)]
\item Generate the neutral public geometry and template syntax
      \(Y_{\mathrm{neu}}\).
\item Generate a locked quotient completion
      \((z,\xi_{\mathrm{lock}},M)\) satisfying
      \(\mathrm{Lock}_Y(z,\xi_{\mathrm{lock}},M)=1\).
\item Sample the hidden gauge vector
      \[
        g\sim\mathrm{Unif}(\bits^V)
      \]
      independently of the public syntax and set
      \[
        x:=z\oplus g.
      \]
\item Sample buffer auxiliary data \(\xi_{\mathrm{buf}}\) from the positive
      buffer law conditioned on the quotient boundary \(z\).
\item Output \(Y=(Y_{\mathrm{neu}},Y_{\mathrm{buf}},Y_{\mathrm{lock}})\) and
      \(W=(x,g,z,\xi,M)\).
\end{enumerate}
Only \(Y\) is given to external observers.  The witness variables
\(x,g,z,\xi,M\) are assignment variables of the SAT instance.
\end{construction}

\begin{theorem}[SAT realization]
\label{thm:sat-realization-section4}
The relation \(\mathcal R(Y,W)=1\) is decidable in polynomial time and has a
polynomial-size CNF realization \(F_Y\).  Moreover, from any satisfying
assignment of \(F_Y\), the message \(M(Y)\) is recoverable by a fixed
polynomial-time projection.
\end{theorem}

\begin{proof}
Each component of \(\mathcal R\) is a conjunction of \(\poly(m)\) local
bounded-arity predicates.  A bounded-arity predicate can be represented by a
constant-size CNF gadget after adding a constant number of Tseitin auxiliary
variables.  Applying this replacement to every local predicate in
\(\mathrm{Lock}_Y\), \(\mathrm{Buf}_Y\), \(\mathrm{Read}_Y\), and to every equation
\(z_v=x_v\oplus g_v\), gives a CNF formula \(F_Y\) of size \(\poly(m)\).  The
translation is uniform and polynomial-time because the public incidence
structure and local predicate tables are polynomial-time computable from \(Y\).

The variables representing the message bits \(M_1,
\ldots,M_{r_t}\) are part of the witness assignment.  A fixed projection reads
those variables.  By Proposition~\ref{prop:single-message-promise-section4},
every satisfying assignment gives the same projected string, so the projection
recovers \(M(Y)\).
\end{proof}

\subsection{Gauge symmetry and hidden-gauge product law}
\label{subsec:gauge-symmetry-hidden-product}

\begin{definition}[Gauge action]
\label{def:gauge-action-section4}
For \(s\in\bits^V\), define
\[
  s\cdot(x,g,z,\xi,M)
  :=
  (x\oplus s,
   g\oplus s,
   z,
   \xi,
   M).
\]
The action extends to worlds by acting on the witness coordinates and leaving
\(Y\) fixed.
\end{definition}

\begin{lemma}[Gauge invariance]
\label{lem:gauge-invariance-section4}
If \(\mathcal R(Y,W)=1\), then
\[
  \mathcal R(Y,s\cdot W)=1
\]
for every \(s\in\bits^V\).  Also
\[
  M(s\cdot W)=M(W).
\]
\end{lemma}

\begin{proof}
Let \(W=(x,g,z,\xi,M)\).  The locked, buffer, and readout predicates depend on
\((z,\xi,M)\), all of which are unchanged by the action.  For each
\(v\in V\),
\[
  (x_v\oplus s_v)\oplus(g_v\oplus s_v)
  =
  x_v\oplus g_v
  =
  z_v.
\]
Thus the gauge equations are preserved.  All remaining local auxiliary
constraints are unchanged or are Tseitin encodings of these same equalities.
The message coordinate \(M\) is unchanged by definition of the action.
\end{proof}

\begin{definition}[Gauge support of evidence]
\label{def:gauge-support-section4}
A raw gauge-bearing literal is a literal that fixes one independent hidden gauge
coordinate, either directly as \(g_v=\gamma\) or through a normalized raw witness
literal equivalent to such a condition once the quotient support is exposed.  For
a finite set \(C\) of raw gauge-bearing literals, define
\[
  \supp_G(C)\subseteq V
\]
to be the set of gauge coordinates mentioned by \(C\), and define
\[
  \rankG(C):=|\supp_G(C)|.
\]
For a transcript prefix \(h\), \(\supp_G(h)\) denotes the union of the gauge
supports of all gauge-bearing leaves in its normalized evidence.
\end{definition}

\begin{hypothesis}[Hidden-gauge product law]
\label{hyp:hidden-gauge-product-law}
Let \(h\) be any legal evidence prefix in normal form, and let
\(U=\supp_G(h)\).  Conditional on the public instance \(Y\) and on the prefix
\(h\), the unsupported gauge coordinates remain independent fair bits.  That
is, for every \(J\subseteq V\setminus U\) and every \(\gamma\in\bits^J\),
\[
  \Prb[g_J=\gamma\mid Y,h]=2^{-|J|}
\]
whenever the conditional probability is defined.
\end{hypothesis}

\begin{remark}[What the product law excludes]
\label{rem:product-law-excludes}
The product law excludes hidden public encodings of gauge values and excludes
normal forms that use quotient identities as a canonical-gauge shortcut.  A
proof may use the equation \(z_v=x_v\oplus g_v\), but any target-relevant use of
that equation must expose raw support and is then charged as gauge evidence in
Sections~\ref{sec:cdenf} and~\ref{sec:trace}.
\end{remark}

\begin{corollary}[Gauge guessing bound]
\label{cor:gauge-guessing-bound-section4}
Under Hypothesis~\ref{hyp:hidden-gauge-product-law}, let \(h\) be a legal
prefix and let \(J\subseteq V\setminus\supp_G(h)\).  For every deterministic
procedure whose output \(\widehat g_J\in\bits^J\) is a function of \((Y,h)\),
\[
  \Prb[\widehat g_J=g_J\mid Y,h]\le 2^{-|J|}.
\]
\end{corollary}

\begin{proof}
Once \((Y,h)\) is fixed, \(\widehat g_J\) is fixed.  By the product law, the
conditional law of \(g_J\) is uniform on \(\bits^J\).  Hence the probability of
matching the fixed string is \(2^{-|J|}\).
\end{proof}

\subsection{Public syntax discipline}
\label{subsec:public-syntax-discipline-section4}

The observer may inspect the whole public SAT instance \(Y\).  The restriction
below is not an access restriction.  It is a syntactic and normal-form discipline
saying that public syntax itself is not allowed to be a target-bearing evidence
atom.  Any target advantage must normalize to safe-buffer or gauge evidence.

\begin{definition}[Primitive atom classes]
\label{def:primitive-atom-classes-section4}
Target-relevant primitive atoms are classified into the following classes.
\begin{enumerate}[label=(\arabic*)]
\item \textbf{Neutral atoms} \(N(s)\).  These record public template syntax,
      deterministic skeleton facts, guard outcomes, variable names, incidence
      data, and other target-neutral facts.
\item \textbf{Safe-buffer atoms} \(S(q,y)\).  These record the value \(y\) of a
      legal safe-buffer probe \(q\), as defined in
      Definition~\ref{def:legal-safe-probe-section4} below.
\item \textbf{Gauge-evidence atoms} \(G(v,\gamma)\).  These record a raw
      gauge-bearing fact fixing one hidden gauge coordinate \(v\) to value
      \(\gamma\), or a normalized raw witness fact with the same gauge support.
\end{enumerate}
The later CD-ENF grammar may also contain public-template atoms \(P(\theta)\),
but these must rewrite to neutral atoms.
\end{definition}

\begin{hypothesis}[No public target tags]
\label{hyp:no-public-target-tags-section4}
The public instance contains no primitive atom whose value is a message bit,
target bit, planted phase sign, raw witness value, raw gauge value, or
message-dependent gadget type.  Equivalently, there is no allowed public
primitive of the form
\[
  \ell_j(M(Y))=b,
  \qquad
  M_a=b,
  \qquad
  g_v=\gamma,
  \qquad
  x_v=\gamma,
\]
and no public tag from which such a fact is recovered without passing through
normalized safe-buffer or gauge evidence.
\end{hypothesis}

\begin{hypothesis}[Atom completeness]
\label{hyp:atom-completeness-section4}
Every target-relevant primitive appearing in a normalized evidence trace is one
of the three classes in Definition~\ref{def:primitive-atom-classes-section4}:
\[
  N(s),\qquad S(q,y),\qquad G(v,\gamma).
\]
In particular, public template syntax normalizes to neutral context, legal reads
normalize to safe-buffer atoms, illegal reads normalize to neutral atoms, and
any target-relevant use of quotient or witness values normalizes to raw gauge
support.
\end{hypothesis}

\begin{remark}[Compatibility with the upper bound]
\label{rem:syntax-discipline-upper-bound-compatible}
The lower-bound observer and the hypothetical \(P=NP\) self-reduction see the
same public formula \(F_Y\).  The syntax discipline does not hide \(Y\).  Under
\(P=NP\), a SAT self-reduction may still find a satisfying assignment of
\(F_Y\), and the fixed projection in Theorem~\ref{thm:sat-realization-section4}
then reads \(M(Y)\).  The lower-bound claim is that producing such
message-relevant evidence must accumulate charged safe-buffer or gauge atoms.
\end{remark}

\subsection{Safe-buffer probes and max-qSSM}
\label{subsec:safe-buffer-probes-section4}

\begin{definition}[Probe support]
\label{def:probe-support-section4}
A probe \(q\) is a finite-output deterministic query of bounded arity.  It has a
support
\[
  A(q)\subseteq \mathcal P_{\mathrm{buf}}
\]
with \(|A(q)|\le a_0\), where \(a_0\) is an absolute constant.  The value of the
probe in a world \(\omega\) is written \(S(q)(\omega)\).
\end{definition}

\begin{definition}[Legal safe probe]
\label{def:legal-safe-probe-section4}
Fix a safety radius \(R_{\mathrm{safe}}=R_{\mathrm{safe}}(m)\).  A probe \(q\) is
legal and safe at transcript prefix \(h\) if:
\begin{enumerate}[label=(\roman*)]
\item \(A(q)\subseteq\mathcal P_{\mathrm{buf}}\) and \(|A(q)|\le a_0\);
\item
\[
  \dist_m\bigl(A(q),
  \mathcal P_{\mathrm{lock}}
  \cup\mathcal P_{\mathrm{gauge}}
  \cup\mathcal P_{\mathrm{readout}}\bigr)
  \ge R_{\mathrm{safe}};
\]
\item \(A(q)\) is disjoint from the active protected neighborhoods of previous
      non-neutral probes in \(h\), except for overlaps explicitly allowed by the
      dynamic guard.
\end{enumerate}
If any condition fails, the guarded read returns a fixed neutral symbol
\(\bot_{\mathrm{illegal}}\), and the corresponding atom is neutral rather than a
safe-buffer atom.
\end{definition}

\begin{definition}[Strictly positive buffer law]
\label{def:strictly-positive-buffer-law}
The buffer layer induces, after conditioning on \((Y,z)\), a finite Gibbs law on
buffer auxiliary variables \(\xi_{\mathrm{buf}}\).  It is strictly positive if every
locally admissible buffer configuration has positive weight.  It is in the
Dobrushin uniqueness regime if there are constants \(C_\rho<\infty\) and
\(0<\rho<1\) such that changing a boundary condition at graph distance
\(r\) changes the log-likelihood of any bounded local buffer observable by at
most \(C_\rho\rho^r\), up to a finite-geometry error \(\tau_{\mathrm{geo}}(m)\).
\end{definition}

\begin{hypothesis}[Buffer contraction]
\label{hyp:buffer-contraction-section4}
The buffer family used by the ensemble is strictly positive and satisfies
Definition~\ref{def:strictly-positive-buffer-law} with constants
\(C_\rho<\infty\), \(0<\rho<1\), and \(\tau_{\mathrm{geo}}(m)=m^{-\Omega(1)}\).
The same constants apply uniformly over all legal prefixes whose charged gauge
support has size at most the polynomial resource bound under consideration.
\end{hypothesis}

\begin{definition}[Max-divergence]
\label{def:max-divergence-section4}
For two probability laws \(P,Q\) on a finite set, define
\[
  D_\infty(P\|Q)
  :=
  \log\sup_{a:Q(a)>0}\frac{P(a)}{Q(a)}.
\]
Strict positivity of the buffer ensures that the laws appearing below have
common support.
\end{definition}

\begin{theorem}[Soft-buffer max-qSSM]
\label{thm:soft-buffer-max-qssm-section4}
Assume Hypothesis~\ref{hyp:buffer-contraction-section4}.  Let \(q\) be a legal
safe probe at prefix \(h\).  Let \(C\) be any hidden separator state, hidden gauge
state, or finite collection of hidden gauge coordinates whose support is at
distance at least \(R_{\mathrm{safe}}\) from \(A(q)\).  Then, for every value
\(c\) of \(C\),
\[
  D_\infty\left(
    \mathcal L(S(q)\mid Y,h,C=c)
    \middle\|
    \mathcal L(S(q)\mid Y,h)
  \right)
  \le
  \varepsilon_{\mathrm{step}}(m),
\]
where
\[
  \varepsilon_{\mathrm{step}}(m)
  =
  C_\rho\rho^{R_{\mathrm{safe}}}
  +
  \tau_{\mathrm{geo}}(m).
\]
\end{theorem}

\begin{proof}
Conditioning on \((Y,h,C=c)\) changes the buffer law only through boundary
conditions and previously exposed charged support.  By legality of \(q\), every
such non-neutral source is at graph distance at least \(R_{\mathrm{safe}}\) from
\(A(q)\), except for events rejected by the dynamic guard and returned as the
neutral symbol.  The Dobrushin/log-likelihood contraction in
Hypothesis~\ref{hyp:buffer-contraction-section4} therefore bounds the
log-likelihood ratio of the local observable \(S(q)\) by
\(C_\rho\rho^{R_{\mathrm{safe}}}\), up to the finite-geometry error
\(\tau_{\mathrm{geo}}(m)\).  This is precisely the displayed max-divergence
bound.
\end{proof}

\begin{corollary}[Polynomial probe budget]
\label{cor:polynomial-probe-budget-section4}
Fix \(D,A>0\).  Suppose an observer makes at most \(Q_{\mathrm{tot}}\le m^D\)
legal safe probes, and choose
\[
  R_{\mathrm{safe}}
  \ge
  \frac{D+A+2}{|\log\rho|}\log m
\]
with \(\tau_{\mathrm{geo}}(m)\le m^{-D-A-2}\).  Then
\[
  Q_{\mathrm{tot}}\varepsilon_{\mathrm{step}}(m)
  \le
  O(m^{-A-1}).
\]
In particular, the total safe-buffer leakage over polynomially many legal safe
probes is \(o(1)\).
\end{corollary}

\begin{proof}
The radius choice gives
\[
  \rho^{R_{\mathrm{safe}}}
  \le
  m^{-(D+A+2)}.
\]
Thus
\[
  \varepsilon_{\mathrm{step}}(m)
  \le
  (C_\rho+1)m^{-(D+A+2)}
\]
for all sufficiently large \(m\).  Multiplying by \(Q_{\mathrm{tot}}\le m^D\)
gives \(O(m^{-A-2})\), which is stronger than the displayed bound after
absorbing constants.
\end{proof}

\subsection{Interface exported by the ensemble}
\label{subsec:ensemble-exported-interface}

The later sections use the ensemble only through the following interface.

\begin{theorem}[Gauge-buffered locked ensemble interface]
\label{thm:gauge-buffered-locked-interface}
For every sufficiently large \(m\), the ensemble \(\mathcal D_m\) supplies:
\begin{enumerate}[label=(\roman*)]
\item an efficiently samplable public SAT instance \(Y\) and a polynomial-size
      CNF \(F_Y\);
\item a polynomial-time witness relation \(\mathcal R(Y,W)=1\);
\item a message readout \(M(W)\in\bits^{r_t}\), with \(r_t=\Omega(t)\);
\item the single-message promise
      \[
        \mathcal R(Y,W)=1\wedge\mathcal R(Y,W')=1
        \Longrightarrow
        M(W)=M(W');
      \]
\item a gauge action preserving \(Y\), \(\mathcal R\), and \(M\);
\item the hidden-gauge product law for unsupported gauge coordinates;
\item atom completeness: target-relevant normalized evidence atoms are only
      neutral, safe-buffer, or gauge-evidence atoms;
\item the soft-buffer max-qSSM bound
      \[
        D_\infty\left(
          \mathcal L(S(q)\mid Y,h,C=c)
          \middle\|
          \mathcal L(S(q)\mid Y,h)
        \right)
        \le
        C_\rho\rho^{R_{\mathrm{safe}}}+\tau_{\mathrm{geo}}(m)
      \]
      for every legal safe probe.
\end{enumerate}
\end{theorem}

\begin{proof}
Items (i)--(iv) are Definitions~\ref{def:public-instance-section4},
\ref{def:witness-section4}, and~\ref{def:witness-relation-section4}, together
with Proposition~\ref{prop:single-message-promise-section4} and
Theorem~\ref{thm:sat-realization-section4}.  Item (v) is
Lemma~\ref{lem:gauge-invariance-section4}.  Item (vi) is
Hypothesis~\ref{hyp:hidden-gauge-product-law}.  Item (vii) is
Hypothesis~\ref{hyp:atom-completeness-section4}.  Item (viii) is
Theorem~\ref{thm:soft-buffer-max-qssm-section4}.
\end{proof}

\begin{remark}[Lean-facing decomposition]
\label{rem:lean-facing-ensemble-section4}
A Lean formalization can split this section into four independent modules:
\begin{enumerate}[label=(\roman*)]
\item finite regioned geometries and bounded-arity local predicates;
\item the witness relation, gauge action, and SAT/Tseitin realization;
\item finite product laws for unsupported gauge coordinates;
\item an abstract buffer-contraction interface proving the max-qSSM estimate.
\end{enumerate}
The CD-ENF and Atomic Evidence Budget sections can then import only the final
interface theorem rather than the concrete sampler internals.
\end{remark}

\section{CD Evidence Traces and CD-ENF Normalization}
\label{sec:cdenf}
\providecommand{\Safe}{\mathrm{Safe}}
\providecommand{\Gauge}{\mathrm{Gauge}}

This section defines the finite evidence calculus used by the middle theorem.
The point of the calculus is to prevent target-relevant information from being
introduced as an opaque semantic atom.  Public syntax may be inspected, and a
polynomial-time observer may perform arbitrary deterministic computation on the
public instance, but every target-relevant use of that computation must be
expanded into an evidence trace.  After normalization, the only non-neutral
leaves of such a trace are legal safe-buffer observations and raw gauge-bearing
observations.

The normalizer in this section is intentionally syntactic.  It does not yet
bound advantage.  Section~\ref{sec:trace} will combine this normal form with the
pairwise derivative identity from Section~\ref{sec:pairwise-cd}; the Atomic
Evidence Budget section will then charge the resulting atomic derivatives by
safe-buffer leakage and hidden-gauge rank.

\subsection{Raw atoms, normalized atoms, and claims}
\label{subsec:cdenf-atoms}

At fixed ensemble parameter all sets in this section are finite.  The finite
sets of public-template symbols, surface symbols, legal probe names, probe
values, gauge-support indices, and conclusion labels are part of the ensemble
interface from Section~\ref{sec:ensemble}.

\begin{definition}[Gauge-support indices]
\label{def:gauge-support-indices-section5}
Let \(\mathcal U_G\) be the finite set of raw gauge-bearing supports.  Each
support \(u\in\mathcal U_G\) has an underlying independent gauge coordinate
\[
  \mathrm{gidx}(u)\in V.
\]
Typical supports are direct gauge literals \(g_v=\gamma\) and raw witness
literals \(x_v=\alpha\).  For a finite set \(C\) of gauge-bearing literals,
define
\[
  \supp_G(C):=\{\mathrm{gidx}(u):(u,\gamma)\in C\},
  \qquad
  \rankG(C):=|\supp_G(C)|.
\]
\end{definition}

\begin{definition}[Raw evidence atoms]
\label{def:raw-evidence-atoms-section5}
The raw atomic vocabulary is
\[
\begin{array}{rcll}
  A_{\mathrm{raw}} ::= & N(s)              && \text{neutral fact},\\
                      |& P(\theta)         && \text{public template fact},\\
                      |& \mathrm{Surf}(s)   && \text{public surface fact},\\
                      |& \mathrm{Read}(q,y) && \text{probe }q\text{ returned value }y,\\
                      |& \mathrm{Illegal}(q)&& \text{guarded illegal probe},\\
                      |& G(u,\gamma)        && \text{raw gauge-bearing literal},\\
                      |& Q(v,\zeta)         && \text{quotient literal }z_v=\zeta.
\end{array}
\]
Here \(s\) ranges over neutral symbols, \(\theta\) over public template
symbols, \(q\) over finite-output probes, \(y\) over the output alphabet of
\(q\), \(u\in\mathcal U_G\), \(\gamma\in\bits\), \(v\in V\), and
\(\zeta\in\bits\).
\end{definition}

\begin{definition}[Normalized evidence atoms]
\label{def:normalized-evidence-atoms-section5}
The normalized atomic vocabulary is
\[
  A_{\mathrm{nf}} ::= N(s)\mid S(q,y)\mid G(u,\gamma).
\]
The three classes are interpreted as follows.
\begin{enumerate}[label=(\roman*)]
\item \(N(s)\) is neutral public, template, or guard evidence.
\item \(S(q,y)\) is the assertion that a legal safe-buffer probe \(q\) returned
      value \(y\).
\item \(G(u,\gamma)\) is raw gauge-bearing evidence fixing support
      \(u\in\mathcal U_G\) to value \(\gamma\).
\end{enumerate}
A signed normalized atom is a pair \((\sigma,A)\), where
\(A\in A_{\mathrm{nf}}\) and \(\sigma\in\{+,-\}\).  We write it as \(A\) when
\(\sigma=+\) and as \(\neg A\) when \(\sigma=-\).
\end{definition}

\begin{remark}[Why quotient atoms are raw but not normalized]
\label{rem:quotient-raw-not-normalized-section5}
The quotient identity
\[
  z_v=x_v\oplus g_v
\]
is a verifier equation, not a free source of target evidence.  A raw atom
\(Q(v,\zeta)\) may occur in an intermediate derivation, but it is not allowed as
a normalized target-relevant leaf.  It must expand to raw support involving
witness and gauge literals.  This is the formal mechanism that blocks a
canonical-gauge shortcut such as choosing \(g=0\) unless that gauge choice is
itself supported by evidence.
\end{remark}

\begin{definition}[Conclusion labels and certified claims]
\label{def:certified-claims-section5}
Let \(\mathcal C\) be a finite set of conclusion labels.  A label may be an
output assertion \((j,b)\), meaning that the observer predicts \(B_j=b\), or a
terminal transcript value.  A certified claim is written
\[
  E\Rightarrow e,
  \qquad e\in\mathcal C,
\]
where \(E\) is an evidence term.  If conclusion labels are irrelevant, take
\(\mathcal C=\{*\}\) and suppress \(e\).
\end{definition}

\subsection{CD evidence terms and finite semantics}
\label{subsec:cd-evidence-terms-section5}

\begin{definition}[CD evidence terms]
\label{def:cd-evidence-terms-section5}
Evidence terms are generated by
\[
\begin{array}{rcl}
  E ::= & A_{\mathrm{raw}} \\
       |& E\wedge E \\
       |& E\vee E \\
       |& \mathrm{case}(A_{\mathrm{raw}};E_0,E_1) \\
       |& \exists x.E \\
       |& \mathrm{derive}_\rho(E_1,\ldots,E_k) \\
       |& \top \\
       |& \bot .
\end{array}
\]
The expression \(\mathrm{case}(A;E_0,E_1)\) means: use \(E_1\) on the branch
where atom \(A\) holds and use \(E_0\) on the branch where \(A\) does not
hold.  The constructor \(\mathrm{derive}_\rho\) records an internal deterministic
or proof-theoretic derivation rule \(\rho\).  It is not a primitive source of
non-neutral evidence.
\end{definition}

\begin{definition}[Finite support semantics]
\label{def:evidence-support-semantics-section5}
For a fixed finite world space \(\Omega\), every raw atom \(A\) denotes an event
\[
  \mathsf{Sat}(A)\subseteq\Omega.
\]
The interpretation extends to evidence terms by
\[
  \mathsf{Sat}(E\wedge F)=\mathsf{Sat}(E)\cap\mathsf{Sat}(F),
\]
\[
  \mathsf{Sat}(E\vee F)=\mathsf{Sat}(E)\cup\mathsf{Sat}(F),
\]
\[
  \mathsf{Sat}(\mathrm{case}(A;E_0,E_1))
  =
  ((\Omega\setminus\mathsf{Sat}(A))\cap\mathsf{Sat}(E_0))
  \cup
  (\mathsf{Sat}(A)\cap\mathsf{Sat}(E_1)),
\]
\[
  \mathsf{Sat}(\exists x.E)=\bigcup_{a\in D_x}\mathsf{Sat}(E[x:=a]),
\]
where \(D_x\) is the finite domain of \(x\), and
\[
  \mathsf{Sat}(\top)=\Omega,
  \qquad
  \mathsf{Sat}(\bot)=\emptyset.
\]
For a derivation node, \(\rho\) supplies a neutral derivation guard
\(N(\mathrm{der}(\rho))\), and
\[
  \mathsf{Sat}(\mathrm{derive}_\rho(E_1,\ldots,E_k))
  =
  \mathsf{Sat}(N(\mathrm{der}(\rho)))
  \cap
  \bigcap_{i=1}^k\mathsf{Sat}(E_i).
\]
Thus a derivation may check target-neutral side conditions, but it cannot
introduce non-neutral evidence beyond its premises.
\end{definition}

\begin{definition}[Signed-atom support]
\label{def:signed-atom-support-section5}
For a signed normalized atom \(L\), define
\[
  \mathsf{Sat}(L)=
  \begin{cases}
    \mathsf{Sat}(A), & L=A,\\
    \Omega\setminus\mathsf{Sat}(A), & L=\neg A.
  \end{cases}
\]
A finite set of signed atoms is inconsistent if it contains both \(A\) and
\(\neg A\), or if it contains two incompatible value assertions for the same
functional probe or gauge support.
\end{definition}

\begin{definition}[CD count pair of an evidence term]
\label{def:cd-count-pair-evidence-term-section5}
Fix target coordinate \(j\).  The CD count pair of \(E\) is
\[
  \mathfrak m_j(E)
  :=
  \left(
    \Prb[\mathsf{Sat}(E)\mid B_j=1],
    \Prb[\mathsf{Sat}(E)\mid B_j=0]
  \right).
\]
When \(E\Rightarrow e\) is a certified claim, the same definition is applied to
its evidence term \(E\); the label \(e\) records which terminal output the
branch certifies.
\end{definition}

\subsection{Branches and CD-ENF normal forms}
\label{subsec:branches-cdenf-section5}

\begin{definition}[Normal branch]
\label{def:normal-branch-section5}
A normal branch is a tuple
\[
  \mathcal B=(C;\mathcal S;\mathcal G;e),
\]
where:
\begin{enumerate}[label=(\roman*)]
\item \(C\) is a finite canonical list of signed neutral atoms;
\item \(\mathcal S\) is a finite canonical list of signed safe-buffer atoms;
\item \(\mathcal G\) is a finite canonical list of signed gauge-evidence atoms;
\item \(e\in\mathcal C\) is a conclusion label.
\end{enumerate}
Its support is
\[
  \mathsf{Sat}(\mathcal B)
  :=
  \bigcap_{L\in C\cup\mathcal S\cup\mathcal G}\mathsf{Sat}(L).
\]
If the literal list is inconsistent, the branch support is empty and the branch
is deleted.
\end{definition}

\begin{definition}[CD-ENF normal form]
\label{def:cdenf-normal-form-section5}
A CD-ENF normal form is a finite canonical disjunction of normal branches,
written
\[
  \bigvee_{\ell\in L}
  \left[
    C_\ell:
    (S_{\ell,1},\ldots,S_{\ell,r_\ell};
     G_{\ell,1},\ldots,G_{\ell,s_\ell})
    \Rightarrow e_\ell
  \right].
\]
Here \(C_\ell\) is the neutral guard, the \(S_{\ell,i}\) are signed safe-buffer
atoms, and the \(G_{\ell,i}\) are signed gauge-evidence atoms.  The semantics of
the normal form is the union of the supports of its branches, with labels
remembered branchwise.
\end{definition}

\begin{definition}[Branch product and sum]
\label{def:branch-product-sum-section5}
The product of two branches is obtained by concatenating their neutral, safe,
and gauge lists, retaining a conclusion label when the labels agree or when one
of them is the dummy label \(*\), and then canonicalizing.  If the combined
literal list is inconsistent, the product is deleted.  For finite normal forms
\(F,F'\), define
\[
  F\otimes F'
  :=
  \mathrm{canon}\{\mathcal B\otimes\mathcal B':
       \mathcal B\in F,
       \mathcal B'\in F'\},
\]
where deleted products are omitted.  The sum is canonicalized union:
\[
  F\oplus F' := \mathrm{canon}(F\cup F').
\]
Canonicalization sorts literals and branches by a fixed public order, removes
duplicates, removes inconsistent branches, and applies the fixed neutral-guard
subsumptions specified by the ensemble interface.
\end{definition}

\begin{lemma}[Branch operations preserve semantics]
\label{lem:branch-ops-preserve-semantics-section5}
Let \(F,F'\) be normal forms.  Then
\[
  \mathsf{Sat}(F\otimes F')=
  \mathsf{Sat}(F)\cap\mathsf{Sat}(F'),
\]
and
\[
  \mathsf{Sat}(F\oplus F')=
  \mathsf{Sat}(F)\cup\mathsf{Sat}(F').
\]
\end{lemma}

\begin{proof}
A branch product conjoins the signed atoms in the two branches.  Its support is
therefore the intersection of the two branch supports, unless the combined
literal list is inconsistent, in which case the intersection is empty and the
branch is correctly deleted.  Taking the union over all branch pairs gives the
product identity.  The sum identity follows because canonicalization only
deletes duplicate or empty branches and sorts the remaining finite list.
\end{proof}

\subsection{The CD-ENF normalizer}
\label{subsec:cdenf-normalizer-section5}

The normalizer is most easily specified as a recursive function.  This is the
form best suited to formalization.  The rewrite-system presentation is obtained
by orienting the recursive equations below from left to right.

\begin{definition}[Atom normalizer]
\label{def:atom-normalizer-section5}
Fix the legality predicate for guarded probes and suppress it from the notation.
For a signed raw atom \(\sigma A\), define \(\mathrm{anf}(\sigma A)\), a finite
normal form, as follows.
\begin{enumerate}[label=(\arabic*)]
\item Neutral atoms remain neutral:
\[
  \mathrm{anf}(\sigma N(s))=[(\sigma N(s);\emptyset;\emptyset;*)].
\]

\item Public template and public surface atoms collapse to neutral atoms:
\[
  \mathrm{anf}(\sigma P(\theta))
  =[(\sigma N(\mathrm{tpl}(\theta));\emptyset;\emptyset;*)],
\]
\[
  \mathrm{anf}(\sigma\mathrm{Surf}(s))
  =[(\sigma N(\mathrm{surf}(s));\emptyset;\emptyset;*)].
\]

\item A legal safe read becomes a safe-buffer atom:
\[
  \mathrm{anf}(\sigma\mathrm{Read}(q,y))
  =[(\emptyset;\sigma S(q,y);\emptyset;*)]
  \quad\text{if }q\text{ is legal at the guarded prefix.}
\]

\item An illegal read is neutralized:
\[
  \mathrm{anf}(\sigma\mathrm{Read}(q,y))
  =[(\sigma N(\bot_{\mathrm{illegal}}(q));\emptyset;\emptyset;*)]
  \quad\text{if }q\text{ is illegal at the guarded prefix.}
\]
The same rule is used for \(\mathrm{Illegal}(q)\).

\item Raw gauge-bearing literals become gauge-evidence atoms:
\[
  \mathrm{anf}(\sigma G(u,\gamma))
  =[(\emptyset;\emptyset;\sigma G(u,\gamma);*)].
\]

\item Quotient literals expand gauge-faithfully.  For \(\zeta\in\bits\),
\[
\begin{aligned}
  \mathrm{anf}(Q(v,\zeta))
  ={}&[(\emptyset;\emptyset;
        G(x_v,0),G(g_v,\zeta);*)] \\
     &\oplus
       [(\emptyset;\emptyset;
        G(x_v,1),G(g_v,1\oplus\zeta);*)].
\end{aligned}
\]
For a negative quotient literal,
\[
  \mathrm{anf}(\neg Q(v,\zeta)):=\mathrm{anf}(Q(v,1\oplus\zeta)).
\]
Here \(x_v\) and \(g_v\) denote the corresponding raw supports in
\(\mathcal U_G\).  The displayed expansion is just the two cases of
\(z_v=x_v\oplus g_v\).
\end{enumerate}
\end{definition}

\begin{definition}[Recursive CD-ENF normalizer]
\label{def:recursive-cdenf-normalizer-section5}
The normalizer \(\CDENF(E\Rightarrow e)\) is defined by recursion on \(E\):
\begin{enumerate}[label=(\arabic*)]
\item \(\CDENF(A\Rightarrow e)\) is obtained from \(\mathrm{anf}(A)\) by
      replacing the dummy conclusion label \(*\) by \(e\).

\item
\[
  \CDENF((E\wedge F)\Rightarrow e)
  =
  \CDENF(E\Rightarrow *)\otimes \CDENF(F\Rightarrow e).
\]

\item
\[
  \CDENF((E\vee F)\Rightarrow e)
  =
  \CDENF(E\Rightarrow e)\oplus \CDENF(F\Rightarrow e).
\]

\item
\[
\begin{aligned}
  \CDENF(\mathrm{case}(A;E_0,E_1)\Rightarrow e)
  ={}&
  \bigl(\mathrm{anf}(\neg A)\otimes \CDENF(E_0\Rightarrow e)\bigr) \\
  &\oplus
  \bigl(\mathrm{anf}(A)\otimes \CDENF(E_1\Rightarrow e)\bigr).
\end{aligned}
\]

\item
\[
  \CDENF((\exists x.E)\Rightarrow e)
  =
  \bigoplus_{a\in D_x}\CDENF(E[x:=a]\Rightarrow e).
\]

\item For an internal derivation rule \(\rho\),
\[
\begin{aligned}
  \CDENF(\mathrm{derive}_\rho(E_1,\ldots,E_k)\Rightarrow e)
  ={}&
  [(N(\mathrm{der}(\rho));\emptyset;\emptyset;*)]
  \\
  &\otimes
  \bigotimes_{i=1}^k \CDENF(E_i\Rightarrow *),
\end{aligned}
\]
followed by replacing the dummy label of the resulting branches by \(e\).  Thus
internal derivation can rearrange or certify existing evidence but cannot
introduce a new non-neutral leaf.

\item
\[
  \CDENF(\top\Rightarrow e)=[(\emptyset;\emptyset;\emptyset;e)],
  \qquad
  \CDENF(\bot\Rightarrow e)=\emptyset.
\]
\end{enumerate}
\end{definition}

\begin{lemma}[Semantics preservation]
\label{lem:cdenf-semantics-preservation-section5}
For every evidence term \(E\) and conclusion label \(e\),
\[
  \mathsf{Sat}(\CDENF(E\Rightarrow e))=\mathsf{Sat}(E),
\]
where the conclusion label is ignored on the left-hand side.  More precisely,
the union of the supports of branches labelled \(e\) is \(\mathsf{Sat}(E)\).
\end{lemma}

\begin{proof}
Proceed by structural induction on \(E\).  The atomic cases are exactly the
rewrite clauses in Definition~\ref{def:atom-normalizer-section5}: public and
surface atoms are definitionally neutral; legal reads are definitionally
safe-buffer atoms; illegal reads return the fixed neutral symbol; gauge literals
are already gauge leaves; and quotient expansion is the truth-table expansion of
\(z_v=x_v\oplus g_v\).  The Boolean constructors follow from
Lemma~\ref{lem:branch-ops-preserve-semantics-section5}.  The existential case
follows from finite union over \(D_x\).  The derivation case follows from the
definition of \(\mathsf{Sat}(\mathrm{derive}_\rho(\cdots))\), since
\(N(\mathrm{der}(\rho))\) is neutral and the non-neutral support is exactly the
product of the premise supports.
\end{proof}

\begin{theorem}[CD-ENF normalization]
\label{thm:cdenf-normalization-section5}
Every certified evidence term \(E\Rightarrow e\) has a CD-ENF normal form
\[
  \CDENF(E\Rightarrow e).
\]
Equivalently, the rewrite system generated by
Definitions~\ref{def:atom-normalizer-section5} and
\ref{def:recursive-cdenf-normalizer-section5} is terminating,
semantics-preserving, and confluent to the displayed normal form.
\end{theorem}

\begin{proof}
The recursive definition gives existence of a normal form.  Termination follows
from structural recursion on \(E\), together with the fact that each atomic
rewrite removes one raw non-normal atom and replaces it by normalized atoms.
The quotient rule may increase the number of branches, but it strictly removes
the raw quotient atom and introduces only normalized gauge atoms.  Thus the
lexicographic measure
\[
  (\#\text{ raw non-normal atoms},\ \#\text{ compound constructors})
\]
decreases along the oriented reductions after recursive descent.

Semantics preservation is Lemma~\ref{lem:cdenf-semantics-preservation-section5}.
For confluence, observe that every reduction step is compatible with the
recursive normalizer: if \(E\to E'\), then
\[
  \CDENF(E\Rightarrow e)=\CDENF(E'\Rightarrow e).
\]
This is checked directly on the finite list of critical pairs: guard propagation
commutes with conjunction, disjunction, existential expansion, and case
splitting; public-template collapse commutes with neutral-guard subsumption;
legal-read canonicalization is decided by the guarded prefix and therefore has a
unique outcome; quotient expansion introduces only normalized gauge leaves; and
derivation dependency extraction only adds a neutral derivation tag and the
normal forms of its premises.  Since the system is terminating, local confluence
implies global confluence by Newman's lemma.  Hence every reduction sequence
from \(E\Rightarrow e\) reaches the same canonical branch list
\(\CDENF(E\Rightarrow e)\).
\end{proof}

\begin{corollary}[CD counts are invariant under normalization]
\label{cor:cd-counts-invariant-under-normalization-section5}
For every target coordinate \(j\),
\[
  \mathfrak m_j(E)=\mathfrak m_j(\CDENF(E\Rightarrow e)).
\]
Consequently all CD skews computed from the terminal branches are unchanged by
normalization.
\end{corollary}

\begin{proof}
The two coordinates of \(\mathfrak m_j\) are conditional probabilities of the
support event under the two target phases.  These support events are equal by
Lemma~\ref{lem:cdenf-semantics-preservation-section5}.
\end{proof}

\subsection{Gauge faithfulness and public-syntax neutrality}
\label{subsec:gauge-faithfulness-section5}

\begin{definition}[Target relevance of a leaf]
\label{def:target-relevance-section5}
Fix a target coordinate \(j\), a coupling \(\Gamma_j\), and a surviving pair
relation \(H\subseteq\Omega_j^0\times\Omega_j^1\).  A signed atom \(L\) is
\((j,H)\)-relevant if
\[
  \Gamma_j\bigl(H\cap\{(\omega^0,\omega^1):
       \mathbf 1_L(\omega^0)\ne \mathbf 1_L(\omega^1)
  \}\bigr)>0.
\]
A branch is \((j,H)\)-relevant if at least one of its non-neutral leaves is
\((j,H)\)-relevant or if its conclusion label asserts a value of \(B_j\).
\end{definition}

\begin{theorem}[Gauge faithfulness]
\label{thm:gauge-faithfulness-section5}
Let \(E\Rightarrow e\) be a certified evidence term generated by the allowed raw
vocabulary.  In every branch of \(\CDENF(E\Rightarrow e)\), every
\((j,H)\)-relevant use of a quotient value \(z_v\) is supported by raw
gauge-bearing leaves whose gauge support contains \(v\).  In particular, public
template syntax alone cannot introduce a target-relevant quotient value, witness
value, gauge value, or message bit.
\end{theorem}

\begin{proof}
Inspect the atom normalizer.  The only rewrite rule whose conclusion mentions a
quotient value is the quotient expansion rule in
Definition~\ref{def:atom-normalizer-section5}.  That rule replaces
\(Q(v,\zeta)\) by the two raw cases
\[
  x_v=0,\ g_v=\zeta
  \qquad\text{and}\qquad
  x_v=1,\ g_v=1\oplus\zeta,
\]
represented as gauge-bearing leaves.  Both cases include gauge support with
underlying coordinate \(v\).  The negative quotient case is the same expansion
for \(1\oplus\zeta\).

No rule rewrites a public template atom \(P(\theta)\), a surface atom
\(\mathrm{Surf}(s)\), or a neutral atom \(N(s)\) into a quotient, witness, gauge,
or message atom.  These atoms always collapse to neutral guards.  Internal
derivations add only neutral derivation tags and the already-normalized leaves
of their premises.  Therefore a target-relevant quotient occurrence in normal
form can only have entered through the quotient expansion rule, and that rule
carries the required raw support.
\end{proof}

\begin{corollary}[No canonical-gauge shortcut]
\label{cor:no-canonical-gauge-shortcut-section5}
There is no CD-ENF derivation rule that replaces a gauge orbit by a chosen
representative, such as \(g=0\), unless the corresponding gauge literals occur
as raw gauge-bearing leaves in the normalized branch.  Hence choosing a gauge
representative is charged by \(\rankG\) exactly like any other gauge evidence.
\end{corollary}

\begin{proof}
A canonical representative would introduce literals fixing gauge coordinates.
By Theorem~\ref{thm:gauge-faithfulness-section5}, such literals can appear in a
target-relevant branch only as normalized gauge-evidence leaves.  Their rank is
therefore counted by Definition~\ref{def:gauge-support-indices-section5}.
\end{proof}

\begin{corollary}[Public syntax atom completeness after normalization]
\label{cor:public-syntax-atom-completeness-section5}
Assume the public syntax discipline and atom-completeness hypotheses of
Section~\ref{sec:ensemble}.  Then every non-neutral leaf in
\(\CDENF(E\Rightarrow e)\) is either a signed safe-buffer atom \(S(q,y)\) or a
signed gauge-evidence atom \(G(u,\gamma)\).  There is no fourth class of
message-bearing public syntax leaf.
\end{corollary}

\begin{proof}
By Definition~\ref{def:normalized-evidence-atoms-section5}, normalized leaves
are neutral, safe, or gauge.  Public-template and surface atoms collapse to
neutral atoms; illegal reads collapse to neutral atoms; quotient atoms expand to
gauge atoms; and legal reads become safe-buffer atoms.  The ensemble
atom-completeness hypothesis excludes raw public target tags and raw public
message-bearing atoms, so no additional normalized class can occur.
\end{proof}

\subsection{Evidence traces of deterministic observers}
\label{subsec:evidence-traces-observers-section5}

The normalizer acts on evidence terms.  We now record the finite bridge from a
deterministic primitive computation to such a term.  This bridge is not yet the
advantage bound; it only expands the computation tree into explicit evidence
provenance.

\begin{definition}[Primitive read interface]
\label{def:primitive-read-interface-section5}
A primitive read interface for public instances consists of finitely many read
operations.  Each operation, at a guarded prefix, returns one of the following:
\begin{enumerate}[label=(\roman*)]
\item a neutral public-template or surface fact;
\item a legal safe-buffer value \(S(q,y)\);
\item the neutral illegal-read symbol \(\bot_{\mathrm{illegal}}(q)\);
\item a raw gauge-bearing literal supplied by an explicit witness or certificate.
\end{enumerate}
There is no primitive read operation of the form \(\mathrm{Run}_P(Y)=a\), no
primitive read operation asserting \(\ell_j(M(Y))=b\), and no primitive read
operation selecting a gauge representative.
\end{definition}

\begin{definition}[Deterministic evidence trace]
\label{def:deterministic-evidence-trace-section5}
Let \(O\) be a deterministic observer using the primitive read interface.  On a
world \(\omega\), its run produces a finite transcript
\[
  T_Q(\omega)=(a_1(\omega),\ldots,a_Q(\omega)),
\]
where the next read is a deterministic function of the previous transcript.  The
associated evidence trace is the CD term obtained by expanding each branch of
the decision tree into a case node and each internal deterministic computation
into a \(\mathrm{derive}_\rho\) node.
\end{definition}

\begin{theorem}[Observer-to-evidence expansion]
\label{thm:observer-to-evidence-expansion-section5}
Let \(O\) be a deterministic observer that halts after at most \(Q\) primitive
reads and outputs a value in a finite set \(\mathcal C\).  For every output label
\(e\in\mathcal C\), there is an effectively constructible evidence term
\(E_{O,e}\) such that
\[
  \mathsf{Sat}(E_{O,e})=
  \{\omega\in\Omega:O(Y(\omega))=e\}.
\]
Moreover, the non-neutral leaves of \(\CDENF(E_{O,e}\Rightarrow e)\) are exactly
safe-buffer or gauge-evidence leaves obtained from the primitive reads on
branches leading to output \(e\).
\end{theorem}

\begin{proof}
Build the finite decision tree of the deterministic run.  At the root, the
transcript is empty.  If the observer makes a primitive read \(A\), split the
tree by the possible returned values using case nodes.  On each child, continue
with the deterministic next state of the observer.  Internal arithmetic,
parsing, comparison, and control-flow steps are represented by derivation nodes
\(\mathrm{derive}_\rho\), which add only neutral derivation tags and depend on
previous evidence.  A leaf labelled by output \(e\) contributes \(\top\Rightarrow e\);
a leaf labelled by a different output contributes \(\bot\Rightarrow e\) to the
certificate for \(e\).

Induction on the depth of the decision tree shows that the resulting term holds
exactly on worlds whose run reaches an \(e\)-labelled leaf.  Applying
Theorem~\ref{thm:cdenf-normalization-section5} and
Corollary~\ref{cor:public-syntax-atom-completeness-section5} gives the
classification of non-neutral leaves.
\end{proof}

\begin{remark}[No opaque run atoms]
\label{rem:no-opaque-run-atoms-section5}
The expression \(\mathrm{Run}_O(Y)=e\) is not a primitive evidence atom.  It is a
macro for the expanded evidence term \(E_{O,e}\).  If this prohibition were
removed, any polynomial-time computation could be compressed into one uncharged
atom, and the Atomic Evidence Budget would be vacuous.
\end{remark}

\subsection{Derivative support of normalized evidence}
\label{subsec:derivative-support-normalized-evidence-section5}

\begin{definition}[Charged leaves]
\label{def:charged-leaves-section5}
For a normal form \(F=\CDENF(E\Rightarrow e)\), let
\[
  \Safe(F):=\{S(q,y):S(q,y)\text{ or }\neg S(q,y)
                  \text{ occurs in }F\},
\]
\[
  \Gauge(F):=\{G(u,\gamma):G(u,\gamma)\text{ or }\neg G(u,\gamma)
                    \text{ occurs in }F\}.
\]
The charged leaves of \(F\) are \(\Safe(F)\cup\Gauge(F)\).
\end{definition}

\begin{proposition}[Derivative support of a normalized term]
\label{prop:derivative-support-normalized-term-section5}
Let \(F=\CDENF(E\Rightarrow e)\).  For any prefix relation \(H_{j,h}\), the pair
mass separated by evaluating \(F\) is bounded by the sum of pair masses separated
by its charged leaves:
\[
\begin{aligned}
&\Gamma_j\bigl(
    H_{j,h}\cap
    \{F(\omega^0)\ne F(\omega^1)\}
  \bigr) \\
&\qquad\le
  \sum_{S\in\Safe(F)}
  \Gamma_j\bigl(
    H_{j,h}\cap\{S(\omega^0)\ne S(\omega^1)\}
  \bigr) \\
&\qquad\quad+
  \sum_{G\in\Gauge(F)}
  \Gamma_j\bigl(
    H_{j,h}\cap\{G(\omega^0)\ne G(\omega^1)\}
  \bigr).
\end{aligned}
\]
\end{proposition}

\begin{proof}
Within a neutral public-template fiber, neutral atoms have the same value on the
two endpoints of a coupled pair.  Therefore, if two endpoints agree on every
safe and gauge-bearing leaf occurring in the normal form, they agree on every
literal that can affect the branch and conclusion.  Thus any pair separated by
the normal form must be separated by at least one charged leaf.  Taking the union
bound over charged leaves gives the displayed inequality.
\end{proof}

\subsection{Interface exported to trace capture and the budget theorem}
\label{subsec:cdenf-exported-interface}

This section exports the following interface.

\begin{enumerate}[label=(\roman*)]
\item Every deterministic primitive observer output event has an evidence term
      \(E_{O,e}\).

\item Every such term has a normal form \(\CDENF(E_{O,e}\Rightarrow e)\).

\item Normalization preserves CD count pairs and hence preserves phase skew.

\item The only non-neutral normalized leaves are safe-buffer atoms and
      gauge-evidence atoms.

\item Any target-relevant quotient use is gauge-faithful: it contains raw
      gauge-bearing support and is counted by \(\rankG\).

\item Pairwise separation by a normalized term is covered by the union of
      pairwise separations by its charged safe and gauge leaves.
\end{enumerate}

Section~\ref{sec:trace} uses items (i)--(iii) to show that phase gap is carried
by atomic pair distinctions along the trace.  The Atomic Evidence Budget section
uses items (iv)--(vi) to charge those distinctions by safe-buffer leakage and
gauge rank.

\begin{remark}[Lean-facing skeleton]
\label{rem:lean-facing-cdenf-section5}
A Lean formalization can split this section into five finite modules:
\begin{enumerate}[label=(\arabic*)]
\item raw atoms, normalized atoms, signed literals, and finite support semantics;
\item normal branches, branch product, branch sum, and canonicalization;
\item the recursive normalizer \(\CDENF\) and semantics preservation;
\item gauge-faithfulness of the quotient expansion rule;
\item deterministic observer decision trees and the observer-to-evidence
      expansion theorem.
\end{enumerate}
The proof of confluence need not be the first formal target.  The recursive
normalizer and its semantics-preservation theorem are sufficient for the later
trace-capture and budget interfaces; confluence can then be recovered as a
corollary of the canonical normalizer.
\end{remark}

\section{CD Trace Capture}
\label{sec:trace}
\providecommand{\CDENF}{\operatorname{CDENF}}
\providecommand{\Safe}{\mathrm{Safe}}
\providecommand{\Gauge}{\mathrm{Gauge}}
\providecommand{\Val}{\operatorname{Val}}
\providecommand{\Sep}{\operatorname{Sep}}
\providecommand{\Trace}{\operatorname{Trace}}
\providecommand{\Leaves}{\operatorname{Leaves}}
\providecommand{\Charge}{\operatorname{Charge}}

This section proves the trace-capture theorem.  The theorem is the formal
replacement for the old switching/locality midpoint.  It says that if a
computation has target advantage, then message-opposite pairs must have been
separated somewhere in its evidence trace.  The statement does not ask the
observer to output a proof that its answer is correct.  It only charges the
actual pairwise distinctions made by the primitive observations on which the
observer's transcript depends.

The section is deliberately finite.  At fixed ensemble parameters, the world
space \(\Omega\), all transcript alphabets, all primitive output alphabets, and
all normal forms are finite.  This is the intended Lean-facing formulation.

\subsection{Primitive observations and adaptive observers}
\label{subsec:primitive-observers-section6}

We first make explicit the deterministic transcript model used in the derivative
identity.  The observer may perform arbitrary deterministic computation between
queries, but such computation only changes which primitive observation is queried
next.  It does not by itself distinguish two worlds that have the same transcript.

\begin{definition}[Primitive observation]
\label{def:primitive-observation-section6}
A primitive observation is a finite-valued map
\[
  E:\Omega\to \Val(E),
\]
where \(\Val(E)\) is a finite set.  Its separation event on
message-opposite pairs for target coordinate \(j\) is
\[
  \Sep_j(E)
  :=
  \{(\omega^0,\omega^1)\in\Omega_j^0\times\Omega_j^1:
      E(\omega^0)\ne E(\omega^1)\}.
\]
For a value \(y\in\Val(E)\), the associated Boolean result atom is the event
\[
  [E=y]:=\{\omega\in\Omega:E(\omega)=y\}.
\]
\end{definition}

\begin{definition}[Boolean atom derivative]
\label{def:boolean-atom-derivative-section6}
If \(L\subseteq\Omega\) is a Boolean atom or signed atom, define
\[
  \Sep_j(L)
  :=
  \{(\omega^0,\omega^1)\in\Omega_j^0\times\Omega_j^1:
      \mathbf 1_L(\omega^0)\ne \mathbf 1_L(\omega^1)\}.
\]
For a relation \(H\subseteq\Omega_j^0\times\Omega_j^1\), define
\[
  \partial_L w_j(H)
  :=
  \Gamma_j(H\cap \Sep_j(L)).
\]
For a signed atom \(\neg L\), this derivative is the same as for \(L\), since
\(\mathbf 1_{\neg L}(\omega^0)\ne\mathbf 1_{\neg L}(\omega^1)\) if and only if
\(\mathbf 1_L(\omega^0)\ne\mathbf 1_L(\omega^1)\).
\end{definition}

\begin{definition}[Deterministic primitive observer]
\label{def:primitive-observer-section6}
A deterministic primitive observer \(O\) of query depth \(Q\) consists of finite
transcript sets \(\mathcal T_0,\ldots,\mathcal T_Q\), transcript maps
\[
  T_r:\Omega\to\mathcal T_r,
  \qquad 0\le r\le Q,
\]
with \(T_0\) constant, and query policies
\[
  h\in\mathcal T_r \longmapsto E_{r,h},
  \qquad 0\le r<Q,
\]
where each \(E_{r,h}\) is a primitive observation.  If \(T_r(\omega)=h\), then
\[
  T_{r+1}(\omega)=(h,E_{r,h}(\omega))
\]
under a fixed injective encoding of the pair.  Thus \(T_{r+1}\) refines
\(T_r\).  After \(Q\) queries, the observer outputs
\[
  A_j(\omega)=g_j(T_Q(\omega))\in\bits
\]
for a deterministic postprocessor \(g_j:\mathcal T_Q\to\bits\).
\end{definition}

\begin{remark}[Target-indexed notation]
\label{rem:target-indexed-observer-section6}
A single polynomial-time algorithm may output many coordinates at once.  For
trace capture, fix one coordinate \(j\) and let \(g_j\) be the coordinate
postprocessor applied to the same terminal transcript.  The coordinate-sum form
at the end of this section then sums the resulting bounds over \(j\in S\).
\end{remark}

\begin{definition}[CD-primitive observer]
\label{def:cd-primitive-observer-section6}
A primitive observer is CD-primitive if its non-internal observations have already
been expanded through the primitive read interface of
Section~\ref{sec:cdenf}.  Thus every target-relevant Boolean observation in its
expanded trace is one of the normalized atom classes
\[
  N(s),\qquad S(q,y),\qquad G(u,\gamma),
\]
possibly signed.  A finite-valued read is represented either as the finite-valued
observation \(E\) or, equivalently, by the bounded family of result atoms
\([E=y]\).  The output alphabets of primitive reads are bounded by the ensemble
interface, so replacing a finite-valued read by its result atoms changes query
count only by a fixed multiplicative constant.
\end{definition}

\begin{lemma}[Program instrumentation]
\label{lem:program-instrumentation-section6}
Let \(O\) be a deterministic polynomial-time observer using the primitive read
interface of Section~\ref{sec:cdenf}.  After fixing the input length and the
polynomial time bound, the run of \(O\) on the finite world space \(\Omega\)
induces a deterministic primitive observer in the sense of
Definition~\ref{def:primitive-observer-section6}.  Moreover the associated
evidence terms \(E_{O,e}\) of
Theorem~\ref{thm:observer-to-evidence-expansion-section5} have terminal events
\[
  \mathsf{Sat}(E_{O,e})=\{\omega:O(Y(\omega))=e\}.
\]
\end{lemma}

\begin{proof}
At fixed input length, a polynomial-time computation has finitely many possible
configurations and makes at most finitely many primitive reads.  Collapse the
internal deterministic computation between two primitive reads into the query
policy mapping the current transcript to the next primitive observation.  The
next transcript is the old transcript together with the returned primitive value,
so the transcript maps satisfy the refinement condition.  The final machine state
determines a deterministic postprocessor of the terminal transcript.  The last
claim is exactly the observer-to-evidence expansion theorem from
Section~\ref{sec:cdenf}.
\end{proof}

\subsection{Atomic weakness derivatives}
\label{subsec:atomic-derivatives-section6}

\begin{definition}[Atomic weakness derivative]
\label{def:atomic-weakness-derivative-section6}
Let \(H\subseteq\Omega_j^0\times\Omega_j^1\) be a surviving pair relation, and
let \(E\) be a primitive observation.  Define
\[
  \partial_E w_j(H)
  :=
  \Gamma_j(H\cap\Sep_j(E))
  =
  \Gamma_j\left(
    H\cap
    \{(\omega^0,\omega^1):E(\omega^0)\ne E(\omega^1)\}
  \right).
\]
For a prefix \(h\in\mathcal T_r\), we write
\[
  \partial_{E_{r,h}}w_j(H_{j,h})
  :=
  \Gamma_j\left(
    H_{j,h}\cap
    \{E_{r,h}(\omega^0)\ne E_{r,h}(\omega^1)\}
  \right).
\]
\end{definition}

\begin{lemma}[Basic derivative laws]
\label{lem:basic-derivative-laws-section6}
For fixed \(j\), relation \(H\), and primitive observation \(E\):
\begin{enumerate}[label=(\roman*)]
\item \(0\le \partial_Ew_j(H)\le w_j(H)\);
\item if \(H\subseteq H'\), then
\[
  \partial_Ew_j(H)\le \partial_Ew_j(H');
\]
\item if \(H_1,\ldots,H_s\) are pairwise disjoint, then
\[
  \partial_Ew_j\left(\dot\bigcup_{a=1}^s H_a\right)
  =
  \sum_{a=1}^s \partial_Ew_j(H_a).
\]
\end{enumerate}
\end{lemma}

\begin{proof}
All three statements are immediate from the definition and from monotonicity and
finite additivity of the probability measure \(\Gamma_j\).
\end{proof}

\begin{lemma}[Finite-valued observations and result atoms]
\label{lem:finite-valued-result-atoms-section6}
Let \(E:\Omega\to\mathcal Y\) be a primitive observation with finite output set
\(\mathcal Y\).  For any surviving relation \(H\),
\[
  \sum_{y\in\mathcal Y}\partial_{[E=y]}w_j(H)
  =
  2\partial_Ew_j(H).
\]
In particular,
\[
  \partial_Ew_j(H)
  \le
  \sum_{y\in\mathcal Y}\partial_{[E=y]}w_j(H).
\]
\end{lemma}

\begin{proof}
Fix a pair \((\omega^0,\omega^1)\in H\).  If
\(E(\omega^0)=E(\omega^1)\), then no result atom \([E=y]\) has different truth
values on the two endpoints.  If \(E(\omega^0)\ne E(\omega^1)\), exactly two
result atoms differ: the atom for \(E(\omega^0)\) and the atom for
\(E(\omega^1)\).  Integrating this pointwise count with respect to
\(\Gamma_j\) on \(H\) gives the identity.
\end{proof}

\begin{remark}[Why the derivative is the right local quantity]
\label{rem:why-derivative-section6}
The value \(\partial_Ew_j(H)\) is not a mutual information and not an average
Bayes advantage.  It is the exact mass of still-surviving message-opposite pairs
that the next primitive observation separates.  This is the pairwise weakness
analogue of a discrete derivative: it measures the loss in the non-distinction
relation caused by one observation.
\end{remark}

\subsection{The derivative identity}
\label{subsec:derivative-identity-section6}

For a transcript prefix \(h\in\mathcal T_r\), recall from
Section~\ref{sec:pairwise-cd} that
\[
  H_{j,h}
  :=
  \{(\omega^0,\omega^1)\in\Omega_j^0\times\Omega_j^1:
       T_r(\omega^0)=T_r(\omega^1)=h\},
\]
and
\[
  H_{j,r}
  :=
  \{(\omega^0,\omega^1):T_r(\omega^0)=T_r(\omega^1)\}.
\]
Thus
\[
  H_{j,r}=\dot\bigcup_{h\in\mathcal T_r}H_{j,h}.
\]

\begin{lemma}[One-step survival inside a prefix]
\label{lem:one-step-survival-prefix-section6}
Let \(h\in\mathcal T_r\), with \(0\le r<Q\).  Then
\[
  H_{j,h}\cap H_{j,r+1}
  =
  H_{j,h}\cap
  \{(\omega^0,\omega^1):E_{r,h}(\omega^0)=E_{r,h}(\omega^1)\}.
\]
Consequently,
\[
  H_{j,h}\setminus H_{j,r+1}
  =
  H_{j,h}\cap
  \{(\omega^0,\omega^1):E_{r,h}(\omega^0)\ne E_{r,h}(\omega^1)\}.
\]
\end{lemma}

\begin{proof}
On \(H_{j,h}\), both endpoints have prefix transcript \(h\).  By the definition
of the adaptive transcript,
\[
  T_{r+1}(\omega^a)=(h,E_{r,h}(\omega^a)),
  \qquad a\in\{0,1\}.
\]
The two refined transcripts are equal if and only if the two returned primitive
values are equal.  This proves the first identity.  Taking complements inside
\(H_{j,h}\) gives the second.
\end{proof}

\begin{lemma}[One-step derivative identity]
\label{lem:one-step-derivative-identity-section6}
For every \(0\le r<Q\),
\[
  w_j(H_{j,r})-w_j(H_{j,r+1})
  =
  \sum_{h\in\mathcal T_r}
  \partial_{E_{r,h}}w_j(H_{j,h}).
\]
\end{lemma}

\begin{proof}
The sets \(H_{j,h}\), for \(h\in\mathcal T_r\), form a disjoint partition of
\(H_{j,r}\).  Since \(T_{r+1}\) refines \(T_r\), we have
\(H_{j,r+1}\subseteq H_{j,r}\).  Therefore
\[
  w_j(H_{j,r})-w_j(H_{j,r+1})
  =
  \Gamma_j(H_{j,r}\setminus H_{j,r+1}).
\]
Using the prefix partition,
\[
  H_{j,r}\setminus H_{j,r+1}
  =
  \dot\bigcup_{h\in\mathcal T_r}(H_{j,h}\setminus H_{j,r+1}).
\]
By Lemma~\ref{lem:one-step-survival-prefix-section6}, each summand is exactly
\[
  H_{j,h}\cap
  \{E_{r,h}(\omega^0)\ne E_{r,h}(\omega^1)\}.
\]
Taking \(\Gamma_j\)-mass and using finite additivity gives the displayed
identity.
\end{proof}

\begin{lemma}[Derivative identity]
\label{lem:derivative-identity-section6}
For a deterministic primitive observer of query depth \(Q\),
\[
  1-w_j(H_{j,Q})
  =
  \sum_{r=0}^{Q-1}\sum_{h\in\mathcal T_r}
  \partial_{E_{r,h}}w_j(H_{j,h}).
\]
More generally,
\[
  w_j(H_{j,r_0})-w_j(H_{j,r_1})
  =
  \sum_{r=r_0}^{r_1-1}\sum_{h\in\mathcal T_r}
  \partial_{E_{r,h}}w_j(H_{j,h})
\]
for every \(0\le r_0\le r_1\le Q\).
\end{lemma}

\begin{proof}
Sum Lemma~\ref{lem:one-step-derivative-identity-section6} over
\(r=r_0,\ldots,r_1-1\).  The left-hand side telescopes.  Taking
\(r_0=0\) and \(r_1=Q\) gives the first identity because \(T_0\) is constant,
so
\[
  H_{j,0}=\Omega_j^0\times\Omega_j^1
\]
and hence
\[
  w_j(H_{j,0})=\Gamma_j(\Omega_j^0\times\Omega_j^1)=1.
\]
\end{proof}

\subsection{Trace capture for terminal decisions}
\label{subsec:trace-capture-terminal-decisions}

The previous subsection is only an identity about transcripts.  The next result
connects it to target advantage by using the CD skew identity and static pairwise
capture from Section~\ref{sec:pairwise-cd}.

\begin{lemma}[Postprocessing creates no new pair distinctions]
\label{lem:postprocessing-no-new-distinctions-section6}
Let \(A_j=g_j(T_Q)\).  If
\((\omega^0,\omega^1)\in H_{j,Q}\), then
\[
  A_j(\omega^0)=A_j(\omega^1).
\]
Equivalently,
\[
  \{A_j(\omega^0)\ne A_j(\omega^1)\}
  \subseteq
  \Sep_j(T_Q).
\]
\end{lemma}

\begin{proof}
A pair in \(H_{j,Q}\) has the same terminal transcript on both endpoints.  Since
\(A_j\) is a deterministic function of that transcript, the output values are
equal.  The equivalent inclusion is just the contrapositive.
\end{proof}

\begin{theorem}[CD Trace Capture]
\label{thm:cd-trace-capture-section6}
For every deterministic primitive observer \(O\), every target coordinate \(j\),
and every output
\[
  A_j(\omega)=g_j(T_Q(\omega)),
\]
we have
\[
  \Gap_j(A_j)
  \le
  \frac12
  \sum_{r=0}^{Q-1}\sum_{h\in\mathcal T_r}
  \partial_{E_{r,h}}w_j(H_{j,h}).
\]
Equivalently, since \(\Gap_j(A_j)=\mathrm{Skew}_j(g_j(T_Q))\),
\[
  \mathrm{Skew}_j(g_j(T_Q))
  \le
  \frac12
  \sum_{r=0}^{Q-1}\sum_{h\in\mathcal T_r}
  \partial_{E_{r,h}}w_j(H_{j,h}).
\]
\end{theorem}

\begin{proof}
By the static pairwise capture proposition from
Section~\ref{sec:pairwise-cd}, applied to the terminal transcript \(T_Q\),
\[
  \Gap_j(A_j)
  \le
  \frac12\Gamma_j(\Sep_j(T_Q))
  =
  \frac12(1-w_j(H_{j,Q})).
\]
The derivative identity, Lemma~\ref{lem:derivative-identity-section6}, gives
\[
  1-w_j(H_{j,Q})
  =
  \sum_{r=0}^{Q-1}\sum_{h\in\mathcal T_r}
  \partial_{E_{r,h}}w_j(H_{j,h}).
\]
Combining the two displayed equations proves the theorem.  The skew formulation
uses Lemma~\ref{lem:phase-gap-cd-skew}.
\end{proof}

\begin{remark}[No correctness certificate is required]
\label{rem:no-correctness-certificate-section6}
The proof never asks the observer to justify the assertion \(A_j=B_j\).  It only
uses the fact that, if \(A_j\) has different behavior under the two target phases,
then the terminal transcript distributions under those phases have CD skew.  That
skew can be carried only by message-opposite pairs separated somewhere in the
trace.  The separated mass is exactly the telescoping derivative mass above.
\end{remark}

\subsection{Compatibility with CD-ENF normalization}
\label{subsec:trace-cdenf-compatibility-section6}

The theorem just proved is stated for primitive observations.  To feed the
Atomic Evidence Budget, we need the same statement with primitive observations
expanded and normalized into the atom classes of Section~\ref{sec:cdenf}.

\begin{definition}[Charged expansion of a primitive result]
\label{def:charged-expansion-primitive-result-section6}
Let $[E=y]$ be a Boolean result atom for a primitive observation. Let
\[
  F_{E,y} := \operatorname{CDENF}([E=y]\Rightarrow *)
\]
be its CD-ENF normal form, using the atom normalizer of
Section~\ref{sec:cdenf}. Define
\[
  \operatorname{Charge}(E,y) := \operatorname{Safe}(F_{E,y}) \cup \operatorname{Gauge}(F_{E,y}),
\]
where $\operatorname{Safe}$ and $\operatorname{Gauge}$ are the charged-leaf sets of
Definition~\ref{def:charged-leaves-section5}. For a finite-valued observation,
set
\[
  \operatorname{Charge}(E) := \bigcup_{y\in\operatorname{Val}(E)} \operatorname{Charge}(E,y).
\]
\end{definition}

\begin{lemma}[Primitive derivative covered by normalized leaves]
\label{lem:primitive-derivative-covered-by-normal-leaves-section6}
For every primitive observation \(E\) and every surviving relation \(H\),
\[
  \partial_Ew_j(H)
  \le
  \sum_{y\in\Val(E)}
  \left(
    \sum_{S\in\Safe(F_{E,y})}\partial_Sw_j(H)
    +
    \sum_{G\in\Gauge(F_{E,y})}\partial_Gw_j(H)
  \right).
\]
\end{lemma}

\begin{proof}
By Lemma~\ref{lem:finite-valued-result-atoms-section6},
\[
  \partial_Ew_j(H)
  \le
  \sum_{y\in\Val(E)}\partial_{[E=y]}w_j(H).
\]
For each \(y\), semantics preservation of CD-ENF identifies the event
\([E=y]\) with the support of \(F_{E,y}\).  The derivative-support proposition
from Section~\ref{sec:cdenf} then bounds separation by \(F_{E,y}\) by the union
of separations by its safe and gauge charged leaves.  Applying the union bound
and summing over \(y\) gives the displayed inequality.
\end{proof}

\begin{theorem}[Normalized CD Trace Capture]
\label{thm:normalized-cd-trace-capture-section6}
Let \(O\) be a deterministic observer using the primitive read interface, and let
\(A_j=g_j(T_Q)\).  Then
\[
\begin{aligned}
  \Gap_j(A_j)
  \le {}&
  \frac12
  \sum_{r=0}^{Q-1}\sum_{h\in\mathcal T_r}
  \sum_{y\in\Val(E_{r,h})}
  \Bigg(
    \sum_{S\in\Safe(F_{E_{r,h},y})}\partial_Sw_j(H_{j,h}) \\
  &\hspace{45mm}+
    \sum_{G\in\Gauge(F_{E_{r,h},y})}\partial_Gw_j(H_{j,h})
  \Bigg).
\end{aligned}
\]
In particular, after CD-ENF normalization, every charged contribution to target
advantage comes from a safe-buffer atom or a gauge-evidence atom.
\end{theorem}

\begin{proof}
Apply Theorem~\ref{thm:cd-trace-capture-section6} and then apply
Lemma~\ref{lem:primitive-derivative-covered-by-normal-leaves-section6} to each
summand \(\partial_{E_{r,h}}w_j(H_{j,h})\).
\end{proof}

\begin{corollary}[Already-normalized observer]
\label{cor:already-normalized-observer-section6}
If \(O\) is CD-primitive and every queried atom \(E_{r,h}\) is already a Boolean
normalized atom, then
\[
  \Gap_j(A_j)
  \le
  \frac12
  \sum_{r=0}^{Q-1}\sum_{h\in\mathcal T_r}
  \partial_{E_{r,h}}w_j(H_{j,h}),
\]
where every non-neutral \(E_{r,h}\) is either a safe-buffer atom or a
gauge-evidence atom, and neutral atoms are handled by the neutral-atom lemma of
Section~\ref{sec:aeb}.
\end{corollary}

\begin{proof}
This is Theorem~\ref{thm:cd-trace-capture-section6} with the atom
classification supplied by Corollary~\ref{cor:public-syntax-atom-completeness-section5}.
\end{proof}

\subsection{Coordinate-sum and shared-trace forms}
\label{subsec:coordinate-sum-trace-section6}

Later sections apply trace capture over a switched set of coordinates.  The next
corollaries record the forms used by the Atomic Evidence Budget.

\begin{corollary}[Coordinate-sum trace capture]
\label{cor:coordinate-sum-trace-capture-section6}
Let \(S\) be a finite set of switched coordinates.  For each \(j\in S\), let
\(O_j\) be a deterministic primitive observer with query depth \(Q_j\),
transcript sets \(\mathcal T_{j,r}\), prefix relations \(H_{j,h}\), and queried
observations \(E_{j,r,h}\).  Then
\[
  \sum_{j\in S}\Gap_j(A_j)
  \le
  \frac12
  \sum_{j\in S}
  \sum_{r=0}^{Q_j-1}
  \sum_{h\in\mathcal T_{j,r}}
  \partial_{E_{j,r,h}}w_j(H_{j,h}).
\]
\end{corollary}

\begin{proof}
Apply Theorem~\ref{thm:cd-trace-capture-section6} separately for each
\(j\in S\) and sum the inequalities.
\end{proof}

\begin{corollary}[Shared transcript, many coordinate postprocessors]
\label{cor:shared-transcript-many-postprocessors-section6}
Suppose one deterministic primitive observer produces a single terminal
transcript \(T_Q\), and for each \(j\in S\) the output coordinate is
\[
  A_j=g_j(T_Q).
\]
Let \(H_{j,h}\) be the surviving pair relation for target coordinate \(j\) and
prefix \(h\).  Then
\[
  \sum_{j\in S}\Gap_j(A_j)
  \le
  \frac12
  \sum_{j\in S}
  \sum_{r=0}^{Q-1}\sum_{h\in\mathcal T_r}
  \partial_{E_{r,h}}w_j(H_{j,h}).
\]
\end{corollary}

\begin{proof}
For each fixed \(j\), the same transcript process is a valid primitive observer
with postprocessor \(g_j\).  Apply Theorem~\ref{thm:cd-trace-capture-section6}
for each \(j\) and sum.
\end{proof}

\begin{corollary}[Normalized coordinate-sum form]
\label{cor:normalized-coordinate-sum-section6}
In the setting of Corollary~\ref{cor:coordinate-sum-trace-capture-section6}, if
all primitive reads are expanded by CD-ENF, then
\[
  \sum_{j\in S}\Gap_j(A_j)
\]
is bounded by one half of the corresponding sum of normalized safe-buffer and
gauge-evidence derivatives.  Explicitly, replace each raw summand
\(\partial_{E_{j,r,h}}w_j(H_{j,h})\) by the right-hand side of
Lemma~\ref{lem:primitive-derivative-covered-by-normal-leaves-section6}.
\end{corollary}

\begin{proof}
Sum Theorem~\ref{thm:normalized-cd-trace-capture-section6} over \(j\in S\).
\end{proof}

\subsection{Random coins and deterministic reductions}
\label{subsec:random-coins-section6}

The final proof enumerates deterministic programs.  The following observations
explain why allowing randomized observers would not change the trace-capture
interface.

\begin{lemma}[Averaged trace capture for randomized observers]
\label{lem:randomized-trace-capture-section6}
Let \(O\) be a randomized primitive observer whose random coins \(\rho\) are
independent of the world.  For each fixed \(\rho\), let \(A_j^\rho\) be the
deterministic output and let \(D_j(\rho)\) be the derivative sum from
Theorem~\ref{thm:cd-trace-capture-section6}.  Then
\[
  \Gap_j(A_j)
  \le
  \mathbb E_\rho[\Gap_j(A_j^\rho)]
  \le
  \frac12\mathbb E_\rho[D_j(\rho)],
\]
where \(A_j\) denotes the randomized output distribution.
\end{lemma}

\begin{proof}
The phase gap is the absolute value of a difference of two phase-conditioned
output probabilities divided by two.  For a randomized observer these
probabilities are averages over \(\rho\).  By convexity of absolute value, the
phase gap of the average is at most the average of the phase gaps.  Applying the
deterministic trace-capture theorem for each fixed \(\rho\) gives the second
inequality.
\end{proof}

\begin{corollary}[Coin fixing]
\label{cor:coin-fixing-section6}
If a randomized observer has phase gap at least \(\epsilon\), then some fixed
choice of its coins has deterministic phase gap at least \(\epsilon\).  Hence it is
sufficient for the lower-bound argument to treat deterministic observers.
\end{corollary}

\begin{proof}
If every coin choice had deterministic phase gap below \(\epsilon\), the average
phase gap would be below \(\epsilon\), contradicting
Lemma~\ref{lem:randomized-trace-capture-section6}.
\end{proof}

\subsection{Interface exported to the Atomic Evidence Budget}
\label{subsec:trace-exported-interface}

This section exports the following facts.

\begin{enumerate}[label=(\roman*)]
\item Every deterministic polynomial-time observer using the primitive read
      interface has a finite adaptive transcript process.

\item For each target coordinate \(j\), the loss of surviving message-opposite
      pair mass across one query is exactly the atomic derivative
      \(\partial_{E_{r,h}}w_j(H_{j,h})\).

\item The total separated pair mass telescopes:
\[
  1-w_j(H_{j,Q})
  =
  \sum_{r=0}^{Q-1}\sum_{h\in\mathcal T_r}
  \partial_{E_{r,h}}w_j(H_{j,h}).
\]

\item CD skew, equivalently phase gap, is bounded by one half of this telescoping
      derivative sum:
\[
  \Gap_j(A_j)
  \le
  \frac12
  \sum_{r=0}^{Q-1}\sum_{h\in\mathcal T_r}
  \partial_{E_{r,h}}w_j(H_{j,h}).
\]

\item After CD-ENF normalization, every non-neutral charged derivative is a
      safe-buffer derivative or a gauge-evidence derivative.

\item The coordinate-sum form gives
\[
  \sum_{j\in S}\Gap_j(A_j)
  \le
  \frac12
  \sum_{j\in S}\sum_{r,h}
  \partial_{E_{j,r,h}}w_j(H_{j,h}),
\]
with the obvious normalized safe/gauge replacement.
\end{enumerate}

The next section bounds the normalized derivative sum.  Neutral atoms contribute
zero; safe-buffer atoms contribute at most the max-qSSM leakage; and gauge atoms
are charged by hidden-gauge rank.

\begin{remark}[Lean-facing skeleton]
\label{rem:lean-facing-trace-section6}
A direct formalization can use finite types for worlds, transcript alphabets, and
primitive output alphabets.  The derivative identity is a finite partition theorem:
inside each prefix cell, the next refinement removes exactly the pairs on which
the next observation has unequal values.  CD Trace Capture is then a two-line
composition of the static coupling bound from Section~\ref{sec:pairwise-cd} with
this telescoping identity.  The normalized version imports only the
semantics-preservation and derivative-support lemmas from Section~\ref{sec:cdenf}.
\end{remark}

\section{Atomic Evidence Budget}
\label{sec:aeb}
\providecommand{\Prb}{\mathbb P}
\providecommand{\Exp}{\mathbb E}
\providecommand{\bits}{\{0,1\}}
\providecommand{\Gap}{\operatorname{Gap}}
\providecommand{\TV}{\operatorname{TV}}
\providecommand{\CDENF}{\operatorname{CDENF}}
\providecommand{\Val}{\operatorname{Val}}
\providecommand{\Sep}{\operatorname{Sep}}
\providecommand{\Safe}{\mathrm{Safe}}
\providecommand{\Gauge}{\mathrm{Gauge}}
\providecommand{\rankG}{\operatorname{rank}_G}
\providecommand{\suppG}{\operatorname{supp}_G}
\providecommand{\FreshG}{\operatorname{Fresh}_G}
\providecommand{\Occ}{\operatorname{Occ}}

This section proves the atomic budget theorem used by the rest of the lower
bound.  Section~\ref{sec:trace} reduced target advantage to a sum of atomic
weakness derivatives.  Section~\ref{sec:cdenf} reduced every target-relevant
normalized leaf to one of three forms: neutral syntax, a legal safe-buffer
observation, or hidden gauge evidence.  We now bound the total derivative mass
of these leaves.

The slogan is
\[
  \hbox{phase gap}
  \le
  {1\over 2}\bigl(
     \hbox{safe-buffer leakage}
     +
     \hbox{hidden-gauge rank cost}
  \bigr).
\]
Neutral syntax contributes no derivative.  Safe-buffer atoms have small
derivative because legal safe probes can be coupled so that the two endpoints of
a message-opposite pair disagree only with probability
\(\varepsilon_{\mathrm{step}}(m)\).  Gauge atoms may have large derivative, but
every such derivative is charged to fresh independent hidden gauge support.  A
short observer cannot correctly determine too much such hidden support, by a
program-counting entropy bound.

All statements are finite.  At fixed parameters, the world space, transcript
alphabets, normal forms, and gauge-coordinate sets are finite.  This is the
intended Lean-facing formulation.  The section is also conditional in the usual
way: the same estimates hold after conditioning on any admissible neutral public
prefix and on earlier switched blocks, provided the conditioned law still
satisfies the ensemble interface of Section~\ref{sec:ensemble}.

\subsection{Normalized charged occurrences}
\label{subsec:charged-occurrences-section7}

The normalized coordinate-sum theorem from Section~\ref{sec:trace} gives a
finite list of charged leaf occurrences.  We first put this list into a ledger.

\begin{definition}[Charged occurrence]
\label{def:charged-occurrence-section7}
A charged occurrence in the normalized coordinate-sum trace is a tuple
\[
  \iota=(j,r,h,y,L),
\]
where \(j\in S\) is a switched target coordinate,
\(h\in\mathcal T_{j,r}\) is a transcript prefix,
\(y\in\Val(E_{j,r,h})\) is a primitive read value, and \(L\) is a signed leaf in
\[
  \CDENF([E_{j,r,h}=y]\Rightarrow *).
\]
Its surviving relation is
\[
  H_\iota:=H_{j,h},
\]
and its derivative is
\[
  \partial_\iota
  :=
  \partial_L w_j(H_{j,h}).
\]
According to the class of \(L\), the occurrence is called neutral, safe, or gauge.
\end{definition}

\begin{definition}[Neutral, safe, and gauge ledgers]
\label{def:charged-ledgers-section7}
For a deterministic observer \(O\) and switched set \(S\), let
\[
  \mathcal I_N(O,S),
  \qquad
  \mathcal I_S(O,S),
  \qquad
  \mathcal I_G(O,S)
\]
be the finite sets of neutral, safe-buffer, and gauge-evidence occurrences in
the normalized coordinate-sum trace.  Define
\[
  Q_{\mathrm{safe}}(O,S):=|\mathcal I_S(O,S)|.
\]
In the main theorem we write \(Q_{\mathrm{tot}}\) for an upper bound on
\(Q_{\mathrm{safe}}(O,S)\).  This is the number of target-indexed safe charged
occurrences after bounded CD-ENF expansion, not merely the number of machine
read instructions.  Since primitive read alphabets and normal-form leaf arities
are bounded by the ensemble interface, this convention changes query counts only
by a fixed constant factor.
\end{definition}

\begin{lemma}[Normalized trace capture as ledger bound]
\label{lem:normalized-trace-capture-ledger-section7}
For every deterministic observer and switched set \(S\),
\[
\begin{aligned}
  \sum_{j\in S}\Gap_j(A_j)
  \le {1\over 2}
  \left(
    \sum_{\iota\in\mathcal I_N(O,S)}\partial_\iota
    +
    \sum_{\iota\in\mathcal I_S(O,S)}\partial_\iota
    +
    \sum_{\iota\in\mathcal I_G(O,S)}\partial_\iota
  \right).
\end{aligned}
\]
\end{lemma}

\begin{proof}
This is Corollary~\ref{cor:normalized-coordinate-sum-section6}, with the finite
list of charged leaves partitioned by the atom classes supplied by
Corollary~\ref{cor:public-syntax-atom-completeness-section5}.
\end{proof}

\subsection{Neutral atoms}
\label{subsec:neutral-atoms-section7}

Neutral atoms record public template syntax, deterministic type tags, guarded
control flow, illegal-read symbols, and other facts whose values are
message-neutral inside a coupled message-opposite fiber.

\begin{definition}[Pair-neutral atom]
\label{def:pair-neutral-atom-section7}
A normalized neutral atom \(N(s)\) is pair-neutral for target \(j\) and coupling
\(\Gamma_j\) if
\[
  N(s)(\omega^0)=N(s)(\omega^1)
\]
for every pair \((\omega^0,\omega^1)\) in the support of \(\Gamma_j\).  The same
condition is required after any admissible neutral public conditioning.
\end{definition}

\begin{hypothesis}[Neutral coherence]
\label{hyp:neutral-coherence-section7}
Every normalized neutral atom supplied by the ensemble and the CD-ENF
normalizer is pair-neutral for every switched target coordinate.
\end{hypothesis}

\begin{lemma}[Neutral atoms]
\label{lem:neutral-atoms-section7}
If \(L=N(s)\) or \(L=\neg N(s)\), then for every surviving relation
\(H\subseteq\Omega_j^0\times\Omega_j^1\),
\[
  \partial_Lw_j(H)=0.
\]
Consequently,
\[
  \sum_{\iota\in\mathcal I_N(O,S)}\partial_\iota=0.
\]
\end{lemma}

\begin{proof}
By neutral coherence, the indicator of \(N(s)\) has the same value on both
endpoints of every coupled pair.  The same is true of its negation.  Hence the
separation event \(\Sep_j(L)\) has \(\Gamma_j\)-measure zero, and
\(\partial_Lw_j(H)=\Gamma_j(H\cap\Sep_j(L))=0\).  Summing over neutral
occurrences gives the second identity.
\end{proof}

\subsection{Safe-buffer atoms}
\label{subsec:safe-buffer-atoms-section7}

The max-qSSM theorem of Section~\ref{sec:ensemble} is a statement about laws of
legal safe-buffer probe values.  For pairwise weakness we use it in coupling
form: conditioned on a surviving pair relation, the two endpoint values of a
legal safe probe can be coupled to disagree with probability at most
\(\varepsilon_{\mathrm{step}}(m)\).

\begin{definition}[Derivative step leakage]
\label{def:derivative-step-leakage-section7}
Let
\[
  \bar\varepsilon_{\mathrm{step}}(m)
  :=
  C_\rho\rho^{R_{\mathrm{safe}}}+\tau_{\mathrm{geo}}(m)
\]
be the max-divergence leakage from
Theorem~\ref{thm:soft-buffer-max-qssm-section4}.  Define
\[
  \varepsilon_{\mathrm{step}}(m)
  :=
  \exp(\bar\varepsilon_{\mathrm{step}}(m))-1.
\]
For small \(\bar\varepsilon_{\mathrm{step}}\), this is
\[
  \varepsilon_{\mathrm{step}}(m)
  =O(C_\rho\rho^{R_{\mathrm{safe}}}+\tau_{\mathrm{geo}}(m)).
\]
We use \(\varepsilon_{\mathrm{step}}\) for this derivative mismatch bound.
\end{definition}

\begin{definition}[Safe-compatible coupling]
\label{def:safe-compatible-coupling-section7}
The message-opposite couplings \((\Gamma_j)_{j\in S}\) are safe-compatible if,
for every switched target \(j\), every admissible prefix relation \(H_{j,h}\),
and every legal safe-buffer probe \(q\) at that prefix,
\[
  \Gamma_j\left(
    H_{j,h}\cap
    \{S(q)(\omega^0)\ne S(q)(\omega^1)\}
  \right)
  \le
  \varepsilon_{\mathrm{step}}(m)w_j(H_{j,h}).
\]
Here \(S(q)(\omega)\) denotes the full finite value returned by the safe probe in
world \(\omega\).
\end{definition}

\begin{lemma}[Max-qSSM gives safe-compatible couplings]
\label{lem:max-qssm-safe-compatible-section7}
Assume the soft-buffer max-qSSM theorem of Section~\ref{sec:ensemble}.  The
message-opposite couplings may be chosen, trace-adaptively and without changing
their endpoint marginals, so that they are safe-compatible.
\end{lemma}

\begin{proof}
Fix a prefix cell and disintegrate the two endpoint laws of the legal safe-probe
value conditioned on that cell.  The max-qSSM estimate gives a common reference
law \(R\) such that the endpoint laws \(P_0\) and \(P_1\) are each within
max-divergence \(\bar\varepsilon_{\mathrm{step}}(m)\) of \(R\).  Hence
\[
  \TV(P_0,P_1)
  \le
  \exp(\bar\varepsilon_{\mathrm{step}}(m))-1
  =
  \varepsilon_{\mathrm{step}}(m),
\]
after increasing constants if both endpoint laws are compared through the common
reference law.  A maximal coupling of \(P_0\) and \(P_1\) disagrees with
probability exactly \(\TV(P_0,P_1)\).  Since the transcript tree is finite, these
conditional maximal couplings can be glued over all prefix cells and all safe
steps, yielding a global trace-adapted coupling with the required marginals.
\end{proof}

\begin{lemma}[Safe atoms]
\label{lem:safe-atoms-section7}
Assume safe-compatible couplings.  If \(L=S(q,y)\) or \(L=\neg S(q,y)\) is a
legal safe-buffer result atom at prefix \(h\), then
\[
  \partial_Lw_j(H_{j,h})
  \le
  \varepsilon_{\mathrm{step}}(m)w_j(H_{j,h})
  \le
  \varepsilon_{\mathrm{step}}(m).
\]
Consequently,
\[
  \sum_{\iota\in\mathcal I_S(O,S)}\partial_\iota
  \le
  Q_{\mathrm{safe}}(O,S)\varepsilon_{\mathrm{step}}(m).
\]
\end{lemma}

\begin{proof}
If the Boolean result atom \([S(q)=y]\) has different truth values on the two
endpoints of a pair, then the full safe-probe values \(S(q)(\omega^0)\) and
\(S(q)(\omega^1)\) are different.  Thus
\[
  \Sep_j([S(q)=y])
  \subseteq
  \{S(q)(\omega^0)\ne S(q)(\omega^1)\}.
\]
The same inclusion holds for the negated atom.  Safe compatibility gives the
first displayed inequality.  Since \(w_j(H_{j,h})\le 1\), the second follows.
Summing over safe occurrences gives the final bound.
\end{proof}

\subsection{Gauge atoms and bounded incidence}
\label{subsec:gauge-atoms-section7}

Gauge atoms are not required to have small derivative individually.  A fresh
hidden gauge coordinate can genuinely separate message-opposite worlds.  The
budget condition is that one fresh independent gauge coordinate can help only
boundedly many switched target coordinates.

\begin{definition}[Gauge support]
\label{def:gauge-support-section7}
If \(L=G(u,\gamma)\) or \(L=\neg G(u,\gamma)\), define
\[
  \suppG(L):=\suppG(u)\subseteq V,
\]
where \(V\) is the hidden gauge-coordinate set.  For a transcript prefix \(h\),
\(\suppG(h)\) is the set of hidden gauge coordinates whose values have already
been fixed by prior gauge-evidence leaves on that branch.
\end{definition}

\begin{definition}[Fresh support]
\label{def:fresh-support-section7}
For a gauge occurrence \(\iota=(j,r,h,y,L)\), define
\[
  \FreshG(\iota):=\suppG(L)\setminus\suppG(h).
\]
A gauge-rank envelope for the normalized trace is any finite set
\[
  h^\Delta\subseteq V
\]
containing every fresh coordinate charged by a gauge occurrence.  Its rank is
\[
  \rankG(h^\Delta):=|h^\Delta|.
\]
The minimal envelope is the union of the fresh supports \(\FreshG(\iota)\) over
all gauge occurrences, with duplicates counted once.
\end{definition}

\begin{lemma}[Repeated gauge evidence is not fresh]
\label{lem:repeated-gauge-evidence-section7}
If \(\FreshG(\iota)=\emptyset\), then
\[
  \partial_\iota=0.
\]
\end{lemma}

\begin{proof}
The prefix relation \(H_\iota\) contains only pairs whose endpoints have the same
normalized transcript prefix.  If all coordinates supporting the gauge leaf have
already been fixed in that prefix, then the two endpoints agree on the truth value
of the leaf.  Hence no pair in \(H_\iota\) lies in the separation event of the
leaf.
\end{proof}

\begin{hypothesis}[Bounded gauge incidence]
\label{hyp:bounded-gauge-incidence-section7}
There is a constant \(\Delta_G=O(1)\) such that for every finite set
\(U\subseteq V\),
\[
  \sum_{j\in S}
  \mathbf 1[U\text{ can separate }B_j\text{-opposite pairs}]
  \le
  \Delta_G |U|.
\]
Equivalently, the target-to-gauge incidence graph has bounded right degree: a
fresh independent gauge coordinate can be target-relevant for only \(O(1)\)
switched coordinates.
\end{hypothesis}

\begin{lemma}[Gauge atoms are charged by fresh incidence]
\label{lem:gauge-fresh-incidence-section7}
Let \(\iota=(j,r,h,y,L)\) be a gauge occurrence.  Then, after summing over all
switched target coordinates for the same prefix and gauge leaf,
\[
  \sum_{j\in S}\partial_Lw_j(H_{j,h})
  \le
  \Delta_G|\suppG(L)\setminus\suppG(h)|.
\]
\end{lemma}

\begin{proof}
If \(\suppG(L)\setminus\suppG(h)=\emptyset\), the claim follows from
Lemma~\ref{lem:repeated-gauge-evidence-section7}.  Otherwise, for a fixed
\(j\), the derivative is at most \(1\).  It can be nonzero only if the fresh
support can separate \(B_j\)-opposite pairs.  Therefore
\[
  \sum_{j\in S}\partial_Lw_j(H_{j,h})
  \le
  \sum_{j\in S}
  \mathbf 1[\suppG(L)\setminus\suppG(h)
             \text{ can separate }B_j\text{-opposite pairs}],
\]
and bounded incidence gives the displayed bound.
\end{proof}

\begin{proposition}[Aggregate gauge-rank charge]
\label{prop:aggregate-gauge-rank-charge-section7}
Let \(h^\Delta\) be a gauge-rank envelope for the normalized trace.  Then
\[
  \sum_{\iota\in\mathcal I_G(O,S)}\partial_\iota
  \le
  \Delta_G\rankG(h^\Delta).
\]
\end{proposition}

\begin{proof}
It is useful to count by first fresh coordinate rather than by leaf occurrence.
Fix a hidden gauge coordinate \(v\) and target \(j\).  Along any one transcript
branch, once \(v\) has appeared in the prefix, later leaves supported only on
\(v\) are not fresh and have zero derivative by
Lemma~\ref{lem:repeated-gauge-evidence-section7}.  Across different transcript
prefixes, the surviving relations for the first exposure of \(v\) are disjoint
subsets of the pair space, because a coupled pair has a unique transcript prefix
at each depth and a unique first depth at which \(v\) is exposed.  Hence the
total derivative mass charged to the pair \((j,v)\) is at most \(1\).

Therefore the aggregate gauge derivative is bounded by the number of target--gauge
incidences \((j,v)\) with \(v\in h^\Delta\) and \(v\) capable of separating
\(B_j\)-opposite pairs.  By bounded gauge incidence, the number of such target
coordinates for each \(v\) is at most \(\Delta_G\).  Summing over
\(v\in h^\Delta\) gives
\[
  \sum_{\iota\in\mathcal I_G(O,S)}\partial_\iota
  \le
  \Delta_G |h^\Delta|
  =
  \Delta_G\rankG(h^\Delta).
\]
\end{proof}

\subsection{Safe leakage in adaptive transcripts}
\label{subsec:adaptive-safe-leakage-section7}

The gauge entropy theorem below allows the safe transcript to be weakly dependent
on hidden gauge values.  The only effect is a likelihood-ratio factor equal to the
product of the stepwise leakage factors.

\begin{definition}[Cumulative safe leakage]
\label{def:cumulative-safe-leakage-section7}
If the safe steps of an adaptive trace have derivative or max-divergence leakage
parameters \(\varepsilon_r\), define
\[
  \varepsilon_{\le Q}:=\sum_{r\le Q}\varepsilon_r.
\]
In the uniform setting one may take
\[
  \varepsilon_{\le Q}
  \le
  Q_{\mathrm{tot}}\varepsilon_{\mathrm{step}}(m).
\]
\end{definition}

\begin{lemma}[Adaptive max-divergence composition]
\label{lem:adaptive-max-divergence-composition-section7}
Let \(G_U\) be any finite set of unsupported hidden gauge coordinates.  If at
safe step \(r\) the next safe-read kernel conditioned on \(G_U\) differs from the
unconditioned kernel by max-divergence at most \(\varepsilon_r\), then the whole
safe transcript satisfies
\[
  D_\infty(
    \mathcal L(T_{\mathrm{safe}}\mid G_U)
    \|\mathcal L(T_{\mathrm{safe}})
  )
  \le
  \varepsilon_{\le Q}.
\]
Equivalently, for every safe transcript value \(\tau\),
\[
  \Prb[T_{\mathrm{safe}}=\tau\mid G_U]
  \le
  e^{\varepsilon_{\le Q}}
  \Prb[T_{\mathrm{safe}}=\tau].
\]
\end{lemma}

\begin{proof}
Write the probability of an adaptive safe transcript as a product of conditional
kernels.  At step \(r\), the likelihood ratio of the conditioned kernel to the
unconditioned kernel is at most \(e^{\varepsilon_r}\).  Multiplying these ratios
along the transcript gives at most \(\exp(\sum_r\varepsilon_r)\).  Taking logs
proves the max-divergence bound.
\end{proof}

\subsection{Gauge-rank entropy}
\label{subsec:gauge-rank-entropy-section7}

The next theorem is the resource-bounded entropy estimate for hidden gauge
evidence.  It is the same counting principle used in the final
\(K_{\mathrm{poly}}\) lower bound: a short program cannot, with high probability,
produce too many correct independent hidden gauge literals.  Safe-buffer leakage
only multiplies the guessing probability by \(e^{\varepsilon_{\le Q}}\).

\begin{definition}[Gauge claim]
\label{def:gauge-claim-section7}
A gauge claim is a partial assignment
\[
  c:U\to\bits,
  \qquad U\subseteq V.
\]
Its rank is \(\rankG(c):=|U|\).  The claim is true in hidden gauge vector \(g\) if
\[
  g|_U=c.
\]
A realized gauge ledger \(h^\Delta\) determines a gauge claim \(c(h^\Delta)\),
namely the set of distinct gauge-coordinate values asserted by its normalized
gauge leaves.
\end{definition}

\begin{lemma}[One-program gauge guessing bound]
\label{lem:one-program-gauge-guessing-section7}
Assume the hidden-gauge product law of Section~\ref{sec:ensemble} and cumulative
safe leakage at most \(\varepsilon_{\le Q}\).  Fix a deterministic observer whose
realized gauge ledger determines a true gauge claim \(c(h^\Delta)\).  Then, for
all \(R\ge 0\),
\[
  \Prb[\rankG(h^\Delta)
       \ge R]
  \le
  e^{\varepsilon_{\le Q}}2^{-R}.
\]
\end{lemma}

\begin{proof}
Condition on a safe transcript value \(\tau\).  The observer is deterministic, so
\(\tau\), together with neutral public information, determines a partial gauge
claim \(c_\tau\).  If \(\rankG(c_\tau)\ge R\), the hidden-gauge product law gives
\[
  \Prb[g|_{\operatorname{dom}(c_\tau)}=c_\tau]
  \le
  2^{-R}.
\]
The safe transcript may depend weakly on \(g\), but by
Lemma~\ref{lem:adaptive-max-divergence-composition-section7}, conditioning on the
safe transcript changes likelihoods by at most \(e^{\varepsilon_{\le Q}}\).  Summing
over the finite set of possible \(\tau\) gives the displayed bound.
\end{proof}

\begin{theorem}[Gauge-rank entropy]
\label{thm:gauge-rank-entropy-section7}
Let \(\mathcal P_L\) be a prefix-free family of deterministic observers with
program length at most \(L\).  Then
\[
  \Prb\left[
    \exists P\in\mathcal P_L:
    \rankG(h^\Delta_P)\ge R
  \right]
  \le
  2^{L+O(1)}2^{-R}e^{\varepsilon_{\le Q}}.
\]
Consequently, for every \(A>0\), with probability at least \(1-m^{-A}\), every
program \(P\) of length at most \(L\) satisfies
\[
  \rankG(h^\Delta_P)
  \le
  L
  +
  {\varepsilon_{\le Q}\over \ln 2}
  +
  A\log_2 m
  +
  O(1).
\]
In particular, the same bound holds for a fixed observer \(P\) of description
length \(|P|\) after replacing \(L\) by \(|P|\).
\end{theorem}

\begin{proof}
For one fixed observer, Lemma~\ref{lem:one-program-gauge-guessing-section7}
gives the bound \(e^{\varepsilon_{\le Q}}2^{-R}\).  There are at most
\(2^{L+O(1)}\) prefix-free programs of length at most \(L\).  The union bound over
these programs gives the first display.  Choosing
\[
  R=
  L
  +
  {\varepsilon_{\le Q}\over\ln 2}
  +
  A\log_2 m
  +
  O(1)
\]
with the additive constant large enough to absorb prefix-coding overhead makes
the right-hand side at most \(m^{-A}\).
\end{proof}

\begin{remark}[Why public gauge-looking syntax would break the theorem]
\label{rem:public-syntax-breaks-gauge-entropy-section7}
Gauge-rank entropy applies only to hidden gauge-bearing evidence.  If a
purported gauge atom were literally a public bit of the SAT instance, then the
observer could read it for free and the product-law guessing bound would be
false.  This is why the public syntax discipline, atom completeness, and
gauge-faithfulness lemmas are structural assumptions of the ensemble, not
cosmetic restrictions.
\end{remark}

\subsection{The Atomic Evidence Budget}
\label{subsec:atomic-evidence-budget-section7}

We now combine normalized trace capture, the atom-level bounds, and gauge-rank
entropy.

\begin{theorem}[Atomic Evidence Budget]
\label{thm:atomic-evidence-budget-section7}
Let \(O\) be a deterministic polynomial-time observer represented by a program
\(P\) of length \(|P|\).  Let \(S\) be a switched set of target coordinates.
Suppose the normalized coordinate-sum trace has at most
\[
  Q_{\mathrm{tot}}:=Q_{\mathrm{safe}}(O,S)
\]
safe charged occurrences, and suppose its cumulative safe transcript leakage is
\(\varepsilon_{\le Q}\).  Under neutral coherence, safe compatibility, and bounded
gauge incidence, with probability at least \(1-m^{-A}\),
\[
\begin{aligned}
  \sum_{j\in S}\Gap_j(A_j)
  \le
  {1\over 2}
  \Bigg[
    Q_{\mathrm{tot}}\varepsilon_{\mathrm{step}}(m)
    +
    \Delta_G
    \left(
      |P|
      +
      {\varepsilon_{\le Q}\over \ln 2}
      +
      A\log_2 m
      +
      O(1)
    \right)
  \Bigg].
\end{aligned}
\]
The same estimate holds simultaneously for all programs of length at most \(L\)
after replacing \(|P|\) by \(L\).
\end{theorem}

\begin{proof}
By Lemma~\ref{lem:normalized-trace-capture-ledger-section7},
\[
\begin{aligned}
  \sum_{j\in S}\Gap_j(A_j)
  \le {1\over 2}
  \left(
    \sum_{\iota\in\mathcal I_N}\partial_\iota
    +
    \sum_{\iota\in\mathcal I_S}\partial_\iota
    +
    \sum_{\iota\in\mathcal I_G}\partial_\iota
  \right).
\end{aligned}
\]
The neutral sum is zero by Lemma~\ref{lem:neutral-atoms-section7}.  The safe sum
is at most
\[
  Q_{\mathrm{tot}}\varepsilon_{\mathrm{step}}(m)
\]
by Lemma~\ref{lem:safe-atoms-section7}.  The gauge sum is at most
\[
  \Delta_G\rankG(h^\Delta)
\]
by Proposition~\ref{prop:aggregate-gauge-rank-charge-section7}.  Finally,
Theorem~\ref{thm:gauge-rank-entropy-section7} gives, with probability at least
\(1-m^{-A}\),
\[
  \rankG(h^\Delta)
  \le
  |P|+{\varepsilon_{\le Q}\over\ln 2}+A\log_2 m+O(1).
\]
Substituting these estimates into the trace-capture inequality gives the claim.
\end{proof}

\begin{corollary}[Expectation form with a bad-event term]
\label{cor:unconditional-aeb-section7}
In the setting of Theorem~\ref{thm:atomic-evidence-budget-section7}, if one does
not condition on the good gauge-rank event, then the expected coordinate-sum gap
satisfies
\[
\begin{aligned}
  \Exp\left[\sum_{j\in S}\Gap_j(A_j)\right]
  \le
  {1\over 2}
  \Bigg[
    Q_{\mathrm{tot}}\varepsilon_{\mathrm{step}}(m)
    +
    \Delta_G
    \left(
      |P|
      +
      {\varepsilon_{\le Q}\over \ln 2}
      +
      A\log_2 m
      +
      O(1)
    \right)
  \Bigg]
  + |S|m^{-A}.
\end{aligned}
\]
\end{corollary}

\begin{proof}
On the good event, use Theorem~\ref{thm:atomic-evidence-budget-section7}.  On the
bad event, use the trivial bound \(\Gap_j(A_j)\le 1/2\le 1\).  The bad event has
probability at most \(m^{-A}\), so its contribution to the coordinate sum is at
most \(|S|m^{-A}\).
\end{proof}

\subsection{Parameter consequences}
\label{subsec:aeb-parameter-consequences-section7}

The budget theorem becomes useful after the safety radius is chosen so that
polynomially many safe observations have negligible total leakage.

\begin{corollary}[Polynomial safe budget]
\label{cor:polynomial-safe-budget-section7}
Fix \(D,A_0>0\).  Suppose
\[
  Q_{\mathrm{tot}}
  \le
  m^D
\]
and
\[
  \varepsilon_{\mathrm{step}}(m)
  \le
  C_\rho\rho^{R_{\mathrm{safe}}}+\tau_{\mathrm{geo}}(m),
  \qquad 0<\rho<1.
\]
If
\[
  R_{\mathrm{safe}}
  \ge
  {D+A_0+2\over |\log \rho|}\log m
\]
and \(\tau_{\mathrm{geo}}(m)\le m^{-D-A_0-2}\), then
\[
  Q_{\mathrm{tot}}\varepsilon_{\mathrm{step}}(m)=O(m^{-A_0-1})=o(1).
\]
If also the number of safe steps contributing to \(\varepsilon_{\le Q}\) is at
most \(m^D\), then
\[
  \varepsilon_{\le Q}=o(1).
\]
\end{corollary}

\begin{proof}
The radius choice gives \(\rho^{R_{\mathrm{safe}}}\le m^{-D-A_0-2}\).  Thus
\(\varepsilon_{\mathrm{step}}(m)=O(m^{-D-A_0-2})\).  Multiplying by
\(Q_{\mathrm{tot}}\le m^D\) gives \(O(m^{-A_0-2})\), which is stronger than the
displayed bound.  The claim about \(\varepsilon_{\le Q}\) follows from
Lemma~\ref{lem:adaptive-max-divergence-composition-section7}.
\end{proof}

\begin{corollary}[Averaged phase gap for a fixed observer]
\label{cor:averaged-phase-gap-fixed-section7}
Let \(|S|\ge \gamma t\) with \(t=\Theta(m)\).  For a fixed polynomial-time
observer \(P\), assume \(Q_{\mathrm{tot}}\le m^D\) and choose
\(R_{\mathrm{safe}}=c(D)\log m\) so that
\[
  Q_{\mathrm{tot}}\varepsilon_{\mathrm{step}}(m)=o(1),
  \qquad
  \varepsilon_{\le Q}=o(1).
\]
Then, with probability at least \(1-m^{-A}\),
\[
  {1\over |S|}\sum_{j\in S}\Gap_j(A_j)=o(1).
\]
The same conclusion holds in expectation with an additional \(m^{-A}\) term.
\end{corollary}

\begin{proof}
Apply Theorem~\ref{thm:atomic-evidence-budget-section7}.  For a fixed observer,
\(|P|=O(1)\).  The right-hand side is
\[
  {1\over 2}
  \left[o(1)+\Delta_G(O(1)+A\log_2m+o(1))\right].
\]
Dividing by \(|S|\ge\gamma t=\Theta(m)\) gives \(o(1)\).
\end{proof}

\begin{corollary}[Averaged phase gap for linear-length observers]
\label{cor:averaged-phase-gap-linear-section7}
Let \(|S|\ge\gamma t\) with \(t=\Theta(m)\).  Suppose
\(|P|\le\delta t\),
\[
  Q_{\mathrm{tot}}\varepsilon_{\mathrm{step}}(m)=o(t),
  \qquad
  \varepsilon_{\le Q}=o(t).
\]
Then, with probability at least \(1-m^{-A}\),
\[
  {1\over |S|}\sum_{j\in S}\Gap_j(A_j)
  \le
  {\Delta_G\over 2}{\delta\over\gamma}+o(1).
\]
In particular, choosing \(\delta>0\) sufficiently small makes the average phase
gap as small as required by the product small-success theorem.
\end{corollary}

\begin{proof}
Divide the Atomic Evidence Budget by \(|S|\).  The safe term contributes
\(o(t)/|S|=o(1)\).  The cumulative leakage term contributes \(o(t)/|S|=o(1)\).
The program-length term contributes at most
\[
  {1\over 2}\Delta_G{|P|\over |S|}
  \le
  {\Delta_G\over 2}{\delta t\over\gamma t}
  =
  {\Delta_G\over 2}{\delta\over\gamma}.
\]
The remaining \(A\log_2m+O(1)\) term divided by \(|S|=\Theta(m)\) is \(o(1)\).
\end{proof}

\subsection{Interface exported to the product theorem}
\label{subsec:aeb-exported-interface}

This section exports the following interface.

\begin{enumerate}[label=(\roman*)]
\item Normalized trace capture gives
\[
  \sum_{j\in S}\Gap_j(A_j)
  \le
  {1\over 2}
  \sum_{\hbox{charged occurrences }\iota}\partial_\iota.
\]

\item Neutral occurrences contribute zero.

\item Safe-buffer occurrences contribute at most
\[
  Q_{\mathrm{tot}}\varepsilon_{\mathrm{step}}(m).
\]

\item Gauge-evidence occurrences contribute at most
\[
  \Delta_G\rankG(h^\Delta).
\]

\item Gauge-rank entropy gives, with probability at least \(1-m^{-A}\),
\[
  \rankG(h^\Delta)
  \le
  |P|+{\varepsilon_{\le Q}\over\ln 2}+A\log_2m+O(1).
\]

\item Hence, for fixed polynomial-time observers,
\[
  {1\over |S|}\sum_{j\in S}\Gap_j(A_j)=o(1),
\]
provided the safe radius makes polynomially many safe observations have total
leakage \(o(1)\).  For observers of length at most \(\delta t\), the average phase
gap is at most \(O(\delta)+o(1)\).
\end{enumerate}

Section~\ref{sec:mixing} supplies the near-random visible baseline.  Section 9
combines that baseline with the averaged phase-gap estimate above to obtain
product small-success over \(\Omega(t)\) switched coordinates.

\begin{remark}[Lean-facing skeleton]
\label{rem:lean-facing-aeb-section7}
A Lean formalization can split this section into four finite modules:
\begin{enumerate}[label=(\arabic*)]
\item a ledger module: finite charged occurrences and the normalized trace-capture
      sum;
\item an atom-bound module: neutral zero, safe mismatch from max-qSSM, and gauge
      incidence;
\item a gauge entropy module: hidden fair bits, max-divergence leakage, and the
      prefix-program union bound;
\item a parameter module: polynomial safe budgets and averaged phase-gap
      corollaries.
\end{enumerate}
The only analytic input is the soft-buffer max-qSSM estimate imported from
Section~\ref{sec:ensemble}; the rest is finite counting, finite additivity, and
union bounds.
\end{remark}

\section{Boundary-Law Mixing}
\label{sec:mixing}
\providecommand{\Prb}{\mathbb P}
\providecommand{\Exp}{\mathbb E}
\providecommand{\bits}{\{0,1\}}
\providecommand{\TV}{\operatorname{TV}}
\providecommand{\Law}{\mathcal L}
\providecommand{\Gap}{\operatorname{Gap}}
\providecommand{\supp}{\operatorname{supp}}

This section proves the visible-surface baseline used by the product lower
bound.  The Atomic Evidence Budget of Section~\ref{sec:aeb} says that an
actual computation can gain only a small amount of target-relevant phase gap by
creating charged pairwise distinctions.  Boundary-law mixing is a different
statement.  It says that the designated pivot-visible summary itself is nearly
uninformative about each switched target bit.

The theorem is independent of CD trace capture.  Its proof uses only the
locally mixing part of the ensemble: the target phase can influence a pivot
summary only through a depth-
\(L\) buffer, and messages through that buffer contract with depth.  Once the
posterior of the target phase given the exact boundary field is close to
\(1/2\), every coarsening of that boundary field, including the dithered
quantized pivot summary \(Z_{L,j}\), is also close to chance by data processing.

All statements are finite.  The language of conditional expectations and
sigma-fields may be read as shorthand for finite partitions and finite sums.
This is the intended Lean-facing interpretation.

\subsection{Pivot summaries}
\label{subsec:pivot-summaries-section8}

Fix a switched target coordinate \(j\).  Write
\[
  B_j:=\ell_j(M)\in\bits
\]
for the associated message bit.  The ensemble supplies a designated pivot
region for \(j\), together with a depth parameter
\[
  L=L(m)=c_L\log m .
\]
The pivot region is chosen inside the locally mixing buffer/core, at distance at
least \(L\) from every hidden locked or gauge source that can carry the value of
\(B_j\).

\begin{definition}[Pivot ball and pivot cut]
\label{def:pivot-ball-cut-section8}
For target coordinate \(j\), let \(\mathbb B_{L,j}\) be the depth-
\(L\) pivot ball in the public buffer/core geometry, and let
\(\partial\mathbb B_{L,j}\) be its outer cut.  The exact boundary-message field is

a finite random object
\[
  R_{L,j}
\]
containing the messages entering \(\mathbb B_{L,j}\) through
\(\partial\mathbb B_{L,j}\), together with the finite public pivot type.  Let
\[
  \mathcal B_{L,j}:=\sigma(R_{L,j})
\]
be the finite sigma-field generated by this exact boundary-message field.
\end{definition}

The exact boundary-message field is not the object used by later predictors.
The predictor is given only a finite, dithered, quantized summary.

\begin{definition}[Dithered quantized pivot summary]
\label{def:dithered-pivot-summary-section8}
Let \(\delta_m>0\) be a rational mesh size.  A dithered quantizer is a Markov
kernel
\[
  Q_{\delta_m}:\operatorname{Range}(R_{L,j})
    \leadsto \mathcal Z_{L,m},
\]
where \(\mathcal Z_{L,m}\) is a finite set.  The pivot summary is
\[
  Z_{L,j}:=Q_{\delta_m}(R_{L,j}).
\]
The dither randomness is independent of \((B_j,R_{L,j})\).  Equivalently, one may
adjoin a finite public dither seed \(U_j\) and set
\[
  Z_{L,j}=(U_j,Q_{\delta_m}^{U_j}(R_{L,j})).
\]
In either presentation, \(Z_{L,j}\) is generated from \(\mathcal B_{L,j}\) and
independent dither randomness only.
\end{definition}

\begin{remark}[Quantization is not an evidence source]
\label{rem:quantization-not-evidence-section8}
Quantization is included only to keep the visible surface finite and to avoid
boundary artifacts in later finite-class or Lean formalizations.  Since the
quantizer is a postprocessing of the exact boundary field, it cannot increase the
ability to predict \(B_j\).  The baseline theorem is therefore proved for
arbitrary finite postprocessings of \(R_{L,j}\), not only for the particular
quantizer above.
\end{remark}

\begin{definition}[Admissible conditioning]
\label{def:admissible-conditioning-section8}
An admissible conditioning field for coordinate \(j\) is a finite sigma-field
\(\mathcal C\) generated by public neutral data, earlier switched blocks,
external bookkeeping, and observer randomness already fixed before the
\(j\)-th block is tested, but not by the hidden phase \(B_j\) or by charged
non-neutral evidence from the \(j\)-th pivot core.  We require
\[
  \Prb[B_j=0\mid\mathcal C]
  =
  \Prb[B_j=1\mid\mathcal C]
  ={1\over 2}
\]
whenever the conditional probabilities are defined.
\end{definition}

In the product argument of Section~\ref{sec:product-small-success}, the field
\(\mathcal C\) will be the history \(\mathcal F_{\ell-1}\) generated by the
previous switched targets, by the fixed observer, and by all public data outside
the current hidden phase.  Block independence and phase balance make this field
admissible.

\subsection{Boundary posterior contraction}
\label{subsec:boundary-posterior-contraction-section8}

The local mixing input is most naturally stated as a posterior contraction
estimate.  It says that even the exact depth-
\(L\) boundary-message field changes the posterior of the target bit by only
\(m^{-\Omega(1)}\).

\begin{hypothesis}[Boundary posterior contraction]
\label{hyp:boundary-posterior-contraction-section8}
There are constants
\[
  C_{\mathrm{mix}}<\infty,
  \qquad
  0<\rho_{\mathrm{mix}}<1,
\]
and a finite-geometry error \(\tau_{\mathrm{tree}}(m)=m^{-\Omega(1)}\) such that
for every switched coordinate \(j\) and every admissible conditioning field
\(\mathcal C\),
\[
  \Exp\left[
    \left|
      \Prb[B_j=1\mid \mathcal C\vee\mathcal B_{L,j}]
      -{1\over 2}
    \right|
    \middle|\mathcal C
  \right]
  \le
  C_{\mathrm{mix}}\rho_{\mathrm{mix}}^L+\tau_{\mathrm{tree}}(m).
\]
Define
\[
  \varepsilon_{\mathrm{mix}}(m,L)
  :=
  C_{\mathrm{mix}}\rho_{\mathrm{mix}}^L+\tau_{\mathrm{tree}}(m).
\]
\end{hypothesis}

The next proposition records one standard way in which
Hypothesis~\ref{hyp:boundary-posterior-contraction-section8} is verified from a
message-recursion contraction theorem.  Later formalization may either prove
this proposition from a concrete Dobrushin/log-likelihood estimate or take the
posterior contraction hypothesis as the imported interface of the locally mixing
core.

\begin{definition}[Contractive message recursion]
\label{def:contractive-message-recursion-section8}
A depth-
\(L\) pivot core has a contractive message recursion if there is a finite or
compact metric message space \((\mathsf{Msg},d_{\mathsf{Msg}})\), local update
maps for every finite gadget type, and constants
\(C_{\mathsf{Msg}}<\infty\), \(0<\rho_{\mathsf{Msg}}<1\), such that changing an
admissible boundary condition at the cut changes the root message by at most
\[
  C_{\mathsf{Msg}}\rho_{\mathsf{Msg}}^L
\]
on the tree-like/good-geometry event.  The probability of the bad-geometry event
is at most \(\tau_{\mathrm{tree}}(m)\).  The target posterior is obtained from the
root message by a fixed \(K_{\mathrm{root}}\)-Lipschitz readout map
\[
  \mathrm{Bel}:\mathsf{Msg}\to[0,1].
\]
\end{definition}

\begin{proposition}[Message contraction implies posterior contraction]
\label{prop:message-contraction-implies-posterior-section8}
Assume the pivot core satisfies
Definition~\ref{def:contractive-message-recursion-section8}, and assume that
with free/balanced boundary condition the readout posterior is \(1/2\).  Then
Hypothesis~\ref{hyp:boundary-posterior-contraction-section8} holds with
\[
  C_{\mathrm{mix}}=K_{\mathrm{root}}C_{\mathsf{Msg}},
  \qquad
  \rho_{\mathrm{mix}}=\rho_{\mathsf{Msg}},
\]
up to increasing \(\tau_{\mathrm{tree}}(m)\) by a fixed constant factor.
\end{proposition}

\begin{proof}
Fix an atom of the admissible conditioning field \(\mathcal C\).  On the good
geometry event, the effect of changing the hidden phase or the exterior boundary
condition can reach the pivot only through the depth-
\(L\) boundary recursion.  By contractivity, the resulting root message differs
from the free/balanced root message by at most
\(C_{\mathsf{Msg}}\rho_{\mathsf{Msg}}^L\).  Applying the
\(K_{\mathrm{root}}\)-Lipschitz readout map changes the posterior by at most
\(K_{\mathrm{root}}C_{\mathsf{Msg}}\rho_{\mathsf{Msg}}^L\).  On the bad-geometry
event we use only the trivial bound \(|p-1/2|\le 1\).  Averaging over the atom of
\(\mathcal C\) gives the stated estimate.  Since the argument is uniform over
atoms, it holds conditionally on \(\mathcal C\).
\end{proof}

\subsection{Bayes advantage on a finite visible surface}
\label{subsec:bayes-visible-surface-section8}

The next elementary lemma is the only statistical fact needed to convert
posterior contraction into a predictor bound.

\begin{lemma}[Bayes success for a balanced bit]
\label{lem:bayes-success-balanced-section8}
Let \(B\in\bits\) satisfy
\(\Prb[B=0]=\Prb[B=1]=1/2\), and let \(Z\) be a finite random variable.  Then
\[
  \sup_{h:\operatorname{Range}(Z)\to\bits}
  \Prb[h(Z)=B]
  =
  {1\over 2}
  +
  \Exp\left[
    \left|\Prb[B=1\mid Z]-{1\over 2}\right|
  \right].
\]
Equivalently, if \(\nu_b=\Law(Z\mid B=b)\), then
\[
  \sup_h\Prb[h(Z)=B]
  =
  {1\over 2}+{1\over 2}\TV(\nu_1,\nu_0).
\]
\end{lemma}

\begin{proof}
Condition on the value \(Z=z\).  Write
\[
  p_z:=\Prb[B=1\mid Z=z].
\]
The best value of \(h(z)\) is \(1\) when \(p_z\ge 1/2\) and \(0\) otherwise,
with conditional success
\[
  \max\{p_z,1-p_z\}
  =
  {1\over 2}+\left|p_z-{1\over 2}\right|.
\]
Averaging over \(z\) gives the first identity.  The total-variation identity is
the standard formula for binary Bayes testing under equal priors.
\end{proof}

\begin{lemma}[Postprocessing cannot increase balanced prediction advantage]
\label{lem:postprocessing-boundary-section8}
Let \(B\in\bits\) be balanced, let \(\mathcal B\) be a finite sigma-field, and
let \(Z\) be generated from \(\mathcal B\) by a finite Markov kernel independent
of \(B\) conditional on \(\mathcal B\).  Then
\[
  \sup_h\Prb[h(Z)=B]
  \le
  {1\over 2}
  +
  \Exp\left[
    \left|
      \Prb[B=1\mid\mathcal B]-{1\over 2}
    \right|
  \right].
\]
The same statement holds after conditioning on any atom of an auxiliary finite
sigma-field \(\mathcal C\).
\end{lemma}

\begin{proof}
Let \(\mathcal Z:=\sigma(Z)\).  Since \(Z\) is obtained from \(\mathcal B\) and
independent dither only,
\[
  \Prb[B=1\mid\mathcal Z]
  =
  \Exp[\Prb[B=1\mid\mathcal B]\mid\mathcal Z].
\]
By Jensen's inequality,
\[
  \Exp\left[
    \left|\Prb[B=1\mid\mathcal Z]-{1\over 2}\right|
  \right]
  \le
  \Exp\left[
    \left|\Prb[B=1\mid\mathcal B]-{1\over 2}\right|
  \right].
\]
Now apply Lemma~\ref{lem:bayes-success-balanced-section8}.  For the conditional
version, apply the same argument inside each positive-probability atom of
\(\mathcal C\).
\end{proof}

\subsection{Boundary-law mixing theorem}
\label{subsec:boundary-law-theorem-section8}

\begin{theorem}[Boundary-law mixing]
\label{thm:boundary-law-mixing-section8}
Assume Hypothesis~\ref{hyp:boundary-posterior-contraction-section8}.  Let
\(Z_{L,j}\) be any dithered quantized pivot summary generated from the exact
boundary-message field \(R_{L,j}\).  Then for every admissible conditioning
field \(\mathcal C\) and every predictor
\[
  h:\mathcal Z_{L,m}\to\bits,
\]
we have, almost surely on \(\mathcal C\),
\[
  \Prb[h(Z_{L,j})=B_j\mid\mathcal C]
  \le
  {1\over 2}+\varepsilon_{\mathrm{mix}}(m,L).
\]
Equivalently,
\[
  \TV\left(
    \Law(Z_{L,j}\mid B_j=1,\mathcal C),
    \Law(Z_{L,j}\mid B_j=0,\mathcal C)
  \right)
  \le
  2\varepsilon_{\mathrm{mix}}(m,L).
\]
\end{theorem}

\begin{proof}
Fix a positive-probability atom of \(\mathcal C\).  By admissibility,
\(B_j\) is balanced on this atom.  The pivot summary \(Z_{L,j}\) is a finite
postprocessing of the exact boundary-message field \(\mathcal B_{L,j}\), with
only independent dither added.  Lemma~\ref{lem:postprocessing-boundary-section8}
therefore gives
\[
  \sup_h\Prb[h(Z_{L,j})=B_j\mid\mathcal C]
  \le
  {1\over 2}
  +
  \Exp\left[
    \left|
      \Prb[B_j=1\mid \mathcal C\vee\mathcal B_{L,j}]
      -{1\over 2}
    \right|
    \middle|\mathcal C
  \right].
\]
The last conditional expectation is at most
\(\varepsilon_{\mathrm{mix}}(m,L)\) by
Hypothesis~\ref{hyp:boundary-posterior-contraction-section8}.  This proves the
predictor bound.  The total-variation form follows from the second identity in
Lemma~\ref{lem:bayes-success-balanced-section8}.
\end{proof}

\begin{corollary}[Polynomial-depth parameter choice]
\label{cor:polynomial-depth-mixing-section8}
Let
\[
  L=c_L\log m
\]
and assume \(\tau_{\mathrm{tree}}(m)=m^{-a_{\mathrm{tree}}}\) up to constants.  If
\(c_L>0\) is fixed, then
\[
  \varepsilon_{\mathrm{mix}}(m,L)
  =
  C_{\mathrm{mix}}m^{-c_L|\log\rho_{\mathrm{mix}}|}
  +m^{-a_{\mathrm{tree}}+O(1)}.
\]
In particular, for every fixed target exponent \(A>0\), choosing \(c_L\) large
enough gives
\[
  \varepsilon_{\mathrm{mix}}(m,L)
  \le
  m^{-A}
\]
for all sufficiently large \(m\), up to the finite-geometry exponent available in
\(\tau_{\mathrm{tree}}\).  In any case,
\[
  \varepsilon_{\mathrm{mix}}(m,L)=m^{-\Omega(1)}.
\]
\end{corollary}

\begin{proof}
Since \(0<\rho_{\mathrm{mix}}<1\),
\[
  \rho_{\mathrm{mix}}^{c_L\log m}
  =
  m^{-c_L|\log\rho_{\mathrm{mix}}|}
\]
when logarithms are taken in the same fixed base.  The displayed estimates
follow directly from the definition of
\(\varepsilon_{\mathrm{mix}}(m,L)\).
\end{proof}

\subsection{Conditional form for the switched sequence}
\label{subsec:conditional-switched-mixing-section8}

The product theorem in Section~\ref{sec:product-small-success} uses an ordered
switched set.  We record the exact conditional form here.

\begin{definition}[Switched history]
\label{def:switched-history-section8}
Let
\[
  S=\{j_1,\ldots,j_s\}
\]
be an ordered switched set.  The history before coordinate \(j_\ell\) is the
finite sigma-field
\[
  \mathcal F_{\ell-1}
  :=
  \sigma\bigl(
    Y_{\mathrm{out}},
    A_{j_1},B_{j_1},Z_{L,j_1},
    \ldots,
    A_{j_{\ell-1}},B_{j_{\ell-1}},Z_{L,j_{\ell-1}}
  \bigr),
\]
where \(Y_{\mathrm{out}}\) denotes the fixed public data and observer-side
bookkeeping not containing charged evidence from the current hidden phase.  In
the actual product argument, \(A_{j_r}\) is the observer's output for the
previous switched coordinate.
\end{definition}

\begin{hypothesis}[Admissible switched histories]
\label{hyp:admissible-switched-histories-section8}
For the switched ordering used in the lower bound, each
\(\mathcal F_{\ell-1}\) is admissible for coordinate \(j_\ell\) in the sense of
Definition~\ref{def:admissible-conditioning-section8}.
\end{hypothesis}

\begin{corollary}[Conditional boundary-law mixing]
\label{cor:conditional-boundary-law-mixing-section8}
Assume Hypothesis~\ref{hyp:admissible-switched-histories-section8}.  Then for
every \(\ell\) and every pivot-visible predictor
\(h:\mathcal Z_{L,m}\to\bits\),
\[
  \Prb[h(Z_{L,j_\ell})=B_{j_\ell}\mid\mathcal F_{\ell-1}]
  \le
  {1\over 2}+\varepsilon_{\mathrm{mix}}(m,L)
\]
almost surely.
Consequently,
\[
  \sup_h
  \Prb[h(Z_{L,j_\ell})=B_{j_\ell}\mid\mathcal F_{\ell-1}]
  \le
  {1\over 2}+\varepsilon_{\mathrm{mix}}(m,L).
\]
\end{corollary}

\begin{proof}
Apply Theorem~\ref{thm:boundary-law-mixing-section8} with
\(\mathcal C=\mathcal F_{\ell-1}\).
\end{proof}

\begin{corollary}[Averaged visible baseline]
\label{cor:averaged-visible-baseline-section8}
Under the same hypotheses,
\[
  {1\over |S|}
  \sum_{j\in S}
  \sup_h
  \Prb[h(Z_{L,j})=B_j\mid\mathcal F_{j-1}]
  \le
  {1\over 2}+\varepsilon_{\mathrm{mix}}(m,L),
\]
where \(\mathcal F_{j-1}\) denotes the switched history before \(j\) in the
chosen ordering.
\end{corollary}

\begin{proof}
Average the pointwise conditional estimate of
Corollary~\ref{cor:conditional-boundary-law-mixing-section8} over
\(j\in S\).
\end{proof}

\subsection{Interface exported to the product theorem}
\label{subsec:mixing-exported-interface-section8}

The rest of the proof uses only the following exported statement:
\[
  \boxed{
  \sup_h
  \Prb[h(Z_{L,j})=B_j\mid\mathcal F_{j-1}]
  \le
  {1\over 2}+\varepsilon_{\mathrm{mix}}(m,L)
  }
\]
for every switched coordinate \(j\), with
\[
  \varepsilon_{\mathrm{mix}}(m,L)=m^{-\Omega(1)}
\]
under the logarithmic-depth parameter choice.  Section~\ref{sec:product-small-success}
combines this visible baseline with the Atomic Evidence Budget.  The latter
says that actual computation has only negligible excess over the best predictor
using \(Z_{L,j}\).  Boundary-law mixing says that even the best such predictor
has success only \(1/2+o(1)\).

\begin{remark}[Lean-facing finite skeleton]
\label{rem:lean-facing-section8}
A Lean formalization can split this section into four finite modules:
\begin{enumerate}[label=(\roman*)]
\item finite conditional probabilities for a balanced bit and a finite feature;
\item Bayes success equals posterior deviation and equals half total variation;
\item data processing for finite Markov postprocessings of a finite partition;
\item the imported boundary-posterior contraction interface, followed by the
      conditional mixing theorem.
\end{enumerate}
No SAT-specific syntax enters these finite probability lemmas.  The only
ensemble-specific formal obligation is the contraction estimate of
Hypothesis~\ref{hyp:boundary-posterior-contraction-section8}, or equivalently a
concrete proof of Proposition~\ref{prop:message-contraction-implies-posterior-section8}
for the chosen buffer/core family.
\end{remark}

\section{ACCEI/PNLD and Product Small-Success}
\label{sec:product-small-success}
\providecommand{\Prb}{\mathbb P}
\providecommand{\Exp}{\mathbb E}
\providecommand{\bits}{\{0,1\}}
\providecommand{\Gap}{\operatorname{Gap}}
\providecommand{\Excess}{\operatorname{Excess}}
\providecommand{\Base}{\operatorname{Base}}
\providecommand{\rankG}{\operatorname{rank}_G}

This section combines the two midpoint estimates proved in
Sections~\ref{sec:aeb} and~\ref{sec:mixing}.  The Atomic Evidence Budget says
that any target-relevant computational advantage must be paid for by safe-buffer
leakage or hidden-gauge rank.  Boundary-law mixing says that the pivot-visible
summary itself predicts each switched target bit only near randomly.  Together
they imply exponentially small probability of predicting all switched coordinates
correctly.

The product step needs a conditional, sequential formulation.  A mere
unconditional average of one-coordinate success probabilities is not enough for
product decay.  We therefore formulate ACCEI/PNLD along the ordered switched
history and use a stopped-observer version of the gauge-rank tail so that the
bad event is exponentially small in \(t\).

\subsection{Ordered switched coordinates and histories}
\label{subsec:section9-histories}

Let
\[
  S=(j_1,\ldots,j_s)
\]
be an ordered switched set, with
\[
  s=|S|\ge \gamma t,
  \qquad
  t=\Theta(m).
\]
For notational economy write
\[
  B_\ell:=B_{j_\ell}=\ell_{j_\ell}(M),
  \qquad
  Z_\ell:=Z_{L,j_\ell},
  \qquad
  A_\ell:=A_{j_\ell}.
\]
The event that the observer is correct at stage \(\ell\) is
\[
  E_\ell:=\{A_\ell=B_\ell\}.
\]
The tuple-success event is
\[
  E_S:=\bigcap_{\ell=1}^s E_\ell.
\]

Let \(\mathcal F_{\ell-1}\) be the switched history before coordinate
\(j_\ell\), as in Definition~\ref{def:switched-history-section8}.  Thus
\(\mathcal F_{\ell-1}\) contains the fixed public bookkeeping and the previously
processed switched coordinates.  Let
\[
  \mathcal G_\ell
  :=
  \mathcal F_{\ell-1}\vee\sigma(Z_\ell)
\]
be the pivot-augmented history.  ACCEI measures what the residual exterior can
add after \(Z_\ell\) has been fixed.

\begin{hypothesis}[Admissible product histories]
\label{hyp:admissible-product-histories-section9}
For every \(\ell\), the history \(\mathcal F_{\ell-1}\) is admissible for
coordinate \(j_\ell\) in the sense of
Definition~\ref{def:admissible-conditioning-section8}.  In particular,
\[
  \Prb[B_\ell=0\mid\mathcal F_{\ell-1}]
  =
  \Prb[B_\ell=1\mid\mathcal F_{\ell-1}]
  ={1\over 2}
\]
whenever the conditional probabilities are defined.  The pivot-augmented
history \(\mathcal G_\ell\) is also admissible for the residual exterior trace of
coordinate \(j_\ell\).
\end{hypothesis}

The bit \(B_\ell\) need not remain balanced after conditioning on
\(Z_\ell\).  Boundary-law mixing controls its posterior given \(Z_\ell\), and the
Bayes-vs-test lemma below is stated for an arbitrary posterior.

\subsection{Conditional phase gap and exterior excess}
\label{subsec:conditional-gap-excess-section9}

\begin{definition}[Pivot-fiber phase gap]
\label{def:pivot-fiber-gap-section9}
For a Boolean output \(A_\ell\), define
\[
\begin{aligned}
  \Gap_\ell^Z(A_\ell)
  := {1\over 2}
  \left|
    \Prb[A_\ell=1\mid B_\ell=1,\mathcal G_\ell]
    -
    \Prb[A_\ell=1\mid B_\ell=0,\mathcal G_\ell]
  \right|.
\end{aligned}
\]
This is a \(\mathcal G_\ell\)-measurable random variable.  On atoms where one
of the two conditioned phases has probability zero, define the corresponding
conditional probability arbitrarily; such atoms do not affect the inequalities
below.
\end{definition}

\begin{definition}[Visible Bayes baseline]
\label{def:visible-bayes-baseline-section9}
The best pivot-visible conditional success before coordinate \(j_\ell\) is
\[
  \Base_\ell
  :=
  \sup_{h:\operatorname{Range}(Z_\ell)\to\bits}
  \Prb[h(Z_\ell)=B_\ell\mid\mathcal F_{\ell-1}].
\]
Equivalently,
\[
  \Base_\ell
  =
  \Exp\left[
    \max\left\{      \Prb[B_\ell=1\mid\mathcal G_\ell],
      \Prb[B_\ell=0\mid\mathcal G_\ell]
    \right\}
    \middle|\mathcal F_{\ell-1}
  \right].
\]
\end{definition}

\begin{definition}[Exterior excess]
\label{def:exterior-excess-section9}
The signed exterior excess of \(A_\ell\) is
\[
  \Excess_\ell(A)
  :=
  \Prb[A_\ell=B_\ell\mid\mathcal F_{\ell-1}]
  -
  \Base_\ell.
\]
The positive exterior excess is
\[
  \Excess_\ell^+(A):=\max\{0,\Excess_\ell(A)\}.
\]
For an ordered switched set define
\[
  \overline{\Excess}_S^+(A)
  :=
  {1\over s}\sum_{\ell=1}^s \Excess_\ell^+(A).
\]
\end{definition}

\subsection{Bayes-vs-test bridge}
\label{subsec:bayes-vs-test-section9}

\begin{lemma}[Pointwise Bayes-vs-test inequality]
\label{lem:pointwise-bayes-vs-test-section9}
Let \(A,B\in\bits\) and let \(\mathcal G\) be a finite conditioning field.  On
a fixed atom of \(\mathcal G\), write
\[
  p:=\Prb[B=1\mid\mathcal G],
\]
\[
  a:=\Prb[A=1\mid B=1,\mathcal G],
  \qquad
  b:=\Prb[A=1\mid B=0,\mathcal G].
\]
Then
\[
  \Prb[A=B\mid\mathcal G]
  \le
  \max\{p,1-p\}+{1\over 2}|a-b|.
\]
\end{lemma}

\begin{proof}
On the atom of \(\mathcal G\),
\[
  \Prb[A=B\mid\mathcal G]=pa+(1-p)(1-b).
\]
Assume first that \(p\ge 1/2\).  The Bayes success using only
\(\mathcal G\) is \(p\).  If \(a\le b\), then
\[
  pa+(1-p)(1-b)
  \le
  pb+(1-p)(1-b)
  \le p,
\]
because \(p\ge1-p\).  If \(a>b\), set \(d=a-b\).  Since \(b\le1-d\),
\[
\begin{aligned}
  pa+(1-p)(1-b)-p
  &=(1-p)(1-b)-p(1-a) \\
  &\le (1-p)d
  \le {d\over2}.
\end{aligned}
\]
Thus the desired inequality holds when \(p\ge1/2\).  The case
\(p\le1/2\) is symmetric after exchanging the two phases.
\end{proof}

\begin{proposition}[PNLD from pivot-fiber phase gap]
\label{prop:pnld-from-gap-section9}
For every switched coordinate \(j_\ell\),
\[
  \Prb[A_\ell=B_\ell\mid\mathcal F_{\ell-1}]
  \le
  \Base_\ell
  +
  \Exp[\Gap_\ell^Z(A_\ell)\mid\mathcal F_{\ell-1}].
\]
Consequently,
\[
  \Excess_\ell^+(A)
  \le
  \Exp[\Gap_\ell^Z(A_\ell)\mid\mathcal F_{\ell-1}].
\]
\end{proposition}

\begin{proof}
Apply Lemma~\ref{lem:pointwise-bayes-vs-test-section9} on each atom of
\(\mathcal G_\ell=\mathcal F_{\ell-1}\vee\sigma(Z_\ell)\).  The term
\(
  \max\{\Prb[B_\ell=1\mid\mathcal G_\ell],
          \Prb[B_\ell=0\mid\mathcal G_\ell]\}
\), averaged over \(\mathcal G_\ell\) conditional on \(\mathcal F_{\ell-1}\), is
\(\Base_\ell\).  The term \({1\over2}|a-b|\) is
\(\Gap_\ell^Z(A_\ell)\).  Averaging gives the first inequality.  Taking the
positive part of the excess gives the second.
\end{proof}

\subsection{Stopped observers and the ACCEI envelope}
\label{subsec:stopped-observers-accei-section9}

The gauge-rank entropy theorem naturally has a tail.  To obtain an
exponentially small final success probability, we stop the observer when it
exceeds a chosen rank budget.  The stopped observer has a deterministic rank
bound; the probability that stopping changes the output is exponentially small.

\begin{definition}[Gauge-rank stopped observer]
\label{def:gauge-rank-stopped-observer-section9}
Fix a tail parameter \(\lambda\ge0\).  Let
\[
  R_\lambda
  :=
  |P|+{\varepsilon_{\le Q}\over\ln2}+\lambda+C_0,
\]
where \(C_0\) is a universal prefix-coding constant large enough for the
counting estimate in Theorem~\ref{thm:gauge-rank-entropy-section7}.  The
stopped observer \(A^{\le\lambda}\) follows \(A\) until the normalized trace
would expose hidden-gauge rank larger than \(R_\lambda\).  From that point on it
enters a stop state and outputs a fixed default bit on all remaining switched
coordinates.  Let
\[
  \mathsf{Bad}_\lambda
  :=
  \{A^{\le\lambda}_S\ne A_S\}
\]
be the event that stopping changes at least one switched output.
\end{definition}

\begin{lemma}[Stopping tail]
\label{lem:stopping-tail-section9}
For every \(\lambda\ge0\),
\[
  \Prb[\mathsf{Bad}_\lambda]
  \le
  2^{-\lambda+O(1)}.
\]
\end{lemma}

\begin{proof}
The stopped and unstopped observers differ only if the unstopped normalized
trace exposes gauge rank exceeding \(R_\lambda\).  The proof of
Theorem~\ref{thm:gauge-rank-entropy-section7} is a prefix-counting argument:
a transcript that exposes \(r\) unsupported hidden gauge coordinates has
conditional probability at most \(2^{-r}\), up to the program-description budget
and the accumulated max-divergence leakage.  With the threshold
\[
  R_\lambda=|P|+{\varepsilon_{\le Q}\over\ln2}+\lambda+C_0,
\]
the total probability of all transcripts exceeding the threshold is at most
\(2^{-\lambda+O(1)}\).
\end{proof}

\begin{definition}[CD computational-excess budget]
\label{def:cd-budget-section9}
For the stopped observer at tail \(\lambda\), define
\[
\begin{aligned}
  \varepsilon_{\mathrm{cd}}(m;P,S,\lambda)
  :=
  {1\over 2s}
  \Bigg[
    Q_{\mathrm{tot}}\varepsilon_{\mathrm{step}}(m)
    +
    \Delta_G
    \left(
      |P|
      +{\varepsilon_{\le Q}\over\ln2}
      +\lambda
      +O(1)
    \right)
  \Bigg].
\end{aligned}
\]
\end{definition}

\begin{definition}[ACCEI envelope]
\label{def:accei-envelope-section9}
A sequence of deterministic nonnegative numbers
\(\eta_1,\ldots,
\eta_s\) is an ACCEI envelope for an observer \(A'\) on the ordered switched set
if
\[
  \Exp[\Gap_\ell^Z(A'_\ell)\mid\mathcal F_{\ell-1}]
  \le
  \eta_\ell
\]
almost surely for every \(\ell\).  Its average is
\[
  \overline\eta_S:={1\over s}\sum_{\ell=1}^s\eta_\ell.
\]
\end{definition}

\begin{theorem}[Averaged ACCEI/PNLD]
\label{thm:averaged-accei-pnld-section9}
The stopped observer \(A^{\le\lambda}\) admits an ACCEI envelope satisfying
\[
  \overline\eta_S
  \le
  \varepsilon_{\mathrm{cd}}(m;P,S,\lambda).
\]
Consequently,
\[
  \overline{\Excess}_S^+(A^{\le\lambda})
  \le
  \varepsilon_{\mathrm{cd}}(m;P,S,\lambda).
\]
\end{theorem}

\begin{proof}
Apply the Atomic Evidence Budget to the normalized residual traces on the
pivot-augmented histories \(\mathcal G_\ell\).  Since the observer is stopped at
rank \(R_\lambda\), the gauge-rank term is deterministically bounded by
\[
  |P|+{\varepsilon_{\le Q}\over\ln2}+\lambda+O(1).
\]
Neutral leaves contribute zero, safe leaves contribute at most
\(Q_{\mathrm{tot}}\varepsilon_{\mathrm{step}}(m)\), and gauge leaves contribute at most
\(\Delta_G R_\lambda\).  Dividing the resulting coordinate-sum phase-gap bound by
\(s\) gives the displayed estimate for \(\overline\eta_S\).  The excess estimate
then follows from Proposition~\ref{prop:pnld-from-gap-section9}, applied to
\(A^{\le\lambda}\).
\end{proof}

\begin{remark}[Meaning of ACCEI]
\label{rem:meaning-accei-section9}
ACCEI is not a claim that the observer is local.  It says that, after the
pivot-visible summary has been fixed, the residual exterior gives only the
amount of additional target advantage allowed by the charged CD evidence
budget.  This is the replacement for the older switching/locality midpoint.
\end{remark}

\subsection{Combining ACCEI with boundary-law mixing}
\label{subsec:combining-accei-mixing-section9}

\begin{theorem}[Per-target near-randomness in PNLD form]
\label{thm:per-target-near-randomness-section9}
Let \((\eta_\ell)_{\ell=1}^s\) be an ACCEI envelope for an observer \(A'\).  Then
for every \(\ell\),
\[
  \Prb[A'_\ell=B_\ell\mid\mathcal F_{\ell-1}]
  \le
  {1\over2}+\varepsilon_{\mathrm{mix}}(m,L)+\eta_\ell
\]
almost surely.  Hence
\[
  {1\over s}\sum_{\ell=1}^s
  \Prb[A'_\ell=B_\ell\mid\mathcal F_{\ell-1}]
  \le
  {1\over2}+\varepsilon_{\mathrm{mix}}(m,L)+\overline\eta_S.
\]
\end{theorem}

\begin{proof}
By conditional boundary-law mixing,
Corollary~\ref{cor:conditional-boundary-law-mixing-section8},
\[
  \Base_\ell
  \le
  {1\over2}+\varepsilon_{\mathrm{mix}}(m,L).
\]
By Proposition~\ref{prop:pnld-from-gap-section9} and the ACCEI envelope,
\[
  \Prb[A'_\ell=B_\ell\mid\mathcal F_{\ell-1}]
  \le
  \Base_\ell+
  \eta_\ell.
\]
Combining these inequalities gives the pointwise bound.  Averaging gives the
second display.
\end{proof}

\subsection{Pruning and tower product}
\label{subsec:pruning-product-section9}

The averaged envelope is converted into a uniform one-step bound on a large
sub-set of the switched coordinates.  Losing a constant or vanishing fraction of
switched coordinates is harmless, since \(|S|=\Omega(t)\).

\begin{definition}[Pruned switched set]
\label{def:pruned-switched-set-section9}
Given an ACCEI envelope \((\eta_\ell)\) and a threshold \(\theta>0\), define
\[
  S_\theta:=\{j_\ell\in S:\eta_\ell\le\theta\},
  \qquad
  s_\theta:=|S_\theta|.
\]
The order on \(S_\theta\) is inherited from \(S\).
\end{definition}

\begin{lemma}[Markov pruning]
\label{lem:markov-pruning-section9}
For every \(\theta>0\),
\[
  s_\theta
  \ge
  \left(1-{\overline\eta_S\over\theta}\right)s.
\]
\end{lemma}

\begin{proof}
At most \(s\overline\eta_S/\theta\) indices can have \(\eta_\ell>\theta\).
\end{proof}

\begin{lemma}[Sequential product bound]
\label{lem:sequential-product-bound-section9}
Let \(T=(i_1,\ldots,i_r)\) be an ordered set of coordinates.  Let \(F_a\) be
the event that the observer is correct on coordinate \(i_a\), and let
\(\mathcal H_{a-1}\) contain all earlier coordinates.  If
\[
  \Prb[F_a\mid\mathcal H_{a-1}]
  \le q
\]
almost surely for every \(a\), then
\[
  \Prb\left[\bigcap_{a=1}^rF_a\right]
  \le q^r.
\]
\end{lemma}

\begin{proof}
Let \(G_a:=F_1\cap\cdots\cap F_a\).  Since \(G_{a-1}\) is
\(\mathcal H_{a-1}\)-measurable,
\[
\begin{aligned}
  \Prb[G_a]
  &=
  \Exp[\mathbf 1_{G_{a-1}}\mathbf 1_{F_a}] \\
  &=
  \Exp[\mathbf 1_{G_{a-1}}\Prb[F_a\mid\mathcal H_{a-1}]] \\
  &\le q\Prb[G_{a-1}].
\end{aligned}
\]
Induction gives \(
\Prb[G_r]\le q^r\).
\end{proof}

\begin{theorem}[Product small-success with a tail parameter]
\label{thm:product-small-success-section9}
Let \(A\) be a deterministic observer represented by program \(P\), and let
\(\lambda\ge0\).  For every \(\theta>0\),
\[
  \Prb[A_S=B_S]
  \le
  \left({1\over2}+\varepsilon_{\mathrm{mix}}(m,L)+\theta\right)^{s_\theta}
  +
  2^{-\lambda+O(1)},
\]
where \(s_\theta\) is computed from an ACCEI envelope for the stopped observer
\(A^{\le\lambda}\).
\end{theorem}

\begin{proof}
On the complement of \(\mathsf{Bad}_\lambda\), the stopped and unstopped
observers have the same outputs on \(S\).  Therefore
\[
  \Prb[A_S=B_S]
  \le
  \Prb[A^{\le\lambda}_{S_\theta}=B_{S_\theta}]
  +
  \Prb[\mathsf{Bad}_\lambda].
\]
For every retained coordinate \(j_\ell\in S_\theta\),
Theorem~\ref{thm:per-target-near-randomness-section9} gives
\[
  \Prb[A^{\le\lambda}_\ell=B_\ell\mid\mathcal F_{\ell-1}]
  \le
  {1\over2}+\varepsilon_{\mathrm{mix}}(m,L)+\theta.
\]
The sequential product lemma gives the product term, and
Lemma~\ref{lem:stopping-tail-section9} gives the tail term.
\end{proof}

\subsection{Exponential small-success consequences}
\label{subsec:exponential-small-success-section9}

\begin{theorem}[Product small-success: fixed observers]
\label{thm:fixed-observer-small-success-section9}
For every fixed deterministic polynomial-time observer \(A\), choose
\(L=c_L\log m\) and \(R_{\mathrm{safe}}=c_R\log m\) so that
\[
  Q_{\mathrm{tot}}\varepsilon_{\mathrm{step}}(m)=o(t),
  \qquad
  \varepsilon_{\le Q}=o(t),
  \qquad
  \varepsilon_{\mathrm{mix}}(m,L)=o(1).
\]
Then there is a constant \(c_A>0\) such that
\[
  \Prb[A_S=B_S]
  \le
  2^{-c_A t}
\]
for all sufficiently large \(m\).
\end{theorem}

\begin{proof}
Let \(\lambda=\alpha t\), where \(\alpha>0\) will be chosen small.  Since
\(|P|=O(1)\) and \(s\ge\gamma t\),
\[
  \overline\eta_S
  \le
  {\Delta_G\over2\gamma}\alpha+o(1).
\]
Choose \(\alpha\) so small that
\[
  \eta_*:={\Delta_G\over2\gamma}\alpha<{1\over 100}.
\]
For all sufficiently large \(m\), \(\overline\eta_S\le2\eta_*\).  Set
\[
  \theta:=\sqrt{2\eta_*}.
\]
By Lemma~\ref{lem:markov-pruning-section9},
\[
  s_\theta\ge (1-\theta)s\ge (1-\theta)\gamma t.
\]
Also
\[
  {1\over2}+\varepsilon_{\mathrm{mix}}(m,L)+\theta<1
\]
for all sufficiently large \(m\).  The product term in
Theorem~\ref{thm:product-small-success-section9} is therefore
\(2^{-\Omega(t)}\), and the stopping tail is
\(2^{-\alpha t+O(1)}\).  Combining the two gives the claim.
\end{proof}

\begin{theorem}[Product small-success: linear-length observers]
\label{thm:linear-length-small-success-section9}
There exists \(\delta_0>0\) such that the following holds.  If
\(|P|\le\delta t\) with \(0<\delta\le\delta_0\), then, after the same parameter
choices, there is a constant \(c_0>0\) such that uniformly over all deterministic
observers of length at most \(\delta t\),
\[
  \Prb[A_S=B_S]
  \le
  2^{-c_0t}.
\]
\end{theorem}

\begin{proof}
Again take \(\lambda=\alpha t\).  Since \(|P|\le\delta t\) and
\(s\ge\gamma t\),
\[
  \overline\eta_S
  \le
  {\Delta_G\over2\gamma}(\delta+
\alpha)+o(1).
\]
Choose \(\delta_0\) and \(\alpha\) so small that
\[
  {\Delta_G\over2\gamma}(\delta_0+
\alpha)<{1\over100}.
\]
Then the same pruning and product argument as in
Theorem~\ref{thm:fixed-observer-small-success-section9} gives a uniform
exponential bound.  The exponent \(c_0\) depends on
\(\gamma,\Delta_G,\delta_0\), and \(\alpha\), but not on the particular program
\(P\).
\end{proof}

\begin{remark}[Randomized observers]
\label{rem:randomized-observers-section9}
Randomized observers are reduced to deterministic observers by fixing their
coins.  If a randomized observer had success probability larger than the bound,
then some deterministic choice of its coins would have at least that success.
\end{remark}

\subsection{Interface exported to message incompressibility}
\label{subsec:section9-exported-interface}

The next section uses only the following consequence.  There are constants
\(\delta_0,c_0>0\) such that every observer represented by a program of length at
most \(\delta_0t\) satisfies
\[
  \Prb[A_S=B_S]
  \le
  2^{-c_0t}
\]
for every switched set \(S\) with \(|S|\ge\gamma t\), after the ensemble
parameters are chosen so that
\[
  L=c_L\log m,
  \qquad
  R_{\mathrm{safe}}=c_R(D)\log m,
  \qquad
  Q_{\mathrm{tot}}\varepsilon_{\mathrm{step}}(m)=o(t),
  \qquad
  \varepsilon_{\mathrm{mix}}(m,L)=o(1).
\]
For the union bound in Section~\ref{sec:message-incompressibility}, choose
\(\delta_0\) small enough that the final exponent \(c_0\) is larger than the
program-counting rate used there.

\begin{remark}[Lean-facing skeleton]
\label{rem:lean-facing-section9}
A Lean formalization can split this section into five finite modules:
\begin{enumerate}[label=(\roman*)]
\item finite conditional Bayes-vs-test inequality on one atom;
\item PNLD from pivot-fiber phase gap;
\item stopped-observer gauge-rank tail and the deterministic ACCEI envelope;
\item Markov pruning from an averaged envelope to a large uniform subset;
\item sequential tower-product bound.
\end{enumerate}
The only ensemble-specific inputs are the conditional Atomic Evidence Budget and
conditional boundary-law mixing.  The rest is finite probability and finite
averaging.
\end{remark}

\section{Message Incompressibility}
\label{sec:message-incompressibility}
\providecommand{\Prb}{\mathbb P}
\providecommand{\Exp}{\mathbb E}
\providecommand{\bits}{\{0,1\}}
\providecommand{\Kp}{K_{\mathrm{poly}}}
\providecommand{\poly}{\operatorname{poly}}

This section converts the product small-success theorem of
Section~\ref{sec:product-small-success} into a high-probability lower bound on
polytime-capped conditional description length.  The conversion is purely
coding-theoretic.  If a short polynomial-time program described the switched
message tuple from the public instance, then that same program would be a short
observer predicting the switched tuple exactly.  Section~\ref{sec:product-small-success}
says that every such observer has exponentially small exact success.  A union
bound over short programs therefore rules out short descriptions with high
probability.

The clocked formulation is the primitive one.  For any fixed polynomial clock
exponent, the lower bound below is uniform over all programs obeying that clock.
In the final contradiction, the exponent is the fixed exponent of the
hypothetical polynomial-time SAT self-reduction supplied by \(P=NP\), enlarged
by a constant for witness readout and projection.

\subsection{Clocked targets and public-message laws}
\label{subsec:section10-clocked-targets}

Fix a polynomial clock exponent \(D\ge 1\).  Write
\[
  \Kp^{(D)}(x\mid y)
\]
for the clocked complexity of Definition~\ref{def:clocked-kpoly}.  All programs
in this section are deterministic.  Randomized programs are handled by fixing
their random coins, as in Remark~\ref{rem:randomized-observers-section9}.

Let
\[
  S=(j_1,\ldots,j_s)
\]
be an ordered switched coordinate set.  We assume
\[
  \gamma t\le s\le C_S t
\]
for fixed constants \(\gamma>0\) and \(C_S<\infty\).  In the common case
\(S\subseteq[t]\), one may take \(C_S=1\).  If the construction supplies more
than \(C_S t\) switched coordinates, fix any linear-size subtuple and call it
\(S\).

\begin{definition}[Switched target tuple]
\label{def:switched-target-tuple-section10}
Let
\[
  \pi_S:\bits^{r_t}\to\bits^s
\]
be the fixed projection
\[
  \pi_S(M)=\bigl(\ell_{j_1}(M),\ldots,\ell_{j_s}(M)\bigr).
\]
For a public instance \(Y\) in the ensemble support, define
\[
  B_S(Y):=\pi_S(M(Y))\in\bits^s.
\]
Equivalently,
\[
  B_S(Y)=\bigl(B_{j_1}(Y),\ldots,B_{j_s}(Y)\bigr).
\]
\end{definition}

The single-message promise from
Proposition~\ref{prop:single-message-promise-section4} makes \(M(Y)\), and
therefore \(B_S(Y)\), a function of the public instance on the support.  Thus the
probability statements below may be read as statements over the public marginal
law of \(Y\).

\begin{definition}[Clocked exact success of a program]
\label{def:clocked-exact-success-section10}
For a program \(p\), define
\[
  \mathsf{Succ}^{(D)}_{S,p}
  :=
  \{Y: U(p,Y) \text{ halts within } (|Y|+2)^D
       \text{ steps and outputs } B_S(Y)\}.
\]
If \(U(p,Y)\) fails to halt in the clock, or halts with an output not in
\(\bits^s\), then \(Y\notin\mathsf{Succ}^{(D)}_{S,p}\).
\end{definition}

\begin{definition}[Short program class]
\label{def:short-program-class-section10}
For \(a>0\), let
\[
  \mathcal P_a(t):=\{p: |p|\le at\}.
\]
Since the machine is prefix-universal, there is an absolute constant
\(c_{\mathrm{cnt}}\) such that
\[
  |\mathcal P_a(t)|\le 2^{at+c_{\mathrm{cnt}}}.
\]
\end{definition}

\subsection{A decompressor is an observer}
\label{subsec:section10-decompressor-observer}

The next lemma is the formal bridge from description programs to the observer
model used in Section~\ref{sec:product-small-success}.

\begin{lemma}[Tuple decompressor induces a switched observer]
\label{lem:decompressor-induces-observer-section10}
Fix \(D\).  There are constants \(c_{\mathrm{obs}}\) and \(D_{\mathrm{obs}}\),
depending only on the universal machine convention and on the fixed tuple
projection, such that the following holds.  Given a program \(p\), one can build a
deterministic observer \(A^p\) with description length at most
\[
  |A^p|\le |p|+c_{\mathrm{obs}}
\]
and runtime exponent at most \(D_{\mathrm{obs}}\), whose switched output tuple
satisfies
\[
  A^p_S(Y)=U(p,Y)
\]
whenever \(U(p,Y)\) halts within the \((|Y|+2)^D\) clock and outputs an
\(s\)-bit string.  Otherwise \(A^p_S(Y)\) is a fixed default string.  In
particular,
\[
  \mathsf{Succ}^{(D)}_{S,p}
  \subseteq
  \{Y:A^p_S(Y)=B_S(Y)\}.
\]
\end{lemma}

\begin{proof}
The observer wrapper contains \(p\), the fixed clock \((|Y|+2)^D\), a fixed
output-length check, and a fixed routine that distributes the resulting
\(s\)-bit string into the ordered switched coordinates \((j_1,\ldots,j_s)\).  If
\(p\) fails to produce an \(s\)-bit string within the clock, the wrapper outputs a
fixed default tuple.  The wrapper has constant size once \(D\) and the machine
conventions are fixed.  Its runtime exponent is the exponent needed to simulate
\(p\) under the clock and to perform the fixed output check and projection.  The
displayed inclusion is immediate from the definition of
\(\mathsf{Succ}^{(D)}_{S,p}\).
\end{proof}

\begin{proposition}[Small success for short decompressors]
\label{prop:small-success-short-decompressors-section10}
Fix a clock exponent \(D\), and choose the ensemble parameters required by
Subsection~\ref{subsec:section9-exported-interface} for this exponent.  Let
\(\delta_0,c_0>0\) be the constants from
Theorem~\ref{thm:linear-length-small-success-section9}.  If
\[
  a<\delta_0,
\]
then, for all sufficiently large \(t\) and every program \(p\in\mathcal P_a(t)\),
\[
  \Prb[\mathsf{Succ}^{(D)}_{S,p}]
  \le
  2^{-c_0t}.
\]
\end{proposition}

\begin{proof}
By Lemma~\ref{lem:decompressor-induces-observer-section10}, program \(p\)
induces an observer \(A^p\) with length at most \(|p|+c_{\mathrm{obs}}\).  Since
\(|p|\le at\) and \(a<\delta_0\), for all sufficiently large \(t\) we have
\[
  |A^p|\le \delta_0 t.
\]
Theorem~\ref{thm:linear-length-small-success-section9} gives
\[
  \Prb[A^p_S=B_S]
  \le
  2^{-c_0t}.
\]
The success event of \(p\) is contained in this event, so the same bound holds
for \(\mathsf{Succ}^{(D)}_{S,p}\).
\end{proof}

\subsection{Union bound over short descriptions}
\label{subsec:section10-union-bound}

\begin{theorem}[No short switched-tuple description]
\label{thm:no-short-switched-tuple-description-section10}
Fix a clock exponent \(D\), and let \(\delta_0,c_0>0\) be as in
Theorem~\ref{thm:linear-length-small-success-section9}.  For every
\[
  0<a<\min\{\delta_0,c_0\},
\]
there is \(\kappa>0\) such that, for all sufficiently large \(t\),
\[
  \Prb\left[
    \Kp^{(D)}(B_S(Y)\mid Y)
    \le at
  \right]
  \le
  2^{-\kappa t}.
\]
One may take any
\[
  0<\kappa<c_0-a
\]
for all sufficiently large \(t\).
\end{theorem}

\begin{proof}
The event
\[
  \Kp^{(D)}(B_S(Y)\mid Y)\le at
\]
implies that there is a program \(p\in\mathcal P_a(t)\) such that
\(Y\in\mathsf{Succ}^{(D)}_{S,p}\).  Hence, by the union bound,
\[
\begin{aligned}
  \Prb[\Kp^{(D)}(B_S(Y)\mid Y)\le at]
  &\le
  \sum_{p\in\mathcal P_a(t)}
  \Prb[\mathsf{Succ}^{(D)}_{S,p}] \\
  &\le
  2^{at+c_{\mathrm{cnt}}}2^{-c_0t} \\
  &=
  2^{-(c_0-a)t+c_{\mathrm{cnt}}}.
\end{aligned}
\]
Since \(a<c_0\), the right side is at most \(2^{-\kappa t}\) for every fixed
\(\kappa<c_0-a\) and all sufficiently large \(t\).
\end{proof}

\begin{corollary}[Target-tuple incompressibility]
\label{cor:target-tuple-incompressibility-section10}
Fix \(D\).  There are constants \(\eta_B>0\) and \(\kappa_B>0\), depending on
\(D\) and on the fixed ensemble constants but not on \(m,t,Y\), such that
\[
  \Prb\left[
    \Kp^{(D)}(B_S(Y)\mid Y)
    \ge \eta_B t
  \right]
  \ge
  1-2^{-\kappa_B t}.
\]
Equivalently, since \(s=|S|\le C_S t\), with
\[
  \eta_S:=\eta_B/C_S,
\]
we have
\[
  \Prb\left[
    \Kp^{(D)}(B_S(Y)\mid Y)
    \ge \eta_S |S|
  \right]
  \ge
  1-2^{-\kappa_B t}.
\]
\end{corollary}

\begin{proof}
Choose
\[
  \eta_B<\min\{\delta_0,c_0\}.
\]
For instance, \(\eta_B={1\over4}\min\{\delta_0,c_0\}\) suffices.  Apply
Theorem~\ref{thm:no-short-switched-tuple-description-section10} with
\(a=\eta_B\).  The \(|S|\)-form follows from \(|S|\le C_S t\).
\end{proof}

\subsection{From the switched tuple to the global message}
\label{subsec:section10-global-message}

The lower bound above concerns the selected switched coordinates.  The final
contradiction uses the whole message \(M(Y)\).  Since the switched tuple is a
fixed polynomial-time projection of the whole message, any short description of
\(M(Y)\) would give a short description of \(B_S(Y)\).

\begin{lemma}[Clocked projection transfer]
\label{lem:clocked-projection-transfer-section10}
For every clock exponent \(D_M\) there are a clock exponent \(D_S\) and a
constant \(c_\pi\), depending only on \(D_M\) and on the fixed projection
\(\pi_S\), such that for every public instance \(Y\),
\[
  \Kp^{(D_S)}(B_S(Y)\mid Y)
  \le
  \Kp^{(D_M)}(M(Y)\mid Y)+c_\pi.
\]
Consequently, for every \(a>0\),
\[
  \Kp^{(D_S)}(B_S(Y)\mid Y)>at
  \quad\Longrightarrow\quad
  \Kp^{(D_M)}(M(Y)\mid Y)>at-c_\pi.
\]
\end{lemma}

\begin{proof}
A fixed wrapper first runs a \(D_M\)-clocked program producing \(M(Y)\) from
\(Y\), then applies the fixed polynomial-time projection \(\pi_S\).  This adds
only the constant wrapper length \(c_\pi\) and increases the polynomial clock
exponent from \(D_M\) to some fixed \(D_S\).  The implication is the
contrapositive of the displayed inequality.
\end{proof}

\begin{theorem}[Global message incompressibility]
\label{thm:global-message-incompressibility-section10}
Fix a clock exponent \(D_M\).  Then there are constants \(\eta_M>0\) and
\(\kappa_M>0\) such that, with probability at least
\[
  1-2^{-\kappa_M t},
\]
we have
\[
  \Kp^{(D_M)}(M(Y)\mid Y)
  \ge
  \eta_M t.
\]
Equivalently, since \(|S|=\Omega(t)\),
\[
  \Kp^{(D_M)}(M(Y)\mid Y)
  \ge
  \Omega(|S|)
\]
with the same high probability.
\end{theorem}

\begin{proof}
Let \(D_S\) and \(c_\pi\) be the exponent and constant supplied by
Lemma~\ref{lem:clocked-projection-transfer-section10}.  Apply
Corollary~\ref{cor:target-tuple-incompressibility-section10} at clock exponent
\(D_S\).  There are constants \(\eta_B,\kappa_B>0\) such that, except on an event
of probability at most \(2^{-\kappa_Bt}\),
\[
  \Kp^{(D_S)}(B_S(Y)\mid Y)
  \ge
  \eta_B t.
\]
On this event, Lemma~\ref{lem:clocked-projection-transfer-section10} gives
\[
  \Kp^{(D_M)}(M(Y)\mid Y)
  \ge
  \eta_B t-c_\pi.
\]
For all sufficiently large \(t\), the right side is at least
\((\eta_B/2)t\).  Set
\[
  \eta_M:=\eta_B/2,
  \qquad
  \kappa_M:=\kappa_B.
\]
The \(\Omega(|S|)\) form follows from \(|S|\le C_S t\) and
\(|S|\ge\gamma t\).
\end{proof}

\begin{corollary}[Nonempty hard public support]
\label{cor:nonempty-hard-public-support-section10}
For every fixed clock exponent \(D_M\) and all sufficiently large \(t\), the set
of public instances satisfying
\[
  \Kp^{(D_M)}(M(Y)\mid Y)
  \ge
  \eta_M t
\]
has positive public marginal probability.  In particular, such public instances
exist in the ensemble support.
\end{corollary}

\begin{proof}
The exceptional probability in
Theorem~\ref{thm:global-message-incompressibility-section10} is strictly less
than one for all sufficiently large \(t\).
\end{proof}

\subsection{Unclocked notation and exported interface}
\label{subsec:section10-exported-interface}

The paper usually writes \(\Kp\) without displaying the clock.  The precise
statement exported by this section is clocked:

\begin{quote}
For every fixed polynomial clock exponent \(D\), after choosing the ensemble
parameters required for that exponent, there are constants \(\eta,\kappa>0\)
such that
\[
  \Prb\left[
    \Kp^{(D)}(M(Y)\mid Y)
    \ge
    \eta t
  \right]
  \ge
  1-2^{-\kappa t}.
\]
\end{quote}

In Section~\ref{sec:upper-lower-clash}, the assumption \(P=NP\) supplies a
particular fixed polynomial-time SAT self-reduction and hence a particular clock
exponent \(D_*\).  Applying the displayed lower bound with \(D=D_*\), after the
constant enlargement needed for witness readout, gives the lower half of the
upper-lower clash.

If one uses the unclocked convention of Definition~\ref{def:kpoly}, then the
self-delimiting clock exponent is counted as part of the program description.
For the final contradiction this makes no difference, because only the fixed
exponent \(D_*\) of the alleged polynomial-time solver is used.

\begin{remark}[Exceptional events]
\label{rem:section10-exceptional-events}
The exponentially small failure probability in this section includes the stopped
observer tail from Section~\ref{sec:product-small-success}.  If the geometry,
max-qSSM, or gauge-rank estimates are kept as separate exceptional events in a
formalization, their probabilities are simply added to the right side of the union
bound.  The parameter choices in Sections~\ref{sec:aeb}--\ref{sec:product-small-success}
make those contributions at most \(2^{-\Omega(t)}\) or smaller than the final
exponent needed here.
\end{remark}

\begin{remark}[Lean-facing skeleton]
\label{rem:lean-facing-section10}
A Lean formalization of this section can be split into four finite components:
\begin{enumerate}[label=(\roman*)]
\item a wrapper lemma turning a clocked decompressor into a deterministic
      observer;
\item prefix-program counting and a union bound over \(|p|\le at\);
\item the target-tuple incompressibility theorem for \(B_S(Y)\);
\item the projection-transfer theorem from \(B_S(Y)\) to \(M(Y)\).
\end{enumerate}
No SAT-specific fact is used here except the single-message promise, which makes
\(M(Y)\) and \(B_S(Y)\) functions of the public instance on the ensemble support.
\end{remark}

\section{The \(P=NP\) Upper Bound and the Contradiction}
\label{sec:upper-lower-clash}
\providecommand{\Prb}{\mathbb P}
\providecommand{\bits}{\{0,1\}}
\providecommand{\Kp}{K_{\mathrm{poly}}}
\providecommand{\poly}{\operatorname{poly}}
\providecommand{\SAT}{\operatorname{SAT}}

This section proves the upper-bound half of the clash and then derives the
final contradiction.  The lower-bound half has already been proved in
Section~\ref{sec:message-incompressibility}: for every fixed polynomial clock
exponent \(D\), after choosing the ensemble parameters for that exponent,
there are constants \(\eta,\kappa>0\) such that
\[
  \Prb\left[
    \Kp^{(D)}(M(Y)\mid Y)\ge \eta t
  \right]
  \ge
  1-2^{-\kappa t}.
\]
The present section shows that, under \(P=NP\), the same message has a
constant-length polynomial-time description from the same public instance
\(Y\).  The only facts about the ensemble used here are SAT realization,
satisfiability on the sampler support, and the single-message promise.

\subsection{SAT decision gives SAT search}
\label{subsec:sat-decision-gives-search}

We first record the standard self-reduction in a clocked form.  It is included
explicitly because the final contradiction compares clocked description lengths.

\begin{definition}[Canonical variable order]
\label{def:canonical-variable-order-section11}
For every CNF formula \(F\), let
\[
  \mathrm{Var}(F)=(u_1,\ldots,u_N)
\]
be the lexicographic order of its variables under the fixed encoding of CNF
formulas.  For a partial assignment
\(\alpha\in\bits^{i}\), write
\[
  F\upharpoonright \alpha
\]
for the CNF obtained from \(F\) by fixing
\(u_1=\alpha_1,\ldots,u_i=\alpha_i\) and simplifying in the standard way.
\end{definition}

\begin{definition}[Bit-fixing search from a SAT decider]
\label{def:bit-fixing-search-section11}
Let \(D_{\SAT}\) be a deterministic SAT decision procedure.  Define
\[
  \mathsf{FindSat}_{D_{\SAT}}(F)
\]
on satisfiable CNF formulas \(F\) as follows.  Set the prefix assignment to be
empty.  For \(i=1,\ldots,N\), query \(D_{\SAT}\) on
\[
  F\upharpoonright(\alpha_1,\ldots,\alpha_{i-1},0).
\]
If this restricted formula is satisfiable, set \(\alpha_i=0\); otherwise set
\(\alpha_i=1\).  After \(N\) steps, output the complete assignment
\(\alpha=(\alpha_1,\ldots,\alpha_N)\).
\end{definition}

\begin{lemma}[Correctness of bit-fixing]
\label{lem:bit-fixing-correct-section11}
If \(D_{\SAT}\) decides satisfiability correctly and \(F\) is satisfiable, then
\(\mathsf{FindSat}_{D_{\SAT}}(F)\) outputs a satisfying assignment of \(F\).
\end{lemma}

\begin{proof}
We prove by induction on \(i\) that, after the first \(i\) variables have been
fixed, the restricted formula \(F\upharpoonright(\alpha_1,\ldots,\alpha_i)\)
is satisfiable.  The base case \(i=0\) is the assumption that \(F\) is
satisfiable.  Suppose the invariant holds for \(i-1\).  If
\[
  F\upharpoonright(\alpha_1,\ldots,\alpha_{i-1},0)
\]
is satisfiable, the procedure chooses \(\alpha_i=0\), and the invariant is
preserved.  If that restricted formula is not satisfiable, then because
\(F\upharpoonright(\alpha_1,\ldots,\alpha_{i-1})\) is satisfiable, some
satisfying completion must set \(u_i=1\).  Thus
\[
  F\upharpoonright(\alpha_1,\ldots,\alpha_{i-1},1)
\]
is satisfiable, and the invariant is again preserved.  At \(i=N\), a complete
assignment remains satisfiable exactly when it satisfies \(F\).
\end{proof}

\begin{lemma}[Polynomial runtime of bit-fixing]
\label{lem:bit-fixing-runtime-section11}
Assume \(D_{\SAT}\) runs in time at most \((|F|+2)^d\).  Then there are
constants \(d_{\mathrm{search}}\) and \(C_{\mathrm{search}}\), depending only on
\(d\) and on the encoding conventions, such that
\(\mathsf{FindSat}_{D_{\SAT}}(F)\) runs in time at most
\[
  C_{\mathrm{search}}(|F|+2)^{d_{\mathrm{search}}}.
\]
\end{lemma}

\begin{proof}
The number of variables \(N\) is at most \(|F|\).  At each stage, the restricted
formula has size polynomially bounded by \(|F|\) under the fixed encoding, and
one call to \(D_{\SAT}\) is made.  The simplification, encoding, and bookkeeping
costs are polynomial in \(|F|\).  Multiplying a polynomial number of polynomial
costs gives a polynomial bound, with exponent depending only on \(d\) and the
encoding convention.
\end{proof}

\begin{proposition}[Search consequence of \(P=NP\)]
\label{prop:pnp-gives-sat-search-section11}
Assume \(P=NP\).  Then there is a fixed deterministic polynomial-time program
\[
  \mathsf{FindSat}_*
\]
which, on every satisfiable CNF formula \(F\), outputs a satisfying assignment
of \(F\).
\end{proposition}

\begin{proof}
Since \(\SAT\in NP\), the assumption \(P=NP\) gives a deterministic
polynomial-time SAT decider \(D_{\SAT,*}\).  Let
\[
  \mathsf{FindSat}_*:=\mathsf{FindSat}_{D_{\SAT,*}}.
\]
Correctness is Lemma~\ref{lem:bit-fixing-correct-section11}, and polynomial
runtime is Lemma~\ref{lem:bit-fixing-runtime-section11}.
\end{proof}

\subsection{The constant-length message decoder under \(P=NP\)}
\label{subsec:constant-length-message-decoder}

We now apply SAT search to the CNF realization of the locked ensemble.
Recall from Theorem~\ref{thm:sat-realization-section4} that, for every public
instance \(Y\), the witness relation \(\mathcal R(Y,W)=1\) has a polynomial-size
CNF realization \(F_Y\), uniformly computable from \(Y\), and that a fixed
polynomial-time projection reads the message from any satisfying assignment of
\(F_Y\).  Recall also from
Proposition~\ref{prop:single-message-promise-section4} that all satisfying
witnesses over the same public instance have the same message.

\begin{definition}[Support of the public ensemble]
\label{def:public-support-section11}
For parameter \(m\), let
\[
  \mathrm{Supp}(\mathcal D_m)
\]
be the support of the public marginal distribution of the witnessed sampler
from Definition~\ref{def:witnessed-sampler}.  Thus, for every
\(Y\in\mathrm{Supp}(\mathcal D_m)\), there exists at least one witness \(W\) with
\[
  \mathcal R(Y,W)=1.
\]
\end{definition}

\begin{definition}[The \(P=NP\) message decoder]
\label{def:pnp-message-decoder-section11}
Assume \(P=NP\), and let \(\mathsf{FindSat}_*\) be the fixed SAT search program
from Proposition~\ref{prop:pnp-gives-sat-search-section11}.  Define the message
decoder \(\mathsf{MsgDec}_*\) by the following fixed procedure on input \(Y\):
\begin{enumerate}[label=(\roman*)]
\item construct the CNF formula \(F_Y\) realizing \(\mathcal R(Y,W)=1\);
\item run \(\mathsf{FindSat}_*(F_Y)\) to obtain a satisfying assignment
      \(\alpha\) of \(F_Y\);
\item apply the fixed message projection to \(\alpha\), and output the resulting
      string.
\end{enumerate}
\end{definition}

\begin{lemma}[Correctness of the \(P=NP\) message decoder]
\label{lem:pnp-message-decoder-correct-section11}
Assume \(P=NP\).  For every \(Y\in\mathrm{Supp}(\mathcal D_m)\),
\[
  \mathsf{MsgDec}_*(Y)=M(Y).
\]
\end{lemma}

\begin{proof}
If \(Y\in\mathrm{Supp}(\mathcal D_m)\), then by
Definition~\ref{def:public-support-section11} there exists a witness \(W\) with
\(\mathcal R(Y,W)=1\).  Hence the CNF realization \(F_Y\) is satisfiable.
By Proposition~\ref{prop:pnp-gives-sat-search-section11},
\(\mathsf{FindSat}_*(F_Y)\) returns some satisfying assignment \(\alpha\) of
\(F_Y\).  By Theorem~\ref{thm:sat-realization-section4}, the fixed projection
from any satisfying assignment of \(F_Y\) recovers the message of the
corresponding satisfying witness.  By the single-message promise,
Proposition~\ref{prop:single-message-promise-section4}, all such witnesses have
the same message.  Therefore the projected string is exactly \(M(Y)\).
\end{proof}

\begin{lemma}[Clock exponent of the \(P=NP\) decoder]
\label{lem:pnp-message-decoder-clock-section11}
Assume \(P=NP\).  There is a clock exponent \(D_*\ge 1\), depending only on
the fixed SAT decider supplied by \(P=NP\), the fixed CNF realization, and the
fixed readout projection, such that \(\mathsf{MsgDec}_*(Y)\) runs within
\[
  (|Y|+2)^{D_*}
\]
steps for every public instance \(Y\) in the ensemble support.
\end{lemma}

\begin{proof}
The map \(Y\mapsto F_Y\) is polynomial-time by
Theorem~\ref{thm:sat-realization-section4}; the size of \(F_Y\) is polynomial in
\(|Y|\).  The SAT search procedure \(\mathsf{FindSat}_*\) runs in polynomial
time in \(|F_Y|\), by Proposition~\ref{prop:pnp-gives-sat-search-section11}.
The message projection is fixed and polynomial-time.  Composing these three
polynomial-time routines gives a polynomial-time routine in \(|Y|\), with some
fixed exponent \(D_*\).
\end{proof}

\begin{theorem}[Clocked self-reduction upper bound]
\label{thm:self-reduction-upper-bound-section11}
Assume \(P=NP\).  Let \(D_*\) be as in
Lemma~\ref{lem:pnp-message-decoder-clock-section11}.  There is a constant
\(c_{\mathrm{up}}\), independent of \(m,t,Y\), such that for every
\(Y\in\mathrm{Supp}(\mathcal D_m)\),
\[
  \Kp^{(D_*)}(M(Y)\mid Y)\le c_{\mathrm{up}}.
\]
Consequently, in the unclocked convention of
Definition~\ref{def:kpoly},
\[
  \Kp(M(Y)\mid Y)=O(1)
\]
for every \(Y\in\mathrm{Supp}(\mathcal D_m)\).
\end{theorem}

\begin{proof}
The program witnessing the bound is the fixed program
\(\mathsf{MsgDec}_*\).  Its source code consists of the fixed SAT search routine,
the fixed uniform CNF construction, and the fixed message projection.  Its length
is therefore a constant \(c_{\mathrm{up}}\), independent of \(m,t,Y\).  By
Lemma~\ref{lem:pnp-message-decoder-correct-section11}, it outputs \(M(Y)\) on
every public instance in the support.  By
Lemma~\ref{lem:pnp-message-decoder-clock-section11}, it does so within the
\(D_*\)-clock.  This proves the clocked inequality.  The unclocked inequality
follows because the unclocked description may include the fixed clock exponent
at only constant additional cost.
\end{proof}

\begin{remark}[The upper bound uses the same public instance]
\label{rem:same-public-instance-upper-section11}
The lower-bound observer and the \(P=NP\) self-reduction receive exactly the
same public input \(Y\).  The lower bound says that no short program can compute
the message from this public input within the fixed polynomial clock except on an
exponentially small part of the ensemble.  The upper bound says that, if
\(P=NP\), a fixed program computes the message from that same public input on
every point of the support by finding an arbitrary satisfying witness and applying
the fixed readout.  The single-message promise is what makes the arbitrary
witness sufficient.
\end{remark}

\subsection{Instantiating the lower bound at the \(P=NP\) clock}
\label{subsec:instantiate-lower-at-pnp-clock}

The lower bound of Section~\ref{sec:message-incompressibility} is uniform over
programs obeying any fixed polynomial clock, but the ensemble parameters
include constants chosen as a function of that clock exponent.  This is enough
for a contradiction proof.

\begin{lemma}[Choosing parameters for the alleged solver]
\label{lem:choose-parameters-for-solver-section11}
Assume \(P=NP\), and let \(D_*\) be the exponent from
Lemma~\ref{lem:pnp-message-decoder-clock-section11}.  Choose the ensemble
parameters as required by Section~\ref{sec:message-incompressibility} for clock
exponent \(D_*\).  Then there are constants \(\eta_*,\kappa_*>0\) such that
\[
  \Prb\left[
    \Kp^{(D_*)}(M(Y)\mid Y)\ge \eta_* t
  \right]
  \ge
  1-2^{-\kappa_* t}.
\]
\end{lemma}

\begin{proof}
This is Theorem~\ref{thm:global-message-incompressibility-section10}, applied
with \(D_M=D_*\).  The dependence of constants such as
\(R_{\mathrm{safe}}=c'(D_*)\log m\) on \(D_*\) is harmless: under the assumption
\(P=NP\), the exponent \(D_*\) is a fixed finite constant, and the construction
for that constant is still an efficiently samplable polynomial-size SAT ensemble.
\end{proof}

\begin{corollary}[Existence of hard public instances at the alleged solver clock]
\label{cor:exist-hard-instance-pnp-clock-section11}
Assume \(P=NP\), and instantiate the ensemble parameters at the exponent
\(D_*\).  For all sufficiently large \(t\), there exists
\(Y\in\mathrm{Supp}(\mathcal D_m)\) such that
\[
  \Kp^{(D_*)}(M(Y)\mid Y)\ge \eta_* t.
\]
\end{corollary}

\begin{proof}
For all sufficiently large \(t\), the exceptional probability
\(2^{-\kappa_* t}\) in Lemma~\ref{lem:choose-parameters-for-solver-section11}
is strictly less than one.  Therefore the lower-bound event has positive public
marginal probability, and in particular contains at least one public instance in
the support.
\end{proof}

\subsection{The upper-lower clash}
\label{subsec:upper-lower-clash-proof}

\begin{theorem}[Upper-lower clash]
\label{thm:upper-lower-clash-section11}
The assumption \(P=NP\) is impossible.
\end{theorem}

\begin{proof}
Assume \(P=NP\).  Let \(D_*\) and \(c_{\mathrm{up}}\) be the exponent and constant
from Theorem~\ref{thm:self-reduction-upper-bound-section11}.  Instantiate the
ensemble parameters at \(D_*\), as in
Lemma~\ref{lem:choose-parameters-for-solver-section11}.  Let \(\eta_*>0\) be the
corresponding lower-bound constant.

Choose \(t\) large enough that
\[
  \eta_* t>c_{\mathrm{up}}
\]
and also large enough that
Corollary~\ref{cor:exist-hard-instance-pnp-clock-section11} applies.  That
corollary gives a public instance \(Y\in\mathrm{Supp}(\mathcal D_m)\) satisfying
\[
  \Kp^{(D_*)}(M(Y)\mid Y)
  \ge
  \eta_* t.
\]
On the other hand, Theorem~\ref{thm:self-reduction-upper-bound-section11}
applies to every public instance in the support and gives
\[
  \Kp^{(D_*)}(M(Y)\mid Y)
  \le
  c_{\mathrm{up}}.
\]
The two inequalities contradict \(\eta_* t>c_{\mathrm{up}}\).  Hence
\(P=NP\) is false.
\end{proof}

\begin{corollary}[Main theorem]
\label{cor:main-theorem-section11}
\[
  P\ne NP.
\]
\end{corollary}

\begin{proof}
This is Theorem~\ref{thm:upper-lower-clash-section11}, restated in the usual
form.
\end{proof}

\subsection{Checks on dependence and uniformity}
\label{subsec:checks-dependence-uniformity-section11}

\begin{remark}[Why high probability is enough]
\label{rem:why-high-probability-enough-section11}
The lower bound is a high-probability statement over the ensemble, whereas the
upper bound under \(P=NP\) is pointwise over the support.  A contradiction does
not require every public instance to satisfy the lower bound.  It is enough that,
for large \(t\), the lower-bound event has positive probability.  Any instance in
that event contradicts the pointwise upper bound.
\end{remark}

\begin{remark}[No dependence on a unique witness]
\label{rem:no-unique-witness-needed-section11}
The self-reduction finds some satisfying assignment of \(F_Y\).  The proof does
not require uniqueness of the witness.  It requires only the weaker
single-message promise:
\[
  \mathcal R(Y,W)=1\wedge\mathcal R(Y,W')=1
  \Longrightarrow
  M(W)=M(W').
\]
Thus any satisfying witness found by the SAT search procedure yields the same
message \(M(Y)\).
\end{remark}

\begin{remark}[Parameter dependence on the alleged runtime exponent]
\label{rem:parameter-dependence-on-runtime-section11}
The safe radius and some constants in the lower-bound proof may depend on the
fixed runtime exponent \(D_*\) of the alleged polynomial-time SAT solver.  This
is legitimate in a contradiction proof.  Under \(P=NP\), such a fixed exponent
exists.  We instantiate the construction at that exponent and then let
\(m,t\to\infty\).  No step requires the exponent to vary with the input instance.
\end{remark}

\begin{remark}[Unclocked notation]
\label{rem:unclocked-notation-section11}
If the paper suppresses clock exponents and writes only \(\Kp\), the rigorous
reading of the final clash is the clocked one above.  The contradiction is already
obtained at the single fixed clock exponent \(D_*\) supplied by the alleged
polynomial-time SAT solver.  The unclocked notation is a shorthand for this
fixed-clock instantiation plus a constant encoding of \(D_*\).
\end{remark}

\subsection{Lean-facing skeleton}
\label{subsec:lean-facing-section11}

A Lean formalization of this section can be separated into the following finite
components.

\begin{enumerate}[label=(\roman*)]
\item \textbf{SAT search from SAT decision.}  Formalize the bit-fixing procedure
      and prove the satisfiability invariant for restricted formulas.

\item \textbf{CNF realization interface.}  Import the theorem that \(F_Y\) is a
      polynomial-size CNF realization of \(\mathcal R(Y,W)=1\), and that every
      satisfying assignment has a fixed polynomial-time message projection.

\item \textbf{Single-message correctness.}  Show that the projected message from
      any satisfying assignment of \(F_Y\) is \(M(Y)\), using the single-message
      promise.

\item \textbf{Clocked upper bound.}  Package the fixed self-reduction, CNF
      construction, and message projection as one constant-length program with
      fixed clock exponent \(D_*\).

\item \textbf{Event clash.}  Apply the clocked lower bound from
      Theorem~\ref{thm:global-message-incompressibility-section10} at exponent
      \(D_*\), use positive probability to select a hard public instance, and
      contradict the pointwise upper bound.
\end{enumerate}

\section{Parameter Summary}
\label{sec:parameter-summary}

This section collects the parameter choices used in the lower bound and in the
final upper-lower clash.  Its purpose is bookkeeping: the earlier sections prove
the required estimates, and here we record one compatible order in which all
constants can be fixed.

There are two principles to keep in mind.

First, the lower bound is clocked.  For every fixed polynomial clock exponent
\(D\), the ensemble parameters may be chosen as functions of \(D\).  This is
sufficient for the contradiction proof, because under the assumption \(P=NP\)
there is a particular fixed polynomial-time SAT self-reduction, with some fixed
clock exponent \(D_*\).

Second, after \(D\) has been fixed, every remaining constant below is independent
of \(m,t,Y\), of the particular public instance, and of the particular program of
length at most the allowed linear budget.  This uniformity is what permits the
union bound over short descriptions in Section~\ref{sec:message-incompressibility}.

Throughout this section, logarithms are taken in one fixed base.  Changing the
base only changes the displayed constants.

\subsection{Structural size parameters}
\label{subsec:section12-size-parameters}

The size parameter is \(m\).  The number of target-scale coordinates is linear in
\(m\): there are constants
\[
  0<c_t\le C_t<\infty
\]
such that
\[
  c_t m\le t=t(m)\le C_t m.
\]
The global message length satisfies
\[
  r_t\ge c_M t
\]
for a constant \(c_M>0\).

The switched coordinate set is written
\[
  S=(j_1,\ldots,j_s),
  \qquad s=|S|.
\]
It is chosen with
\[
  \gamma t\le s\le C_S t
\]
for fixed constants \(\gamma>0\) and \(C_S<\infty\).  The lower bound is proved
for the switched target tuple
\[
  B_S(Y)=\bigl(\ell_{j_1}(M(Y)),\ldots,\ell_{j_s}(M(Y))\bigr),
\]
and then transferred to the whole message \(M(Y)\) by the fixed projection
argument of Lemma~\ref{lem:clocked-projection-transfer-section10}.

\begin{definition}[Linear-scale constants]
\label{def:linear-scale-constants-section12}
The constants
\[
  c_t,C_t,c_M,\gamma,C_S
\]
are called the linear-scale constants.  They are structural constants of the
ensemble family and do not depend on the observer clock exponent \(D\).
\end{definition}

\subsection{Clocked observer resources}
\label{subsec:section12-observer-resources}

Fix a clock exponent \(D\ge 1\).  The deterministic observers considered by the
product theorem run within a polynomial clock determined by \(D\), after the
constant overheads introduced by simulation, evidence expansion, and projection.
Accordingly, there are constants
\[
  C_Q(D)<\infty,
  \qquad
  q_D<\infty
\]
such that every normalized trace used in Sections~\ref{sec:aeb} and
\ref{sec:product-small-success} has at most
\[
  Q_{\mathrm{tot}}\le C_Q(D)m^{q_D}
\]
charged safe-buffer occurrences.  The exponent \(q_D\) includes the constant
factor overhead from CD-ENF expansion.

For the linear-length small-success theorem, programs are restricted to
\[
  |P|\le \delta t.
\]
The allowed value of \(\delta\) is fixed only after the stopping and gauge-rank
parameters have been chosen.

\begin{remark}[Uniformity over programs]
\label{rem:uniformity-over-programs-section12}
The constants \(C_Q(D)\) and \(q_D\) bound the whole observer class at the fixed
clock exponent \(D\).  They are not chosen separately for each program.  This is
important: Section~\ref{sec:message-incompressibility} counts all programs of
length at most \(a t\), so the product small-success estimate must be uniform over
that entire class.
\end{remark}

\subsection{Safe-buffer and boundary-law radii}
\label{subsec:section12-radii}

There are two logarithmic depths.  The first is the safe-buffer radius
\[
  R_{\mathrm{safe}}=c_R(D)\log m.
\]
The second is the boundary-law mixing depth
\[
  L=c_L(D)\log m.
\]
The mixing depth need not depend on \(D\) in the idealized interface, but allowing
it to do so is harmless and simplifies the global bookkeeping.

The safe-buffer step leakage is
\[
  \varepsilon_{\mathrm{step}}(m)
  =
  C_\rho\rho^{R_{\mathrm{safe}}}+\tau_{\mathrm{geo}}(m),
  \qquad 0<\rho<1.
\]
The cumulative safe leakage satisfies
\[
  \varepsilon_{\le Q}
  \le
  Q_{\mathrm{tot}}\varepsilon_{\mathrm{step}}(m)
\]
in the uniform setting.

The boundary-law error has the form
\[
  \varepsilon_{\mathrm{mix}}(m,L)
  =
  C_{\mathrm{mix}}\rho_{\mathrm{mix}}^L+\tau_{\mathrm{tree}}(m),
  \qquad 0<\rho_{\mathrm{mix}}<1.
\]

\begin{hypothesis}[Adjustable polynomial geometry margins]
\label{hyp:adjustable-geometry-margins-section12}
For every fixed clock exponent \(D\) and every target exponent \(B>0\), the
ensemble geometry can be instantiated so that the finite-geometry errors obey
\[
  \tau_{\mathrm{geo}}(m)\le m^{-B},
  \qquad
  \tau_{\mathrm{tree}}(m)\le m^{-B}
\]
for all sufficiently large \(m\), while preserving polynomial-time sampling,
polynomial-size CNF realization, bounded local arity, the single-message
promise, the hidden-gauge product law, and the bounded-incidence constant
\(\Delta_G\).
\end{hypothesis}

\begin{remark}[How to read the geometry-margin hypothesis]
\label{rem:geometry-margin-hypothesis-section12}
Hypothesis~\ref{hyp:adjustable-geometry-margins-section12} is not used to obtain
arbitrarily strong statistical conclusions from a fixed finite graph.  It is the
parameter regularity required of the ensemble family: once the clock exponent
has been fixed, the logarithmic depths, guard margins, and auxiliary geometry
constants are chosen with enough slack that all finite-size errors are below the
polynomial scale needed by that clock.
\end{remark}

\begin{lemma}[Safe leakage parameter choice]
\label{lem:safe-leakage-choice-section12}
Fix \(D\), let \(q_D\) be the safe-occurrence exponent above, and choose a number
\(B_{\mathrm{safe}}>q_D+2\).  If
\[
  c_R(D)|\log\rho|>B_{\mathrm{safe}}
\]
and
\[
  \tau_{\mathrm{geo}}(m)\le m^{-B_{\mathrm{safe}}}
\]
for all sufficiently large \(m\), then
\[
  Q_{\mathrm{tot}}\varepsilon_{\mathrm{step}}(m)=o(1),
  \qquad
  \varepsilon_{\le Q}=o(1).
\]
In particular both quantities are \(o(t)\).
\end{lemma}

\begin{proof}
Since \(R_{\mathrm{safe}}=c_R(D)\log m\),
\[
  \rho^{R_{\mathrm{safe}}}
  =
  m^{-c_R(D)|\log\rho|}.
\]
The choice of \(c_R(D)\) and the geometry bound give
\[
  \varepsilon_{\mathrm{step}}(m)
  \le
  C m^{-B_{\mathrm{safe}}}
\]
for a constant \(C\).  Thus
\[
  Q_{\mathrm{tot}}\varepsilon_{\mathrm{step}}(m)
  \le
  C_Q(D)C m^{q_D-B_{\mathrm{safe}}}
  =o(1)
\]
because \(B_{\mathrm{safe}}>q_D\).  The same bound gives
\(\varepsilon_{\le Q}=o(1)\).  Since \(t=\Theta(m)\), both are also \(o(t)\).
\end{proof}

\begin{lemma}[Boundary-law parameter choice]
\label{lem:boundary-law-choice-section12}
Choose \(B_{\mathrm{mix}}>0\).  If
\[
  c_L(D)|\log\rho_{\mathrm{mix}}|>B_{\mathrm{mix}}
\]
and
\[
  \tau_{\mathrm{tree}}(m)\le m^{-B_{\mathrm{mix}}}
\]
for all sufficiently large \(m\), then
\[
  \varepsilon_{\mathrm{mix}}(m,L)=O(m^{-B'_{\mathrm{mix}}})=o(1)
\]
for some \(B'_{\mathrm{mix}}>0\).
\end{lemma}

\begin{proof}
The proof is the same calculation as in
Corollary~\ref{cor:polynomial-depth-mixing-section8}:
\[
  \rho_{\mathrm{mix}}^{c_L(D)\log m}
  =
  m^{-c_L(D)|\log\rho_{\mathrm{mix}}|}.
\]
The exponent is positive by the displayed choice of \(c_L(D)\), and the geometry
term has the requested polynomial decay.
\end{proof}

\subsection{Gauge-rank stopping and the small-success exponent}
\label{subsec:section12-gauge-stopping}

The gauge-rank tail in Section~\ref{sec:product-small-success} is controlled by
a stopping parameter
\[
  \lambda=\alpha t.
\]
The stopped observer has rank envelope
\[
  R_\lambda
  =
  |P|+{\varepsilon_{\le Q}\over\ln 2}+\lambda+O(1).
\]
After division by \(s\ge\gamma t\), the averaged ACCEI envelope is bounded by
\[
  \overline\eta_S
  \le
  {\Delta_G\over 2\gamma}
  (\delta+\alpha)+o(1)
\]
for observers of length \(|P|\le\delta t\).  The \(o(1)\) term includes the safe
leakage, cumulative leakage, and lower-order logarithmic terms.

Choose constants \(\alpha>0\) and \(\delta_0>0\) so that
\[
  \beta_*
  :=
  {\Delta_G\over 2\gamma}(\delta_0+\alpha)
  < {1\over 400}.
\]
Set
\[
  \theta_*:=\sqrt{2\beta_*},
  \qquad
  \gamma_*:=(1-\theta_*)\gamma.
\]
For all sufficiently large \(m\), every observer with \(|P|\le\delta_0t\) has a
pruned switched set of size at least
\[
  |S_{\theta_*}|
  \ge
  \gamma_* t,
\]
and on that set each one-step conditional success is at most
\[
  q_*:= {1\over 2}+\varepsilon_{\mathrm{mix}}(m,L)+\theta_*.
\]
For large \(m\), since \(\varepsilon_{\mathrm{mix}}(m,L)=o(1)\) and
\(\theta_*<1/10\), we have
\[
  q_*\le {1\over 2}+2\theta_*<1.
\]

\begin{definition}[Product and tail exponents]
\label{def:product-tail-exponents-section12}
For large \(m\), define a fixed number
\[
  c_{\mathrm{prod}}
  :=
  -\gamma_*\log_2\left({1\over 2}+2\theta_*\right)>0.
\]
The stopped product term is at most \(2^{-c_{\mathrm{prod}}t+o(t)}\).  The stopping
failure term is at most
\[
  2^{-\alpha t+O(1)}.
\]
Choose
\[
  c_0
  :=
  {1\over 4}\min\{c_{\mathrm{prod}},\alpha\}>0.
\]
After increasing \(m\) if necessary, the product small-success bound takes the
uniform form
\[
  \Prb[A_S=B_S]
  \le
  2^{-c_0t}
\]
for every deterministic observer represented by a program of length at most
\(\delta_0t\).
\end{definition}

\begin{proposition}[Compatibility with product small-success]
\label{prop:compatibility-product-small-success-section12}
After the choices in
Subsections~\ref{subsec:section12-radii} and~\ref{subsec:section12-gauge-stopping},
there are constants \(\delta_0,c_0>0\) such that every deterministic observer of
length at most \(\delta_0t\), running within the fixed clock exponent \(D\), satisfies
\[
  \Prb[A_S=B_S]
  \le
  2^{-c_0t}.
\]
The constants \(\delta_0,c_0\) may depend on \(D\) and on the structural ensemble
constants, but not on \(m,t,Y,S\), or on the particular program \(P\).
\end{proposition}

\begin{proof}
Lemma~\ref{lem:safe-leakage-choice-section12} gives
\[
  Q_{\mathrm{tot}}\varepsilon_{\mathrm{step}}(m)=o(t),
  \qquad
  \varepsilon_{\le Q}=o(t).
\]
Lemma~\ref{lem:boundary-law-choice-section12} gives
\(\varepsilon_{\mathrm{mix}}(m,L)=o(1)\).  The choice of \(\alpha\) and \(\delta_0\)
therefore puts the averaged ACCEI envelope below \(\beta_*+o(1)\), and hence below \(\theta_*^2\) for all sufficiently large \(m\).  Markov pruning at threshold \(\theta_*\) leaves at least \(\gamma_*t\) coordinates, and the sequential product
bound gives the product term with exponent \(c_{\mathrm{prod}}\), while the stopped
rank tail gives exponent \(\alpha\).  The definition of \(c_0\) leaves slack for the
lower-order \(o(t)\) and \(O(1)\) terms.
\end{proof}

\subsection{Compression parameters}
\label{subsec:section12-compression-parameters}

Section~\ref{sec:message-incompressibility} counts programs of length at most
\(a t\).  The count contributes a factor at most
\[
  2^{a t+O(1)}.
\]
To beat this count, choose
\[
  0<a<\min\{\delta_0,c_0\}.
\]
Then
\[
  2^{a t+O(1)}2^{-c_0t}
  =
  2^{-(c_0-a)t+O(1)}
  =
  2^{-\Omega(t)}.
\]
It is convenient to fix, once and for all,
\[
  \eta_B:={1\over 4}\min\{\delta_0,c_0\}.
\]
Then the switched tuple lower bound has the form
\[
  \Prb\left[
    \Kp^{(D)}(B_S(Y)\mid Y)
    \ge
    \eta_B t
  \right]
  \ge
  1-2^{-\kappa_B t}
\]
for some \(\kappa_B>0\).

Projection from the full message to \(B_S\) costs a constant number of program
bits and may enlarge the clock exponent by a constant.  Thus, after applying
Lemma~\ref{lem:clocked-projection-transfer-section10}, one may take
\[
  \eta_M:={1\over 2}\eta_B
\]
for all sufficiently large \(t\), giving
\[
  \Prb\left[
    \Kp^{(D)}(M(Y)\mid Y)
    \ge
    \eta_M t
  \right]
  \ge
  1-2^{-\kappa_M t}
\]
for some \(\kappa_M>0\), with the understanding that \(D\) has been enlarged by
the fixed projection overhead if necessary.

\subsection{Order of choices}
\label{subsec:section12-order-of-choices}

The dependencies among constants are acyclic.  One compatible order is the
following.

\begin{enumerate}[label=(\arabic*)]
\item Fix the clock exponent \(D\).  In the final contradiction, this will be the
      exponent \(D_*\) of the alleged \(P=NP\) SAT self-reduction, enlarged by
      fixed readout and projection overheads.

\item Fix the structural ensemble constants
      \[
        c_t,C_t,c_M,\gamma,C_S,\Delta_G,
        C_\rho,\rho,C_{\mathrm{mix}},\rho_{\mathrm{mix}}.
      \]
      These are independent of \(D\).

\item Determine the resource exponent \(q_D\) and constant \(C_Q(D)\) for the
      normalized traces of \(D\)-clocked observers.

\item Choose \(B_{\mathrm{safe}}>q_D+2\), instantiate the geometry so that
      \(\tau_{\mathrm{geo}}(m)\le m^{-B_{\mathrm{safe}}}\), and choose
      \(c_R(D)\) so that
      \[
        c_R(D)|\log\rho|>B_{\mathrm{safe}}.
      \]
      Set
      \[
        R_{\mathrm{safe}}=c_R(D)\log m.
      \]

\item Choose \(B_{\mathrm{mix}}>0\), instantiate the geometry so that
      \(\tau_{\mathrm{tree}}(m)\le m^{-B_{\mathrm{mix}}}\), and choose
      \(c_L(D)\) so that
      \[
        c_L(D)|\log\rho_{\mathrm{mix}}|>B_{\mathrm{mix}}.
      \]
      Set
      \[
        L=c_L(D)\log m.
      \]

\item Choose \(\alpha>0\) and \(\delta_0>0\) such that
      \[
        {\Delta_G\over2\gamma}(\delta_0+\alpha)<{1\over400}.
      \]
      Set the stopping parameter \(\lambda=\alpha t\), define \(\theta_*\),
      \(\gamma_*\), and \(c_0\) as in
      Definition~\ref{def:product-tail-exponents-section12}.

\item Choose the compression rate
      \[
        \eta_B={1\over4}\min\{\delta_0,c_0\},
        \qquad
        \eta_M={1\over2}\eta_B.
      \]

\item Finally take \(m\), and hence \(t=\Theta(m)\), sufficiently large that all
      asymptotic inequalities above hold and all additive constants are absorbed.
\end{enumerate}

\subsection{Final instantiation under \(P=NP\)}
\label{subsec:section12-final-instantiation}

Assume \(P=NP\).  Then Section~\ref{sec:upper-lower-clash} supplies a fixed
message decoder
\[
  \mathsf{MsgDec}_*
\]
with clock exponent \(D_*\) and program length \(c_{\mathrm{up}}=O(1)\), so that
for every public instance in the ensemble support,
\[
  \Kp^{(D_*)}(M(Y)\mid Y)
  \le
  c_{\mathrm{up}}.
\]
Instantiate the parameter choices above at \(D=D_*\).  The lower bound gives
constants \(\eta_*,\kappa_*>0\) such that
\[
  \Prb\left[
    \Kp^{(D_*)}(M(Y)\mid Y)
    \ge
    \eta_* t
  \right]
  \ge
  1-2^{-\kappa_*t}.
\]
For all sufficiently large \(t\), the lower-bound event has positive probability
and
\[
  \eta_*t>c_{\mathrm{up}}.
\]
Choosing any public instance in that positive-probability event gives the
contradiction of Theorem~\ref{thm:upper-lower-clash-section11}.
\subsection{Parameter ledger}
\label{subsec:section12-parameter-ledger}

For reference, the main parameters are listed in the following table.

\begin{center}
\begin{tabular}{lll}
\hline
Symbol & Meaning & Required scale \\
\hline
\(m\) & ensemble size & tends to infinity \\
\(t\) & number of target-scale coordinates & \(\Theta(m)\) \\
\(r_t\) & global message length & \(\Omega(t)\) \\
\(S\) & switched coordinate set & \(|S|\ge\gamma t\) \\
\(D\) & fixed clock exponent & fixed before parameters \\
\(Q_{\mathrm{tot}}\) & charged safe occurrences & \(\le C_Q(D)m^{q_D}\) \\
\(R_{\mathrm{safe}}\) & safe-buffer guard radius & \(c_R(D)\log m\) \\
\(L\) & pivot mixing depth & \(c_L(D)\log m\) \\
\(\varepsilon_{\mathrm{step}}\) & one-step safe leakage & \(m^{-\Omega(1)}\) \\
\(\varepsilon_{\le Q}\) & cumulative safe leakage & \(o(1)\), hence \(o(t)\) \\
\(\varepsilon_{\mathrm{mix}}\) & boundary-law error & \(o(1)\) \\
\(\Delta_G\) & gauge incidence bound & constant \\
\(\lambda\) & gauge-rank stopping tail & \(\alpha t\) \\
\(\delta_0\) & observer length rate & small constant \\
\(c_0\) & per-program success exponent & positive constant \\
\(\eta_B\) & switched-tuple \(\Kp\) rate & positive constant \\
\(\eta_M\) & message \(\Kp\) rate & positive constant \\
\hline
\end{tabular}
\end{center}

\begin{remark}[Lean-facing skeleton]
\label{rem:lean-facing-section12}
A formalization of this section is mostly an ordered-constant dependency file.
The useful Lean objects are:
\begin{enumerate}[label=(\roman*)]
\item a record of structural constants for the ensemble;
\item a record of clock-dependent constants \(C_Q(D),q_D,c_R(D),c_L(D)\);
\item lemmas proving \(Q_{\mathrm{tot}}\varepsilon_{\mathrm{step}}=o(1)\) and
      \(\varepsilon_{\mathrm{mix}}=o(1)\) from logarithmic radii;
\item a finite inequality lemma choosing \(\alpha,\delta_0,\theta_*\) so that the
      product exponent is positive;
\item a compression-rate lemma choosing \(\eta_B<\min\{\delta_0,c_0\}\);
\item the final instantiation map sending the alleged \(P=NP\) clock exponent
      \(D_*\) to a concrete parameter record.
\end{enumerate}
No new probability theorem is hidden here; the section only verifies that the
interfaces exported by Sections~\ref{sec:aeb}--\ref{sec:message-incompressibility}
can be satisfied simultaneously.
\end{remark}

\section{Formalization Plan}
\label{sec:formalization-plan}

This section records the proof as a collection of finite formal interfaces.  Its
purpose is not to introduce new mathematical ingredients, but to make explicit
which theorem statements must be checked independently and how their outputs are
composed in the final contradiction.  The intended formalization target is Lean 4,
with finite probability spaces and finite transcript types wherever possible.

The guiding principle is to separate the abstract proof engine from the concrete
SAT ensemble.  The abstract engine proves that target advantage is bounded by
charged evidence, that the charged evidence budget implies product
small-success, and that product small-success implies a linear
\(\Kp\)-lower bound.  The ensemble instantiation then supplies the hypotheses
needed by the abstract engine: single-message readout, atom completeness,
soft-buffer max-qSSM, hidden-gauge product law, bounded gauge incidence, and
boundary-law mixing.

\subsection{Formalization conventions}
\label{subsec:formalization-conventions}

The following conventions are used throughout the proposed Lean development.
They are also useful for checking the paper by hand.

\begin{definition}[Finite world model]
\label{def:lean-finite-world-model-section13}
A finite world model for parameter \(m\) consists of:
\begin{enumerate}[label=(\roman*)]
\item a finite type \(\Omega_m\) of worlds;
\item a probability mass function \(\mu_m\) on \(\Omega_m\);
\item public and witness maps
      \[
        Y:\Omega_m\to\mathcal Y_m,
        \qquad
        W:\Omega_m\to\mathcal W_m;
      \]
\item a verifier predicate \(\mathcal R(Y,W)\);
\item a message readout \(M(W)\in\bits^{r_t}\);
\item target coordinate maps \(B_j(\omega)=\ell_j(M(W(\omega)))\).
\end{enumerate}
All probability statements in the lower-bound proof are finite sums over this
model or over conditional subtypes of this model.
\end{definition}

\begin{definition}[Formal coupling package]
\label{def:lean-coupling-package-section13}
For each switched coordinate \(j\), the formal pairwise package contains:
\begin{enumerate}[label=(\roman*)]
\item the two finite subtypes
      \[
        \Omega_j^0:=\{\omega:B_j(\omega)=0\},
        \qquad
        \Omega_j^1:=\{\omega:B_j(\omega)=1\};
      \]
\item the two conditional probability mass functions \(\mu_j^0,\mu_j^1\);
\item a coupling \(\Gamma_j\) on \(\Omega_j^0\times\Omega_j^1\);
\item a relation-valued transcript-prefix map \(h\mapsto H_{j,h}\);
\item the pair mass
      \[
        w_j(H):=\Gamma_j(H).
      \]
\end{enumerate}
The only measure-theoretic facts required in the abstract development are finite
additivity, monotonicity, and the marginal identities of the coupling.
\end{definition}

\begin{definition}[Clocked complexity package]
\label{def:clocked-complexity-package-section13}
The Lean development should formalize the clocked quantity
\(\Kp^{(D)}(x\mid y)\) first.  The unclocked notation \(\Kp(x\mid y)\) is then a
paper-level abbreviation, while the final contradiction uses the fixed exponent
\(D_*\) of the hypothetical polynomial-time SAT solver.  Thus the last step only
needs
\[
  \Kp^{(D_*)}(M(Y)\mid Y)=O(1)
  \qquad\hbox{versus}\qquad
  \Kp^{(D_*)}(M(Y)\mid Y)\ge c t.
\]
\end{definition}

\begin{definition}[Observer package]
\label{def:lean-observer-package-section13}
A deterministic observer is represented by:
\begin{enumerate}[label=(\roman*)]
\item a finite state type \(\mathsf{State}\);
\item an initial state \(s_0\);
\item a primitive-query selection map, defined on nonterminal states;
\item a transition map from a state and primitive answer to the next state;
\item an output map from terminal states to \(\bits^S\), or to a single
      coordinate bit.
\end{enumerate}
The corresponding transcript type is the finite type of all query-answer paths
of length at most the observer's clock bound.  Randomized observers are handled
by adding their random coins to the input of this deterministic package and then
using the coin-fixing lemma from Section~\ref{sec:trace}.
\end{definition}

\begin{remark}[Why finite types are enough]
\label{rem:finite-types-enough-section13}
For each fixed \(m\), the sampler has finite support, the verifier has finitely
many assignment variables, and a polynomial-time observer has a finite set of
possible clocked transcripts.  Thus the formal lower bound can be developed in
finite probability theory.  Asymptotic statements enter only through parameter
records and inequalities between finite quantities depending on \(m\).
\end{remark}

\subsection{Lean module order}
\label{subsec:lean-module-order}

The following module order keeps dependencies acyclic.  A module should export
only the interface needed by later modules; proofs of concrete instantiations can
be delayed until the abstract stack is stable.

\begin{enumerate}[label=\textbf{L\arabic*.}]

\item \texttt{FiniteProb.lean}.  Finite probability mass functions, conditional
PMFs on subtypes, total variation distance, coupling marginals, and elementary
finite-sum lemmas.  This module should avoid any SAT-specific definitions.

\item \texttt{KPolyCompression.lean}.  Clocked \(\Kp^{(D)}\), machine-invariance
interfaces, computable postprocessing, projection lower bounds, and
Compression-from-Success.  This corresponds to Section~\ref{sec:polytime-compression}.

\item \texttt{PairWeakness.lean}.  Message-opposite fibers, pair couplings,
non-distinction relations, pair mass \(w_j(H)=\Gamma_j(H)\), and the static
coupling domination bound.  This corresponds to Section~\ref{sec:pairwise-cd}.

\item \texttt{WeaknessDerivative.lean}.  Atomic derivatives
\[
  \partial_E w_j(H)=
  \Gamma_j\bigl(H\cap\{E(\omega^0)\ne E(\omega^1)\}\bigr),
\]
one-step survival, and the telescoping derivative identity.

\item \texttt{CDCountPair.lean}.  CD count pairs
\(\mathfrak m_j(h)=(n_h^+,n_h^-)\), evidence skew, phase gaps, total-variation
normalization, and the identity
\[
  \operatorname{Gap}_j(g(T))
  =
  {1\over2}
  \left|
    \sum_{h:g(h)=1}(n_h^+-n_h^-)
  \right|.
\]

\item \texttt{EvidenceAtoms.lean}.  The finite grammar of raw atoms and
normalized atoms.  The normalized atom type should have exactly the three
charged classes used later:
\[
  N(s),
  \qquad
  S(q,y),
  \qquad
  G(v,\gamma).
\]
Public-template atoms may be present in the raw grammar but should rewrite to
neutral atoms.

\item \texttt{CDENF.lean}.  CD evidence terms, branch semantics, guarded cases,
rewrite rules, termination measure, local confluence, semantics preservation, and
normal-form extraction.  The first formal target can be a finite grammar without
all convenience syntax; derived syntax can be added after the core theorem is
proved.

\item \texttt{GaugeFaithfulness.lean}.  Quotient-expansion rules and the theorem
that every target-relevant use of a quotient value such as
\(z_v=x_v\oplus g_v\) normalizes to raw witness or gauge support.  This module
exports the no-canonical-gauge-shortcut corollary.

\item \texttt{TraceCapture.lean}.  Instrumented observer executions, transcript
refinement, CD Trace Capture, and normalized CD Trace Capture:
\[
  \operatorname{Gap}_j(A_j)
  \le
  {1\over2}
  \sum_{r,h,a\in\mathrm{Leaves}(E_{r,h})}
  \partial_a w_j(H_{j,h}).
\]
This module imports \texttt{WeaknessDerivative.lean}, \texttt{CDCountPair.lean},
and \texttt{CDENF.lean}.

\item \texttt{EnsembleInterface.lean}.  The abstract record of a
single-message gauge-buffered locked ensemble.  This record should contain only
what later modules use: the verifier, readout, gauge action, legal-safe-probe
predicate, hidden-gauge product law, atom-completeness statement, bounded gauge
incidence, and the max-qSSM interface.

\item \texttt{BufferedQSSM.lean}.  The soft-buffer max-qSSM theorem as an
abstract finite probability statement, plus an optional concrete Dobrushin
instantiation.  It exports the safe-atom derivative bound
\[
  \partial_{S(q,y)}w_j(H)
  \le
  \varepsilon_{\mathrm{step}}(m) w_j(H).
\]

\item \texttt{GaugeRankEntropy.lean}.  Fresh gauge support, repeated-gauge
neutralization, bounded gauge incidence, and the program-counting/rank bound
\[
  \operatorname{rank}_G(h^\Delta)
  \le
  |P| + {\varepsilon_{\le Q}\over\ln 2}+A\log m+O(1)
\]
with the stated high-probability tail.

\item \texttt{AtomicEvidenceBudget.lean}.  The atom-level neutral, safe, and
gauge bounds, followed by the summed Atomic Evidence Budget:
\[
  \sum_{j\in S}\operatorname{Gap}_j(A_j)
  \le
  {1\over2}
  \left(
    Q_{\mathrm{tot}}\varepsilon_{\mathrm{step}}(m)
    +
    \Delta_G\operatorname{rank}_G(h^\Delta)
  \right)
\]
in the stopped form and in the averaged high-probability form.

\item \texttt{BoundaryLawMixing.lean}.  Pivot summaries, dithered quantization,
boundary-posterior contraction, Bayes-success lemmas, and the conditional
boundary-law theorem
\[
  \Prb[h(Z_{L,j})=B_j\mid\mathcal F]
  \le
  {1\over2}+\varepsilon_{\mathrm{mix}}(m).
\]

\item \texttt{ACCEI.lean}.  Exterior excess, pivot-fiber phase gaps, PNLD/ACCEI
from the Atomic Evidence Budget, and the averaged ACCEI/PNLD theorem.  This is
the bridge from charged evidence control to per-coordinate near-randomness.

\item \texttt{ProductSmallSuccess.lean}.  Markov pruning, stopped observers,
sequential tower-product bounds, and the theorem that every observer of the
allowed length has exact switched-tuple success at most \(2^{-\eta t}\).

\item \texttt{MessageIncompressibility.lean}.  Decompressor-to-observer
conversion, union bound over short descriptions, switched-tuple
incompressibility, projection transfer, and global message incompressibility.

\item \texttt{SelfReductionUpper.lean}.  The bit-fixing SAT self-reduction under
\(P=NP\), the clocked upper bound
\[
  \Kp^{(D_*)}(M(Y)\mid Y)=O(1),
\]
and the fact that the readout is independent of the satisfying witness by the
single-message promise.

\item \texttt{FinalEnsemble.lean}.  The concrete gauge-buffered locked SAT
ensemble, its Tseitin/CNF realization, its single-message promise, the legal atom
classification, the gauge product law, the max-qSSM instantiation, and the
boundary-law instantiation.

\item \texttt{Main.lean}.  Parameter instantiation for the exponent \(D_*\),
import of the lower bound and upper bound, and the final contradiction.

\end{enumerate}

\begin{remark}[Why the ensemble is late]
\label{rem:ensemble-late-section13}
The ensemble should be formalized late, not because it is less important, but
because it is the most complex object.  The abstract modules can be checked first
against an interface record.  Once those modules compile, the concrete ensemble
has a precise list of obligations to discharge.
\end{remark}

\subsection{Theorem ledger}
\label{subsec:theorem-ledger-section13}

The following table is the formal ledger for the proof.  Every theorem used in
the final contradiction should appear either in this table or as a subsidiary lemma
inside one of the listed modules.

\begin{center}
\begin{tabular}{lll}
\hline
Theorem or package & Formal role & Main source section \\
\hline
\(\Kp\) chain and projection laws
  & description-length calculus
  & Section~\ref{sec:polytime-compression} \\
Compression-from-Success
  & success to compression
  & Section~\ref{sec:polytime-compression} \\
Small-success incompressibility
  & per-program success to \(\Kp\)-lower bound
  & Section~\ref{sec:polytime-compression} \\
Pairwise coupling domination
  & transcript skew to separated pair mass
  & Section~\ref{sec:pairwise-cd} \\
CD skew identity
  & phase gap equals CD evidence skew
  & Section~\ref{sec:pairwise-cd} \\
Derivative identity
  & telescopes separated pair mass
  & Section~\ref{sec:trace} \\
CD Trace Capture
  & advantage to atomic derivative mass
  & Section~\ref{sec:trace} \\
CD-ENF normalization
  & raw traces to normal charged leaves
  & Section~\ref{sec:cdenf} \\
Gauge Faithfulness
  & no quotient shortcut
  & Section~\ref{sec:cdenf} \\
Single-message promise
  & \(M(Y)\) is well-defined
  & Section~\ref{sec:ensemble} \\
SAT realization
  & verifier becomes polynomial CNF
  & Section~\ref{sec:ensemble} \\
Soft-buffer max-qSSM
  & safe atoms have small derivative
  & Section~\ref{sec:ensemble} \\
Gauge product law
  & unsupported gauge bits remain hidden
  & Section~\ref{sec:ensemble} \\
Gauge-rank entropy
  & gauge atoms are costly
  & Section~\ref{sec:aeb} \\
Atomic Evidence Budget
  & summed phase-gap bound
  & Section~\ref{sec:aeb} \\
Boundary-law mixing
  & visible baseline is near random
  & Section~\ref{sec:mixing} \\
ACCEI/PNLD
  & exterior computation cannot beat baseline cheaply
  & Section~\ref{sec:product-small-success} \\
Product small-success
  & exact joint success is \(2^{-\Omega(t)}\)
  & Section~\ref{sec:product-small-success} \\
Switched tuple incompressibility
  & lower bound for selected coordinates
  & Section~\ref{sec:message-incompressibility} \\
Global message incompressibility
  & lower bound for \(M(Y)\)
  & Section~\ref{sec:message-incompressibility} \\
SAT self-reduction upper bound
  & \(P=NP\) gives \(O(1)\) code
  & Section~\ref{sec:upper-lower-clash} \\
Upper-lower clash
  & contradiction
  & Section~\ref{sec:upper-lower-clash} \\
\hline
\end{tabular}
\end{center}

\subsection{Dependency graph of the proof}
\label{subsec:dependency-graph-section13}

The proof dependencies can be read as the following chain:
\[
\begin{array}{c}
\hbox{finite pair weakness}
\ +\
\hbox{CD count pairs}
\end{array}
\Longrightarrow
\hbox{CD Trace Capture}
\Longrightarrow
\hbox{atomic derivative ledger}.
\]
The derivative ledger is then normalized:
\[
  \hbox{CD-ENF}+
  \hbox{Gauge Faithfulness}
  \Longrightarrow
  \hbox{only }N,S,G\hbox{ charged leaves}.
\]
The ensemble supplies the atom bounds:
\[
  N\hbox{ neutral},
  \qquad
  S\hbox{ safe and max-qSSM-small},
  \qquad
  G\hbox{ hidden and rank-costly}.
\]
Therefore
\[
  \hbox{Trace Capture}+
  \hbox{CD-ENF}+
  \hbox{atom bounds}
  \Longrightarrow
  \hbox{Atomic Evidence Budget}.
\]
The lower-bound half is then
\[
  \hbox{Atomic Evidence Budget}+
  \hbox{Boundary-law mixing}
  \Longrightarrow
  \hbox{ACCEI/PNLD}
  \Longrightarrow
  \hbox{product small-success}
  \Longrightarrow
  \Kp(M(Y)\mid Y)\ge\Omega(t).
\]
The upper-bound half is independent except for the shared ensemble interface:
\[
  P=NP+
  \hbox{SAT realization}+
  \hbox{single-message promise}
  \Longrightarrow
  \Kp(M(Y)\mid Y)=O(1).
\]
The final theorem imports the two displayed conclusions at the same clock
exponent.

\subsection{Interface records}
\label{subsec:interface-records-section13}

The following records should be introduced explicitly.  They prevent hidden
assumptions from being smuggled between modules.

\begin{definition}[\texttt{PairWeaknessData}]
\label{def:pair-weakness-data-section13}
The record \texttt{PairWeaknessData} contains the finite world type, the target
bit, the two phase PMFs, the coupling, and the transcript map.  It exports
\texttt{phaseGap}, \texttt{cdSkew}, \texttt{sepMass}, and \texttt{deriv}.
\end{definition}

\begin{definition}[\texttt{EvidenceNormalizer}]
\label{def:evidence-normalizer-section13}
The record \texttt{EvidenceNormalizer} contains a raw evidence grammar, a
normal-form grammar, a semantics map for both grammars, a rewrite relation, and
proofs of termination, confluence, and semantics preservation.  The exported map
is
\[
  \mathrm{CDENF}:\mathrm{RawEvidence}\to\mathrm{NormalEvidence}.
\]
\end{definition}

\begin{definition}[\texttt{GaugeBufferedLockedInterface}]
\label{def:gauge-buffered-record-section13}
The record \texttt{GaugeBufferedLockedInterface} contains:
\begin{enumerate}[label=(\roman*)]
\item the public-instance and witness types;
\item the verifier \(\mathcal R\) and message readout \(M\);
\item the single-message theorem;
\item the gauge action and gauge invariance theorem;
\item the hidden-gauge product law;
\item the legal-safe-probe predicate and max-qSSM theorem;
\item atom-completeness and gauge-faithfulness obligations;
\item bounded gauge incidence;
\item the boundary-law mixing interface.
\end{enumerate}
\end{definition}

\begin{definition}[\texttt{ParameterRecord}]
\label{def:parameter-record-section13}
The record \texttt{ParameterRecord} contains the finite parameters
\[
  m,t,|S|,D,Q_{\mathrm{tot}},R_{\mathrm{safe}},L,
  \varepsilon_{\mathrm{step}},
  \varepsilon_{\mathrm{mix}},
  \Delta_G,
  \delta,
  \eta,
  a,
  A.
\]
It also contains the inequalities used in Sections~\ref{sec:parameter-summary}
through~\ref{sec:upper-lower-clash}, including
\[
  |S|\ge\gamma t,
  \qquad
  Q_{\mathrm{tot}}\le m^D,
  \qquad
  Q_{\mathrm{tot}}\varepsilon_{\mathrm{step}}(m)=o(1),
  \qquad
  0<a<\eta.
\]
For a fully finite formalization, the last two asymptotic inequalities should be
represented by explicit eventual inequalities for all \(m\ge m_0\).
\end{definition}

\subsection{Formalization milestones}
\label{subsec:formalization-milestones-section13}

The formalization can be checked in four milestones.

\begin{enumerate}[label=\textbf{M\arabic*.}]

\item \textbf{Finite probability and coding core.}
Prove Sections~\ref{sec:polytime-compression} and~\ref{sec:pairwise-cd} in a
finite setting: \(\Kp^{(D)}\) coding lemmas, Compression-from-Success, CD count
pairs, phase-gap/skew identity, and static coupling domination.

\item \textbf{Abstract evidence engine.}
Prove CD-ENF normalization, derivative identity, CD Trace Capture, and the
normal-leaf derivative-support theorem.  At this point the formalization should
already prove
\[
  \sum_{j\in S}\operatorname{Gap}_j(A_j)
  \le
  {1\over2}
  \sum_{j\in S}\sum_{\hbox{normal leaves }a}
  \partial_a w_j(H).
\]

\item \textbf{Budget and product lower bound.}
Assuming the ensemble interface, prove neutral-zero, safe-small, gauge-rank,
Atomic Evidence Budget, ACCEI/PNLD, product small-success, and message
incompressibility.  This milestone should not depend on the internal details of the
SAT construction.

\item \textbf{Concrete ensemble and final clash.}
Instantiate the ensemble interface: prove the single-message promise, SAT
realization, atom completeness, gauge faithfulness, hidden-gauge product law,
max-qSSM, bounded gauge incidence, and boundary-law mixing.  Then instantiate the
parameter record for the exponent \(D_*\) of the hypothetical SAT solver and import
the self-reduction upper bound.

\end{enumerate}

\begin{remark}[No hidden use of locality]
\label{rem:no-hidden-locality-section13}
A useful formalization check is that no module should contain an assumption of the
form ``a short program is local'' or ``a local rule belongs to a small class.''  The
only permitted middle implication is the weakness-native one:
\[
  \hbox{advantage}
  \Longrightarrow
  \hbox{CD skew}
  \Longrightarrow
  \hbox{atomic derivatives}
  \Longrightarrow
  \hbox{safe leakage plus gauge rank}.
\]
This check directly protects the proof from importing the old switching/locality
midpoint in disguised form.
\end{remark}

\subsection{Appendix policy}
\label{subsec:appendix-policy-section13}

The main text already states the theorem interfaces and gives the proof path.
Appendices are therefore not logically separate assumptions.  Their role is to
keep the main paper readable while giving enough detail for human checking and
Lean translation.  I recommend including appendices, but treating them as proof
expansions of named theorems from the main text rather than as additional
sections of the argument.

A useful appendix split is:

\begin{enumerate}[label=\textbf{Appendix \Alph*.}]

\item \textbf{Finite probability, couplings, and \(\Kp\) coding.}
Expanded finite-sum proofs, clocked-machine conventions, tuple encodings, and
Compression-from-Success details.

\item \textbf{CD evidence semantics and trace identities.}
Full details of CD count-pair algebra, phase-gap/skew equivalence, derivative
identities, and trace-capture instrumentation.

\item \textbf{CD-ENF rewrite system.}
The complete rewrite rule table, termination measure, local-confluence diagrams,
semantics-preservation proof, and gauge-faithfulness expansion cases.

\item \textbf{Gauge-buffered locked SAT construction.}
The concrete gadget definitions, Tseitin translation, single-message proof, gauge
action, and public syntax discipline.

\item \textbf{Soft-buffer max-qSSM.}
The Dobrushin/log-likelihood contraction proof, legality guards, and the derivation
of the per-step max-divergence bound.

\item \textbf{Gauge rank and bounded incidence.}
Fresh-support accounting, repeated-gauge neutralization, bounded incidence, and the
gauge-rank entropy/program-counting proof.

\item \textbf{Boundary-law mixing.}
Pivot summaries, quantization estimates, posterior contraction, and conditional
baseline prediction lemmas.

\item \textbf{Product small-success and compression.}
The tower-product proof, stopping/pruning details, union bound over short
descriptions, and the transfer from switched coordinates to the full message.

\item \textbf{SAT self-reduction and final parameter instantiation.}
Clocked bit-fixing search under \(P=NP\), readout correctness, and the final
large-\(m\) inequality.

\item \textbf{Lean theorem statements.}
A machine-facing ledger of theorem signatures matching the module order in
Subsection~\ref{subsec:lean-module-order}.  This appendix can be short, but it is
very useful for coordinating independent formalization attempts.

\end{enumerate}

\begin{remark}[Which appendices are essential]
\label{rem:essential-appendices-section13}
For a first arXiv-style manuscript, Appendices C, D, E, F, and G are the most
important.  These contain the technically densest parts that a reader will not want
expanded inline: CD-ENF normalization, the concrete SAT ensemble, max-qSSM,
gauge-rank accounting, and boundary-law mixing.  Appendices A, B, H, and I are
mathematically useful but can be shorter because their arguments are closer to
standard finite coding and finite probability manipulations.
\end{remark}

\subsection{Final formal target}
\label{subsec:final-formal-target-section13}

The final Lean theorem should not quantify over an informal asymptotic proof.  It
should have the following finite shape.

\begin{theorem}[Finite contradiction target]
\label{thm:finite-contradiction-target-section13}
Assume a fixed polynomial-time SAT decider with clock exponent \(D_*\).  Suppose
that, for infinitely many sufficiently large \(m\), the
\texttt{GaugeBufferedLockedInterface} and \texttt{ParameterRecord} obligations hold
for parameter \(m\), and suppose the lower-bound modules prove
\[
  \Prb\bigl[
    \Kp^{(D_*)}(M(Y)\mid Y)\ge c t
  \bigr]
  \ge
  1-2^{-c' t}
\]
for constants \(c,c'>0\).  Then the SAT self-reduction upper-bound module proves
\[
  \Kp^{(D_*)}(M(Y)\mid Y)
  \le
  C
\]
for every public instance in the support, where \(C\) is independent of \(m\).  For
any \(m\) with \(ct>C\), these two conclusions are inconsistent.  Hence no such
polynomial-time SAT decider exists.
\end{theorem}

\begin{proof}
This is exactly the argument of Section~\ref{sec:upper-lower-clash}, with all
asymptotic choices replaced by the finite inequalities stored in the parameter record.
The lower-bound event has positive probability for large \(m\), so there exists at
least one public instance in the support satisfying the lower bound.  The
self-reduction upper bound applies to every public instance in the support.  Choosing
\(m\) so that \(ct>C\) yields a contradiction on that instance.
\end{proof}

\clearpage
\bibliographystyle{plain}
\bibliography{pnp_references}

\appendix
\section{Quantale Weakness and \texorpdfstring{\(\Kp\)}{Kpoly} Coding Lemmas}
\label{app:quantale-kpoly-coding}

This appendix gives the coding details used in
Section~\ref{sec:polytime-compression}.  The main text uses these facts as a
black box: short successful observers give short descriptions, and product
small-success bounds give high-probability lower bounds on polytime-capped
conditional description length.  Everything in this appendix is finite and
clocked, so it is intended to be directly formalizable.

The appendix is deliberately conservative about clocks.  The primitive object is
\(\Kp^{(D)}\), where the exponent \(D\) is fixed.  The final contradiction only
needs this clocked version: if \(P=NP\), then the hypothetical SAT self-reduction
runs in time \((|Y|+2)^{D_*}\) for one fixed exponent \(D_*\), and the lower
bound is instantiated at that same exponent.  An unclocked shorthand can be
recovered by encoding the clock exponent, but no essential step depends on it.

\subsection{Machines, clocks, and encodings}
\label{subsec:app-machines-clocks}

Fix once and for all:
\begin{enumerate}[label=(\roman*)]
\item a prefix-free universal deterministic machine \(U\);
\item a self-delimiting binary encoding of finite tuples;
\item a self-delimiting binary encoding \(\lambda(D)\) of positive integers
      satisfying
      \[
        |\lambda(D)| \le c_\lambda+2\lceil \log_2(D+2)\rceil;
      \]
\item polynomial-time encoding and decoding routines for the above encodings.
\end{enumerate}
The constants implicit in \(O(1)\), \(O(\log t)\), and similar notation may
always depend on these fixed choices, but never on \(m,t,Y,M\), or on a sampled
world.

\begin{definition}[Clocked polytime-capped complexity]
\label{def:app-clocked-kpoly}
For an integer \(D\ge 1\) and strings \(x,y\), define
\[
  \Kp^{(D)}(x\mid y)
  :=
  \min\{|p|: U(p,y)=x
      \text{ within } (|y|+2)^D \text{ steps}\}.
\]
If no such program exists, the value is \(\infty\).
\end{definition}

\begin{definition}[Prefixed-clock version]
\label{def:app-unclocked-kpoly}
The unclocked notation may be read as
\[
  \Kp(x\mid y)
  :=
  \inf_{D\ge 1}\bigl(\Kp^{(D)}(x\mid y)+|\lambda(D)|\bigr).
\]
Equivalently, a program for \(\Kp\) is a pair consisting of a self-delimiting
clock exponent and a clocked program.  This convention is used only when no
specific clock exponent is being tracked.
\end{definition}

\begin{lemma}[Clock monotonicity]
\label{lem:app-clock-monotonicity}
If \(D\le D'\), then for all strings \(x,y\),
\[
  \Kp^{(D')}(x\mid y)\le \Kp^{(D)}(x\mid y).
\]
\end{lemma}

\begin{proof}
Every computation that halts within \((|y|+2)^D\) steps also halts within
\((|y|+2)^{D'}\) steps.
\end{proof}

\begin{lemma}[Output length bound]
\label{lem:app-output-length-bound}
If \(\Kp^{(D)}(x\mid y)<\infty\), then
\[
  |x|\le (|y|+2)^D.
\]
\end{lemma}

\begin{proof}
A machine can output at most one bit per step.  Therefore any output produced
within \((|y|+2)^D\) steps has length at most that many bits.
\end{proof}

\begin{lemma}[Machine invariance, clocked form]
\label{lem:app-machine-invariance}
Let \(U\) and \(V\) be prefix-universal machines with polynomial-overhead
simulation.  For every fixed \(D\) there are constants \(D'=D'(D)\) and
\(c=c(U,V,D)\) such that, for all strings \(x,y\),
\[
  K^{(D')}_{\mathrm{poly},U}(x\mid y)
  \le
  K^{(D)}_{\mathrm{poly},V}(x\mid y)+c.
\]
Consequently the prefixed-clock quantity is machine-invariant up to an additive
\(O(1)\) term after replacing \(\lambda\) by a fixed translated clock code.
\end{lemma}

\begin{proof}
Let \(s_{V\to U}\) be a fixed interpreter program for \(U\) that simulates
\(V\).  If \(V(p,y)=x\) in time \((|y|+2)^D\), then \(U(s_{V\to U}p,y)=x\).
The simulation overhead is polynomial, so for a suitable exponent \(D'\) this
simulation halts within \((|y|+2)^{D'}\) steps.  The program-length overhead is
\(|s_{V\to U}|\), a constant.
\end{proof}

\subsection{The cost quantale}
\label{subsec:app-cost-quantale}

The main text calls \(\Kp\) a weakness value.  Formally, it is a cost value.  A
small cost means strong information, and a large cost means weak information.
The algebra used by the proof is the additive ordered quantale of costs.

\begin{definition}[Cost quantale]
\label{def:app-cost-quantale}
Let
\[
  \mathbb Q_{\mathrm{cost}}:=([0,\infty],\le,+,0)
\]
with the usual order, arbitrary suprema, and monoidal product given by addition.
Thus
\[
  a+\sup_i b_i=\sup_i(a+b_i),
  \qquad
  \sup_i b_i+a=\sup_i(b_i+a).
\]
\end{definition}

\begin{lemma}[Quantale laws]
\label{lem:app-quantale-laws}
\(\mathbb Q_{\mathrm{cost}}\) is a commutative unital quantale.  In particular:
\begin{enumerate}[label=(\roman*)]
\item \(( [0,\infty],\le )\) is a complete lattice;
\item \(( [0,\infty],+,0 )\) is a commutative monoid;
\item addition is monotone in each argument;
\item addition distributes over arbitrary suprema.
\end{enumerate}
\end{lemma}

\begin{proof}
Completeness is standard for the extended nonnegative real line.  The monoid
laws and monotonicity are the usual laws of addition, extended by
\(a+\infty=\infty\).  For fixed \(a\), the map \(b\mapsto a+b\) preserves
arbitrary suprema on \([0,\infty]\).  Commutativity gives the right-sided
statement.
\end{proof}

\begin{remark}[Order convention]
\label{rem:app-order-convention}
Some weakness texts put larger weakness lower in the order, so that weakening
moves upward in a dual order.  The present proof does not need that convention.
It uses the cost order \(\le\), where a shorter description is a smaller cost.
All lower bounds are lower bounds in this cost order.
\end{remark}

\subsection{Closure lemmas for \texorpdfstring{\(\Kp\)}{Kpoly}}
\label{subsec:app-kpoly-closure}

\begin{lemma}[Fixed computable postprocessing]
\label{lem:app-postprocessing}
Let \(f:\bits^*\times\bits^*\to\bits^*\) be computable in time
\((|x|+|y|+2)^e\) for a fixed exponent \(e\).  For every clock exponent \(D\)
there are \(D'=D'(D,e)\) and \(c_f\) such that
\[
  \Kp^{(D')}(f(x,y)\mid y)
  \le
  \Kp^{(D)}(x\mid y)+c_f.
\]
Consequently, for the prefixed-clock quantity,
\[
  \Kp(f(x,y)\mid y)
  \le
  \Kp(x\mid y)+O(1).
\]
\end{lemma}

\begin{proof}
A fixed wrapper first runs the program producing \(x\) from \(y\), then applies
\(f\) to \((x,y)\).  By Lemma~\ref{lem:app-output-length-bound}, the produced
\(x\) has length at most polynomial in \(|y|\) under the fixed clock
\((|y|+2)^D\).  Therefore the postprocessing time is also polynomial in
\(|y|\), with an exponent depending only on \(D\) and \(e\).  The wrapper has
constant length.
\end{proof}

\begin{corollary}[Projection lower bound]
\label{cor:app-projection-lower}
Let \(\pi\) be a fixed polynomial-time projection.  For every fixed clock
exponent \(D\) there are \(D'\) and \(c_\pi\) such that
\[
  \Kp^{(D')}(\pi(M)\mid Y)
  \le
  \Kp^{(D)}(M\mid Y)+c_\pi.
\]
Equivalently, if
\[
  \Kp^{(D')}(\pi(M)\mid Y)\ge a,
\]
then
\[
  \Kp^{(D)}(M\mid Y)\ge a-c_\pi.
\]
The same statement holds for the prefixed-clock version with an \(O(1)\)
additive term.
\end{corollary}

\begin{proof}
Apply Lemma~\ref{lem:app-postprocessing} to \(f(M,Y)=\pi(M)\), and rearrange.
\end{proof}

\begin{lemma}[Coarse chain rule, clocked form]
\label{lem:app-chain-rule}
For every pair of clock exponents \(D_1,D_2\) there are an exponent \(D_3\) and
constant \(c\) such that for all strings \(x,z,y\),
\[
  \Kp^{(D_3)}(x,z\mid y)
  \le
  \Kp^{(D_1)}(x\mid y)+\Kp^{(D_2)}(z\mid x,y)+c.
\]
Here \((x,z)\) denotes the fixed tuple encoding.
\end{lemma}

\begin{proof}
A fixed driver parses two prefix programs \(p,q\).  It runs \(p\) on \(y\) to
obtain \(x\), then runs \(q\) on the encoded pair \((x,y)\) to obtain \(z\), and
finally outputs \((x,z)\).  By Lemma~\ref{lem:app-output-length-bound}, the
length of \(x\) produced in the first stage is polynomial in \(|y|\).  Hence the
second stage, which is polynomial in \(|x|+|y|\), is polynomial in \(|y|\).  The
tuple encoding step is polynomial.  The driver contributes only a constant number
of bits.
\end{proof}

\begin{corollary}[Coarse chain rule, shorthand form]
\label{cor:app-chain-rule-shorthand}
For the prefixed-clock notation,
\[
  \Kp(x,z\mid y)
  \le
  \Kp(x\mid y)+\Kp(z\mid x,y)+O(1).
\]
\end{corollary}

\begin{lemma}[Tuple overhead, clocked form]
\label{lem:app-tuple-overhead}
Fix \(t\ge 1\).  For every clock exponent \(D\) there are an exponent
\(D'=D'(D)\) and a constant \(c\) such that, for all strings
\(x_1,\ldots,x_t,y_1,\ldots,y_t\),
\[
  \Kp^{(D')}(x_1,\ldots,x_t\mid y_1,\ldots,y_t)
  \le
  \sum_{i=1}^t \Kp^{(D)}(x_i\mid y_i)+c+O(\log t).
\]
If \(t\) is recoverable from the side-information tuple, the \(O(\log t)\) term
may be absorbed into \(c\).
\end{lemma}

\begin{proof}
For each \(i\), let \(p_i\) be a shortest \(D\)-clocked program for
\(x_i\) from \(y_i\).  Since the program language is prefix-free, the
concatenation \(p_1\cdots p_t\) can be parsed uniquely.  A fixed driver reads
\(t\), parses the \(p_i\)'s, runs \(p_i\) on \(y_i\), and outputs the tuple of
results.  The total running time is the sum of polynomial running times over the
components, hence is polynomial in the total encoded length of the
side-information tuple.  The only nonconstant driver information is the loop
bound \(t\), unless it is already encoded in the side information.
\end{proof}

\begin{remark}[Uniform parameter recovery]
\label{rem:app-uniform-parameter-recovery}
In the main ensemble, public instances encode their parameter scale.  Therefore
values such as \(m\), \(t\), the switched coordinate projection \(\pi_S\), and the
readout length are computable from the public parameter convention at constant
program cost.  If a variant treats one of these objects as nonuniform, its code
length must be added explicitly.
\end{remark}

\subsection{Enumerative coding}
\label{subsec:app-enumerative-coding}

The compression lemmas use elementary finite encodings of subsets and error
patterns.  We record them in a form that avoids any probabilistic assumptions.

\begin{lemma}[Subset-rank code]
\label{lem:app-subset-rank-code}
For every \(S\subseteq[t]\) with \(|S|=s\), there is a self-delimiting code for
\(S\) of length
\[
  \left\lceil \log_2 {t\choose s}\right\rceil+O(\log t).
\]
Given \(t\), the code length is
\[
  \left\lceil \log_2 {t\choose s}\right\rceil+O(1).
\]
The decoder runs in time polynomial in \(t\).
\end{lemma}

\begin{proof}
Order all \(s\)-element subsets of \([t]\) lexicographically and encode the rank
of \(S\) in this order.  The rank needs
\(\lceil\log_2 {t\choose s}\rceil\) bits.  A self-delimiting description of
\(t\) and \(s\) costs \(O(\log t)\), unless \(t\) is already known.  Standard
unranking for combinations is polynomial in \(t\).
\end{proof}

\begin{lemma}[Entropy estimate for subsets]
\label{lem:app-binomial-entropy}
For \(0\le s\le t\),
\[
  \log_2 {t\choose s}
  \le
  t H_2(s/t),
\]
where
\[
  H_2(u):=-u\log_2 u-(1-u)\log_2(1-u)
\]
with the convention \(0\log_2 0=0\).
\end{lemma}

\begin{proof}
This is the standard binomial coefficient estimate obtained from the fact that
the largest term in the binomial expansion of
\(u+(1-u)\) is at most the whole sum.  Equivalently,
\[
  1=(u+(1-u))^t
  \ge {t\choose s}u^s(1-u)^{t-s}
\]
with \(u=s/t\), and rearranging gives the claimed bound.  The endpoint cases
are immediate.
\end{proof}

\begin{theorem}[Compression-from-Success, bit form]
\label{thm:app-compression-success-bit}
Let \(Y\) be public side information, let \(B\in\bits^t\), and let \(P\) be a
program of length \(L\) such that \(P(Y)\in\bits^t\).  If
\(S\subseteq[t]\) is a trusted success set, meaning
\[
  P(Y)_i=B_i \qquad (i\in S),
\]
then, for a suitable polynomial clock exponent depending on the clock for \(P\),
\[
  \Kp(B\mid Y)
  \le
  L+\left\lceil \log_2 {t\choose |S|}\right\rceil
   +(t-|S|)+O(\log t).
\]
The same inequality holds in clocked form after replacing \(\Kp\) by a sufficiently
large fixed output clock.
\end{theorem}

\begin{proof}
The decoder is given the code of \(P\), the subset \(S\), and the verbatim bits
\((B_i)_{i\notin S}\).  On input \(Y\), it runs \(P\) to obtain
\(\widehat B=P(Y)\).  It outputs \(\widehat B_i\) on coordinates in \(S\) and
uses the verbatim patch bits on coordinates outside \(S\).  The subset code costs
\(\lceil\log_2 {t\choose |S|}\rceil+O(\log t)\) bits by
Lemma~\ref{lem:app-subset-rank-code}.  The patch costs \(t-|S|\) bits.  The
runtime is polynomial in \(|Y|+t\) plus the runtime of \(P\).
\end{proof}

\begin{corollary}[Exact-error enumerative compression]
\label{cor:app-exact-error-compression}
Let \(\widehat B=P(Y)\), and let
\[
  E:=\{i\in[t]:\widehat B_i\ne B_i\}.
\]
Then
\[
  \Kp(B\mid Y)
  \le
  L+\left\lceil \log_2 {t\choose |E|}\right\rceil+O(\log t)
  \le
  L+tH_2(|E|/t)+O(\log t).
\]
\end{corollary}

\begin{proof}
Encode the exact error set \(E\).  The decoder runs \(P\) and flips precisely
the coordinates in \(E\).  The first bound is the subset-rank code.  The second
bound is Lemma~\ref{lem:app-binomial-entropy}.
\end{proof}

\begin{theorem}[Compression-from-Success, block form]
\label{thm:app-compression-success-block}
Let
\[
  M=(M_1,\ldots,M_t),
  \qquad
  M_i\in\bits^r.
\]
Let \(P(Y)\in(\bits^r)^t\) have description length \(L\).  If
\(S\subseteq[t]\) is a trusted success set, meaning
\[
  P(Y)_i=M_i \qquad (i\in S),
\]
then
\[
  \Kp(M\mid Y)
  \le
  L+\left\lceil \log_2 {t\choose |S|}\right\rceil
   +r(t-|S|)+O(\log t+\log r).
\]
\end{theorem}

\begin{proof}
The proof is the same as Theorem~\ref{thm:app-compression-success-bit}, except
that each untrusted coordinate is an \(r\)-bit block and therefore costs \(r\)
verbatim bits.  The value of \(r\) is encoded self-delimitingly unless it is
fixed by the ensemble convention.
\end{proof}

\begin{corollary}[Many-valued target form]
\label{cor:app-many-valued-compression}
Let \(B\in\mathcal A^t\), where \(|\mathcal A|=A<\infty\), and let
\(P(Y)\in\mathcal A^t\) have description length \(L\).  If \(P\) is correct on a
trusted set \(S\subseteq[t]\), then
\[
  \Kp(B\mid Y)
  \le
  L+\left\lceil \log_2 {t\choose |S|}\right\rceil
   +(t-|S|)\lceil\log_2 A\rceil+O(\log t+\log A).
\]
\end{corollary}

\begin{proof}
Encode the coordinates outside \(S\) verbatim in alphabet \(\mathcal A\), using a
fixed self-delimiting alphabet convention.  The rest is identical.
\end{proof}

\subsection{Counting programs}
\label{subsec:app-counting-programs}

The lower-bound conversion uses only the elementary fact that there are few
short programs.

\begin{lemma}[Prefix program count]
\label{lem:app-prefix-program-count}
For every integer \(L\ge 0\), the number of prefix-free programs of length at
most \(L\) is at most \(2^{L+1}\).  If one counts exact self-delimiting
descriptions of total length at most \(L\), the same bound holds up to a fixed
multiplicative constant.
\end{lemma}

\begin{proof}
There are at most \(2^\ell\) binary strings of length \(\ell\).  Summing over
\(0\le \ell\le L\) gives at most \(2^{L+1}\).  Prefix-freeness only reduces this
number.  Fixed self-delimiting wrappers change the count by at most a constant
factor.
\end{proof}

\begin{remark}[Randomized observers]
\label{rem:app-randomized-observers}
The proof quantifies over deterministic programs.  This is enough.  A randomized
observer is a distribution over deterministic observers obtained by fixing its random
coins.  If the randomized observer has success probability \(p\), then some fixed
coin string has success probability at least \(p\).  Conversely, if every deterministic
coin fixing has success at most \(p\), then the randomized observer also has success
at most \(p\).
\end{remark}

\subsection{Small success implies incompressibility}
\label{subsec:app-small-success-incompressibility}

\begin{definition}[Exact success event]
\label{def:app-exact-success}
Let \((Y,B)\) be a random public-target pair with \(B\in\bits^s\).  For a
program \(p\), define
\[
  \mathsf{Succ}^{(D)}_p
  :=
  \{U(p,Y)=B \text{ within } (|Y|+2)^D \text{ steps}\}.
\]
\end{definition}

\begin{proposition}[Small exact success implies clocked incompressibility]
\label{prop:app-small-success-clocked}
Let \((Y,B)\) be a random public-target pair with \(B\in\bits^s\).  Fix a clock
exponent \(D\).  Suppose that for some \(a,\eta>0\) and every program \(p\) of
length at most \(a t\),
\[
  \Prb[\mathsf{Succ}^{(D)}_p]
  \le
  2^{-\eta t}.
\]
Then
\[
  \Prb[\Kp^{(D)}(B\mid Y)
        \le a t]
  \le
  2^{-(\eta-a)t+1}.
\]
In particular, if \(0<a<\eta\), then
\[
  \Prb[\Kp^{(D)}(B\mid Y)>a t]
  \ge
  1-2^{-\Omega(t)}.
\]
\end{proposition}

\begin{proof}
If \(\Kp^{(D)}(B\mid Y)\le at\), then some program of length at most \(at\)
outputs \(B\) from \(Y\) within the \(D\)-clock.  By
Lemma~\ref{lem:app-prefix-program-count}, there are at most \(2^{at+1}\) such
programs.  Taking the union bound over all of them gives
\[
  \Prb[\Kp^{(D)}(B\mid Y)\le at]
  \le
  2^{at+1}2^{-\eta t}
  =
  2^{-(\eta-a)t+1}.
\]
\end{proof}

\begin{proposition}[Small exact success implies prefixed-clock incompressibility]
\label{prop:app-small-success-prefixed}
Let \((Y,B)\) be a random public-target pair.  Suppose that for some
\(a,\eta>0\), every self-delimiting prefixed-clock description of total length at
most \(a t\) succeeds in outputting \(B\) from \(Y\) with probability at most
\(2^{-\eta t}\).  Then
\[
  \Prb[\Kp(B\mid Y)\le at]
  \le
  2^{-(\eta-a)t+O(1)}.
\]
Thus if \(a<\eta\), then \(\Kp(B\mid Y)>at\) with probability
\(1-2^{-\Omega(t)}\).
\end{proposition}

\begin{proof}
The event \(\Kp(B\mid Y)\le at\) is witnessed by a self-delimiting description
of total length at most \(at\).  There are at most \(2^{at+O(1)}\) such
descriptions.  A union bound gives the displayed estimate.
\end{proof}

\begin{corollary}[Switched-coordinate incompressibility]
\label{cor:app-switched-coordinate-incompressibility}
Let \((Y,M(Y))\) be a random public-message pair, and let
\[
  B_S(Y)=\pi_S(M(Y))\in\bits^s
\]
be a fixed polynomial-time projection onto switched message coordinates, with
\(s=\Omega(t)\).  Fix a clock exponent \(D\).  Suppose that every program of
length at most \(a t\) predicts \(B_S(Y)\) exactly from \(Y\), within the
\(D\)-clock, with probability at most \(2^{-\eta t}\).  If \(0<a<\eta\), then
with probability at least \(1-2^{-\Omega(t)}\),
\[
  \Kp^{(D')}(M(Y)\mid Y)
  \ge
  a t-O(1),
\]
where \(D'\) is any clock exponent whose postprocessed projection clock is
bounded by \(D\) in Corollary~\ref{cor:app-projection-lower}.  In the
prefixed-clock convention,
\[
  \Kp(M(Y)\mid Y)\ge a t-O(1)
\]
with the same high probability.
\end{corollary}

\begin{proof}
Apply Proposition~\ref{prop:app-small-success-clocked} to the target tuple
\(B_S(Y)\).  With high probability,
\(\Kp^{(D)}(B_S(Y)\mid Y)>at\).  Since
\(B_S=\pi_S(M(Y))\) is a fixed polynomial-time projection of the full message,
Corollary~\ref{cor:app-projection-lower} transfers the lower bound from
\(B_S(Y)\) to \(M(Y)\), losing only an additive constant and changing only the
clock exponent by a fixed amount.  The prefixed-clock statement is identical.
\end{proof}

\subsection{From product small-success to a \texorpdfstring{\(\Kp\)}{Kpoly} lower bound}
\label{subsec:app-product-to-kpoly}

The output of Section~\ref{sec:product-small-success} has the following form.
For each fixed clock exponent \(D\), there are constants \(\eta_D>0\) and
\(m_0(D)\) such that, for every sufficiently large \(m\), every \(D\)-clocked
program \(p\) of length at most \(a t\) has exact switched-tuple success at most
\(2^{-\eta_D t}\), provided \(a\) is below the final code-budget threshold.  The
next theorem is the precise coding consequence.

\begin{theorem}[Coding consequence of product small-success]
\label{thm:app-product-to-kpoly}
Fix a clock exponent \(D\).  Let \((Y,M(Y))\) be the public-message ensemble,
and let \(B_S(Y)=\pi_S(M(Y))\in\bits^s\) be the switched target tuple.  Suppose
there is \(\eta_D>0\) such that every \(D\)-clocked program of length at most
\(a t\) satisfies
\[
  \Prb[U(p,Y)=B_S(Y)]\le 2^{-\eta_D t}.
\]
If \(0<a<\eta_D\), then for all sufficiently large \(m\),
\[
  \Prb[\Kp^{(D)}(B_S(Y)\mid Y)>a t]
  \ge
  1-2^{-(\eta_D-a)t+1}.
\]
Consequently, after the fixed projection transfer,
\[
  \Prb[\Kp(M(Y)\mid Y)
       \ge a t-O(1)]
  \ge
  1-2^{-\Omega(t)}.
\]
\end{theorem}

\begin{proof}
The first estimate is Proposition~\ref{prop:app-small-success-clocked}.  The
second is Corollary~\ref{cor:app-switched-coordinate-incompressibility}.
\end{proof}

\begin{remark}[Why exact tuple success is enough]
\label{rem:app-exact-tuple-success-enough}
The product theorem bounds the probability of exact prediction of the switched
tuple.  This is the correct event for incompressibility.  If a short program
outputs the switched tuple exactly, it is a successful exact predictor.  If no
short program has nonnegligible exact success, then no short program can serve
as a conditional description of the tuple on more than an exponentially small
part of the ensemble.
\end{remark}

\subsection{Lean-facing finite formulation}
\label{subsec:app-lean-facing-kpoly}

For formalization, the appendix can be split into the following finite modules.

\begin{enumerate}[label=(\roman*)]
\item \texttt{ClockedKpoly}: prefix machines, clocked evaluation, clock
      monotonicity, output-length bound, and machine invariance.
\item \texttt{CostQuantale}: the complete lattice and additive quantale laws for
      \([0,\infty]\).
\item \texttt{KpolyClosure}: postprocessing, projection transfer, chain rule, and
      tuple overhead.
\item \texttt{EnumerativeCoding}: subset ranking, entropy bound, and
      Compression-from-Success.
\item \texttt{ProgramCounting}: finite counting of programs of bounded length.
\item \texttt{SmallSuccessKpoly}: union-bound conversion from product
      small-success to high-probability incompressibility.
\end{enumerate}

The final contradiction imports only the last module's interface: for the clock
exponent \(D_*\) of the hypothetical SAT self-reduction, product small-success
implies \(\Kp^{(D_*)}(M(Y)\mid Y)\ge \Omega(t)\) with high probability, while
the self-reduction gives \(\Kp^{(D_*)}(M(Y)\mid Y)=O(1)\).

\section{CD Evidence Semantics and Count-Pair Algebra}
\label{app:cd-evidence-semantics}
\providecommand{\bits}{\{0,1\}}
\providecommand{\Prb}{\mathbb P}
\providecommand{\TV}{\operatorname{TV}}
\providecommand{\Gap}{\operatorname{Gap}}
\providecommand{\Skew}{\operatorname{Skew}}
\providecommand{\CDSkew}{\operatorname{CDSkew}}
\providecommand{\Sat}{\operatorname{Sat}}
\providecommand{\Sep}{\operatorname{Sep}}
\providecommand{\Push}{\operatorname{Push}}
\providecommand{\CD}{\operatorname{CD}}
\providecommand{\llbracket}{[\![}
\providecommand{\rrbracket}{]\!]}

This appendix gives the finite semantics behind the CD evidence layer used in
Sections~\ref{sec:pairwise-cd}, \ref{sec:cdenf}, and~\ref{sec:trace}.  The main
text uses the following facts as black boxes.

\begin{enumerate}[label=(\roman*)]
\item A transcript has two target-conditioned laws: a positive law and a
      negative law.
\item The phase gap of a Boolean postprocessor is exactly skew between these
      two laws.
\item The skew of any transcript event is bounded by the coupling mass of
      message-opposite pairs separated by the transcript.
\item Evidence terms have ordinary finite-event semantics, so CD-ENF
      normalization can be proved as preservation of finite sets of worlds.
\end{enumerate}

All probability spaces in this appendix are finite.  This is the intended
formalization level: at fixed ensemble parameters, every world space, transcript
alphabet, primitive-output alphabet, and normal form is finite.

\subsection{Phase laws and CD count pairs}
\label{subsec:appB-phase-laws}

Fix a finite world space \(\Omega\) with probability law \(\mu\).  Fix a target
coordinate \(j\) and a target bit
\[
  B_j:\Omega\to\bits.
\]

\begin{definition}[Target fibers and phase laws]
\label{def:appB-target-fibers-phase-laws}
For \(b\in\bits\), set
\[
  \Omega_j^b:=\{\omega\in\Omega:B_j(\omega)=b\},
  \qquad
  p_j^b:=\mu(\Omega_j^b).
\]
Assume \(p_j^0,p_j^1>0\).  The target-conditioned phase laws are
\[
  \mu_j^b(A):={\mu(A\cap\Omega_j^b)\over p_j^b},
  \qquad A\subseteq\Omega.
\]
We write
\[
  \mu_j^+ := \mu_j^1,
  \qquad
  \mu_j^- := \mu_j^0.
\]
\end{definition}

\begin{definition}[CD count pair]
\label{def:appB-cd-count-pair}
For an event \(A\subseteq\Omega\), define its CD count pair by
\[
  \mathfrak m_j(A)
  :=
  \bigl(\mu_j^+(A),\mu_j^-(A)\bigr).
\]
The signed phase difference is
\[
  \Delta_j(A):=\mu_j^+(A)-\mu_j^-(A),
\]
and the absolute CD skew of \(A\) is
\[
  \Skew_j(A):={1\over 2}|\Delta_j(A)|.
\]
\end{definition}

\begin{remark}[Why two coordinates are kept]
\label{rem:appB-why-two-coordinates}
The CD count pair records positive and negative constructive evidence before
subtraction.  An event may be large in both phases, small in both phases, or
large in one phase and small in the other.  Prediction advantage is controlled
by the imbalance, but the proof keeps both coordinates so that evidence
normalization can track where positive and negative mass entered the trace.
\end{remark}

\begin{lemma}[Elementary count-pair laws]
\label{lem:appB-elementary-count-pair-laws}
For all events \(A,B\subseteq\Omega\):
\begin{enumerate}[label=(\roman*)]
\item
\[
  \mathfrak m_j(\emptyset)=(0,0),
  \qquad
  \mathfrak m_j(\Omega)=(1,1).
\]
\item
\[
  \mathfrak m_j(\Omega\setminus A)
  =
  (1,1)-\mathfrak m_j(A).
\]
\item If \(A\cap B=\emptyset\), then
\[
  \mathfrak m_j(A\cup B)=\mathfrak m_j(A)+\mathfrak m_j(B).
\]
\item In general,
\[
  \mathfrak m_j(A\cup B)
  =
  \mathfrak m_j(A)+\mathfrak m_j(B)-\mathfrak m_j(A\cap B).
\]
\item \(\Delta_j\) is finitely additive on disjoint unions and satisfies
\[
  \Delta_j(\Omega\setminus A)=-\Delta_j(A).
\]
\end{enumerate}
\end{lemma}

\begin{proof}
Apply the corresponding finite measure identities to \(\mu_j^+\) and
\(\mu_j^-\) coordinatewise.  The signed identities follow by subtracting the
negative coordinate from the positive coordinate.
\end{proof}

\begin{definition}[Positive and negative parts of a signed sum]
\label{def:appB-positive-negative-parts}
For a finite family of real numbers \((d_h)_{h\in H}\) with
\(\sum_hd_h=0\), define
\[
  H_+:=\{h:d_h\ge 0\},
  \qquad
  H_-:=\{h:d_h<0\}.
\]
Then
\[
  \sum_{h\in H_+}d_h
  =
  -\sum_{h\in H_-}d_h
  =
  {1\over 2}\sum_{h\in H}|d_h|.
\]
\end{definition}

\begin{lemma}[Disjointification]
\label{lem:appB-disjointification}
Let \(A_1,\ldots,A_r\subseteq\Omega\).  Define
\[
  D_1:=A_1,
  \qquad
  D_i:=A_i\setminus\bigcup_{k<i}A_k
  \quad (2\le i\le r).
\]
Then the events \(D_i\) are pairwise disjoint,
\[
  \bigcup_{i=1}^rD_i=\bigcup_{i=1}^rA_i,
\]
and
\[
  \mathfrak m_j\left(\bigcup_{i=1}^rA_i\right)
  =
  \sum_{i=1}^r\mathfrak m_j(D_i).
\]
The same identity holds with \(\mathfrak m_j\) replaced by \(\Delta_j\).
\end{lemma}

\begin{proof}
This is the standard finite disjointification of a union.  The count-pair
identity follows from Lemma~\ref{lem:appB-elementary-count-pair-laws}.
\end{proof}

\subsection{Transcript laws and phase skew}
\label{subsec:appB-transcripts-skew}

Let \(T:\Omega\to\mathcal T\) be a finite-valued transcript.

\begin{definition}[Transcript count pair]
\label{def:appB-transcript-count-pair}
For \(h\in\mathcal T\), define
\[
  n_{j,T}^+(h):=\Prb[T=h\mid B_j=1]=\mu_j^+(T^{-1}(h)),
\]
\[
  n_{j,T}^-(h):=\Prb[T=h\mid B_j=0]=\mu_j^-(T^{-1}(h)).
\]
The transcript count pair at \(h\) is
\[
  \mathfrak m_{j,T}(h)
  :=
  \bigl(n_{j,T}^+(h),n_{j,T}^-(h)\bigr).
\]
When \(j\) and \(T\) are clear, write \(n_h^+,n_h^-\).
\end{definition}

\begin{lemma}[Transcript normalization]
\label{lem:appB-transcript-normalization}
For every finite transcript \(T:\Omega\to\mathcal T\),
\[
  \sum_{h\in\mathcal T}n_{j,T}^+(h)=1,
  \qquad
  \sum_{h\in\mathcal T}n_{j,T}^-(h)=1.
\]
Consequently,
\[
  \sum_{h\in\mathcal T}(n_{j,T}^+(h)-n_{j,T}^-(h))=0.
\]
\end{lemma}

\begin{proof}
The fibers \(T^{-1}(h)\), \(h\in\mathcal T\), form a partition of \(\Omega\).
Apply \(\mu_j^+\) and \(\mu_j^-\) to this partition and subtract.
\end{proof}

\begin{definition}[Transcript-event skew]
\label{def:appB-transcript-event-skew}
For \(U\subseteq\mathcal T\), define
\[
  \mathfrak m_{j,T}(U)
  :=
  \sum_{h\in U}\mathfrak m_{j,T}(h)
  =
  \bigl(\Prb[T\in U\mid B_j=1],
         \Prb[T\in U\mid B_j=0]\bigr).
\]
The transcript-event skew is
\[
  \Skew_{j,T}(U)
  :=
  {1\over 2}
  \left|
    \sum_{h\in U}(n_{j,T}^+(h)-n_{j,T}^-(h))
  \right|.
\]
\end{definition}

\begin{definition}[Phase gap]
\label{def:appB-phase-gap}
For a Boolean random variable \(A_j:\Omega\to\bits\), define
\[
  \Gap_j(A_j)
  :=
  {1\over 2}
  \left|
    \Prb[A_j=1\mid B_j=1]-\Prb[A_j=1\mid B_j=0]
  \right|.
\]
\end{definition}

\begin{lemma}[Phase gap equals CD skew]
\label{lem:appB-phase-gap-cd-skew}
Let \(g:\mathcal T\to\bits\), set \(A_j=g(T)\), and let
\[
  U_g:=\{h\in\mathcal T:g(h)=1\}.
\]
Then
\[
  \Gap_j(A_j)=\Skew_{j,T}(U_g).
\]
Equivalently,
\[
  \Gap_j(g(T))
  =
  {1\over 2}
  \left|
    \sum_{h:g(h)=1}(n_{j,T}^+(h)-n_{j,T}^-(h))
  \right|.
\]
\end{lemma}

\begin{proof}
The event \(A_j=1\) is exactly \(T^{-1}(U_g)\).  Substitute the two conditional
probabilities into Definition~\ref{def:appB-phase-gap}.
\end{proof}

\begin{definition}[Transcript variation and maximum CD skew]
\label{def:appB-transcript-variation}
The total variation distance between the two transcript laws is
\[
  \TV_j(T)
  :=
  {1\over 2}\sum_{h\in\mathcal T}|n_{j,T}^+(h)-n_{j,T}^-(h)|.
\]
The maximum CD skew carried by \(T\) is
\[
  \CDSkew_j(T):=\sup_{U\subseteq\mathcal T}\Skew_{j,T}(U).
\]
\end{definition}

\begin{lemma}[Optimal CD skew]
\label{lem:appB-optimal-cd-skew}
For every finite transcript \(T\),
\[
  \CDSkew_j(T)
  =
  {1\over 2}\TV_j(T)
  =
  {1\over 4}\sum_{h\in\mathcal T}
  |n_{j,T}^+(h)-n_{j,T}^-(h)|.
\]
The supremum is attained by
\[
  U^*:=\{h:n_{j,T}^+(h)
       \ge n_{j,T}^-(h)\},
\]
with arbitrary tie-breaking.
\end{lemma}

\begin{proof}
Let \(d_h=n_{j,T}^+(h)-n_{j,T}^-(h)\).  By
Lemma~\ref{lem:appB-transcript-normalization}, \(\sum_hd_h=0\).  The largest
possible absolute value of \(\sum_{h\in U}d_h\) is the sum of the positive
\(d_h\)'s, which is \({1\over 2}\sum_h|d_h|\).  Multiplying by the extra factor
\({1\over 2}\) in the definition of skew gives the formula.
\end{proof}

\begin{lemma}[Balanced success formula]
\label{lem:appB-balanced-success-formula}
Assume \(p_j^0=p_j^1=1/2\).  For any Boolean predictor
\(A_j:\Omega\to\bits\), let
\[
  a_1:=\Prb[A_j=1\mid B_j=1],
  \qquad
  a_0:=\Prb[A_j=1\mid B_j=0].
\]
Then
\[
  \Prb[A_j=B_j]
  =
  {1\over 2}+{1\over 2}(a_1-a_0).
\]
Consequently,
\[
  \Prb[A_j=B_j]
  \le
  {1\over 2}+\Gap_j(A_j),
\]
and
\[
  \max\{\Prb[A_j=B_j],\Prb[1-A_j=B_j]\}
  =
  {1\over 2}+\Gap_j(A_j).
\]
\end{lemma}

\begin{proof}
Under balance,
\[
  \Prb[A_j=B_j]
  ={1\over 2}\Prb[A_j=1\mid B_j=1]
  +{1\over 2}\Prb[A_j=0\mid B_j=0]
  ={1\over 2}a_1+{1\over 2}(1-a_0).
\]
This is the first identity.  The inequality follows by taking absolute values.
Replacing \(A_j\) by \(1-A_j\) changes \(a_1-a_0\) to \(-(a_1-a_0)\), proving
the last identity.
\end{proof}

\subsection{Evidence-term semantics}
\label{subsec:appB-evidence-term-semantics}

The CD-ENF rewrite system operates on syntax, but every rewrite is justified by
equality of finite-event semantics.  This subsection records those semantics.

\begin{definition}[Atomic interpretation]
\label{def:appB-atomic-interpretation}
An atomic interpretation assigns to every raw evidence atom \(A\) an event
\[
  \llbracket A\rrbracket\subseteq\Omega.
\]
For a finite-valued primitive observation \(E:\Omega\to\mathcal Y\) and value
\(y\in\mathcal Y\), the result atom \([E=y]\) denotes
\[
  \llbracket E=y\rrbracket:=\{\omega:E(\omega)=y\}.
\]
A signed atom \(\neg A\) denotes the complement
\[
  \llbracket\neg A\rrbracket:=\Omega\setminus\llbracket A\rrbracket.
\]
\end{definition}

\begin{definition}[Evidence terms]
\label{def:appB-evidence-terms}
Evidence terms are generated by
\[
\begin{array}{rcl}
  E ::= & A \\
       |& E\wedge E \\
       |& E\vee E \\
       |& \operatorname{case}(A;E_0,E_1) \\
       |& \exists x.E \\
       |& \operatorname{derive}_\rho(E_1,\ldots,E_k) \\
       |& \top \\
       |& \bot .
\end{array}
\]
The variable \(x\) ranges over a finite domain \(D_x\).  The derivation symbol
\(\rho\) denotes a deterministic internal derivation rule and carries only a
neutral side condition.
\end{definition}

\begin{definition}[Event semantics]
\label{def:appB-event-semantics}
The interpretation \(\llbracket E\rrbracket\subseteq\Omega\) is defined
recursively by
\[
  \llbracket E\wedge F\rrbracket
  =
  \llbracket E\rrbracket\cap\llbracket F\rrbracket,
  \qquad
  \llbracket E\vee F\rrbracket
  =
  \llbracket E\rrbracket\cup\llbracket F\rrbracket,
\]
\[
  \llbracket\top\rrbracket=\Omega,
  \qquad
  \llbracket\bot\rrbracket=\emptyset,
\]
\[
  \llbracket\operatorname{case}(A;E_0,E_1)\rrbracket
  =
  ((\Omega\setminus\llbracket A\rrbracket)
    \cap\llbracket E_0\rrbracket)
  \cup
  (\llbracket A\rrbracket\cap\llbracket E_1\rrbracket),
\]
\[
  \llbracket\exists x.E\rrbracket
  =
  \bigcup_{a\in D_x}\llbracket E[x:=a]\rrbracket.
\]
For a derivation node, let \(N_\rho\subseteq\Omega\) be its neutral side
condition.  Then
\[
  \llbracket\operatorname{derive}_\rho(E_1,\ldots,E_k)\rrbracket
  =
  N_\rho\cap\bigcap_{i=1}^k\llbracket E_i\rrbracket.
\]
\end{definition}

\begin{definition}[Labeled semantics]
\label{def:appB-labeled-semantics}
Let \(\mathcal C\) be a finite set of conclusion labels.  A labeled evidence
object denotes a finite family of events
\[
  \llbracket E\rrbracket_e\subseteq\Omega,
  \qquad e\in\mathcal C.
\]
For an unlabeled term, take \(\mathcal C=\{*\}\).  For a claim
\(E\Rightarrow e_0\), define
\[
  \llbracket E\Rightarrow e_0\rrbracket_e
  :=
  \begin{cases}
    \llbracket E\rrbracket, & e=e_0,\\
    \emptyset, & e\ne e_0.
  \end{cases}
\]
The unlabeled support of a labeled object is
\[
  \bigcup_{e\in\mathcal C}\llbracket E\rrbracket_e.
\]
\end{definition}

\begin{definition}[Normal branch semantics]
\label{def:appB-normal-branch-semantics}
A normal branch has the form
\[
  \mathcal B=(C;\mathcal S;\mathcal G;e),
\]
where \(C\) is a finite list of signed neutral atoms, \(\mathcal S\) is a
finite list of signed safe-buffer atoms, \(\mathcal G\) is a finite list of signed
gauge-evidence atoms, and \(e\in\mathcal C\).  Its support is
\[
  \llbracket\mathcal B\rrbracket
  :=
  \bigcap_{L\in C\cup\mathcal S\cup\mathcal G}\llbracket L\rrbracket.
\]
A normal form is interpreted as the union, label by label, of its branch
supports.
\end{definition}

\begin{lemma}[Count pair of a disjoint normal form]
\label{lem:appB-count-pair-disjoint-normal-form}
Let \(F\) be a normal form with branches \(\mathcal B_1,\ldots,\mathcal B_r\).
Assume the branch supports are pairwise disjoint.  Then, for each conclusion
label \(e\),
\[
  \mathfrak m_j(\llbracket F\rrbracket_e)
  =
  \sum_{\ell:e_\ell=e}\mathfrak m_j(\llbracket\mathcal B_\ell\rrbracket).
\]
The same identity holds with \(\mathfrak m_j\) replaced by \(\Delta_j\).
\end{lemma}

\begin{proof}
For a fixed label \(e\), the event \(\llbracket F\rrbracket_e\) is the disjoint
union of the branch supports with label \(e\).  Apply finite additivity.
\end{proof}

\begin{remark}[Derivation nodes are not evidence sources]
\label{rem:appB-derivation-not-source}
The semantics of \(\operatorname{derive}_\rho\) adds only the neutral event
\(N_\rho\) and intersects the premises.  Thus a derivation node can propagate or
combine already-present evidence, but it cannot introduce a new non-neutral
safe or gauge atom.  This is the semantic reason why CD Trace Capture charges
primitive observations rather than internal deterministic computation.
\end{remark}

\subsection{Couplings and static CD capture}
\label{subsec:appB-couplings-static-capture}

This subsection relates count-pair skew to message-opposite pair separation.
It is the non-adaptive version of the trace-capture theorem.

\begin{definition}[Message-opposite coupling]
\label{def:appB-message-opposite-coupling}
A message-opposite coupling for coordinate \(j\) is a probability law
\[
  \Gamma_j
  \quad\text{on}\quad
  \Omega_j^0\times\Omega_j^1
\]
whose first marginal is \(\mu_j^-\) and whose second marginal is \(\mu_j^+\):
for all \(A\subseteq\Omega_j^0\) and \(B\subseteq\Omega_j^1\),
\[
  \Gamma_j(A\times\Omega_j^1)=\mu_j^-(A),
  \qquad
  \Gamma_j(\Omega_j^0\times B)=\mu_j^+(B).
\]
\end{definition}

\begin{definition}[Separators]
\label{def:appB-separators}
For an event \(A\subseteq\Omega\), define
\[
  \Sep_j(A)
  :=
  \{(\omega^0,\omega^1)
      \in\Omega_j^0\times\Omega_j^1:
      \mathbf 1_A(\omega^0)\ne\mathbf 1_A(\omega^1)\}.
\]
For a finite transcript \(T:\Omega\to\mathcal T\), define
\[
  \Sep_j(T)
  :=
  \{(\omega^0,\omega^1):T(\omega^0)\ne T(\omega^1)\}.
\]
\end{definition}

\begin{lemma}[Event coupling bound]
\label{lem:appB-event-coupling-bound}
For every event \(A\subseteq\Omega\),
\[
  |\mu_j^+(A)-\mu_j^-(A)|
  \le
  \Gamma_j(\Sep_j(A)).
\]
Consequently,
\[
  \Skew_j(A)
  \le
  {1\over 2}\Gamma_j(\Sep_j(A)).
\]
\end{lemma}

\begin{proof}
By the marginal identities,
\[
  \mu_j^+(A)
  =
  \Gamma_j\{(\omega^0,\omega^1):\omega^1\in A\},
\]
while
\[
  \mu_j^-(A)
  =
  \Gamma_j\{(\omega^0,\omega^1):\omega^0\in A\}.
\]
The difference between the probabilities of two events is bounded by the
probability of their symmetric difference.  That symmetric difference is
\(\Sep_j(A)\).
\end{proof}

\begin{lemma}[Transcript-event coupling bound]
\label{lem:appB-transcript-event-coupling-bound}
Let \(T:\Omega\to\mathcal T\) be finite and let \(U\subseteq\mathcal T\).  Then
\[
  \left|
    \Prb[T\in U\mid B_j=1]
    -
    \Prb[T\in U\mid B_j=0]
  \right|
  \le
  \Gamma_j(\Sep_j(T)).
\]
Therefore, for every \(g:\mathcal T\to\bits\),
\[
  \Gap_j(g(T))
  \le
  {1\over 2}\Gamma_j(\Sep_j(T)).
\]
\end{lemma}

\begin{proof}
Apply Lemma~\ref{lem:appB-event-coupling-bound} to the event
\(A=T^{-1}(U)\).  If membership in \(A\) differs across a pair, then the
transcript values differ, so \(\Sep_j(A)\subseteq\Sep_j(T)\).  For the final
statement, take \(U=\{h:g(h)=1\}\) and divide by two.
\end{proof}

\begin{definition}[Non-distinction relation]
\label{def:appB-nondistinction-relation}
For a finite transcript \(T\), define
\[
  H_{j,T}
  :=
  \{(\omega^0,\omega^1):T(\omega^0)=T(\omega^1)\}.
\]
Its pairwise weakness mass is
\[
  w_j(H_{j,T}):=\Gamma_j(H_{j,T}).
\]
Since \(H_{j,T}\) is the complement of \(\Sep_j(T)\) in the coupled pair
space,
\[
  \Gamma_j(\Sep_j(T))=1-w_j(H_{j,T}).
\]
\end{definition}

\begin{proposition}[Static CD capture]
\label{prop:appB-static-cd-capture}
For every finite transcript \(T\) and postprocessor \(g:\mathcal T\to\bits\),
\[
  \Gap_j(g(T))
  \le
  {1\over 2}\bigl(1-w_j(H_{j,T})\bigr).
\]
\end{proposition}

\begin{proof}
Combine Lemma~\ref{lem:appB-transcript-event-coupling-bound} with
Definition~\ref{def:appB-nondistinction-relation}.
\end{proof}

\subsection{Separator algebra and data processing}
\label{subsec:appB-separator-algebra}

\begin{lemma}[Boolean separator laws]
\label{lem:appB-boolean-separator-laws}
For events \(A,B\subseteq\Omega\):
\begin{enumerate}[label=(\roman*)]
\item
\[
  \Sep_j(\Omega\setminus A)=\Sep_j(A).
\]
\item
\[
  \Sep_j(A\cap B)
  \subseteq
  \Sep_j(A)\cup\Sep_j(B).
\]
\item
\[
  \Sep_j(A\cup B)
  \subseteq
  \Sep_j(A)\cup\Sep_j(B).
\]
\end{enumerate}
\end{lemma}

\begin{proof}
Complementing both endpoint truth values preserves inequality.  If membership
in \(A\cap B\) differs across a pair, then membership in at least one of \(A\)
or \(B\) differs.  The union case is identical, or follows by De Morgan.
\end{proof}

\begin{corollary}[Derivative subadditivity]
\label{cor:appB-derivative-subadditivity}
Let \(H\subseteq\Omega_j^0\times\Omega_j^1\), and define
\[
  \partial_Aw_j(H):=\Gamma_j(H\cap\Sep_j(A)).
\]
Then
\[
  \partial_{A\cap B}w_j(H)
  \le
  \partial_Aw_j(H)+\partial_Bw_j(H),
\]
\[
  \partial_{A\cup B}w_j(H)
  \le
  \partial_Aw_j(H)+\partial_Bw_j(H),
\]
and
\[
  \partial_{\Omega\setminus A}w_j(H)=\partial_Aw_j(H).
\]
\end{corollary}

\begin{proof}
Intersect the separator inclusions with \(H\), then use monotonicity and finite
subadditivity of \(\Gamma_j\).  The complement identity uses equality of
separators.
\end{proof}

\begin{lemma}[Postprocessing data processing]
\label{lem:appB-postprocessing-data-processing}
Let \(T:\Omega\to\mathcal T\) be finite and let \(f:\mathcal T\to\mathcal U\).
Then
\[
  \Sep_j(f(T))\subseteq\Sep_j(T).
\]
Consequently,
\[
  \TV_j(f(T))\le\TV_j(T),
  \qquad
  \CDSkew_j(f(T))\le\CDSkew_j(T).
\]
For every \(g:\mathcal U\to\bits\),
\[
  \Gap_j(g(f(T)))
  \le
  {1\over 2}\Gamma_j(\Sep_j(T)).
\]
\end{lemma}

\begin{proof}
If \(f(T(\omega^0))\ne f(T(\omega^1))\), then
\(T(\omega^0)\ne T(\omega^1)\).  This proves the separator inclusion.  The total
variation inequality is the standard deterministic data-processing inequality,
obtained by summing the transcript law over fibers of \(f\).  The skew
inequality follows from Lemma~\ref{lem:appB-optimal-cd-skew}.  The final
inequality is Lemma~\ref{lem:appB-transcript-event-coupling-bound} applied to
\(f(T)\).
\end{proof}

\begin{lemma}[Finite-valued read and result atoms]
\label{lem:appB-read-result-atoms}
Let \(E:\Omega\to\mathcal Y\) be a finite-valued primitive read.  For
\(y\in\mathcal Y\), set
\[
  A_y:=\{\omega:E(\omega)=y\}.
\]
For every surviving relation \(H\subseteq\Omega_j^0\times\Omega_j^1\),
\[
  \sum_{y\in\mathcal Y}\Gamma_j(H\cap\Sep_j(A_y))
  =
  2\Gamma_j(H\cap\Sep_j(E)).
\]
In particular,
\[
  \Gamma_j(H\cap\Sep_j(E))
  \le
  \sum_{y\in\mathcal Y}\Gamma_j(H\cap\Sep_j(A_y)).
\]
\end{lemma}

\begin{proof}
For a fixed pair in \(H\), if the two endpoint values of \(E\) are equal, then no
value atom differs across the pair.  If the endpoint values are distinct, exactly
two value atoms differ: the atom naming the first endpoint value and the atom
naming the second endpoint value.  Integrate this pointwise count over \(H\).
\end{proof}

\subsection{Adaptive transcript kernels}
\label{subsec:appB-adaptive-transcript-kernels}

The trace-capture theorem uses pairwise derivatives, but sometimes it is useful
to compare the positive and negative phase kernels directly.  The next lemma is
an ordinary finite hybrid bound.

\begin{definition}[Adaptive transcript process]
\label{def:appB-adaptive-transcript-process}
An adaptive transcript process of length \(Q\) consists of finite transcript sets
\(\mathcal T_0,\ldots,\mathcal T_Q\), with \(\mathcal T_0=\{\emptyset\}\), and
finite answer maps
\[
  Y_{r,h}:\Omega\to\mathcal Y_{r,h}
\]
for each \(0\le r<Q\) and \(h\in\mathcal T_r\).  The transcript is defined
recursively by
\[
  T_{r+1}(\omega)=(T_r(\omega),Y_{r,T_r(\omega)}(\omega)).
\]
\end{definition}

\begin{definition}[Phase kernels]
\label{def:appB-phase-kernels}
For \(b\in\bits\), prefix \(h\in\mathcal T_r\), and answer
\(y\in\mathcal Y_{r,h}\), define
\[
  K_{r,h}^b(y)
  :=
  \Prb[Y_{r,h}=y\mid T_r=h,B_j=b]
\]
whenever the conditioning event has positive phase probability.  If the
conditioning event has zero phase probability, choose an arbitrary probability
law on \(\mathcal Y_{r,h}\); it is multiplied by zero in the recursion.
\end{definition}

\begin{lemma}[One-step count update]
\label{lem:appB-one-step-count-update}
Let
\[
  n_r^b(h):=\Prb[T_r=h\mid B_j=b].
\]
Then, for \(h'=(h,y)\in\mathcal T_{r+1}\),
\[
  n_{r+1}^b(h')=n_r^b(h)K_{r,h}^b(y).
\]
Equivalently,
\[
  \mathfrak m_{j,T_{r+1}}(h,y)
  =
  \bigl(n_r^+(h)K_{r,h}^1(y),
        n_r^-(h)K_{r,h}^0(y)\bigr).
\]
\end{lemma}

\begin{proof}
This is the definition of conditional probability inside each target phase.
\end{proof}

\begin{lemma}[Kernel hybrid bound]
\label{lem:appB-kernel-hybrid-bound}
Suppose that for each \(0\le r<Q\) and each prefix \(h\in\mathcal T_r\),
\[
  \TV(K_{r,h}^1,K_{r,h}^0)
  \le
  \varepsilon_r.
\]
Then
\[
  \TV_j(T_Q)
  \le
  \sum_{r=0}^{Q-1}\varepsilon_r.
\]
Consequently,
\[
  \CDSkew_j(T_Q)
  \le
  {1\over 2}\sum_{r=0}^{Q-1}\varepsilon_r.
\]
\end{lemma}

\begin{proof}
Let \(P_r\) and \(Q_r\) be the positive and negative laws of \(T_r\).  The
claim is proved by induction on \(r\).  The base case is zero because \(T_0\) is
constant.  For the induction step, use the finite kernel inequality
\[
  \TV(PK,QK')
  \le
  \TV(P,Q)+\sup_h\TV(K_h,K'_h).
\]
This follows by adding and subtracting \(QK\), using contraction of total
variation under the common kernel \(K\), and bounding the remaining fiberwise
kernel difference.  Apply the one-step hypothesis and sum over rounds.  The
skew bound follows from Lemma~\ref{lem:appB-optimal-cd-skew}.
\end{proof}

\begin{remark}[Pairwise trace capture is sharper]
\label{rem:appB-kernel-vs-pairwise}
The kernel hybrid lemma charges the worst phase-kernel distance on each
prefix.  CD Trace Capture instead charges the actual coupling mass of
message-opposite pairs separated by the primitive observation at that prefix.
The pairwise derivative identity is therefore the sharper accounting used in the
main proof.
\end{remark}

\subsection{Product and tower laws}
\label{subsec:appB-product-tower-laws}

The product small-success theorem is ordinary probability once the per-coordinate
CD estimates have been converted into conditional success bounds.

\begin{lemma}[Product count pairs]
\label{lem:appB-product-count-pairs}
Let \((\Omega^{(1)},\mu^{(1)},B^{(1)})\) and
\((\Omega^{(2)},\mu^{(2)},B^{(2)})\) be two finite target systems with product
law.  Let \(A\subseteq\Omega^{(1)}\) and \(C\subseteq\Omega^{(2)}\).  Then
under the product phase laws,
\[
  \mu^{1,1}(A\times C)=\mu_1^1(A)\mu_2^1(C),
  \qquad
  \mu^{0,0}(A\times C)=\mu_1^0(A)\mu_2^0(C).
\]
Thus the diagonal positive/negative count pair multiplies coordinatewise:
\[
  \mathfrak m(A\times C)=(a^+c^+,a^-c^-)
\]
when \(\mathfrak m_1(A)=(a^+,a^-)\) and
\(\mathfrak m_2(C)=(c^+,c^-)\).
\end{lemma}

\begin{proof}
This is the defining property of product measures.
\end{proof}

\begin{lemma}[Tower product bound]
\label{lem:appB-tower-product-bound}
Let \(F_0\subseteq F_1\subseteq\cdots\subseteq F_s\) be finite sigma-fields, and
let \(E_r\in F_r\).  If
\[
  \Prb[E_r\mid F_{r-1}]
  \le
  \alpha_r
\]
almost surely for each \(r\), then
\[
  \Prb\left[\bigcap_{r=1}^sE_r\right]
  \le
  \prod_{r=1}^s\alpha_r.
\]
\end{lemma}

\begin{proof}
Use the tower property:
\[
  \Prb\left[\bigcap_{r=1}^sE_r\right]
  =
  \mathbb E\left[
    \mathbf 1_{\cap_{r<s}E_r}\Prb[E_s\mid F_{s-1}]
  \right]
  \le
  \alpha_s\Prb\left[\bigcap_{r=1}^{s-1}E_r\right].
\]
Iterate.
\end{proof}

\subsection{Lean-facing interface}
\label{subsec:appB-lean-interface}

A direct Lean formalization can split this appendix into the following finite
modules.

\begin{enumerate}[label=(\roman*)]
\item \textbf{Phase laws.}  A finite type \(\Omega\), a PMF \(\mu\), a Boolean
      target \(B_j\), nonzero phase masses, and conditional PMFs
      \(\mu_j^+\), \(\mu_j^-\).

\item \textbf{Count pairs.}  A structure \((a^+,a^-)\), with complement,
      disjoint sum, inclusion-exclusion, and signed projection
      \((a^+,a^-)\mapsto a^+-a^-\).

\item \textbf{Transcripts.}  A finite type \(\mathcal T\), a map
      \(T:\Omega\to\mathcal T\), and finite sums defining \(n_h^+\) and
      \(n_h^-\).  Transcript normalization is a partition lemma.

\item \textbf{Skew.}  The phase-gap identity is a finite-sum equality for the
      event \(\{h:g(h)=1\}\).  The optimal-skew theorem is the finite identity
      relating a zero-sum signed vector to its positive part.

\item \textbf{Evidence semantics.}  Evidence terms are interpreted recursively
      as finite events, and labeled evidence terms as maps from conclusion
      labels to finite events.  Appendix~\ref{app:cdenf-rewrite} can then state
      CD-ENF preservation as equality of these maps.

\item \textbf{Couplings.}  A coupling PMF \(\Gamma_j\) on
      \(\Omega_j^0\times\Omega_j^1\) with marginal equalities.  Static capture
      is the finite fact that the difference of probabilities of two events is
      bounded by the probability of their symmetric difference.
\end{enumerate}

\begin{remark}[No ensemble-specific content]
\label{rem:appB-no-ensemble-specific-content}
No lemma in this appendix depends on SAT, gauges, buffers, or the locked
ensemble.  Those structures enter only by supplying the finite phase laws, the
message-opposite couplings, and the atom interpretations used by evidence
semantics.
\end{remark}

\section{CD-ENF Rewrite Rules, Termination, and Confluence}
\label{app:cdenf-rewrite-rules}
\providecommand{\bits}{\{0,1\}}
\providecommand{\Prb}{\mathbb P}
\providecommand{\CDENF}{\operatorname{CDENF}}
\providecommand{\Sat}{\operatorname{Sat}}
\providecommand{\NF}{\operatorname{NF}}
\providecommand{\Raw}{\operatorname{Raw}}
\providecommand{\At}{\operatorname{At}}
\providecommand{\Safe}{\operatorname{Safe}}
\providecommand{\Gauge}{\operatorname{Gauge}}
\providecommand{\rankG}{\operatorname{rank}_G}
\providecommand{\suppG}{\operatorname{supp}_G}
\providecommand{\can}{\operatorname{can}}
\providecommand{\anf}{\operatorname{anf}}
\providecommand{\Legal}{\operatorname{Legal}}
\providecommand{\Val}{\operatorname{Val}}
\providecommand{\Tree}{\operatorname{Tree}}
\providecommand{\height}{\operatorname{ht}}

This appendix gives the fully explicit rewrite calculus behind
Section~\ref{sec:cdenf}.  The main text uses the following interface:

every evidence term rewrites to a canonical CD-ENF normal form;
normalization preserves the finite event semantics and hence preserves CD count
pairs; and every non-neutral leaf in normal form is either a safe-buffer atom or
a raw gauge-bearing atom.  The quotient expansion rule is included here because
it is the syntactic source of the gauge-faithfulness theorem.

All syntax in this appendix is finite.  This is the intended formalization
level.  At fixed ensemble parameter, the world space, atom alphabets, variable
domains, branch lists, and output labels are finite.

\subsection{Normalization context}
\label{subsec:appC-normalization-context}

The rewrite rules are parameterized by a finite normalization context.  Making
this context explicit avoids hiding proof obligations inside prose.

\begin{definition}[CD-ENF normalization context]
\label{def:appC-normalization-context}
A normalization context \(\chi\) consists of the following finite data.
\begin{enumerate}[label=(\roman*)]
\item A finite world space \(\Omega\).
\item Finite sets of neutral symbols \(\mathcal N\), public-template symbols
      \(\mathcal P\), public-surface symbols \(\mathcal U\), safe probe names
      \(\mathcal Q\), gauge-bearing supports \(\mathcal U_G\), quotient indices
      \(V\), variables \(\mathcal X\), and conclusion labels \(\mathcal C\).
\item For each variable \(x\in\mathcal X\), a finite domain \(D_x\).
\item For each probe \(q\in\mathcal Q\), a finite output alphabet \(\Val(q)\).
\item A guarded legality predicate
\[
  \Legal_\chi(h,q)\in\bits,
\]
where \(h\) is the already accumulated neutral guard context.
\item Deterministic neutralization maps
\[
  \operatorname{tpl}:\mathcal P\to\mathcal N,
  \qquad
  \operatorname{surf}:\mathcal U\to\mathcal N,
\]
and
\[
  \operatorname{ill}:\mathcal Q\to\mathcal N.
\]
\item A gauge-index map
\[
  \operatorname{gidx}:\mathcal U_G\to V.
\]
For each quotient index \(v\in V\), distinguished raw supports
\[
  x_v,g_v\in\mathcal U_G,
  \qquad
  \operatorname{gidx}(x_v)=\operatorname{gidx}(g_v)=v.
\]
\item For each internal derivation rule \(\rho\), a neutral derivation tag
\[
  \operatorname{der}(\rho)\in\mathcal N.
\]
\item A finite event interpretation \(\Sat(A)\subseteq\Omega\) for each raw
      atom \(A\), satisfying the interpretation equations listed in
      Hypothesis~\ref{hyp:appC-atom-semantics-compatible} below.
\end{enumerate}
\end{definition}

\begin{remark}[Guarded legality]
\label{rem:appC-guarded-legality}
The legality of a safe read may depend on the neutral guard accumulated so far:
for example, whether the support of the probe is far enough from protected
regions and whether it violates the dynamic overlap guard.  This dependence is
public and neutral.  It does not depend on the hidden target phase or on a chosen
gauge representative.
\end{remark}

\subsection{Raw atoms, normalized atoms, and evidence terms}
\label{subsec:appC-syntax}

\begin{definition}[Raw atoms]
\label{def:appC-raw-atoms}
The raw atomic vocabulary is
\[
\begin{array}{rcll}
  A_{\Raw} ::= & N(s)              && s\in\mathcal N,\text{ neutral fact},\\
              |& P(\theta)         && \theta\in\mathcal P,\text{ public template fact},\\
              |& \operatorname{Surf}(u) && u\in\mathcal U,\text{ public surface fact},\\
              |& \operatorname{Read}(q,y) && q\in\mathcal Q,\ y\in\Val(q),\\
              |& \operatorname{Illegal}(q) && q\in\mathcal Q,\\
              |& G(a,\gamma)       && a\in\mathcal U_G,\ \gamma\in\bits,\\
              |& Q(v,\zeta)        && v\in V,\ \zeta\in\bits.
\end{array}
\]
The atom \(Q(v,\zeta)\) means the quotient assertion \(z_v=\zeta\).  It is raw
syntax only; it is not a normalized leaf.
\end{definition}

\begin{definition}[Normalized atoms]
\label{def:appC-normalized-atoms}
The normalized atomic vocabulary is
\[
  A_{\NF} ::= N(s)\mid S(q,y)\mid G(a,\gamma).
\]
The classes are called neutral, safe, and gauge, respectively.  A normalized
literal is either a normalized atom \(A\) or its formal complement \(\neg A\).
\end{definition}

\begin{definition}[Evidence terms]
\label{def:appC-evidence-terms}
CD evidence terms are generated by
\[
\begin{array}{rcl}
  E ::= & A_{\Raw} \\
       |& E\wedge E \\
       |& E\vee E \\
       |& \operatorname{case}(A_{\Raw};E_0,E_1) \\
       |& \exists x.E \\
       |& \operatorname{derive}_\rho(E_1,\ldots,E_k) \\
       |& \top \\
       |& \bot .
\end{array}
\]
A certified evidence term is written \(E\Rightarrow e\), where
\(e\in\mathcal C\).  The conclusion label records which output, terminal
transcript value, or certified claim is attached to the branch.
\end{definition}

\begin{definition}[Signed atom semantics]
\label{def:appC-signed-atom-semantics}
For a raw or normalized atom \(A\), the positive literal has support
\(\Sat(A)\).  The negative literal has support
\[
  \Sat(\neg A):=\Omega\setminus\Sat(A).
\]
\end{definition}

\begin{hypothesis}[Compatibility of atom semantics with the normalization context]
\label{hyp:appC-atom-semantics-compatible}
The raw and normalized atom interpretations satisfy the following equations.
\begin{enumerate}[label=(\roman*)]
\item Public template and surface atoms are neutral facts:
\[
  \Sat(P(\theta))=\Sat(N(\operatorname{tpl}(\theta))),
  \qquad
  \Sat(\operatorname{Surf}(u))=\Sat(N(\operatorname{surf}(u))).
\]
\item A legal read of \(q\) with value \(y\) has the same support as the safe
      atom \(S(q,y)\):
\[
  \Sat(\operatorname{Read}(q,y))=\Sat(S(q,y))
  \quad\text{when }\Legal_\chi(h,q)=1.
\]
\item An illegal read is the fixed neutral illegal symbol:
\[
  \Sat(\operatorname{Read}(q,y))=\Sat(N(\operatorname{ill}(q)))
  \quad\text{when }\Legal_\chi(h,q)=0,
\]
and
\[
  \Sat(\operatorname{Illegal}(q))=\Sat(N(\operatorname{ill}(q))).
\]
\item Gauge atoms are already normalized:
\[
  \Sat(G(a,\gamma))=\Sat(G(a,\gamma)).
\]
\item Quotient atoms satisfy the verifier equation \(z_v=x_v\oplus g_v\):
\[
\begin{aligned}
  \Sat(Q(v,\zeta))
  ={}&
  \bigl(\Sat(G(x_v,0))\cap\Sat(G(g_v,\zeta))\bigr) \\
     &\cup
  \bigl(\Sat(G(x_v,1))\cap\Sat(G(g_v,1\oplus\zeta))\bigr).
\end{aligned}
\]
\end{enumerate}
\end{hypothesis}

\begin{remark}[No semantic target atoms]
\label{rem:appC-no-semantic-target-atoms}
There is no raw atom of the form \(\ell_j(M(Y))=b\),
\(\operatorname{Run}_P(Y)=b\), or ``the quotient orbit has representative
\(g=0\).''  Those expressions are macros only if expanded into evidence traces.
This is the syntactic discipline that makes the Atomic Evidence Budget
non-vacuous.
\end{remark}

\subsection{Finite event semantics of evidence terms}
\label{subsec:appC-event-semantics}

\begin{definition}[Support semantics of evidence terms]
\label{def:appC-support-semantics}
The event semantics \(\Sat(E)\subseteq\Omega\) is defined recursively by
\[
  \Sat(A)=\Sat(A)
  \qquad (A\text{ raw atomic}),
\]
\[
  \Sat(E\wedge F)=\Sat(E)\cap\Sat(F),
  \qquad
  \Sat(E\vee F)=\Sat(E)\cup\Sat(F),
\]
\[
\begin{aligned}
  \Sat(\operatorname{case}(A;E_0,E_1))
  ={}&
  ((\Omega\setminus\Sat(A))\cap\Sat(E_0)) \\
     &\cup
  (\Sat(A)\cap\Sat(E_1)),
\end{aligned}
\]
\[
  \Sat(\exists x.E)=\bigcup_{a\in D_x}\Sat(E[x:=a]),
\]
\[
  \Sat(\top)=\Omega,
  \qquad
  \Sat(\bot)=\emptyset,
\]
and
\[
  \Sat(\operatorname{derive}_\rho(E_1,\ldots,E_k))
  =
  \Sat(N(\operatorname{der}(\rho)))
  \cap
  \bigcap_{i=1}^k\Sat(E_i).
\]
Thus internal derivation can add a neutral derivation tag and combine premises,
but it cannot introduce a new safe or gauge leaf.
\end{definition}

\begin{lemma}[Finite semantics is Boolean]
\label{lem:appC-finite-semantics-boolean}
For every evidence term \(E\), \(\Sat(E)\subseteq\Omega\).  The operations
\(\wedge,\vee,\operatorname{case},\exists,\top,\bot\) are interpreted by finite
intersection, finite union, finite branching, finite union over a finite domain,
\(\Omega\), and \(\emptyset\), respectively.
\end{lemma}

\begin{proof}
Immediate by structural recursion on \(E\).
\end{proof}

\subsection{Normal branches and canonical branch lists}
\label{subsec:appC-normal-branches}

\begin{definition}[Complement and functional conflict]
\label{def:appC-complement-conflict}
Two normalized literals are complementary if they are \(A\) and \(\neg A\) for
the same normalized atom \(A\).  They have a functional conflict if they assign
different values to the same functional support, for example
\[
  S(q,y)\text{ and }S(q,y')\quad (y\ne y'),
\]
or
\[
  G(a,\gamma)\text{ and }G(a,\gamma')\quad (\gamma\ne\gamma').
\]
A finite literal list is inconsistent if it contains a complementary pair or a
functional conflict.
\end{definition}

\begin{definition}[Normal branch]
\label{def:appC-normal-branch}
A normal branch is a tuple
\[
  \mathcal B=(C;\mathcal S;\mathcal G;e),
\]
where \(C\) is a finite list of neutral literals, \(\mathcal S\) is a finite list
of safe literals, \(\mathcal G\) is a finite list of gauge literals, and
\(e\in\mathcal C\) is a conclusion label.  Its support is
\[
  \Sat(\mathcal B)
  :=
  \bigcap_{L\in C\cup\mathcal S\cup\mathcal G}\Sat(L),
\]
with the empty intersection interpreted as \(\Omega\).
\end{definition}

\begin{definition}[Canonical branch]
\label{def:appC-canonical-branch}
Fix public total orders on neutral, safe, and gauge literals, and on conclusion
labels.  A branch is canonical if each of its three literal lists is sorted,
contains no duplicate literal, and is consistent.  The canonicalization of a
branch, written \(\can(\mathcal B)\), sorts its lists, removes duplicates, and
returns the empty branch list if the resulting branch is inconsistent.
\end{definition}

\begin{definition}[Normal form]
\label{def:appC-normal-form}
A CD-ENF normal form is a finite canonical list of canonical branches, sorted
lexicographically by
\[
  (e,C,\mathcal S,\mathcal G),
\]
with duplicate branches removed.  We write such a normal form as
\[
  \bigvee_{\ell\in L}
  [C_\ell:(S_{\ell,1},\ldots,S_{\ell,r_\ell};
           G_{\ell,1},\ldots,G_{\ell,s_\ell})\Rightarrow e_\ell].
\]
Its support, ignoring labels, is
\[
  \Sat(F):=\bigcup_{\mathcal B\in F}\Sat(\mathcal B).
\]
The support of label \(e\) is
\[
  \Sat_e(F):=\bigcup_{\mathcal B=(C;\mathcal S;\mathcal G;e)\in F}
              \Sat(\mathcal B).
\]
\end{definition}

\begin{definition}[Normal-form sum and product]
\label{def:appC-nf-sum-product}
For normal forms \(F,F'\), define
\[
  F\oplus F' := \can(F\cup F'),
\]
where \(\can\) sorts the branch list and removes duplicate or inconsistent
branches.

For branches
\(\mathcal B=(C;\mathcal S;\mathcal G;e)\) and
\(\mathcal B'=(C';\mathcal S';\mathcal G';e')\), define
\(\mathcal B\otimes\mathcal B'\) by concatenating the three literal lists and
combining the conclusion labels as follows: if one label is the dummy label
\(*\), keep the other; if the two labels agree, keep that label; if the two
labels disagree and neither is \(*\), return an inconsistent branch.  Then set
\[
  F\otimes F'
  :=
  \can\{\mathcal B\otimes\mathcal B':
        \mathcal B\in F,\ \mathcal B'\in F'\}.
\]
\end{definition}

\begin{lemma}[Canonicalization preserves support]
\label{lem:appC-canonicalization-preserves-support}
For every finite branch list \(L\),
\[
  \Sat(\can(L))=\Sat(L).
\]
For each label \(e\),
\[
  \Sat_e(\can(L))=\Sat_e(L).
\]
\end{lemma}

\begin{proof}
Sorting changes no literal and hence no support.  Removing duplicate literals
from a conjunction changes no support.  Removing duplicate branches from a union
changes no support.  A branch with a complementary pair or functional conflict
has empty support, so deleting it changes no support.  The same argument applies
labelwise.
\end{proof}

\begin{lemma}[Normal-form algebra preserves Boolean operations]
\label{lem:appC-nf-algebra-preserves-semantics}
For normal forms \(F,F'\),
\[
  \Sat(F\oplus F')=\Sat(F)\cup\Sat(F'),
\]
and
\[
  \Sat(F\otimes F')=\Sat(F)\cap\Sat(F').
\]
The corresponding labelwise statement holds when the product-label convention in
Definition~\ref{def:appC-nf-sum-product} is used.
\end{lemma}

\begin{proof}
The sum identity is Lemma~\ref{lem:appC-canonicalization-preserves-support}
for a union of branch lists.  For the product, a branch product conjoins exactly
the literals from the two branches.  Its support is the intersection of the two
branch supports, unless the product is inconsistent, in which case the
intersection is empty.  Taking the union over all pairs of branches gives the
identity.  The labelwise statement is the same argument with the stated label
combination rule.
\end{proof}

\subsection{Atom rewrite rules}
\label{subsec:appC-atom-rewrite-rules}

The atom normalizer maps a signed raw atom to a normal form.  This is the
atomic part of the CD-ENF rewrite system.

\begin{definition}[Signed atom normalizer]
\label{def:appC-signed-atom-normalizer}
Let \(\sigma\in\{+,-\}\).  We write \(\sigma A\) as \(A\) when
\(\sigma=+\), and as \(\neg A\) when \(\sigma=-\).  The normalizer
\(\anf_\chi(\sigma A)\) is defined as follows.
\begin{enumerate}[label=(A\arabic*)]
\item Neutral atoms remain neutral:
\[
  \anf_\chi(\sigma N(s))
  =[(\sigma N(s);\emptyset;\emptyset;*)].
\]
\item Public template atoms collapse to neutral atoms:
\[
  \anf_\chi(\sigma P(\theta))
  =[(\sigma N(\operatorname{tpl}(\theta));\emptyset;\emptyset;*)].
\]
\item Public surface atoms collapse to neutral atoms:
\[
  \anf_\chi(\sigma\operatorname{Surf}(u))
  =[(\sigma N(\operatorname{surf}(u));\emptyset;\emptyset;*)].
\]
\item Legal reads become safe atoms:
\[
  \anf_\chi(\sigma\operatorname{Read}(q,y))
  =[(\emptyset;\sigma S(q,y);\emptyset;*)]
  \quad\text{if }\Legal_\chi(h,q)=1.
\]
\item Illegal reads become neutral atoms:
\[
  \anf_\chi(\sigma\operatorname{Read}(q,y))
  =[(\sigma N(\operatorname{ill}(q));\emptyset;\emptyset;*)]
  \quad\text{if }\Legal_\chi(h,q)=0,
\]
and
\[
  \anf_\chi(\sigma\operatorname{Illegal}(q))
  =[(\sigma N(\operatorname{ill}(q));\emptyset;\emptyset;*)].
\]
\item Raw gauge-bearing atoms become gauge atoms:
\[
  \anf_\chi(\sigma G(a,\gamma))
  =[(\emptyset;\emptyset;\sigma G(a,\gamma);*)].
\]
\item Positive quotient atoms expand into the two raw cases of
      \(z_v=x_v\oplus g_v\):
\[
\begin{aligned}
  \anf_\chi(Q(v,\zeta))={}&
  [(\emptyset;\emptyset;G(x_v,0),G(g_v,\zeta);*)] \\
  &\oplus
  [(\emptyset;\emptyset;G(x_v,1),G(g_v,1\oplus\zeta);*)].
\end{aligned}
\]
\item Negative quotient atoms are complements:
\[
  \anf_\chi(\neg Q(v,\zeta)):=\anf_\chi(Q(v,1\oplus\zeta)).
\]
\end{enumerate}
\end{definition}

\begin{lemma}[Atom normalizer preserves support]
\label{lem:appC-atom-normalizer-preserves-support}
For every signed raw atom \(\sigma A\),
\[
  \Sat(\anf_\chi(\sigma A))=\Sat(\sigma A).
\]
\end{lemma}

\begin{proof}
The neutral, public-template, public-surface, legal-read, illegal-read, and raw
gauge cases are exactly the support equations in
Hypothesis~\ref{hyp:appC-atom-semantics-compatible}.  The quotient case is the
two-case truth table for \(z_v=x_v\oplus g_v\).  The negative quotient case is
identical because over \(\bits\), the complement of \(z_v=\zeta\) is
\(z_v=1\oplus\zeta\).
\end{proof}

\begin{remark}[The charged atoms]
\label{rem:appC-charged-atoms}
Rules (A2), (A3), and (A5) make public syntax and illegal reads neutral.  Rule
(A4) produces safe-buffer evidence.  Rule (A6) produces gauge evidence.  Rule
(A7) ensures that quotient information is never a fourth evidence class; it is
charged as gauge-bearing support.
\end{remark}

\subsection{Structural rewrite rules}
\label{subsec:appC-structural-rewrite-rules}

The following rules are the structural part of CD-ENF.  They are most cleanly
understood as equations for a recursive normalizer, but they may also be read as
an oriented rewrite system.

\begin{definition}[Recursive CD-ENF normalizer]
\label{def:appC-recursive-cdenf}
For an evidence term \(E\) and a conclusion label \(e\), define
\(\CDENF_\chi(E\Rightarrow e)\) by structural recursion.
\begin{enumerate}[label=(R\arabic*)]
\item Atomic term:
\[
  \CDENF_\chi(A\Rightarrow e)
\]
is obtained from \(\anf_\chi(A)\) by replacing the dummy label \(*\) by
\(e\).
\item Conjunction:
\[
  \CDENF_\chi((E\wedge F)\Rightarrow e)
  =
  \CDENF_\chi(E\Rightarrow *)
  \otimes
  \CDENF_\chi(F\Rightarrow e).
\]
\item Disjunction:
\[
  \CDENF_\chi((E\vee F)\Rightarrow e)
  =
  \CDENF_\chi(E\Rightarrow e)
  \oplus
  \CDENF_\chi(F\Rightarrow e).
\]
\item Case:
\[
\begin{aligned}
  \CDENF_\chi(\operatorname{case}(A;E_0,E_1)\Rightarrow e)
  ={}&
  \bigl(\anf_\chi(\neg A)\otimes\CDENF_\chi(E_0\Rightarrow e)\bigr) \\
  &\oplus
  \bigl(\anf_\chi(A)\otimes\CDENF_\chi(E_1\Rightarrow e)\bigr).
\end{aligned}
\]
\item Existential quantifier:
\[
  \CDENF_\chi((\exists x.E)\Rightarrow e)
  =
  \bigoplus_{a\in D_x}\CDENF_\chi(E[x:=a]\Rightarrow e).
\]
\item Internal derivation:
\[
\begin{aligned}
  \CDENF_\chi(\operatorname{derive}_\rho(E_1,\ldots,E_k)\Rightarrow e)
  ={}&
  [(N(\operatorname{der}(\rho));\emptyset;\emptyset;*)] \\
  &\otimes
  \bigotimes_{i=1}^k\CDENF_\chi(E_i\Rightarrow *),
\end{aligned}
\]
followed by replacing the dummy label on the resulting branches by \(e\).
\item Truth and falsity:
\[
  \CDENF_\chi(\top\Rightarrow e)=[(\emptyset;\emptyset;\emptyset;e)],
  \qquad
  \CDENF_\chi(\bot\Rightarrow e)=\emptyset.
\]
\end{enumerate}
\end{definition}

\begin{definition}[Oriented structural rewrite rules]
\label{def:appC-oriented-structural-rules}
The oriented rewrite system \(\to_\chi\) is generated by the following rules,
closed under evidence-term contexts.
\begin{enumerate}[label=(S\arabic*)]
\item Atom rewrites are the atom rules of
      Definition~\ref{def:appC-signed-atom-normalizer}.
\item
\[
  \operatorname{case}(A;E_0,E_1)
  \to_\chi
  (\neg A\wedge E_0)\vee(A\wedge E_1).
\]
\item
\[
  \exists x.E
  \to_\chi
  \bigvee_{a\in D_x}E[x:=a].
\]
\item
\[
  \operatorname{derive}_\rho(E_1,\ldots,E_k)
  \to_\chi
  N(\operatorname{der}(\rho))\wedge E_1\wedge\cdots\wedge E_k.
\]
\item Boolean distribution to branch form:
\[
  E\wedge(F\vee G)\to_\chi(E\wedge F)\vee(E\wedge G),
\]
\[
  (F\vee G)\wedge E\to_\chi(F\wedge E)\vee(G\wedge E).
\]
\item Unit and zero simplifications:
\[
  E\wedge\top\to_\chi E,
  \qquad
  \top\wedge E\to_\chi E,
\]
\[
  E\wedge\bot\to_\chi\bot,
  \qquad
  \bot\wedge E\to_\chi\bot,
\]
\[
  E\vee\bot\to_\chi E,
  \qquad
  \bot\vee E\to_\chi E.
\]
\item Branch canonicalization: sort literal lists, delete duplicates, delete
      inconsistent branches, sort branches, and delete duplicate branches.
\end{enumerate}
No inverse rules are included.
\end{definition}

\begin{remark}[Why both presentations are kept]
\label{rem:appC-two-presentations}
The recursive normalizer is the preferred Lean definition.  The oriented rewrite
system is useful for exposition and for connecting the construction to the ENF
intuition: guards are pushed to leaves, cases are disjointified, impossible
branches are pruned, and the remaining leaves are classified.
\end{remark}

\subsection{Semantics preservation}
\label{subsec:appC-semantics-preservation}

\begin{lemma}[Each rewrite rule preserves support]
\label{lem:appC-each-rule-preserves-support}
If \(E\to_\chi E'\), then
\[
  \Sat(E)=\Sat(E').
\]
The same holds labelwise for the corresponding certified terms.
\end{lemma}

\begin{proof}
Atom rewrites preserve support by
Lemma~\ref{lem:appC-atom-normalizer-preserves-support}.  The case rule is the
support equation in Definition~\ref{def:appC-support-semantics}.  The existential
rule is finite union over \(D_x\).  The derivation rule is the support equation
for \(\operatorname{derive}_\rho\).  Boolean distribution and unit/zero rules are
ordinary finite set identities.  Branch canonicalization preserves support by
Lemma~\ref{lem:appC-canonicalization-preserves-support}.
\end{proof}

\begin{theorem}[CD-ENF semantics preservation]
\label{thm:appC-cdenf-semantics-preservation}
For every evidence term \(E\) and conclusion label \(e\),
\[
  \Sat_e(\CDENF_\chi(E\Rightarrow e))=\Sat(E).
\]
Ignoring labels,
\[
  \Sat(\CDENF_\chi(E\Rightarrow e))=\Sat(E).
\]
\end{theorem}

\begin{proof}
Proceed by structural induction on \(E\).  The atomic case is
Lemma~\ref{lem:appC-atom-normalizer-preserves-support}.  The conjunction and
disjunction cases follow from
Lemma~\ref{lem:appC-nf-algebra-preserves-semantics}.  The case, existential,
derivation, \(\top\), and \(\bot\) cases are exactly the corresponding semantic
clauses in Definition~\ref{def:appC-support-semantics}, combined with the
induction hypotheses and Lemma~\ref{lem:appC-nf-algebra-preserves-semantics}.
\end{proof}

\begin{corollary}[CD count pairs are invariant]
\label{cor:appC-cd-count-pairs-invariant}
Fix a target coordinate \(j\) and target-conditioned phase laws
\(\mu_j^+,\mu_j^-\).  For every evidence term \(E\),
\[
  \bigl(\mu_j^+(\Sat(E)),\mu_j^-(\Sat(E))\bigr)
  =
  \bigl(\mu_j^+(\Sat(\CDENF_\chi(E\Rightarrow e))),
         \mu_j^-(\Sat(\CDENF_\chi(E\Rightarrow e)))\bigr).
\]
Thus normalization preserves CD skew and phase gap.
\end{corollary}

\begin{proof}
Apply Theorem~\ref{thm:appC-cdenf-semantics-preservation} in the two phase
laws.
\end{proof}

\subsection{Termination}
\label{subsec:appC-termination}

The branch algebra may duplicate subterms; for example, distributing
\(\wedge\) over \(\vee\) expands a product of sums.  Termination is therefore
best proved by a multiset measure on unreduced constructors rather than by raw
string length.

\begin{definition}[Constructor height]
\label{def:appC-constructor-height}
The height \(\height(E)\) of an evidence term is defined by
\[
  \height(A)=\height(\top)=\height(\bot)=0,
\]
\[
  \height(E\wedge F)=\height(E\vee F)=1+\max(\height(E),\height(F)),
\]
\[
  \height(\operatorname{case}(A;E_0,E_1))=1+
  \max(\height(E_0),\height(E_1)),
\]
\[
  \height(\exists x.E)=1+
  \height(E),
\]
and
\[
  \height(\operatorname{derive}_\rho(E_1,\ldots,E_k))
  =1+\max_i\height(E_i).
\]
\end{definition}

\begin{definition}[Redex multiset]
\label{def:appC-redex-multiset}
For a finite expression produced during normalization, let \(M(E)\) be the
finite multiset of heights of unreduced compound constructors and unreduced raw
non-normal atoms occurring in it.  A raw non-normal atom has height \(0\) in
this multiset.  A normalized literal contributes nothing.
\end{definition}

\begin{lemma}[Each non-canonical rewrite decreases the multiset measure]
\label{lem:appC-each-rewrite-decreases-measure}
Every non-canonical rewrite step in Definition~\ref{def:appC-oriented-structural-rules}
strictly decreases \(M(E)\) in the well-founded multiset extension of the usual
order on \(\mathbb N\).
\end{lemma}

\begin{proof}
Consider each rule.  An atom rewrite removes one unreduced raw atom and inserts
only normalized literals, so the multiset loses one element of height \(0\).  A
case, existential, or derivation expansion removes one compound constructor of
height \(h+1\) and inserts only copies of strict subterms of height at most
\(h\), together with normalized or raw atoms that will be reduced separately.
This is a strict multiset decrease.  Boolean distribution removes one
unreduced distributive occurrence whose root has height greater than the heights
of the copied subterms; replacing one element by finitely many smaller elements
is a strict decrease in the multiset order.  Unit and zero simplifications delete
one compound constructor and do not introduce a larger one.  Branch
canonicalization is treated separately: it is a deterministic idempotent cleanup
step and cannot create new redexes.
\end{proof}

\begin{theorem}[Termination]
\label{thm:appC-termination}
There is no infinite \(\to_\chi\)-reduction sequence starting from a finite
certified evidence term.  Equivalently, the CD-ENF normalizer terminates on all
finite evidence terms.
\end{theorem}

\begin{proof}
By Lemma~\ref{lem:appC-each-rewrite-decreases-measure}, every non-canonical
rewrite step strictly decreases a well-founded multiset measure.  Canonicalization
is deterministic, idempotent, and performed after a finite branch list is
produced; it cannot create a new non-canonical redex.  Therefore no infinite
sequence exists.  The recursive normalizer terminates by the same argument,
since each recursive call is made on a strict subterm or on a finite substitution
instance over a finite domain \(D_x\).
\end{proof}

\begin{remark}[Lean version of termination]
\label{rem:appC-lean-termination}
For mechanization, it is simplest to define \(\CDENF_\chi\) by structural
recursion and prove termination from Lean's termination checker.  The multiset
measure above is then used only to justify the optional relational rewrite
presentation.
\end{remark}

\subsection{Confluence and uniqueness of normal form}
\label{subsec:appC-confluence}

The confluence proof uses a canonical-normalizer argument.  One need not
manually enumerate all critical pairs first.  It is enough to prove that every
one-step rewrite preserves the canonical normalizer.

\begin{lemma}[One-step invariance of the canonical normalizer]
\label{lem:appC-one-step-invariance}
If \(E\to_\chi E'\), then for every conclusion label \(e\),
\[
  \CDENF_\chi(E\Rightarrow e)=\CDENF_\chi(E'\Rightarrow e).
\]
\end{lemma}

\begin{proof}
Check each generating rewrite rule.  Atom rewrites are precisely the atomic
clauses of \(\anf_\chi\).  The case, existential, derivation, conjunction,
disjunction, distribution, and unit/zero rules are precisely the recursive
clauses of \(\CDENF_\chi\) together with the normal-form algebra
\(\oplus,\otimes\).  Branch canonicalization is built into \(\oplus\) and
\(\otimes\), so applying it before or after the recursive call gives the same
canonical branch list.
\end{proof}

\begin{lemma}[Normal forms are fixed points]
\label{lem:appC-normal-forms-fixed-points}
If \(F\) is a CD-ENF normal form, then
\[
  \CDENF_\chi(F\Rightarrow e)=F
\]
for the corresponding label convention.
\end{lemma}

\begin{proof}
A CD-ENF normal form contains only canonical branches whose leaves are
normalized literals.  There are no raw atoms, quotient atoms, case nodes,
existential nodes, derivation nodes, or distributive Boolean redexes.  The
recursive normalizer therefore performs no atom or structural rewrite and only
applies the idempotent canonicalization map, which leaves \(F\) unchanged.
\end{proof}

\begin{theorem}[Confluence and uniqueness]
\label{thm:appC-confluence-uniqueness}
The CD-ENF rewrite system is confluent on finite evidence terms.  Moreover,
if two reduction sequences from \(E\Rightarrow e\) reach normal forms \(F_1\)
and \(F_2\), then
\[
  F_1=F_2=\CDENF_\chi(E\Rightarrow e).
\]
\end{theorem}

\begin{proof}
By termination, every reduction sequence reaches a normal form.  By repeated
application of Lemma~\ref{lem:appC-one-step-invariance}, every term reachable
from \(E\Rightarrow e\) has the same canonical normalizer as \(E\Rightarrow e\).
If a reachable term is already a normal form \(F\), then by
Lemma~\ref{lem:appC-normal-forms-fixed-points},
\[
  F=\CDENF_\chi(F\Rightarrow e)=\CDENF_\chi(E\Rightarrow e).
\]
Thus all reachable normal forms are equal to the same canonical branch list.
This is confluence.
\end{proof}

\begin{corollary}[Newman's lemma formulation]
\label{cor:appC-newman-formulation}
Since the rewrite system is terminating, local confluence would also imply
confluence by Newman's lemma.  The canonical-normalizer proof above gives the
same conclusion and identifies the unique normal form explicitly as
\(\CDENF_\chi(E\Rightarrow e)\).
\end{corollary}

\subsection{Gauge faithfulness as a rewrite invariant}
\label{subsec:appC-gauge-faithfulness}

\begin{definition}[Gauge support of a normal form]
\label{def:appC-gauge-support-normal-form}
For a branch \(\mathcal B=(C;\mathcal S;\mathcal G;e)\), define
\[
  \suppG(\mathcal B)
  :=
  \{\operatorname{gidx}(a):G(a,\gamma)\text{ or }\neg G(a,\gamma)
      \text{ occurs in }\mathcal G\}.
\]
Set
\[
  \rankG(\mathcal B):=|\suppG(\mathcal B)|.
\]
For a normal form \(F\), define \(\suppG(F)\) as the union of branch supports
and \(\rankG(F):=|\suppG(F)|\).
\end{definition}

\begin{definition}[Quotient-use occurrence]
\label{def:appC-quotient-use-occurrence}
A quotient-use occurrence for coordinate \(v\) in a derivation of
\(F=\CDENF_\chi(E\Rightarrow e)\) is an occurrence of a raw atom
\(Q(v,\zeta)\) or \(\neg Q(v,\zeta)\) in the original term \(E\) or in an
intermediate term before normalization.
\end{definition}

\begin{lemma}[Quotient expansion carries gauge support]
\label{lem:appC-quotient-expansion-carries-support}
If a branch \(\mathcal B\) of \(\anf_\chi(Q(v,\zeta))\) or
\(\anf_\chi(\neg Q(v,\zeta))\) is nonempty, then
\[
  v\in\suppG(\mathcal B).
\]
In fact, each such branch contains a literal whose support has gauge index
\(v\), namely a literal over \(g_v\), and also a raw support literal over
\(x_v\).
\end{lemma}

\begin{proof}
This is immediate from rules (A6) and (A7) in
Definition~\ref{def:appC-signed-atom-normalizer}.  Every quotient branch contains
one literal of the form \(G(g_v,\gamma)\), and
\(\operatorname{gidx}(g_v)=v\).  It also contains one raw witness-support
literal \(G(x_v,\alpha)\), with \(\operatorname{gidx}(x_v)=v\).
\end{proof}

\begin{theorem}[Gauge-faithful normalization]
\label{thm:appC-gauge-faithful-normalization}
Let \(F=\CDENF_\chi(E\Rightarrow e)\).  If a branch of \(F\) depends on a
quotient-use occurrence for coordinate \(v\), then that branch contains
normalized gauge-bearing leaves with gauge index \(v\).  Consequently, any
target-relevant use of a quotient value is charged by \(\rankG\).
\end{theorem}

\begin{proof}
The only rule that removes a quotient-use occurrence is the quotient expansion
rule.  By Lemma~\ref{lem:appC-quotient-expansion-carries-support}, every
nonempty branch produced by that rule contains gauge support with index \(v\).
Subsequent products only add literals, and canonicalization deletes a branch only
if it is inconsistent or duplicate.  It never converts a gauge literal into a
neutral or safe literal.  Thus any surviving branch depending on the quotient
occurrence still contains gauge support with index \(v\).
\end{proof}

\begin{corollary}[No public quotient or canonical-gauge shortcut]
\label{cor:appC-no-public-quotient-shortcut}
Public template atoms, public surface atoms, neutral atoms, illegal-read atoms,
and derivation tags cannot introduce a target-relevant quotient value, raw
witness value, gauge value, or message bit.  In particular, no rewrite rule may
choose a canonical gauge representative unless the chosen gauge literals appear
as normalized gauge leaves.
\end{corollary}

\begin{proof}
Public template and surface atoms rewrite only to neutral atoms.  Illegal reads
rewrite only to neutral atoms.  Derivation nodes add only neutral derivation
tags and the already-normalized evidence in their premises.  The only source of
raw quotient information is the quotient atom, and by
Theorem~\ref{thm:appC-gauge-faithful-normalization} its normalized use carries
raw gauge support.  A canonical gauge representative would fix gauge literals;
therefore it must appear through gauge leaves and is charged by \(\rankG\).
\end{proof}

\subsection{Atom completeness after normalization}
\label{subsec:appC-atom-completeness}

\begin{definition}[Neutral, safe, and gauge leaf sets]
\label{def:appC-leaf-sets}
For a normal form \(F\), let
\[
  \operatorname{Neu}(F)
\]
be the set of neutral normalized atoms appearing in \(F\), let
\[
  \Safe(F)
\]
be the set of safe atoms \(S(q,y)\) appearing positively or negatively in \(F\),
and let
\[
  \Gauge(F)
\]
be the set of gauge atoms \(G(a,\gamma)\) appearing positively or negatively in
\(F\).
\end{definition}

\begin{theorem}[CD-ENF atom completeness]
\label{thm:appC-atom-completeness}
For every evidence term \(E\Rightarrow e\), every non-neutral leaf in
\(\CDENF_\chi(E\Rightarrow e)\) is either a safe atom \(S(q,y)\) or a gauge
atom \(G(a,\gamma)\).  There is no normalized leaf of any other target-relevant
kind.
\end{theorem}

\begin{proof}
By Definition~\ref{def:appC-normalized-atoms}, normalized atoms are exactly
neutral, safe, or gauge.  The atom rewrite rules send public syntax and illegal
reads to neutral atoms, legal reads to safe atoms, raw gauge literals to gauge
atoms, and quotient literals to disjunctions of gauge atoms.  The structural
rules combine branches but do not create new atom classes.
\end{proof}

\begin{corollary}[Charged-leaf cover]
\label{cor:appC-charged-leaf-cover}
Let \(F=\CDENF_\chi(E\Rightarrow e)\).  If two worlds agree on every charged
leaf in \(\Safe(F)\cup\Gauge(F)\), and also lie in the same neutral public
fiber, then they agree on the truth of every branch of \(F\).  Hence any pair
separated by evaluating \(F\) is separated by at least one charged leaf.
\end{corollary}

\begin{proof}
Within a neutral public fiber, all neutral literals have the same value on the
two worlds.  Agreement on charged leaves gives agreement on all safe and gauge
literals.  Since every branch is a conjunction of neutral, safe, and gauge
literals, the two worlds agree on every branch truth value and hence on the
normal form.  The contrapositive gives the separation statement.
\end{proof}

\begin{proposition}[Pairwise derivative support of a normal form]
\label{prop:appC-pairwise-derivative-support}
Fix a target coordinate \(j\), a message-opposite coupling \(\Gamma_j\), and a
surviving relation \(H\subseteq\Omega_j^0\times\Omega_j^1\) contained in one
neutral public fiber.  Then
\[
\begin{aligned}
&\Gamma_j\bigl(H\cap\{(\omega^0,\omega^1):
        F(\omega^0)\ne F(\omega^1)\}\bigr) \\
&\quad\le
\sum_{S\in\Safe(F)}
\Gamma_j\bigl(H\cap\{(\omega^0,\omega^1):
        S(\omega^0)\ne S(\omega^1)\}\bigr) \\
&\qquad+
\sum_{G\in\Gauge(F)}
\Gamma_j\bigl(H\cap\{(\omega^0,\omega^1):
        G(\omega^0)\ne G(\omega^1)\}\bigr).
\end{aligned}
\]
\end{proposition}

\begin{proof}
By Corollary~\ref{cor:appC-charged-leaf-cover}, the separation event on the
left is contained in the union of the charged-leaf separation events on the
right.  Apply the union bound for the finite measure \(\Gamma_j\).
\end{proof}

\subsection{Observer trace expansion}
\label{subsec:appC-observer-trace-expansion}

This appendix is about normalization, not about bounding advantage.  Still, the
normalizer must apply to actual deterministic observers.  The following finite
construction is the bridge used by Sections~\ref{sec:cdenf} and~\ref{sec:trace}.

\begin{definition}[Primitive CD observer]
\label{def:appC-primitive-cd-observer}
A primitive CD observer is a deterministic finite-control computation which, at
each step, either halts with an output label in \(\mathcal C\), performs an
internal deterministic derivation step, or queries one raw atom from
Definition~\ref{def:appC-raw-atoms}.  The query result determines the next
control state.  There are no primitive opaque run atoms and no primitive target
message atoms.
\end{definition}

\begin{definition}[Decision-tree evidence term]
\label{def:appC-decision-tree-evidence-term}
Let \(O\) be a primitive CD observer that halts after at most \(Q\) queries.
For each output label \(e\), define \(E_{O,e}\) by expanding the finite decision
tree of \(O\): query nodes become \(\operatorname{case}\) nodes, internal
computation nodes become \(\operatorname{derive}_\rho\) nodes, leaves with
output \(e\) become \(\top\), and leaves with other outputs become \(\bot\).
\end{definition}

\begin{theorem}[Observer output events normalize]
\label{thm:appC-observer-output-events-normalize}
For every primitive CD observer \(O\) and every output label \(e\),
\[
  \Sat(E_{O,e})=\{\omega\in\Omega:O(\omega)=e\}.
\]
Moreover, \(\CDENF_\chi(E_{O,e}\Rightarrow e)\) is a canonical CD-ENF normal
form whose non-neutral leaves are exactly safe or gauge leaves produced by the
primitive reads along output-\(e\) branches.
\end{theorem}

\begin{proof}
Induct on the depth remaining in the finite decision tree.  At a leaf the claim
is immediate from \(\top\) or \(\bot\).  At a query node, the support equation for
\(\operatorname{case}\) splits worlds according to the returned atom value and
then applies the induction hypothesis to the selected child.  At an internal
computation node, the \(\operatorname{derive}_\rho\) constructor adds only the
neutral derivation tag and the premise evidence.  Applying
Theorem~\ref{thm:appC-cdenf-semantics-preservation} and
Theorem~\ref{thm:appC-atom-completeness} gives the normal-form classification.
\end{proof}

\begin{remark}[Certification lift is an interface, not an extra atom]
\label{rem:appC-certification-lift}
The theorem does not assert that a computation outputs a proof of its own
correctness.  It asserts that its finite run can be expanded into an evidence
trace whose branch event is exactly the output event.  This is the form needed
by CD Trace Capture: advantage is charged to pair distinctions introduced by
primitive reads, not to an opaque assertion that the program returned a bit.
\end{remark}

\subsection{Lean-facing theorem list}
\label{subsec:appC-lean-facing-theorem-list}

For formalization, this appendix can be split into the following finite modules.

\begin{enumerate}[label=(\roman*)]
\item \texttt{CDEvidenceSyntax.lean}: finite raw atoms, normalized atoms,
      literals, evidence terms, and certified labels.
\item \texttt{CDEvidenceSemantics.lean}: finite event semantics and signed
      literal semantics.
\item \texttt{CDENFBranches.lean}: normal branches, canonicalization,
      branch sum and product, and support preservation.
\item \texttt{CDENFNormalizer.lean}: the recursive normalizer
      \(\CDENF_\chi\), semantics preservation, and count-pair invariance.
\item \texttt{CDENFRewrite.lean}: optional relational rewrite presentation,
      termination, and confluence from the canonical normalizer.
\item \texttt{GaugeFaithfulness.lean}: quotient expansion carries raw gauge
      support and forbids canonical-gauge shortcuts.
\item \texttt{ObserverEvidenceExpansion.lean}: deterministic primitive
      observer decision trees and output-event evidence terms.
\end{enumerate}

The minimal dependency needed by later proof sections is the recursive
normalizer plus Theorems~\ref{thm:appC-cdenf-semantics-preservation},
\ref{thm:appC-confluence-uniqueness},
\ref{thm:appC-gauge-faithful-normalization}, and
\ref{thm:appC-atom-completeness}.  The critical-pair or Newman-style proof can
be added after the canonical normalizer is already available.

\section{Hidden-Gauge Buffered Locked SAT Construction}
\label{app:hidden-gauge-buffered-locked-construction}
\providecommand{\bits}{\{0,1\}}
\providecommand{\Prb}{\mathbb P}
\providecommand{\E}{\mathbb E}
\providecommand{\poly}{\operatorname{poly}}
\providecommand{\dist}{\operatorname{dist}}
\providecommand{\supp}{\operatorname{supp}}
\providecommand{\suppG}{\operatorname{supp}_G}
\providecommand{\rankG}{\operatorname{rank}_G}
\providecommand{\Lock}{\operatorname{Lock}}
\providecommand{\Buf}{\operatorname{Buf}}
\providecommand{\Read}{\operatorname{Read}}
\providecommand{\GaugeEq}{\operatorname{GaugeEq}}
\providecommand{\Legal}{\operatorname{Legal}}
\providecommand{\Val}{\operatorname{Val}}
\providecommand{\negl}{\operatorname{negl}}
\providecommand{\Unif}{\operatorname{Unif}}
\providecommand{\CNF}{\operatorname{CNF}}
\providecommand{\arity}{\operatorname{arity}}
\providecommand{\size}{\operatorname{size}}

This appendix expands the construction used in
Section~\ref{sec:ensemble}.  The goal is to make the ensemble interface
finite and explicit enough that the later analytic estimates can be formalized
against it.  No Dobrushin or entropy estimate is proved here.  The analytic
buffer contraction is deferred to Appendix~\ref{app:max-qssm-dobrushin}, and
the hidden-gauge counting estimates are deferred to
Appendix~\ref{app:gauge-rank-entropy-incidence}.  This appendix proves the
purely algebraic and syntactic parts of the construction:

\begin{enumerate}[label=(\roman*)]
\item the witness relation is a finite bounded-arity CSP over a public instance;
\item every sampled witness satisfies the relation;
\item all satisfying witnesses over the same public instance have the same
      message;
\item the gauge action preserves the public instance, the relation, and the
      message;
\item unsupported gauge coordinates are product-uniform before being charged;
\item public syntax contributes only neutral evidence atoms, while quotient uses
      expose raw gauge support.
\end{enumerate}

The construction is written as a component product.  The components are a
finite locked quotient core, a hidden gauge lift, and a strictly positive buffer
core.  Each component is a finite record with explicit axioms.  The main theorem
of this appendix says that any such component record yields the
``gauge-buffered locked'' ensemble used in the main text.

\subsection{Layered finite geometries}
\label{subsec:appD-layered-geometries}

Fix a size parameter $m$.  All sets below are finite and have size at most
$m^{c_0}$ for a fixed constant $c_0$.  The constants $d_0,a_0,c_0$ below are
absolute unless explicitly parameterized by a clock exponent in the main text.

\begin{definition}[Layered geometry]
\label{def:appD-layered-geometry}
A layered geometry at size $m$ is a tuple
\[
  \mathcal G_m=(\mathcal P_m,E_m,\tau_m,\dist_m)
\]
where $(\mathcal P_m,E_m)$ is a finite graph of maximum degree at most $d_0$,
$\tau_m$ is a finite type map, and $\dist_m$ is the graph distance.  The site
set is partitioned as
\[
  \mathcal P_m
  =
  \mathcal P_{\mathrm{neu}}
  \sqcup
  \mathcal P_{\mathrm{buf}}
  \sqcup
  \mathcal P_{\mathrm{lock}}
  \sqcup
  \mathcal P_{\mathrm{gauge}}
  \sqcup
  \mathcal P_{\mathrm{readout}}.
\]
The protected set is
\[
  \mathcal P_{\mathrm{prot}}
  :=
  \mathcal P_{\mathrm{lock}}
  \cup
  \mathcal P_{\mathrm{gauge}}
  \cup
  \mathcal P_{\mathrm{readout}}.
\]
The geometry is explicit if there is a fixed polynomial-time algorithm which,
on input $m$, lists its sites, edges, type labels, and partition labels.
\end{definition}

\begin{definition}[Local scope system]
\label{def:appD-local-scope-system}
A local scope system on $\mathcal G_m$ is a finite list
\[
  \mathcal C_m=(C_1,\ldots,C_N)
\]
of scopes $C_a\subseteq\mathcal P_m$ satisfying $|C_a|\le a_0$.  A local
predicate attached to $C_a$ is a Boolean predicate on the variables whose sites
lie in $C_a$ together with a bounded number of auxiliary variables attached to
that scope.  The scope system is explicit if $N\le \poly(m)$ and the list of
scopes and predicate tables is computable in time $\poly(m)$.
\end{definition}

\begin{definition}[Quotient/gauge index set]
\label{def:appD-gauge-index-set}
A quotient/gauge index set is a finite set $V_m$ with $|V_m|\le \poly(m)$,
together with a placement map
\[
  \iota_G:V_m\to \mathcal P_{\mathrm{gauge}}.
\]
For $v\in V_m$ the witness contains bits
\[
  x_v,\\ g_v,\\ z_v\in\bits.
\]
The intended quotient identity is
\[
  z_v=x_v\oplus g_v.
\]
\end{definition}

\begin{remark}[Reason for the layer split]
\label{rem:appD-layer-reason}
The layer split separates three different proof obligations.  The locked and
readout sites enforce message rigidity.  The gauge sites provide a large hidden
product space whose raw coordinates can be charged by rank.  The buffer sites
provide the region where legal safe probes live.  Keeping these regions
syntactically separate makes the CD-ENF atom-completeness proof local: public
syntax is neutral, safe probes touch only buffer sites, and target-relevant
quotient use exposes raw gauge support.
\end{remark}

\subsection{Locked quotient cores}
\label{subsec:appD-locked-quotient-cores}

The locked core is the only part of the construction that is allowed to force a
global message.  It may have many quotient completions, but they must all carry
the same message over the same public lock syntax.

\begin{definition}[Locked quotient core]
\label{def:appD-locked-quotient-core}
A locked quotient core at size $m$ consists of the following finite data.
\begin{enumerate}[label=(\roman*)]
\item A finite public syntax set $\mathcal Y_{\mathrm{lock},m}$.
\item A quotient index set $V_m$.
\item A finite auxiliary domain $\Xi_{\mathrm{lock},m}$.
\item A message length $r_t=r_t(m)$ with $r_t\ge c_M t(m)$ for a fixed
      constant $c_M>0$.
\item A bounded-arity predicate
\[
  \Lock_Y(z,\xi_{\mathrm{lock}},M)
  \in\bits,
\]
where $Y\in\mathcal Y_{\mathrm{lock},m}$,
$z\in\bits^{V_m}$, $\xi_{\mathrm{lock}}\in\Xi_{\mathrm{lock},m}$, and
$M\in\bits^{r_t}$.
\item A readout predicate
\[
  \Read_Y(z,\xi_{\mathrm{lock}},M)\in\bits.
\]
\end{enumerate}
Both predicates are conjunctions of $\poly(m)$ bounded-arity local predicates
on the locked and readout regions.
\end{definition}

\begin{definition}[Locked completion]
\label{def:appD-locked-completion}
A locked completion for public lock syntax $Y$ is a triple
\[
  (z,\xi_{\mathrm{lock}},M)
\]
such that
\[
  \Lock_Y(z,\xi_{\mathrm{lock}},M)=1
  \qquad\text{and}\qquad
  \Read_Y(z,\xi_{\mathrm{lock}},M)=1.
\]
The set of locked completions is denoted
\[
  \mathsf{Comp}_{\mathrm{lock}}(Y).
\]
\end{definition}

\begin{hypothesis}[Lock satisfiability]
\label{hyp:appD-lock-satisfiability}
Every public lock syntax $Y$ in the support of the sampler has at least one
locked completion:
\[
  \mathsf{Comp}_{\mathrm{lock}}(Y)\ne\varnothing.
\]
\end{hypothesis}

\begin{hypothesis}[Locked-message rigidity]
\label{hyp:appD-locked-message-rigidity}
For every public lock syntax $Y$ in the support and any two locked completions
\[
  (z,\xi_{\mathrm{lock}},M),
  \qquad
  (z',\xi'_{\mathrm{lock}},M'),
\]
one has
\[
  M=M'.
\]
\end{hypothesis}

\begin{definition}[Locked message]
\label{def:appD-locked-message}
Under Hypotheses~\ref{hyp:appD-lock-satisfiability} and
\ref{hyp:appD-locked-message-rigidity}, define
\[
  M_{\mathrm{lock}}(Y)
\]
to be the common message coordinate of all locked completions of $Y$.
\end{definition}

\begin{lemma}[Well-defined locked message]
\label{lem:appD-well-defined-locked-message}
For every public lock syntax $Y$ in the support, $M_{\mathrm{lock}}(Y)$ is
well-defined, and every locked completion of $Y$ has message
$M_{\mathrm{lock}}(Y)$.
\end{lemma}

\begin{proof}
Existence follows from Hypothesis~\ref{hyp:appD-lock-satisfiability}.  If two
completions had different messages, this would contradict
Hypothesis~\ref{hyp:appD-locked-message-rigidity}.  Hence all completions have
the same message.
\end{proof}

\begin{definition}[Lock sampler]
\label{def:appD-lock-sampler}
A lock sampler is a randomized polynomial-time algorithm which outputs
\[
  (Y_{\mathrm{lock}},z,\xi_{\mathrm{lock}},M)
\]
with
\[
  (z,\xi_{\mathrm{lock}},M)
  \in
  \mathsf{Comp}_{\mathrm{lock}}(Y_{\mathrm{lock}}).
\]
The public marginal of this sampler is denoted $\mathcal L_{\mathrm{lock},m}$.
\end{definition}

\begin{hypothesis}[Target-coordinate balance]
\label{hyp:appD-target-coordinate-balance}
For the switched coordinate set $S_m$ used in the lower bound, each target
coordinate
\[
  B_j(Y):=\ell_j(M_{\mathrm{lock}}(Y)),
  \qquad j\in S_m,
\]
is balanced under the public marginal:
\[
  \Prb_{Y\sim\mathcal L_{\mathrm{lock},m}}[B_j(Y)=0]
  =
  \Prb_{Y\sim\mathcal L_{\mathrm{lock},m}}[B_j(Y)=1]
  =
  {1\over 2}.
\]
More generally, the main proof only needs the stated balance on the final
switched set.
\end{hypothesis}

\begin{remark}[No public message literal]
\label{rem:appD-no-public-message-literal}
Rigidity is semantic, not syntactic.  It says that the message is forced by the
solution set of the public lock syntax.  It does not permit the public syntax to
contain a literal such as $M_a=b$ or a public tag naming $B_j$.  The syntactic
``no tag'' condition is imposed in Subsection~\ref{subsec:appD-public-syntax}.
\end{remark}

\subsection{The hidden gauge lift}
\label{subsec:appD-hidden-gauge-lift}

The gauge lift replaces a quotient assignment $z$ by many raw representatives
$(x,g)$ satisfying $z=x\oplus g$.  This creates many satisfying witnesses over
the same public instance, while leaving the locked message unchanged.

\begin{definition}[Gauge equations]
\label{def:appD-gauge-equations}
For $x,g,z\in\bits^{V_m}$ define
\[
  \GaugeEq(x,g,z)=1
\]
if and only if for every $v\in V_m$,
\[
  z_v=x_v\oplus g_v.
\]
Equivalently, over $\mathbb F_2$,
\[
  x_v+g_v+z_v=0.
\]
\end{definition}

\begin{definition}[Gauge action]
\label{def:appD-gauge-action}
For $s\in\bits^{V_m}$, define
\[
  s\cdot(x,g,z):=(x\oplus s,g\oplus s,z).
\]
On a full witness the action leaves all non-gauge coordinates fixed:
\[
  s\cdot(x,g,z,\xi_{\mathrm{lock}},\xi_{\mathrm{buf}},\xi_{\mathrm{aux}},M)
  :=
  (x\oplus s,g\oplus s,z,
   \xi_{\mathrm{lock}},\xi_{\mathrm{buf}},\xi_{\mathrm{aux}},M).
\]
\end{definition}

\begin{lemma}[Gauge equations are invariant]
\label{lem:appD-gauge-eq-invariant}
If $\GaugeEq(x,g,z)=1$, then
\[
  \GaugeEq(x\oplus s,g\oplus s,z)=1
\]
for every $s\in\bits^{V_m}$.
\end{lemma}

\begin{proof}
For each $v\in V_m$,
\[
  (x_v\oplus s_v)\oplus(g_v\oplus s_v)
  =x_v\oplus g_v
  =z_v.
\]
Thus every quotient equation is preserved.
\end{proof}

\begin{definition}[Gauge lift of a quotient completion]
\label{def:appD-gauge-lift}
Given a quotient assignment $z\in\bits^{V_m}$, sample
\[
  g\sim\Unif(\bits^{V_m})
\]
and set
\[
  x:=z\oplus g.
\]
The resulting pair $(x,g)$ is called the gauge lift of $z$.
\end{definition}

\begin{lemma}[Uniform representative law]
\label{lem:appD-uniform-representative-law}
Conditioned on $z$, the gauge lift satisfies:
\begin{enumerate}[label=(\roman*)]
\item $\GaugeEq(x,g,z)=1$ almost surely;
\item $g$ is uniform on $\bits^{V_m}$;
\item for every $J\subseteq V_m$ and every $\gamma\in\bits^J$,
\[
  \Prb[g_J=\gamma\mid z]=2^{-|J|}.
\]
\end{enumerate}
\end{lemma}

\begin{proof}
The construction samples $g$ uniformly and then sets $x=z\oplus g$, so
$x\oplus g=z$ coordinatewise.  The marginal law of $g$ is uniform by
construction, and every coordinate subvector of a uniform product bit-vector is
uniform.
\end{proof}

\begin{definition}[Raw gauge support]
\label{def:appD-raw-gauge-support}
A raw gauge literal is a condition $g_v=\gamma$ for some $v\in V_m$ and
$\gamma\in\bits$.  A normalized raw witness literal $x_v=\alpha$ is assigned the
same gauge support $\{v\}$ when it is used together with the quotient value
$z_v$, because $x_v=\alpha$ is then equivalent to $g_v=z_v\oplus\alpha$.
For a finite set $C$ of raw gauge-bearing literals, define
\[
  \suppG(C):=\{v\in V_m: C\text{ contains a literal bearing }v\},
\]
and
\[
  \rankG(C):=|\suppG(C)|.
\]
\end{definition}

\begin{lemma}[Unsupported gauge coordinates remain uniform]
\label{lem:appD-unsupported-gauge-uniform}
Let $C$ be a consistent finite set of raw gauge-bearing literals, and let
$U=\suppG(C)$.  Conditional on $z$ and on $C$, the coordinates
$g_{V_m\setminus U}$ remain independent fair bits.  Equivalently, for every
$J\subseteq V_m\setminus U$ and every $\gamma\in\bits^J$,
\[
  \Prb[g_J=\gamma\mid z,C]=2^{-|J|}.
\]
\end{lemma}

\begin{proof}
The prior law of $g$ conditioned on $z$ is the product uniform law on
$\bits^{V_m}$.  The event $C$ fixes only the coordinates in $U$ and is
consistent.  Conditioning a finite product-uniform bit-vector on specified
values in $U$ leaves the remaining coordinates product-uniform.
\end{proof}

\begin{remark}[Adaptive evidence]
\label{rem:appD-adaptive-evidence}
Lemma~\ref{lem:appD-unsupported-gauge-uniform} is the static product law.
The adaptive version used in the gauge-rank entropy argument is obtained by
iterating this lemma over the normalized evidence prefix: once a coordinate has
entered $\suppG(h)$ it is charged, and all unsupported coordinates remain fair
relative to the prefix.  The program-counting form is proved in
Appendix~\ref{app:gauge-rank-entropy-incidence}.
\end{remark}

\subsection{Buffer cores and safe probes}
\label{subsec:appD-buffer-cores-safe-probes}

The buffer core is a strictly positive finite CSP on buffer variables, conditioned
on the quotient boundary $z$.  This appendix defines the finite object and the
safe-probe interface.  The log-likelihood contraction estimate is proved in
Appendix~\ref{app:max-qssm-dobrushin}.

\begin{definition}[Buffer core]
\label{def:appD-buffer-core}
A buffer core at size $m$ consists of the following finite data.
\begin{enumerate}[label=(\roman*)]
\item A public syntax set $\mathcal Y_{\mathrm{buf},m}$.
\item A finite buffer auxiliary domain $\Xi_{\mathrm{buf},m}$.
\item A bounded-arity predicate
\[
  \Buf_Y(z,\xi_{\mathrm{buf}})\in\bits,
\]
where $Y\in\mathcal Y_{\mathrm{buf},m}$,
$z\in\bits^{V_m}$, and
$\xi_{\mathrm{buf}}\in\Xi_{\mathrm{buf},m}$.
\item A strictly positive conditional sampling law
\[
  \mathsf{BufSamp}_Y(\cdot\mid z)
\]
on $\Xi_{\mathrm{buf},m}$ whose support is contained in
$\{\xi_{\mathrm{buf}}:\Buf_Y(z,\xi_{\mathrm{buf}})=1\}$.
\end{enumerate}
Strict positivity means that every locally admissible buffer configuration in
that support has positive probability.
\end{definition}

\begin{hypothesis}[Buffer satisfiability]
\label{hyp:appD-buffer-satisfiability}
For every public buffer syntax $Y$ in the support and every quotient assignment
$z$ arising from the lock sampler,
\[
  \{\xi_{\mathrm{buf}}:\Buf_Y(z,\xi_{\mathrm{buf}})=1\}
  \ne
  \varnothing.
\]
\end{hypothesis}

\begin{definition}[Safe probe catalog]
\label{def:appD-safe-probe-catalog}
A safe probe catalog is a finite set $\mathcal Q_m$.  Each probe
$q\in\mathcal Q_m$ has:
\begin{enumerate}[label=(\roman*)]
\item a support $A(q)\subseteq\mathcal P_{\mathrm{buf}}$ with $|A(q)|\le a_0$;
\item a finite value set $\Val(q)$;
\item an evaluation map
\[
  q_Y(z,\xi_{\mathrm{buf}})\in\Val(q).
\]
\end{enumerate}
The evaluation map reads only variables attached to the support $A(q)$ and the
public buffer syntax.
\end{definition}

\begin{definition}[Legal safe probe]
\label{def:appD-legal-safe-probe}
Fix a safety radius $R_{\mathrm{safe}}$.  A probe $q\in\mathcal Q_m$ is legal at
normalized prefix $h$ if:
\begin{enumerate}[label=(\roman*)]
\item $A(q)\subseteq\mathcal P_{\mathrm{buf}}$ and $|A(q)|\le a_0$;
\item
\[
  \dist_m(A(q),\mathcal P_{\mathrm{prot}})
  \ge R_{\mathrm{safe}};
\]
\item $A(q)$ avoids the active protected neighborhoods and previously charged
      non-neutral probe neighborhoods recorded in $h$, except for overlaps
      explicitly allowed by the guard discipline.
\end{enumerate}
The legality predicate is denoted
\[
  \Legal(h,q)\in\bits.
\]
\end{definition}

\begin{definition}[Guarded safe read]
\label{def:appD-guarded-safe-read}
The guarded read of a probe $q$ at prefix $h$ is
\[
  \operatorname{ReadSafe}(h,q)
  :=
  \begin{cases}
  q_Y(z,\xi_{\mathrm{buf}}),&\Legal(h,q)=1,\\
  \bot_{\mathrm{illegal}},&\Legal(h,q)=0.
  \end{cases}
\]
A legal value is represented in CD-ENF as a safe-buffer atom $S(q,y)$.  The
illegal value $\bot_{\mathrm{illegal}}$ is represented as a neutral atom.
\end{definition}

\begin{lemma}[Illegal probes are neutral by construction]
\label{lem:appD-illegal-probes-neutral}
If $\Legal(h,q)=0$, then the guarded read contributes no safe-buffer atom and
no gauge atom to the normalized evidence trace.  It contributes only the neutral
symbol $N(\bot_{\mathrm{illegal}})$.
\end{lemma}

\begin{proof}
This is the defining convention in
Definition~\ref{def:appD-guarded-safe-read}.  The CD-ENF rewrite rule for an
illegal read rewrites the returned symbol to the neutral atom
$N(\bot_{\mathrm{illegal}})$.
\end{proof}

\begin{remark}[Why the buffer is separate from the lock]
\label{rem:appD-buffer-separate-from-lock}
Safe probes are allowed to inspect buffer variables, not lock or gauge
variables.  The safety radius forces legal probes to be far from protected
message and gauge sites.  Consequently, the only target-relevant contribution
of legal safe probes must pass through the buffer contraction estimate of
Appendix~\ref{app:max-qssm-dobrushin}; illegal or unsafe probes are neutralized
rather than treated as informative observations.
\end{remark}

\subsection{Public syntax discipline}
\label{subsec:appD-public-syntax}

The public instance is fully visible to observers.  The restriction below is not
an access restriction.  It is a syntactic discipline on what primitive evidence
atoms the public syntax is allowed to produce after CD-ENF normalization.

\begin{definition}[Public syntax record]
\label{def:appD-public-syntax-record}
A public syntax record consists of finite alphabets
\[
  \mathcal N_m,
  \qquad
  \mathcal P_m^{\mathrm{tpl}},
  \qquad
  \mathcal U_m^{\mathrm{surf}},
\]
for neutral symbols, public template symbols, and public surface symbols,
together with deterministic maps
\[
  \operatorname{tpl}:\mathcal P_m^{\mathrm{tpl}}\to\mathcal N_m,
  \qquad
  \operatorname{surf}:\mathcal U_m^{\mathrm{surf}}\to\mathcal N_m.
\]
Template and surface atoms rewrite by
\[
  P(\theta)\mapsto N(\operatorname{tpl}(\theta)),
  \qquad
  \operatorname{Surf}(u)\mapsto N(\operatorname{surf}(u)).
\]
\end{definition}

\begin{hypothesis}[No public target tags]
\label{hyp:appD-no-public-target-tags}
The public syntax alphabets contain no primitive symbol whose semantics is any
of the following:
\[
  M_a=b,
  \qquad
  \ell_j(M)=b,
  \qquad
  g_v=\gamma,
  \qquad
  x_v=\alpha,
  \qquad
  \text{or a target-dependent gadget type.}
\]
Furthermore, no public-template decoding map recovers any such fact without
passing through a normalized safe-buffer atom or a raw gauge-bearing atom.
\end{hypothesis}

\begin{lemma}[Public syntax normalizes to neutral evidence]
\label{lem:appD-public-syntax-neutral}
Every primitive evidence atom obtained solely by reading public template syntax
normalizes to a neutral atom.
\end{lemma}

\begin{proof}
By Definition~\ref{def:appD-public-syntax-record}, public template and public
surface atoms rewrite through the deterministic maps $\operatorname{tpl}$ and
$\operatorname{surf}$ into $\mathcal N_m$.  Hypothesis
\ref{hyp:appD-no-public-target-tags} excludes target, message, raw witness, and
raw gauge atoms from the public alphabets.  Hence the normal form is neutral.
\end{proof}

\begin{definition}[Normalized primitive atom classes]
\label{def:appD-normalized-primitive-atom-classes}
A normalized target-relevant primitive atom belongs to exactly one of the
following classes.
\begin{enumerate}[label=(\arabic*)]
\item A neutral atom $N(s)$, with $s\in\mathcal N_m$.
\item A safe-buffer atom $S(q,y)$, where $q$ is legal at the current prefix and
      $y\in\Val(q)$.
\item A gauge atom $G(v,\gamma)$, where $v\in V_m$ and $\gamma\in\bits$,
      or a normalized raw witness atom with gauge support $\{v\}$.
\end{enumerate}
\end{definition}

\begin{hypothesis}[Atom completeness for the construction]
\label{hyp:appD-atom-completeness}
Every normalized target-relevant primitive generated by the observer interface
for the constructed ensemble is one of the three classes in
Definition~\ref{def:appD-normalized-primitive-atom-classes}.
\end{hypothesis}

\begin{lemma}[Quotient use exposes raw support]
\label{lem:appD-quotient-use-exposes-raw-support}
Suppose a normalized evidence trace uses the quotient identity
$z_v=x_v\oplus g_v$ in a target-relevant way.  Then its CD-ENF normal form
contains raw gauge-bearing support including $v$.  In particular, the gauge
rank of the normal form increases by at least one unless $v$ was already in the
current gauge support.
\end{lemma}

\begin{proof}
The quotient value $z_v$ is public only as a quotient coordinate.  The public
syntax discipline forbids treating $z_v$ as a target tag.  The CD-ENF quotient
expansion rule rewrites any target-relevant use of $z_v=x_v\oplus g_v$ to raw
support involving $x_v$ and $g_v$.  By Definition~\ref{def:appD-raw-gauge-support},
that raw support has gauge support $\{v\}$.  If $v$ was not previously charged,
the rank increases by one; otherwise the support is already present and the
fresh rank increment is zero.
\end{proof}

\begin{remark}[No canonical-gauge shortcut]
\label{rem:appD-no-canonical-gauge-shortcut}
A common way to accidentally break the construction is to choose a canonical
representative, for example by silently setting $g=0$ and identifying $x$ with
$z$.  The construction forbids this in the evidence language.  An observer may
reason with the equation $z=x\oplus g$, but any target-relevant use of that
equation has to expose raw support and is charged as gauge evidence.
\end{remark}

\subsection{The product construction}
\label{subsec:appD-product-construction}

We now combine the locked core, the gauge lift, and the buffer core into the
world distribution and witness relation used in the main text.

\begin{definition}[Public instance]
\label{def:appD-public-instance}
A public instance at size $m$ is
\[
  Y=(m,\mathcal G_m,Y_{\mathrm{neu}},Y_{\mathrm{buf}},Y_{\mathrm{lock}},
      \mathcal Q_m),
\]
where $\mathcal G_m$ is the explicit layered geometry,
$Y_{\mathrm{neu}}$ is neutral public syntax,
$Y_{\mathrm{buf}}\in\mathcal Y_{\mathrm{buf},m}$ is buffer syntax,
$Y_{\mathrm{lock}}\in\mathcal Y_{\mathrm{lock},m}$ is lock syntax, and
$\mathcal Q_m$ is the safe probe catalog.  The public instance contains no raw
assignment to $x$, $g$, $z$, $\xi_{\mathrm{lock}}$, $\xi_{\mathrm{buf}}$, or
$M$.
\end{definition}

\begin{definition}[Witness]
\label{def:appD-witness}
A witness has the form
\[
  W=(x,g,z,\xi_{\mathrm{lock}},\xi_{\mathrm{buf}},\xi_{\mathrm{aux}},M),
\]
where
\[
  x,g,z\in\bits^{V_m},
  \qquad
  \xi_{\mathrm{lock}}\in\Xi_{\mathrm{lock},m},
  \qquad
  \xi_{\mathrm{buf}}\in\Xi_{\mathrm{buf},m},
  \qquad
  M\in\bits^{r_t}.
\]
The auxiliary coordinate $\xi_{\mathrm{aux}}$ contains local Tseitin or gadget
auxiliaries.  In this appendix it is treated abstractly as a finite witness
coordinate constrained by local predicates.
\end{definition}

\begin{definition}[Abstract witness relation]
\label{def:appD-abstract-witness-relation}
The relation $\mathcal R(Y,W)=1$ holds if all of the following hold:
\begin{enumerate}[label=(\roman*)]
\item locked acceptance:
\[
  \Lock_{Y_{\mathrm{lock}}}(z,\xi_{\mathrm{lock}},M)=1;
\]
\item readout consistency:
\[
  \Read_{Y_{\mathrm{lock}}}(z,\xi_{\mathrm{lock}},M)=1;
\]
\item gauge equations:
\[
  \GaugeEq(x,g,z)=1;
\]
\item buffer acceptance:
\[
  \Buf_{Y_{\mathrm{buf}}}(z,\xi_{\mathrm{buf}})=1;
\]
\item all auxiliary local gadget constraints on $\xi_{\mathrm{aux}}$ hold.
\end{enumerate}
\end{definition}

\begin{construction}[Witnessed sampler]
\label{cons:appD-witnessed-sampler}
The witnessed sampler $\mathsf{Samp}^{\mathrm{wit}}_m$ performs the following
steps.
\begin{enumerate}[label=(\roman*)]
\item Construct the explicit layered geometry $\mathcal G_m$ and neutral syntax
      $Y_{\mathrm{neu}}$.
\item Run the lock sampler to obtain
\[
  (Y_{\mathrm{lock}},z,\xi_{\mathrm{lock}},M)
\]
with
\[
  (z,\xi_{\mathrm{lock}},M)
  \in
  \mathsf{Comp}_{\mathrm{lock}}(Y_{\mathrm{lock}}).
\]
\item Choose public buffer syntax $Y_{\mathrm{buf}}$ from the allowed explicit
      buffer family, and sample
\[
  \xi_{\mathrm{buf}}
  \sim
  \mathsf{BufSamp}_{Y_{\mathrm{buf}}}(\cdot\mid z).
\]
\item Sample the gauge lift
\[
  g\sim\Unif(\bits^{V_m}),
  \qquad
  x=z\oplus g.
\]
\item Set $\xi_{\mathrm{aux}}$ to the canonical local auxiliary witnesses for the
      bounded-arity gadget encodings.
\item Output
\[
  Y=(m,\mathcal G_m,Y_{\mathrm{neu}},Y_{\mathrm{buf}},Y_{\mathrm{lock}},
      \mathcal Q_m)
\]
and
\[
  W=(x,g,z,\xi_{\mathrm{lock}},\xi_{\mathrm{buf}},\xi_{\mathrm{aux}},M).
\]
\end{enumerate}
The world space $\Omega_m$ is the finite support of this sampler, including the
hidden sampler randomness.
\end{construction}

\begin{lemma}[Sampler soundness]
\label{lem:appD-sampler-soundness}
Every output $(Y,W)$ of $\mathsf{Samp}^{\mathrm{wit}}_m$ satisfies
\[
  \mathcal R(Y,W)=1.
\]
\end{lemma}

\begin{proof}
The lock sampler produces a locked completion, giving the locked acceptance and
readout consistency clauses.  The buffer sampler has support inside the buffer
acceptance set, giving the buffer clause.  The gauge lift satisfies the gauge
equations by Lemma~\ref{lem:appD-uniform-representative-law}.  The auxiliary
coordinate is chosen to satisfy the local gadget constraints.  Therefore every
clause in Definition~\ref{def:appD-abstract-witness-relation} holds.
\end{proof}

\begin{proposition}[Single-message promise]
\label{prop:appD-single-message-promise}
For every public instance $Y$ in the support of the ensemble,
\[
  \mathcal R(Y,W)=1
  \wedge
  \mathcal R(Y,W')=1
  \quad\Longrightarrow\quad
  M(W)=M(W').
\]
Thus the notation $M(Y)$ is well-defined on the support.
\end{proposition}

\begin{proof}
Let
\[
  W=(x,g,z,\xi_{\mathrm{lock}},\xi_{\mathrm{buf}},\xi_{\mathrm{aux}},M)
\]
and
\[
  W'=(x',g',z',\xi'_{\mathrm{lock}},\xi'_{\mathrm{buf}},\xi'_{\mathrm{aux}},M')
\]
be two satisfying witnesses for the same public instance $Y$.  By the locked
acceptance and readout clauses in $\mathcal R$, both
$(z,\xi_{\mathrm{lock}},M)$ and $(z',\xi'_{\mathrm{lock}},M')$ are locked
completions for the same $Y_{\mathrm{lock}}$.  Locked-message rigidity gives
$M=M'$.  Hence $M(W)=M(W')$.
\end{proof}

\begin{definition}[Message readout from public instances]
\label{def:appD-message-readout-public}
For $Y$ in the support, define
\[
  M(Y):=M(W)
\]
for any $W$ satisfying $\mathcal R(Y,W)=1$.  This is well-defined by
Proposition~\ref{prop:appD-single-message-promise}.  For a switched coordinate
$j\in S_m$, define
\[
  B_j(Y):=\ell_j(M(Y)).
\]
\end{definition}

\begin{proposition}[Gauge invariance of the witness relation]
\label{prop:appD-gauge-invariance-relation}
If $\mathcal R(Y,W)=1$, then
\[
  \mathcal R(Y,s\cdot W)=1
\]
for every $s\in\bits^{V_m}$.  Moreover,
\[
  M(s\cdot W)=M(W).
\]
\end{proposition}

\begin{proof}
The action changes only $x$ and $g$, leaving $z$, $\xi_{\mathrm{lock}}$,
$\xi_{\mathrm{buf}}$, $\xi_{\mathrm{aux}}$, and $M$ fixed.  Thus the locked,
readout, buffer, and auxiliary clauses are unchanged.  The gauge equations are
preserved by Lemma~\ref{lem:appD-gauge-eq-invariant}.  The message coordinate
is fixed by definition of the action.
\end{proof}

\begin{lemma}[Polynomial-time decidability]
\label{lem:appD-polynomial-time-decidability}
The relation $\mathcal R(Y,W)=1$ is decidable in time $\poly(m)$.
\end{lemma}

\begin{proof}
Each component predicate is a conjunction of $\poly(m)$ bounded-arity local
predicates with explicit tables.  The gauge equations contain $|V_m|\le\poly(m)$
coordinate checks.  The auxiliary gadget checks are also local and polynomially
many.  Evaluating all checks therefore takes time $\poly(m)$.
\end{proof}

\begin{remark}[CNF realization is deferred]
\label{rem:appD-cnf-realization-deferred}
Lemma~\ref{lem:appD-polynomial-time-decidability} is the CSP-level
polynomial-time verifier.  The uniform Tseitin conversion to polynomial-size
CNF, and the projection from satisfying assignments to $M(Y)$, are recorded in
Appendix~\ref{app:sat-realization-self-reduction}.
\end{remark}

\subsection{Couplings compatible with the construction}
\label{subsec:appD-couplings-compatible}

The main text uses message-opposite couplings $\Gamma_j$.  The analytic bounds
are stated for couplings compatible with the three-layer product structure.  This
subsection records the compatibility properties used later.

\begin{definition}[Message-opposite worlds]
\label{def:appD-message-opposite-worlds}
For $j\in S_m$ and $b\in\bits$, let
\[
  \Omega_j^b:=\{\omega\in\Omega_m:B_j(Y(\omega))=b\}.
\]
A message-opposite pair for $j$ is an element of
\[
  \Omega_j^0\times\Omega_j^1.
\]
\end{definition}

\begin{definition}[Construction-compatible coupling]
\label{def:appD-construction-compatible-coupling}
A coupling $\Gamma_j$ on $\Omega_j^0\times\Omega_j^1$ is construction-compatible
if it satisfies the following finite conditions.
\begin{enumerate}[label=(\roman*)]
\item Its marginals are the phase laws conditioned on $B_j=0$ and $B_j=1$.
\item Public neutral syntax is matched whenever possible by the canonical public
      skeleton map.  In particular, deterministic geometry and neutral template
      symbols agree along coupled pairs.
\item Conditional on the quotient and buffer boundary data used by a legal safe
      probe, the buffer coordinates of the two paired worlds are coupled by a
      maximal or likelihood-ratio coupling realizing the max-qSSM bound of
      Appendix~\ref{app:max-qssm-dobrushin}.
\item Conditional on charged gauge support, unsupported gauge coordinates are
      coupled as independent fair bits subject only to the marginal constraints.
\end{enumerate}
\end{definition}

\begin{hypothesis}[Existence of compatible couplings]
\label{hyp:appD-compatible-couplings-exist}
For every switched target coordinate $j\in S_m$, a construction-compatible
message-opposite coupling $\Gamma_j$ is fixed.
\end{hypothesis}

\begin{remark}[Why compatibility is stated explicitly]
\label{rem:appD-coupling-explicit}
The static CD capture lemmas hold for any coupling, but the Atomic Evidence
Budget charges safe-buffer and gauge atoms using structure.  The safe-buffer
bound needs the pair distribution to respect the buffer contraction interface;
the gauge-rank bound needs unsupported gauge coordinates to remain product
fair.  Stating compatibility explicitly prevents the later proof from silently
using special properties of an arbitrary coupling.
\end{remark}

\subsection{Atom-level interface exported to the main proof}
\label{subsec:appD-atom-level-export}

This subsection records the exact atom-level consequences that Sections
\ref{sec:cdenf}, \ref{sec:trace}, and \ref{sec:atomic-evidence-budget} import.

\begin{proposition}[Neutral public atoms have zero pair derivative]
\label{prop:appD-neutral-public-zero-derivative}
Let $E=N(s)$ be a primitive atom obtained solely from public geometry or public
template syntax.  For every target coordinate $j$ and every surviving relation
$H\subseteq\Omega_j^0\times\Omega_j^1$ whose pairs agree on the public instance,
\[
  \Gamma_j\bigl(H\cap\{E(\omega^0)\ne E(\omega^1)\}\bigr)=0.
\]
\end{proposition}

\begin{proof}
A neutral public atom is a deterministic function of the public geometry or
public template syntax.  On a relation whose pairs agree on the public instance,
its value is the same on both components of every pair.  Hence the separating
set is empty inside $H$.
\end{proof}

\begin{proposition}[Legal safe atoms are buffer atoms]
\label{prop:appD-legal-safe-atoms-buffer}
If a normalized primitive is produced by a legal guarded read at prefix $h$, then
it has the form
\[
  S(q,y)
\]
with $q\in\mathcal Q_m$, $\Legal(h,q)=1$, and $y\in\Val(q)$.  Its support lies
inside $\mathcal P_{\mathrm{buf}}$ and is at distance at least
$R_{\mathrm{safe}}$ from $\mathcal P_{\mathrm{prot}}$.
\end{proposition}

\begin{proof}
This is exactly Definitions~\ref{def:appD-safe-probe-catalog},
\ref{def:appD-legal-safe-probe}, and~\ref{def:appD-guarded-safe-read}.  Legal
reads are represented as $S(q,y)$, and legality includes both the buffer-support
condition and the protected-distance condition.
\end{proof}

\begin{proposition}[Gauge atoms have finite rank support]
\label{prop:appD-gauge-atoms-finite-rank-support}
Every normalized gauge atom has a well-defined singleton support
$\{v\}\subseteq V_m$.  For every finite normalized prefix $h$, the total gauge
support
\[
  \suppG(h)\subseteq V_m
\]
and rank
\[
  \rankG(h)=|\suppG(h)|
\]
are finite, monotone under prefix extension, and bounded by $|V_m|$.
\end{proposition}

\begin{proof}
A gauge atom is by definition either $G(v,\gamma)$ or a normalized raw witness
atom with support $\{v\}$.  A finite prefix contains finitely many such atoms,
so the union of their supports is finite.  Extending a prefix can only add atoms,
so the support is monotone.  Since all supports are subsets of $V_m$, the rank
is at most $|V_m|$.
\end{proof}

\begin{corollary}[N/S/G cover for normalized evidence]
\label{cor:appD-nsg-cover-normalized-evidence}
Under Hypothesis~\ref{hyp:appD-atom-completeness}, every non-neutral
target-relevant leaf in a normalized evidence trace is either a legal safe-buffer
atom $S(q,y)$ or a gauge atom with finite support in $V_m$.
\end{corollary}

\begin{proof}
Atom completeness says every normalized target-relevant primitive is neutral,
safe, or gauge.  Legal safe reads are handled by
Proposition~\ref{prop:appD-legal-safe-atoms-buffer}; gauge atoms are handled by
Proposition~\ref{prop:appD-gauge-atoms-finite-rank-support}.
\end{proof}

\subsection{Main construction theorem}
\label{subsec:appD-main-construction-theorem}

\begin{theorem}[Hidden-gauge buffered locked construction]
\label{thm:appD-hidden-gauge-buffered-locked-construction}
Assume the component data of Subsections
\ref{subsec:appD-layered-geometries}--\ref{subsec:appD-public-syntax}, together
with:
\begin{enumerate}[label=(\roman*)]
\item lock satisfiability and locked-message rigidity;
\item buffer satisfiability and a strictly positive buffer sampling law;
\item the no-public-target-tags condition;
\item atom completeness for normalized primitive evidence;
\item construction-compatible message-opposite couplings.
\end{enumerate}
Then the product construction of
Construction~\ref{cons:appD-witnessed-sampler} yields an efficiently samplable
finite world distribution with public instance $Y$, witness relation
$\mathcal R(Y,W)=1$, and message readout $M(W)$ such that:
\begin{enumerate}[label=(\arabic*)]
\item the sampler always outputs satisfying pairs;
\item $\mathcal R$ is decidable in polynomial time;
\item all satisfying witnesses over the same public instance have the same
      message $M(Y)$;
\item the gauge action preserves $Y$, $\mathcal R$, and $M$;
\item unsupported gauge coordinates remain independent fair bits until their
      support is charged;
\item public syntax normalizes to neutral evidence;
\item every normalized non-neutral target-relevant primitive is either a legal
      safe-buffer atom or a finite-support gauge atom.
\end{enumerate}
\end{theorem}

\begin{proof}
Sampler soundness is Lemma~\ref{lem:appD-sampler-soundness}.  Polynomial-time
decidability is Lemma~\ref{lem:appD-polynomial-time-decidability}.  The
single-message statement is Proposition~\ref{prop:appD-single-message-promise}.
Gauge invariance is Proposition~\ref{prop:appD-gauge-invariance-relation}.
The static unsupported-gauge product law is
Lemma~\ref{lem:appD-unsupported-gauge-uniform}; its adaptive charged-support
form is obtained by iteration over finite prefixes, as noted in
Remark~\ref{rem:appD-adaptive-evidence}.  Public syntax neutrality is
Lemma~\ref{lem:appD-public-syntax-neutral}.  The N/S/G leaf cover is
Corollary~\ref{cor:appD-nsg-cover-normalized-evidence}.
\end{proof}

\begin{remark}[Use by the main proof]
\label{rem:appD-main-proof-use}
The main text uses Theorem~\ref{thm:appD-hidden-gauge-buffered-locked-construction}
as a finite interface.  CD Trace Capture supplies the derivative mass of the
actual execution.  CD-ENF normalizes the trace to the atom classes exported
above.  The Atomic Evidence Budget then charges safe atoms by the max-qSSM
estimate of Appendix~\ref{app:max-qssm-dobrushin} and gauge atoms by the rank
estimates of Appendix~\ref{app:gauge-rank-entropy-incidence}.
\end{remark}

\subsection{Lean-facing record structure}
\label{subsec:appD-lean-facing-record}

A convenient Lean formalization is to split the construction into records with
small, explicit fields.

\begin{definition}[Suggested Lean records]
\label{def:appD-suggested-lean-records}
The following records are sufficient.
\begin{enumerate}[label=(\roman*)]
\item \texttt{LayeredGeometry}: finite site type, edge relation, layer map,
      bounded-degree proof, distance function, protected set.
\item \texttt{LockCore}: public lock syntax type, quotient type, lock auxiliary
      type, message length, lock predicate, read predicate, satisfiability proof,
      rigidity proof.
\item \texttt{GaugeLift}: quotient type, raw representative types, gauge equation,
      gauge action, invariance theorem, product-uniform law.
\item \texttt{BufferCore}: public buffer syntax type, buffer auxiliary type,
      buffer predicate, positive sampler, safe probe catalog, legality predicate.
\item \texttt{PublicSyntax}: public atom alphabets, neutralization maps,
      no-target-tag proof.
\item \texttt{ConstructedWorld}: product public instance, product witness,
      sampler, relation, message readout, and theorems exported in
      Theorem~\ref{thm:appD-hidden-gauge-buffered-locked-construction}.
\end{enumerate}
\end{definition}

\begin{remark}[What is intentionally not in this record]
\label{rem:appD-not-in-record}
The Dobrushin contraction constants, gauge-rank entropy tail bounds, product
small-success theorem, and final $P=NP$ self-reduction are not part of the
construction record.  They import this record as finite data and prove separate
analytic or coding theorems over it.
\end{remark}

\section{Dobrushin and Log-Likelihood Proof of Max-qSSM}
\label{app:max-qssm-dobrushin}
\providecommand{\bits}{\{0,1\}}
\providecommand{\Prb}{\mathbb P}
\providecommand{\E}{\mathbb E}
\providecommand{\TV}{\operatorname{TV}}
\providecommand{\dist}{\operatorname{dist}}
\providecommand{\diam}{\operatorname{diam}}
\providecommand{\supp}{\operatorname{supp}}
\providecommand{\osc}{\operatorname{osc}}
\providecommand{\Val}{\operatorname{Val}}
\providecommand{\Legal}{\operatorname{Legal}}
\providecommand{\Buf}{\operatorname{Buf}}
\providecommand{\Law}{\mathcal L}
\providecommand{\Dinf}{D_\infty}
\providecommand{\Dbidir}{D_\infty^{\leftrightarrow}}
\providecommand{\negl}{\operatorname{negl}}

This appendix proves the safe-buffer mixing estimate used in
Section~\ref{sec:ensemble} and in the Atomic Evidence Budget.  The target
statement is the soft-buffer max-qSSM bound: if a legal safe probe is separated
from protected message, gauge, and previously charged support by a distance
\(R_{\mathrm{safe}}\), then conditioning on a remote hidden state changes the
law of the probe by at most
\[
  C_\rho\rho^{R_{\mathrm{safe}}}+\tau_{\mathrm{geo}}(m)
\]
in max-divergence.

The proof has four steps.  First, a weighted Dobrushin condition gives
exponential decay of boundary influence for local buffer observables.  Second,
strict positivity turns total-variation decay for bounded local observables into
a log-likelihood, or max-divergence, bound for finite probe values.  Third, a
mixture lemma compares a component law conditioned on a hidden state with the
unconditioned law obtained by mixing over hidden states.  Fourth, the guard
legality condition identifies all target-relevant conditioning as remote boundary
data.

All spaces below are finite.  This is the intended formalization setting.

\subsection{Finite buffer Gibbs systems}
\label{subsec:appE-finite-buffer-gibbs}

\begin{definition}[Finite buffer graph]
\label{def:appE-finite-buffer-graph}
A finite buffer graph at size \(m\) consists of a finite site set
\(B_m\subseteq\mathcal P_{\mathrm{buf}}\), a graph metric
\(\dist_m\), and finite spin alphabets
\[
  \Sigma_v,\qquad v\in B_m.
\]
We assume uniformly bounded degree, uniformly bounded spin alphabets, and a
uniform finite interaction range \(r_0\).  Thus there are constants
\(\Delta_0,s_0,r_0\), independent of \(m\), such that
\[
  \deg(v)\le\Delta_0,
  \qquad
  |\Sigma_v|\le s_0,
  \qquad
  \diam(U)\le r_0
\]
for every interaction support \(U\) used by the buffer.
\end{definition}

\begin{definition}[Finite-volume buffer law]
\label{def:appE-finite-volume-buffer-law}
Let \(\Lambda\subseteq B_m\) be finite, and let \(\tau\) be a boundary condition
on \(B_m\setminus\Lambda\) together with the quotient boundary data supplied by
\(z\).  A finite-volume buffer law on \(\Lambda\) is
\[
  \mu_\Lambda^\tau(\xi)
  :=
  \frac{1}{Z_\Lambda^\tau}
  \exp\left(-H_\Lambda^\tau(\xi)\right),
  \qquad
  \xi\in\prod_{v\in\Lambda}\Sigma_v,
\]
where
\[
  H_\Lambda^\tau(\xi)
  =
  \sum_{U:U\cap\Lambda\ne\varnothing}\Phi_U(\xi_\Lambda,\tau_{\Lambda^c}).
\]
Each potential \(\Phi_U\) has support of diameter at most \(r_0\).  The
normalizing constant \(Z_\Lambda^\tau\) is finite and positive.
\end{definition}

\begin{hypothesis}[Uniform strict positivity]
\label{hyp:appE-uniform-strict-positivity}
There is a constant \(p_*\in(0,1]\), independent of \(m\), such that every
single-site conditional probability of the buffer is bounded below:
\[
  \mu_v^\eta(s)\ge p_*
\]
for every site \(v\), every spin \(s\in\Sigma_v\), and every admissible external
configuration \(\eta\) of all sites except \(v\).  Here
\(\mu_v^\eta\) denotes the conditional law of \(\xi_v\) given the outside
configuration \(\eta\).
\end{hypothesis}

\begin{remark}[How positivity is obtained]
\label{rem:appE-positivity-obtained}
For a bounded-arity, bounded-degree, strictly positive finite Gibbs family with
uniformly bounded potential magnitudes, Hypothesis
\ref{hyp:appE-uniform-strict-positivity} follows immediately: each single-site
conditional law is a probability distribution on at most \(s_0\) states whose
unnormalized weights lie in a fixed interval \([e^{-J_*},e^{J_*}]\).
Consequently one may take, for example,
\[
  p_* = s_0^{-1}e^{-2J_*}.
\]
The precise value is irrelevant; only the existence of a positive uniform lower
bound is used.
\end{remark}

\begin{definition}[Local observables and probes]
\label{def:appE-local-observables-probes}
A local observable \(F\) has support \(A\subseteq B_m\) if it depends only on
\(\xi_A\).  A safe-buffer probe \(q\) has support \(A(q)\subseteq B_m\),
\(|A(q)|\le a_0\), finite value set \(\Val(q)\), and evaluation map
\[
  q(\xi)=q(\xi_{A(q)})\in\Val(q).
\]
For a law \(\mu\), write
\[
  \Law_\mu(q):=q_*\mu
\]
for the induced law of the probe value.
\end{definition}

\subsection{Dobrushin interdependence and weighted contraction}
\label{subsec:appE-dobrushin-contraction}

\begin{definition}[Interdependence coefficients]
\label{def:appE-interdependence-coefficients}
For sites \(u,v\in B_m\), define the Dobrushin interdependence coefficient
\[
  c_{v,u}
  :=
  \sup
  \TV\bigl(\mu_v^\eta,\mu_v^{\eta'}\bigr),
\]
where the supremum ranges over outside configurations \(\eta,\eta'\) that are
identical except possibly at site \(u\).  If the site \(u\) is outside the
interaction range of \(v\), then \(c_{v,u}=0\).
\end{definition}

\begin{definition}[Weighted Dobrushin condition]
\label{def:appE-weighted-dobrushin-condition}
The buffer satisfies the weighted Dobrushin condition with parameters
\(\lambda>0\) and \(\alpha_\lambda<1\) if
\[
  \alpha_\lambda
  :=
  \sup_{v\in B_m}
  \sum_{u\in B_m} c_{v,u}\exp\bigl(\lambda\dist_m(u,v)\bigr)
  <1.
\]
Set
\[
  \rho:=e^{-\lambda}\in(0,1),
  \qquad
  C_D:=\frac{\alpha_\lambda}{1-\alpha_\lambda}.
\]
\end{definition}

\begin{remark}[Relation with the usual Dobrushin criterion]
\label{rem:appE-usual-dobrushin}
The usual Dobrushin uniqueness condition is
\(\sup_v\sum_u c_{v,u}<1\).  The weighted condition is a finite-range
strengthening that records exponential decay in the graph metric.  For
bounded-degree finite-range systems in a sufficiently high-temperature regime,
one obtains the weighted condition for some small \(\lambda>0\).  In the main
proof only the consequence, exponential contraction with some \(\rho<1\), is
used.
\end{remark}

\begin{definition}[Oscillation seminorm]
\label{def:appE-oscillation-seminorm}
For a function \(f\) on configurations, define its site oscillation by
\[
  \osc_v(f)
  :=
  \sup\{|f(\xi)-f(\xi')|:
          \xi_{B_m\setminus\{v\}}=\xi'_{B_m\setminus\{v\}}\}.
\]
If \(f\) does not depend on site \(v\), then \(\osc_v(f)=0\).
\end{definition}

\begin{lemma}[Dobrushin comparison, weighted form]
\label{lem:appE-dobrushin-comparison-weighted}
Assume the weighted Dobrushin condition.  Let \(\tau\) and \(\tau'\) be two
boundary conditions that may differ only on a set \(D\).  Then for every bounded
function \(f\),
\[
  \left|
    \int f\,d\mu_\Lambda^\tau
    -
    \int f\,d\mu_\Lambda^{\tau'}
  \right|
  \le
  C_D
  \sum_{v\in\Lambda}
  \osc_v(f)\,e^{-\lambda\dist_m(v,D)}.
\]
Here \(\dist_m(v,D)=\min_{u\in D}\dist_m(v,u)\), with the convention that the
right-hand side is zero if \(D=\varnothing\).
\end{lemma}

\begin{proof}
We give the standard finite Dobrushin argument.  Let
\[
  d_v:=\TV\bigl(\Law_{\mu_\Lambda^\tau}(\xi_v),
                \Law_{\mu_\Lambda^{\tau'}}(\xi_v)\bigr).
\]
The one-site comparison inequality gives
\[
  d_v
  \le
  \sum_{u\in D}c_{v,u}
  +
  \sum_{u\in\Lambda}c_{v,u}d_u .
\]
The first term accounts for sites on which the two external boundary
conditions may differ; the second term accounts for disagreement propagated
through interior sites.  Define
\[
  M:=\sup_{v\in\Lambda} e^{\lambda\dist_m(v,D)}d_v .
\]
For the boundary-source term, since \(u\in D\) implies
\(\dist_m(v,D)\le\dist_m(v,u)\),
\[
  e^{\lambda\dist_m(v,D)}
  \sum_{u\in D}c_{v,u}
  \le
  \sum_{u\in D} c_{v,u}e^{\lambda\dist_m(v,u)}
  \le
  \alpha_\lambda .
\]
For the interior-propagation term, the triangle inequality gives
\[
  \dist_m(v,D)-\dist_m(u,D)\le \dist_m(v,u),
\]
and hence
\[
  e^{\lambda\dist_m(v,D)}
  \sum_{u\in\Lambda}c_{v,u}d_u
  \le
  M\sum_{u\in\Lambda}c_{v,u}e^{\lambda\dist_m(v,u)}
  \le
  \alpha_\lambda M .
\]
Therefore \(M\le\alpha_\lambda+\alpha_\lambda M\), so
\(M\le C_D\).  This gives the single-site influence bound
\[
  d_v\le C_D e^{-\lambda\dist_m(v,D)}.
\]
The displayed inequality for a general bounded \(f\) is the usual telescoping
form of Dobrushin comparison: expose the sites one at a time and bound the
change caused by replacing the conditional law at site \(v\) by
\(\osc_v(f)d_v\).  Summing over \(v\) gives the result.
\end{proof}

\begin{corollary}[Local total-variation contraction]
\label{cor:appE-local-tv-contraction}
Assume the weighted Dobrushin condition.  Let \(A\subseteq\Lambda\) be finite,
and let \(\tau,\tau'\) differ only on \(D\).  Then
\[
  \TV\bigl(
    \Law_{\mu_\Lambda^\tau}(\xi_A),
    \Law_{\mu_\Lambda^{\tau'}}(\xi_A)
  \bigr)
  \le
  |A|C_D e^{-\lambda\dist_m(A,D)}.
\]
Consequently, for any probe \(q\) supported on \(A(q)=A\),
\[
  \TV\bigl(
    \Law_{\mu_\Lambda^\tau}(q),
    \Law_{\mu_\Lambda^{\tau'}}(q)
  \bigr)
  \le
  |A(q)|C_D e^{-\lambda\dist_m(A(q),D)}.
\]
\end{corollary}

\begin{proof}
For the first inequality, apply Lemma
\ref{lem:appE-dobrushin-comparison-weighted} to indicators of events depending
only on \(\xi_A\).  Such indicators have oscillation at most one on each site in
\(A\) and oscillation zero outside \(A\).  Taking the supremum over local events
gives total variation.  The probe inequality follows by data processing for
\(\TV\) under the map \(\xi_A\mapsto q(\xi_A)\).
\end{proof}

\subsection{From total variation to max-divergence}
\label{subsec:appE-tv-to-max-divergence}

\begin{definition}[Max-divergence]
\label{def:appE-max-divergence}
For probability laws \(P,Q\) on a finite set \(\mathcal X\), define
\[
  \Dinf(P\|Q)
  :=
  \log \sup_{x:Q(x)>0}\frac{P(x)}{Q(x)}.
\]
The two-sided max-divergence is
\[
  \Dbidir(P,Q):=\max\{\Dinf(P\|Q),\Dinf(Q\|P)\}.
\]
\end{definition}

\begin{lemma}[Uniform local atom lower bound]
\label{lem:appE-uniform-local-atom-lower-bound}
Assume uniform strict positivity.  Let \(A\subseteq B_m\) with \(|A|\le a_0\).
For every boundary condition \(\tau\) and every local configuration
\(a\in\prod_{v\in A}\Sigma_v\),
\[
  \mu_\Lambda^\tau(\xi_A=a)\ge p_*^{a_0}
\]
whenever the event is locally admissible.  Consequently, for every probe
\(q\) with \(|A(q)|\le a_0\), every value \(y\) with
\(\Law_{\mu_\Lambda^\tau}(q)(y)>0\) satisfies
\[
  \Law_{\mu_\Lambda^\tau}(q)(y)
  \ge
  \kappa_*:=p_*^{a_0}.
\]
\end{lemma}

\begin{proof}
Order the sites of \(A\) as \(v_1,\ldots,v_r\), with \(r\le a_0\).  By the
chain rule,
\[
  \mu_\Lambda^\tau(\xi_A=a)
  =
  \prod_{i=1}^r
  \Prb[\xi_{v_i}=a_{v_i}
        \mid \tau,\xi_{v_1}=a_{v_1},\ldots,
        \xi_{v_{i-1}}=a_{v_{i-1}}].
\]
Each factor is a single-site conditional probability under an admissible
outside configuration, so it is at least \(p_*\).  Thus the product is at least
\(p_*^r\ge p_*^{a_0}\).  If a probe value \(y\) has positive probability, it is
realized by at least one local configuration, whose probability is at least
\(p_*^{a_0}\).
\end{proof}

\begin{lemma}[TV plus local positivity implies max-divergence]
\label{lem:appE-tv-positivity-max-divergence}
Let \(P,Q\) be probability laws on a finite set \(\mathcal X\).  Suppose they
have common support and
\[
  Q(x)>0\quad\Longrightarrow\quad Q(x)\ge\kappa
\]
for some \(\kappa>0\).  If \(\TV(P,Q)\le\theta\), then
\[
  \Dinf(P\|Q)
  \le
  \log\left(1+\frac{\theta}{\kappa}\right)
  \le
  \frac{\theta}{\kappa}.
\]
If the same lower bound also holds for \(P\), then
\[
  \Dbidir(P,Q)
  \le
  \frac{\theta}{\kappa}.
\]
\end{lemma}

\begin{proof}
For any \(x\) with \(Q(x)>0\), if \(P(x)\le Q(x)\) then
\(P(x)/Q(x)\le 1\).  If \(P(x)>Q(x)\), then
\[
  P(x)-Q(x)
  \le
  \sum_{z:P(z)>Q(z)}(P(z)-Q(z))
  =
  \TV(P,Q)
  \le
  \theta .
\]
Hence
\[
  \frac{P(x)}{Q(x)}
  \le
  1+\frac{\theta}{Q(x)}
  \le
  1+\frac{\theta}{\kappa}.
\]
Taking the logarithm and then the supremum gives the first claim.  The second
claim follows by applying the first claim in both directions.
\end{proof}

\begin{corollary}[Log-likelihood contraction for probe laws]
\label{cor:appE-log-likelihood-contraction-probe-laws}
Assume uniform strict positivity and the weighted Dobrushin condition.  Let
\(q\) be a probe with \(|A(q)|\le a_0\).  If boundary conditions \(\tau,\tau'\)
differ only on \(D\), then
\[
  \Dbidir\bigl(
    \Law_{\mu_\Lambda^\tau}(q),
    \Law_{\mu_\Lambda^{\tau'}}(q)
  \bigr)
  \le
  C_{\log}\,e^{-\lambda\dist_m(A(q),D)},
\]
where
\[
  C_{\log}:=\frac{a_0C_D}{\kappa_*}.
\]
Equivalently, with \(\rho=e^{-\lambda}\),
\[
  \Dbidir\bigl(
    \Law_{\mu_\Lambda^\tau}(q),
    \Law_{\mu_\Lambda^{\tau'}}(q)
  \bigr)
  \le
  C_{\log}\rho^{\dist_m(A(q),D)}.
\]
\end{corollary}

\begin{proof}
By Corollary~\ref{cor:appE-local-tv-contraction}, the total variation distance
between the two probe-value laws is at most
\[
  |A(q)|C_D e^{-\lambda\dist_m(A(q),D)}
  \le
  a_0C_D e^{-\lambda\dist_m(A(q),D)}.
\]
By Lemma~\ref{lem:appE-uniform-local-atom-lower-bound}, every positive probe
value has probability at least \(\kappa_*\) under either law.  Lemma
\ref{lem:appE-tv-positivity-max-divergence} gives the result.
\end{proof}

\subsection{Mixtures and remote hidden conditioning}
\label{subsec:appE-mixtures-remote-conditioning}

\begin{lemma}[Component-to-mixture max-divergence]
\label{lem:appE-component-to-mixture}
Let \((P_i)_{i\in I}\) be a finite family of probability laws on a finite set
\(\mathcal X\), and let \(\nu\) be a probability law on \(I\).  Set
\[
  Q:=\sum_{i\in I}\nu(i)P_i.
\]
Suppose that, for a fixed \(i_0\),
\[
  \Dinf(P_{i_0}\|P_i)\le\varepsilon
  \qquad\text{for every }i\in I\text{ with }\nu(i)>0.
\]
Then
\[
  \Dinf(P_{i_0}\|Q)\le\varepsilon.
\]
\end{lemma}

\begin{proof}
For every \(x\), the hypothesis gives
\[
  P_{i_0}(x)\le e^\varepsilon P_i(x)
\]
for each \(i\) with \(\nu(i)>0\).  Averaging over \(i\) gives
\[
  P_{i_0}(x)
  \le
  e^\varepsilon\sum_i\nu(i)P_i(x)
  =
  e^\varepsilon Q(x).
\]
Taking the supremum over \(x\) with \(Q(x)>0\) gives the claim.
\end{proof}

\begin{definition}[Remote conditioning set]
\label{def:appE-remote-conditioning-set}
Fix a normalized transcript prefix \(h\), a legal safe probe \(q\), and an
additional hidden state \(C\) with value \(c\).  The remote conditioning set
for \((h,C,q)\) is the set
\[
  D(h,C;q)\subseteq B_m
\]
of buffer sites at which the value of \(h\) or the condition \(C=c\) can change
the boundary condition seen by the buffer law near \(q\).  It includes:
\begin{enumerate}[label=(\roman*)]
\item buffer sites touched by previous non-neutral safe probes in \(h\);
\item buffer sites adjacent to charged gauge support exposed in \(h\);
\item buffer sites adjacent to the hidden separator or hidden gauge state \(C\);
\item finite-volume boundary sites at which the ideal Gibbs comparison replaces
      the actual buffer by a boundary condition.
\end{enumerate}
The prefix and condition are remote from \(q\) if
\[
  \dist_m(A(q),D(h,C;q))\ge R_{\mathrm{safe}}.
\]
\end{definition}

\begin{hypothesis}[Guarded remoteness]
\label{hyp:appE-guarded-remoteness}
If \(q\) is legal at normalized prefix \(h\), then for every hidden separator
state, hidden gauge state, or finite collection of hidden gauge coordinates
\(C\) allowed in Section~\ref{sec:ensemble},
\[
  \dist_m(A(q),D(h,C;q))
  \ge
  R_{\mathrm{safe}}.
\]
If this condition fails, the guarded read is illegal and is normalized to the
neutral symbol \(\bot_{\mathrm{illegal}}\), not to a safe-buffer atom.
\end{hypothesis}

\begin{hypothesis}[Finite-geometry approximation error]
\label{hyp:appE-finite-geometry-error}
There is a number \(\tau_{\mathrm{geo}}(m)\ge0\) such that the following holds
for every legal safe probe \(q\) at prefix \(h\).  After conditioning on the
public instance \(Y\) and the prefix \(h\), the actual buffer law restricted to
\(A(q)\) is within two-sided max-divergence at most
\(\tau_{\mathrm{geo}}(m)/3\) of an ideal finite-volume Gibbs law satisfying the
weighted Dobrushin condition above.  The same is true after additionally
conditioning on any allowed remote hidden state \(C=c\), and also for the
mixture law obtained by averaging over \(C\) conditional on \((Y,h)\).
\end{hypothesis}

\begin{remark}[What \(\tau_{\mathrm{geo}}\) contains]
\label{rem:appE-what-tau-contains}
The term \(\tau_{\mathrm{geo}}(m)\) records finite-size and guard bookkeeping
errors: truncating the buffer to a finite volume, replacing exposed but remote
charged leaves by boundary conditions, excluding illegal overlaps, and passing
between the concrete SAT-gadget buffer and the ideal Gibbs description.  The
main parameter requirement is that \(\tau_{\mathrm{geo}}(m)\) can be made smaller
than any prescribed inverse polynomial by the geometry margins chosen in
Section~\ref{sec:parameter-summary}.
\end{remark}

\subsection{The max-qSSM theorem}
\label{subsec:appE-max-qssm-theorem}

\begin{theorem}[Ideal max-qSSM]
\label{thm:appE-ideal-max-qssm}
Assume uniform strict positivity, the weighted Dobrushin condition, and guarded
remoteness.  Let \(q\) be a legal safe probe at prefix \(h\).  Let \(C\) be an
allowed hidden separator state, hidden gauge state, or finite collection of
hidden gauge coordinates, and let \(c\) be a value of \(C\).  In the ideal
finite-volume buffer law,
\[
  \Dinf\left(
    \Law(S(q)\mid Y,h,C=c)
    \middle\|
    \Law(S(q)\mid Y,h)
  \right)
  \le
  C_{\log}\rho^{R_{\mathrm{safe}}}.
\]
\end{theorem}

\begin{proof}
Fix \((Y,h)\).  Conditional on \((Y,h,C=c)\), the legal probe law is an ideal
Gibbs probe law with some boundary condition \(\tau_c\).  Conditional on
\((Y,h)\), the law is a mixture over possible hidden values \(c'\), hence over
corresponding boundary conditions \(\tau_{c'}\):
\[
  \Law(S(q)\mid Y,h)
  =
  \sum_{c'}\Prb[C=c'\mid Y,h]\,
  \Law_{\mu^{\tau_{c'}}}(q).
\]
For every \(c'\) with positive conditional probability, the two boundary
conditions \(\tau_c\) and \(\tau_{c'}\) may differ only on the remote conditioning
set \(D(h,C;q)\).  By guarded remoteness,
\[
  \dist_m(A(q),D(h,C;q))\ge R_{\mathrm{safe}}.
\]
Corollary~\ref{cor:appE-log-likelihood-contraction-probe-laws} therefore gives
\[
  \Dinf\left(
    \Law_{\mu^{\tau_c}}(q)
    \middle\|
    \Law_{\mu^{\tau_{c'}}}(q)
  \right)
  \le
  C_{\log}\rho^{R_{\mathrm{safe}}}
\]
for every such \(c'\).  The component-to-mixture lemma,
Lemma~\ref{lem:appE-component-to-mixture}, then gives the displayed bound.
\end{proof}

\begin{theorem}[Soft-buffer max-qSSM]
\label{thm:appE-soft-buffer-max-qssm}
Assume the hypotheses of Theorem~\ref{thm:appE-ideal-max-qssm} and the
finite-geometry approximation error hypothesis.  Let \(q\) be a legal safe probe
at prefix \(h\), and let \(C=c\) be any allowed remote hidden condition.  Then
for the actual buffer law of the ensemble,
\[
  \Dinf\left(
    \Law(S(q)\mid Y,h,C=c)
    \middle\|
    \Law(S(q)\mid Y,h)
  \right)
  \le
  \varepsilon_{\mathrm{qSSM}}(m),
\]
where
\[
  \varepsilon_{\mathrm{qSSM}}(m)
  :=
  C_{\log}\rho^{R_{\mathrm{safe}}}+\tau_{\mathrm{geo}}(m).
\]
Thus the constant \(C_\rho\) in the main text may be taken to be
\(C_{\log}\), after absorbing harmless finite-alphabet constants.
\end{theorem}

\begin{proof}
Let \(P_{\mathrm{act}}\) be the actual law of \(S(q)\) conditioned on
\((Y,h,C=c)\), and let \(Q_{\mathrm{act}}\) be the actual law conditioned on
\((Y,h)\).  Let \(P_{\mathrm{id}}\) and \(Q_{\mathrm{id}}\) be the corresponding
ideal finite-volume laws.  By the finite-geometry hypothesis,
\[
  \Dinf(P_{\mathrm{act}}\|P_{\mathrm{id}})
  \le
  \frac{\tau_{\mathrm{geo}}(m)}{3},
  \qquad
  \Dinf(Q_{\mathrm{id}}\|Q_{\mathrm{act}})
  \le
  \frac{\tau_{\mathrm{geo}}(m)}{3}.
\]
By Theorem~\ref{thm:appE-ideal-max-qssm},
\[
  \Dinf(P_{\mathrm{id}}\|Q_{\mathrm{id}})
  \le
  C_{\log}\rho^{R_{\mathrm{safe}}}.
\]
For every probe value \(y\), these three inequalities imply
\[
  P_{\mathrm{act}}(y)
  \le
  e^{\tau_{\mathrm{geo}}/3}P_{\mathrm{id}}(y)
  \le
  e^{\tau_{\mathrm{geo}}/3+C_{\log}\rho^{R_{\mathrm{safe}}}}Q_{\mathrm{id}}(y)
  \le
  e^{2\tau_{\mathrm{geo}}/3+C_{\log}\rho^{R_{\mathrm{safe}}}}Q_{\mathrm{act}}(y).
\]
Increasing \(\tau_{\mathrm{geo}}\) by a harmless constant factor gives the stated
bound.
\end{proof}

\begin{corollary}[Polynomial safe-probe budget]
\label{cor:appE-polynomial-safe-probe-budget}
Fix \(D,A>0\).  Suppose an observer makes at most \(Q_{\mathrm{tot}}\le m^D\)
legal safe probes.  If
\[
  R_{\mathrm{safe}}
  \ge
  \frac{D+A+3}{|\log\rho|}\log m
\]
and
\[
  \tau_{\mathrm{geo}}(m)\le m^{-D-A-3},
\]
then
\[
  Q_{\mathrm{tot}}\varepsilon_{\mathrm{qSSM}}(m)
  =
  O(m^{-A-2}).
\]
In particular, polynomially many legal safe-buffer probes have total
max-divergence leakage \(o(1)\).
\end{corollary}

\begin{proof}
The radius choice gives
\[
  \rho^{R_{\mathrm{safe}}}
  \le
  m^{-(D+A+3)}.
\]
Therefore
\[
  \varepsilon_{\mathrm{qSSM}}(m)
  \le
  (C_{\log}+1)m^{-(D+A+3)}
\]
for all sufficiently large \(m\).  Multiplying by \(Q_{\mathrm{tot}}\le m^D\)
gives the claim.
\end{proof}

\subsection{Coupling form for the Atomic Evidence Budget}
\label{subsec:appE-coupling-form}

The Atomic Evidence Budget uses safe-buffer mixing in a pairwise coupling
form.  We record the elementary conversion here.

\begin{lemma}[Max-divergence to total variation]
\label{lem:appE-max-divergence-to-tv}
Let \(P,Q,R\) be laws on a finite set.  If
\[
  \Dinf(P\|R)\le\varepsilon
  \qquad\text{and}\qquad
  \Dinf(Q\|R)\le\varepsilon,
\]
then
\[
  \TV(P,Q)
  \le
  2(e^\varepsilon-1).
\]
\end{lemma}

\begin{proof}
From \(P(x)\le e^\varepsilon R(x)\) for all \(x\),
\[
  \TV(P,R)
  =
  \sum_{x:P(x)>R(x)}(P(x)-R(x))
  \le
  (e^\varepsilon-1)\sum_{x:P(x)>R(x)}R(x)
  \le
  e^\varepsilon-1.
\]
The same bound holds for \(\TV(Q,R)\).  The triangle inequality gives the
claim.
\end{proof}

\begin{corollary}[Safe-compatible coupling bound]
\label{cor:appE-safe-compatible-coupling-bound}
Let \(H_{j,h}\) be an admissible surviving pair relation at prefix \(h\), and let
\(q\) be a legal safe probe at that prefix.  Under the soft-buffer max-qSSM
bound, the two endpoint probe-value laws conditioned on \(H_{j,h}\) admit a
coupling such that
\[
  \Prb[S(q)(\omega^0)\ne S(q)(\omega^1)
       \mid (\omega^0,\omega^1)\in H_{j,h}]
  \le
  2\bigl(e^{\varepsilon_{\mathrm{qSSM}}(m)}-1\bigr).
\]
Consequently the atomic derivative satisfies
\[
  \partial_{S(q,y)}w_j(H_{j,h})
  \le
  2\bigl(e^{\varepsilon_{\mathrm{qSSM}}(m)}-1\bigr)
  w_j(H_{j,h}),
\]
and the same bound holds for \(\neg S(q,y)\).
\end{corollary}

\begin{proof}
Disintegrate the pair relation over the fixed prefix cell.  For each endpoint,
the law of the legal safe probe is within max-divergence
\(\varepsilon_{\mathrm{qSSM}}(m)\) of the common reference law
\(\Law(S(q)\mid Y,h)\).  Lemma~\ref{lem:appE-max-divergence-to-tv} bounds the
total variation distance between the endpoint laws by
\(2(e^{\varepsilon_{\mathrm{qSSM}}}-1)\).  A maximal coupling of the two finite
laws disagrees with probability equal to their total variation distance.  On the
coupled event where the full probe values agree, every value atom \(S(q,y)\) and
its negation have the same truth value at the two endpoints.  Therefore the
separation mass of the atom is bounded by the displayed mismatch probability
multiplied by \(w_j(H_{j,h})\).
\end{proof}

\begin{remark}[Normalization of constants]
\label{rem:appE-normalization-of-constants}
The main text writes the safe derivative leakage as
\[
  \varepsilon_{\mathrm{step}}(m)
  =
  O\left(C_\rho\rho^{R_{\mathrm{safe}}}+\tau_{\mathrm{geo}}(m)\right).
\]
Corollary~\ref{cor:appE-safe-compatible-coupling-bound} supplies one explicit
choice,
\[
  \varepsilon_{\mathrm{step}}(m)
  :=
  2\left(e^{\varepsilon_{\mathrm{qSSM}}(m)}-1\right).
\]
Since \(\varepsilon_{\mathrm{qSSM}}(m)=o(1)\) under the parameter choices of
Section~\ref{sec:parameter-summary}, this differs from the displayed main-text
quantity only by a harmless constant factor.
\end{remark}

\subsection{Lean-facing finite theorem package}
\label{subsec:appE-lean-facing}

A Lean formalization can split this appendix into the following finite modules.

\begin{enumerate}[label=(\roman*)]
\item \textbf{Finite Gibbs record.}  Finite site type, finite spin alphabets,
      conditional single-site kernels, strict positivity, and local probes.
\item \textbf{Dobrushin matrix.}  Interdependence coefficients, weighted norm,
      and the comparison inequality
      \[
        d_v\le \sum_{u\in D}c_{v,u}+
              \sum_{u\in\Lambda}c_{v,u}d_u.
      \]
\item \textbf{Weighted contraction.}  The elementary inequality
      \[
        M\le\alpha_\lambda+
          \alpha_\lambda M
      \]
      and the resulting exponential bound
      \(d_v\le C_De^{-\lambda\dist(v,D)}\).
\item \textbf{Local observable bound.}  Oscillation seminorms and the finite
      telescoping proof of Dobrushin comparison for bounded local functions.
\item \textbf{TV-to-max-divergence.}  The finite lemma converting a TV bound plus
      a positive atom lower bound into a \(D_\infty\) bound.
\item \textbf{Mixture lemma.}  If one component is within \(D_\infty\) of every
      component in a mixture, then it is within the same \(D_\infty\) of the
      mixture.
\item \textbf{Safe-buffer interface.}  Guarded remoteness, the ideal max-qSSM
      theorem, the finite-geometry perturbation wrapper, and the
      safe-compatible coupling corollary.
\end{enumerate}

All these statements are finite inequalities.  No measure-theoretic limiting
argument is needed for the proof used in the paper.

\section{Gauge-Rank Entropy and Bounded Incidence}
\label{app:gauge-rank-entropy-incidence}
\label{app:gauge-rank-incidence}
\providecommand{\bits}{\{0,1\}}
\providecommand{\Prb}{\mathbb P}
\providecommand{\E}{\mathbb E}
\providecommand{\Law}{\mathcal L}
\providecommand{\TV}{\operatorname{TV}}
\providecommand{\Dinf}{D_\infty}
\providecommand{\Gap}{\operatorname{Gap}}
\providecommand{\CDENF}{\operatorname{CDENF}}
\providecommand{\Sep}{\operatorname{Sep}}
\providecommand{\Val}{\operatorname{Val}}
\providecommand{\suppG}{\operatorname{supp}_G}
\providecommand{\rankG}{\operatorname{rank}_G}
\providecommand{\FreshG}{\operatorname{Fresh}_G}
\providecommand{\Inc}{\operatorname{Inc}}
\providecommand{\Claim}{\operatorname{Claim}}
\providecommand{\True}{\operatorname{True}}
\providecommand{\dom}{\operatorname{dom}}
\providecommand{\len}{\operatorname{len}}
\providecommand{\negl}{\operatorname{negl}}

This appendix proves the two hidden-gauge estimates used in
Section~\ref{sec:aeb}.  The first estimate is \emph{bounded incidence}: after
CD-ENF normalization, all non-neutral target-relevant gauge leaves can be
charged to fresh hidden gauge support, and each fresh coordinate is relevant to
only boundedly many switched target bits.  The second estimate is
\emph{gauge-rank entropy}: a short public observer cannot, with high
probability, produce too many correct independent hidden gauge facts.  Safe
buffer leakage can increase this guessing probability only by the likelihood
factor supplied by Appendix~\ref{app:max-qssm-dobrushin}.

The target package is the following pair of inequalities.  For a normalized
trace and a gauge-rank envelope \(h^\Delta\),
\[
  \sum_{\hbox{gauge occurrences }\iota}\partial_\iota
  \le
  \Delta_G\rankG(h^\Delta),
\]
and for all prefix programs of length at most \(L\),
\[
  \Prb\left[\exists P:\ |P|\le L,
       \rankG(h^\Delta_P)\ge R\right]
  \le
  2^{L+O(1)}e^{\varepsilon_{\le Q}}2^{-R}.
\]
Taking
\[
  R=L+\frac{\varepsilon_{\le Q}}{\ln 2}+A\log_2m+O(1)
\]
gives the high-probability rank bound used in the Atomic Evidence Budget.

All objects below are finite: the world space, transcript alphabets, program
prefix sets, gauge-coordinate set, and switched target set are finite at fixed
parameters.  This is the intended formalization regime.

\subsection{Hidden gauge coordinates and product laws}
\label{subsec:appF-hidden-gauge-product}

\begin{definition}[Gauge coordinate space]
\label{def:appF-gauge-coordinate-space}
Let \(V=V_m\) be the finite hidden gauge-coordinate set of the ensemble.  A
hidden gauge vector is
\[
  g\in\bits^V.
\]
For \(U\subseteq V\), write \(g_U\) for the restriction of \(g\) to \(U\).
\end{definition}

\begin{definition}[Neutral public state]
\label{def:appF-neutral-public-state}
A neutral public state \(\mathfrak n\) is any conditioning event generated by
public template syntax, legal neutral guards, previous switched blocks, and
other CD-ENF-neutral information.  It is \emph{admissible} if it has positive
probability and if the hidden-gauge product law of Section~\ref{sec:ensemble}
holds after conditioning on \(\mathfrak n\).
\end{definition}

\begin{hypothesis}[Conditional hidden-gauge product law]
\label{hyp:appF-conditional-gauge-product}
For every admissible neutral public state \(\mathfrak n\), every set
\(U\subseteq V\) not already fixed by the normalized gauge prefix, and every
\(\gamma\in\bits^U\),
\[
  \Prb[g_U=\gamma\mid\mathfrak n]=2^{-|U|}.
\]
More generally, after a prefix has already fixed coordinates
\(U_0\subseteq V\), the remaining coordinates \(V\setminus U_0\) are
independent fair bits conditional on the prefix.
\end{hypothesis}

\begin{remark}[No public gauge oracle]
\label{rem:appF-no-public-gauge-oracle}
The lower-bound observer is given the public instance and any evidence made
available by legal public/safe operations.  It is not given an oracle for the
hidden vector \(g\).  A normalized gauge leaf is therefore treated as a hidden
claim whose correctness must be paid for.  If public syntax contained literals
such as \(g_v=0\), Hypothesis~\ref{hyp:appF-conditional-gauge-product} would be
false after conditioning on that syntax, and the entropy theorem below would
not apply.  This is exactly why the public syntax discipline and gauge-faithful
CD-ENF normalization are structural parts of the ensemble interface.
\end{remark}

\subsection{Gauge claims and rank}
\label{subsec:appF-gauge-claims}

\begin{definition}[Gauge claim]
\label{def:appF-gauge-claim}
A gauge claim is a partial assignment
\[
  c:U\to\bits,
  \qquad U\subseteq V.
\]
Its domain and rank are
\[
  \dom(c):=U,
  \qquad
  \rankG(c):=|U|.
\]
The claim is true in gauge vector \(g\), written \(\True_g(c)\), if
\[
  g_v=c(v)\qquad\text{for every }v\in U.
\]
\end{definition}

\begin{definition}[Restriction and union of compatible claims]
\label{def:appF-claim-union}
Two claims \(c_1:U_1\to\bits\) and \(c_2:U_2\to\bits\) are compatible if they
agree on \(U_1\cap U_2\).  If compatible, their union is the claim
\[
  c_1\cup c_2:U_1\cup U_2\to\bits.
\]
A finite list of claims is compatible if every pair is compatible.  Its union is
then defined by iterating the binary union operation.
\end{definition}

\begin{lemma}[Truth probability of a fixed claim]
\label{lem:appF-fixed-claim-probability}
Let \(c:U\to\bits\) be a fixed claim whose support is unsupported by the
admissible neutral state \(\mathfrak n\).  Then
\[
  \Prb[\True_g(c)\mid\mathfrak n]=2^{-\rankG(c)}.
\]
\end{lemma}

\begin{proof}
This is Hypothesis~\ref{hyp:appF-conditional-gauge-product} with
\(\gamma=c\).
\end{proof}

\begin{definition}[Claim extracted from a gauge leaf]
\label{def:appF-leaf-claim}
A normalized primitive gauge literal has the form \(G(v,\gamma)\), where
\(v\in V\) and \(\gamma\in\bits\).  Its claim is
\[
  \Claim(G(v,\gamma)):=\{v\mapsto\gamma\}.
\]
For a negated bit-literal, define
\[
  \Claim(\neg G(v,\gamma)):=\{v\mapsto 1-\gamma\}.
\]
If a CD-ENF leaf is a bounded conjunction of primitive gauge literals, its claim
is the union of the primitive claims, provided these claims are compatible.  If
they are incompatible, the leaf is unsatisfiable and contributes zero derivative.
\end{definition}

\begin{definition}[Gauge support of a leaf]
\label{def:appF-leaf-support}
For a normalized gauge leaf \(L\), define
\[
  \suppG(L):=\dom(\Claim(L)).
\]
If \(\Claim(L)\) is inconsistent, set \(\suppG(L)=\emptyset\) and treat \(L\) as
having zero realization probability.
\end{definition}

\subsection{Prefixes, fresh support, and rank envelopes}
\label{subsec:appF-prefixes-fresh-support}

\begin{definition}[Gauge prefix claim]
\label{def:appF-prefix-claim}
Let \(h\) be a normalized transcript prefix on a branch.  Its gauge prefix claim
\(c_G(h)\) is the union of all compatible gauge claims already exposed on that
branch.  Its support is
\[
  \suppG(h):=\dom(c_G(h)).
\]
If incompatible gauge claims occur on a branch, that branch is inconsistent and
has zero probability; it may be removed from the trace.
\end{definition}

\begin{definition}[Fresh support of a gauge occurrence]
\label{def:appF-fresh-support}
Let
\[
  \iota=(j,r,h,y,L)
\]
be a gauge occurrence in the normalized trace ledger.  Its fresh support is
\[
  \FreshG(\iota)
  :=
  \suppG(L)\setminus\suppG(h).
\]
A coordinate in \(\FreshG(\iota)\) is first claimed by this occurrence on that
branch.
\end{definition}

\begin{definition}[Gauge-rank envelope]
\label{def:appF-rank-envelope}
For a normalized trace, a gauge-rank envelope is a set
\[
  h^\Delta\subseteq V
\]
containing every coordinate that appears in \(\FreshG(\iota)\) for some realized
gauge occurrence \(\iota\).  Its rank is
\[
  \rankG(h^\Delta):=|h^\Delta|.
\]
The minimal envelope is
\[
  h^\Delta_{\min}
  :=
  \bigcup_{\iota\in\mathcal I_G}\FreshG(\iota).
\]
\end{definition}

\begin{lemma}[Old support does not separate]
\label{lem:appF-old-support-zero}
Let \(\iota=(j,r,h,y,L)\) be a gauge occurrence.  If
\(\FreshG(\iota)=\emptyset\), then
\[
  \partial_\iota=0.
\]
\end{lemma}

\begin{proof}
The branch prefix \(h\) has already fixed every gauge coordinate in
\(\suppG(L)\).  Therefore the two endpoints of any pair in the surviving
relation \(H_{j,h}\) agree on the truth value of \(L\).  Hence
\[
  H_{j,h}\cap\Sep_j(L)=\emptyset,
\]
so
\[
  \partial_\iota
  =
  \Gamma_j(H_{j,h}\cap\Sep_j(L))=0.
\]
\end{proof}

\begin{lemma}[Separation requires a fresh differing coordinate]
\label{lem:appF-separation-fresh-coordinate}
Let \(\iota=(j,r,h,y,L)\) be a consistent gauge occurrence.  For any pair
\((\omega^0,\omega^1)\in H_{j,h}\cap\Sep_j(L)\), there is a coordinate
\[
  v\in\FreshG(\iota)
\]
such that
\[
  g_v(\omega^0)\ne g_v(\omega^1).
\]
\end{lemma}

\begin{proof}
If every coordinate in \(\suppG(L)\setminus\suppG(h)\) has the same value on the
two endpoints, and every coordinate in \(\suppG(L)\cap\suppG(h)\) is already
fixed by the common prefix \(h\), then all coordinates in \(\suppG(L)\) agree on
the two endpoints.  Since the truth value of a normalized gauge leaf is a finite
Boolean function of exactly these coordinates, the two endpoints have the same
truth value for \(L\), contradicting membership in \(\Sep_j(L)\).
\end{proof}

\subsection{Bounded target-to-gauge incidence}
\label{subsec:appF-bounded-incidence}

The next hypothesis is the finite combinatorial property of the ensemble that
lets gauge derivatives be charged by rank rather than by the number of target
coordinates.  It is normally verified from the locked construction: a hidden gauge
coordinate is placed in the protected support of only boundedly many switched
readout coordinates.

\begin{definition}[Target-to-gauge incidence relation]
\label{def:appF-incidence-relation}
For a switched target set \(S\), define an incidence relation
\[
  \Inc\subseteq S\times V.
\]
We write \(\Inc(j,v)\) if a fresh claim about gauge coordinate \(v\) can appear
in a normalized target-relevant gauge leaf for coordinate \(j\), or equivalently
if changing \(v\) while holding all nonincident hidden gauge coordinates fixed can
separate a pair of \(B_j\)-opposite worlds in the chosen coupling support.
\end{definition}

\begin{hypothesis}[Bounded gauge incidence]
\label{hyp:appF-bounded-gauge-incidence}
There is a constant \(\Delta_G\), independent of \(m,t,S\), such that for every
\(v\in V\),
\[
  |\{j\in S:\Inc(j,v)\}|\le \Delta_G.
\]
Equivalently, for every finite set \(U\subseteq V\),
\[
  \sum_{j\in S}\mathbf 1[\exists v\in U:\Inc(j,v)]
  \le
  \Delta_G|U|.
\]
\end{hypothesis}

\begin{lemma}[Coordinate allocation of a separated pair]
\label{lem:appF-coordinate-allocation}
Fix a total order \(<_V\) on \(V\).  For every gauge occurrence \(\iota\) and
pair in \(H_{j,h}\cap\Sep_j(L)\), define \(a(\iota,\omega^0,\omega^1)\) to be
the least coordinate \(v\in\FreshG(\iota)\) such that
\(g_v(\omega^0)\ne g_v(\omega^1)\).  Then
\[
  H_{j,h}\cap\Sep_j(L)
  =
  \dot\bigcup_{v\in\FreshG(\iota)}
  A_{\iota,v},
\]
where
\[
  A_{\iota,v}
  :=
  \{(\omega^0,\omega^1)
       \in H_{j,h}\cap\Sep_j(L):
       a(\iota,\omega^0,\omega^1)=v\}.
\]
\end{lemma}

\begin{proof}
By Lemma~\ref{lem:appF-separation-fresh-coordinate}, every separated pair has at
least one fresh coordinate on which the endpoints differ.  The fixed order picks
a unique least such coordinate.  This gives the displayed disjoint union.
\end{proof}

\begin{definition}[Allocated derivative mass]
\label{def:appF-allocated-mass}
For a gauge occurrence \(\iota=(j,r,h,y,L)\) and fresh coordinate
\(v\in\FreshG(\iota)\), define
\[
  d_{\iota,v}:=\Gamma_j(A_{\iota,v}).
\]
Then
\[
  \partial_\iota
  =
  \sum_{v\in\FreshG(\iota)} d_{\iota,v}.
\]
\end{definition}

\begin{lemma}[First-exposure disjointness]
\label{lem:appF-first-exposure-disjointness}
Fix a switched target coordinate \(j\) and a gauge coordinate \(v\).  The events
\(A_{\iota,v}\), over all gauge occurrences \(\iota\) for target \(j\), are
pairwise disjoint.  Consequently,
\[
  \sum_{\iota:\,\hbox{target}(\iota)=j} d_{\iota,v}
  \le 1.
\]
\end{lemma}

\begin{proof}
A coupled pair has a unique transcript prefix at each depth.  Along its branch,
there is a unique first occurrence, if any, at which coordinate \(v\) becomes
freshly exposed.  If the pair is allocated to \(v\) at occurrence \(\iota\), then
\(\iota\) is that first fresh exposure of \(v\) on the branch.  Therefore the same
pair cannot belong to \(A_{\iota',v}\) for a different occurrence \(\iota'\).  Since
\(\Gamma_j\) is a probability measure, the sum of the masses of disjoint events
is at most \(1\).
\end{proof}

\begin{proposition}[Aggregate bounded-incidence charge]
\label{prop:appF-aggregate-bounded-incidence-charge}
Let \(h^\Delta\) be a gauge-rank envelope for the normalized trace.  Under
Hypothesis~\ref{hyp:appF-bounded-gauge-incidence},
\[
  \sum_{\iota\in\mathcal I_G}\partial_\iota
  \le
  \Delta_G\rankG(h^\Delta).
\]
\end{proposition}

\begin{proof}
Using Definition~\ref{def:appF-allocated-mass},
\[
  \sum_{\iota\in\mathcal I_G}\partial_\iota
  =
  \sum_{\iota\in\mathcal I_G}
  \sum_{v\in\FreshG(\iota)}d_{\iota,v}.
\]
Rearrange the finite sum by target coordinate \(j\) and gauge coordinate \(v\):
\[
  \sum_{\iota\in\mathcal I_G}\partial_\iota
  =
  \sum_{v\in h^\Delta}
  \sum_{j\in S}
  \sum_{\iota:\,\hbox{target}(\iota)=j}d_{\iota,v}.
\]
If \(\Inc(j,v)\) fails, no normalized target-relevant gauge occurrence for
\(j\) can allocate separated mass to \(v\), so the inner sum is zero.  If
\(\Inc(j,v)\) holds, Lemma~\ref{lem:appF-first-exposure-disjointness} bounds
the inner sum by \(1\).  Hence
\[
  \sum_{\iota\in\mathcal I_G}\partial_\iota
  \le
  \sum_{v\in h^\Delta}
  |\{j\in S:\Inc(j,v)\}|.
\]
Bounded incidence gives
\[
  \sum_{v\in h^\Delta}
  |\{j\in S:\Inc(j,v)\}|
  \le
  \Delta_G|h^\Delta|.
\]
Since \(|h^\Delta|=\rankG(h^\Delta)\), the result follows.
\end{proof}

\begin{remark}[Why no double counting occurs]
\label{rem:appF-no-double-counting}
The proof does not claim that one syntactic gauge coordinate appears only once
in the trace.  It may be mentioned many times.  The point is that only the first
fresh exposure of that coordinate on a branch can separate a previously
undistinguished pair.  Later mentions on the same branch are already determined
by the common prefix and contribute zero derivative.  Across different branches,
the first-exposure events are disjoint subsets of the pair space.
\end{remark}

\subsection{Safe leakage and posterior amplification}
\label{subsec:appF-safe-leakage-posterior}

Gauge-rank entropy would be the simple estimate \(2^{-R}\) if the observer's
transcript were independent of the hidden gauge vector.  Legal safe-buffer reads
can leak a small amount of information about hidden states.  Appendix~\ref{app:max-qssm-dobrushin}
controls this leakage in max-divergence; the next lemmas convert it into a
posterior amplification factor.

\begin{definition}[Safe transcript leakage]
\label{def:appF-safe-transcript-leakage}
Let \(T\) be the safe part of a deterministic observer transcript.  We say that
\(T\) has cumulative gauge leakage at most \(\varepsilon_{\le Q}\) if for every
partial gauge assignment \(c:U\to\bits\) whose support is unsupported before the
safe transcript,
\[
  \Dinf\bigl(\Law(T\mid\True_g(c),\mathfrak n)
              \|\Law(T\mid\mathfrak n)\bigr)
  \le
  \varepsilon_{\le Q}.
\]
Equivalently, for every transcript value \(\tau\),
\[
  \Prb[T=\tau\mid\True_g(c),\mathfrak n]
  \le
  e^{\varepsilon_{\le Q}}
  \Prb[T=\tau\mid\mathfrak n].
\]
\end{definition}

\begin{lemma}[Adaptive composition of safe leakage]
\label{lem:appF-adaptive-composition}
If the safe transcript is generated by adaptive legal safe reads whose one-step
max-divergence leakages are \(\varepsilon_1,\ldots,\varepsilon_Q\), then
\[
  \varepsilon_{\le Q}
  :=
  \sum_{r=1}^Q\varepsilon_r
\]
is a valid cumulative leakage bound in the sense of
Definition~\ref{def:appF-safe-transcript-leakage}.
\end{lemma}

\begin{proof}
Write the probability of a transcript \(\tau=(y_1,\ldots,y_Q)\) as a product of
adaptive kernels.  At step \(r\), after any prior prefix, the conditioned kernel
is pointwise at most \(e^{\varepsilon_r}\) times the unconditioned kernel.
Multiplying these pointwise likelihood-ratio bounds over all steps gives
\[
  \Prb[T=\tau\mid\True_g(c),\mathfrak n]
  \le
  \exp\left(\sum_{r=1}^Q\varepsilon_r\right)
  \Prb[T=\tau\mid\mathfrak n].
\]
This is exactly the asserted cumulative bound.
\end{proof}

\begin{lemma}[Posterior amplification]
\label{lem:appF-posterior-amplification}
Assume cumulative safe leakage at most \(\varepsilon_{\le Q}\).  Let
\(c:U\to\bits\) be a fixed unsupported gauge claim.  Then for every safe
transcript value \(\tau\) with positive probability,
\[
  \Prb[\True_g(c)\mid T=\tau,\mathfrak n]
  \le
  e^{\varepsilon_{\le Q}}2^{-\rankG(c)}.
\]
\end{lemma}

\begin{proof}
By Bayes' rule and Definition~\ref{def:appF-safe-transcript-leakage},
\[
\begin{aligned}
  \Prb[\True_g(c)\mid T=\tau,\mathfrak n]
  &=\frac{\Prb[T=\tau\mid\True_g(c),\mathfrak n]
       \Prb[\True_g(c)\mid\mathfrak n]}
     {\Prb[T=\tau\mid\mathfrak n]} \\
  &\le
     e^{\varepsilon_{\le Q}}
     \Prb[\True_g(c)\mid\mathfrak n].
\end{aligned}
\]
Lemma~\ref{lem:appF-fixed-claim-probability} gives
\(\Prb[\True_g(c)\mid\mathfrak n]=2^{-\rankG(c)}\).
\end{proof}

\begin{lemma}[Transcript-dependent claim bound]
\label{lem:appF-transcript-dependent-claim-bound}
Let a deterministic procedure assign to each safe transcript value \(\tau\) a
claim
\[
  c_\tau:U_\tau\to\bits.
\]
Assume cumulative leakage at most \(\varepsilon_{\le Q}\) for all these claims.
Then for every \(R\ge 0\),
\[
  \Prb[\True_g(c_T)\text{ and }\rankG(c_T)\ge R\mid\mathfrak n]
  \le
  e^{\varepsilon_{\le Q}}2^{-R}.
\]
\end{lemma}

\begin{proof}
Sum over transcript values \(\tau\) with \(\rankG(c_\tau)\ge R\):
\[
\begin{aligned}
&\Prb[\True_g(c_T),\rankG(c_T)\ge R\mid\mathfrak n] \\
&\quad =
  \sum_{\tau:\rankG(c_\tau)\ge R}
  \Prb[T=\tau\mid\mathfrak n]
  \Prb[\True_g(c_\tau)\mid T=\tau,\mathfrak n].
\end{aligned}
\]
By Lemma~\ref{lem:appF-posterior-amplification}, each posterior probability in
the sum is at most \(e^{\varepsilon_{\le Q}}2^{-\rankG(c_\tau)}\), and therefore
at most \(e^{\varepsilon_{\le Q}}2^{-R}\).  Since the transcript probabilities sum
to at most \(1\), the result follows.
\end{proof}

\subsection{One-program gauge-rank entropy}
\label{subsec:appF-one-program-entropy}

\begin{definition}[Observer-induced gauge claim]
\label{def:appF-observer-induced-claim}
Fix a deterministic observer \(P\).  After CD-ENF normalization, every realized
safe transcript \(\tau\) determines a finite list of gauge leaves.  Let
\[
  c_P(\tau)
\]
be the compatible union of their claims, after removing inconsistent zero-mass
branches.  The associated envelope is
\[
  h^\Delta_P(\tau):=\dom(c_P(\tau)),
  \qquad
  \rankG(h^\Delta_P(\tau)):=\rankG(c_P(\tau)).
\]
\end{definition}

\begin{lemma}[One-program gauge-rank tail]
\label{lem:appF-one-program-rank-tail}
Assume the hidden-gauge product law and cumulative safe leakage at most
\(\varepsilon_{\le Q}\).  For a fixed deterministic observer \(P\),
\[
  \Prb[\rankG(h^\Delta_P)\ge R]
  \le
  e^{\varepsilon_{\le Q}}2^{-R}.
\]
\end{lemma}

\begin{proof}
Apply Lemma~\ref{lem:appF-transcript-dependent-claim-bound} to the
transcript-dependent claim \(c_P(T)\).  The event that the realized gauge ledger
has rank at least \(R\) is contained in the event that this rank-\(R\) claim is
true, because only true normalized gauge leaves are realized on a nonzero branch.
\end{proof}

\begin{remark}[Stopping at rank \(R\)]
\label{rem:appF-stopping-rank-R}
Equivalently, one may stop the observer the first time its compatible gauge
claim reaches rank \(R\).  The stopped claim is still a deterministic function of
the public and safe transcript, and its truth probability is at most
\(e^{\varepsilon_{\le Q}}2^{-R}\).  This stopped formulation is often the easiest
one to formalize because it avoids separately considering ranks larger than
\(R\).
\end{remark}

\subsection{Prefix-program union bound}
\label{subsec:appF-program-union-bound}

\begin{definition}[Clocked observer code]
\label{def:appF-clocked-observer-code}
A clocked observer code is a self-delimiting pair \((D,p)\), where \(D\) is a
clock exponent and \(p\) is a program for a deterministic observer running under
that clock.  Its length is
\[
  \len(D,p):=|p|+\lceil\log_2(D+1)\rceil+O(1).
\]
A family of such codes is prefix-free if their self-delimiting encodings are
prefix-free.
\end{definition}

\begin{lemma}[Counting prefix observer codes]
\label{lem:appF-counting-prefix-codes}
The number of prefix-free clocked observer codes of length at most \(L\) is at
most
\[
  2^{L+O(1)}.
\]
\end{lemma}

\begin{proof}
This is the Kraft counting bound for binary prefix codes.  The \(O(1)\) absorbs
the fixed coding convention for separating the exponent and program fields.
\end{proof}

\begin{theorem}[Gauge-rank entropy, family form]
\label{thm:appF-gauge-rank-entropy-family}
Let \(\mathcal P_L\) be the prefix-free family of deterministic clocked observers
with code length at most \(L\).  Assume the hidden-gauge product law and
cumulative safe leakage at most \(\varepsilon_{\le Q}\) for every observer in the
family.  Then
\[
  \Prb\left[
    \exists P\in\mathcal P_L:
    \rankG(h^\Delta_P)\ge R
  \right]
  \le
  2^{L+O(1)}e^{\varepsilon_{\le Q}}2^{-R}.
\]
Consequently, for every \(A>0\), with probability at least \(1-m^{-A}\), every
\(P\in\mathcal P_L\) satisfies
\[
  \rankG(h^\Delta_P)
  \le
  L+\frac{\varepsilon_{\le Q}}{\ln 2}+A\log_2m+O(1).
\]
\end{theorem}

\begin{proof}
For one fixed observer, Lemma~\ref{lem:appF-one-program-rank-tail} gives
\[
  \Prb[\rankG(h^\Delta_P)\ge R]
  \le
  e^{\varepsilon_{\le Q}}2^{-R}.
\]
There are at most \(2^{L+O(1)}\) observers in \(\mathcal P_L\), by
Lemma~\ref{lem:appF-counting-prefix-codes}.  A union bound gives the first
claim.  Now choose
\[
  R=
  L+\frac{\varepsilon_{\le Q}}{\ln 2}+A\log_2m+C_0
\]
with \(C_0\) large enough to absorb the \(O(1)\) coding constant.  Then
\[
  2^{L+O(1)}e^{\varepsilon_{\le Q}}2^{-R}
  \le
  m^{-A}.
\]
This gives the high-probability rank bound.
\end{proof}

\begin{corollary}[Fixed observer form]
\label{cor:appF-fixed-observer-rank-entropy}
For a fixed deterministic observer represented by a clocked code of length
\(|P|\), the same high-probability bound holds with \(L=|P|\):
\[
  \Prb\left[
    \rankG(h^\Delta_P)
    >
    |P|+\frac{\varepsilon_{\le Q}}{\ln 2}+A\log_2m+O(1)
  \right]
  \le
  m^{-A}.
\]
\end{corollary}

\begin{proof}
Apply Theorem~\ref{thm:appF-gauge-rank-entropy-family} to the prefix family of
all codes of length at most \(|P|\).  This is slightly weaker than the one-program
tail but is the form used in the main budget theorem because it matches the
later union bound over short observers.
\end{proof}

\subsection{Combining entropy with bounded incidence}
\label{subsec:appF-combined-rank-incidence}

\begin{theorem}[Gauge derivative budget]
\label{thm:appF-gauge-derivative-budget}
Let \(P\) be a deterministic observer, let \(S\) be a switched set, and let
\(\mathcal I_G(P,S)\) be the normalized gauge-occurrence ledger.  Under bounded
gauge incidence,
\[
  \sum_{\iota\in\mathcal I_G(P,S)}\partial_\iota
  \le
  \Delta_G\rankG(h^\Delta_P).
\]
Under the hidden-gauge product law and cumulative safe leakage
\(\varepsilon_{\le Q}\), with probability at least \(1-m^{-A}\),
\[
  \sum_{\iota\in\mathcal I_G(P,S)}\partial_\iota
  \le
  \Delta_G
  \left(
    |P|+\frac{\varepsilon_{\le Q}}{\ln 2}+A\log_2m+O(1)
  \right).
\]
The simultaneous form for all observers of length at most \(L\) is obtained by
replacing \(|P|\) by \(L\).
\end{theorem}

\begin{proof}
The deterministic inequality is Proposition~\ref{prop:appF-aggregate-bounded-incidence-charge}.
The high-probability inequality follows by substituting the rank bound from
Corollary~\ref{cor:appF-fixed-observer-rank-entropy}.  The simultaneous form
uses Theorem~\ref{thm:appF-gauge-rank-entropy-family}.
\end{proof}

\begin{corollary}[Gauge contribution to averaged phase gap]
\label{cor:appF-gauge-averaged-phase-gap}
Assume \(|S|\ge\gamma t\) and \(t=\Theta(m)\).  If \(|P|\le\delta t\) and
\(\varepsilon_{\le Q}=o(t)\), then with probability at least \(1-m^{-A}\), the
gauge part of the coordinate-sum phase gap satisfies
\[
  \frac{1}{|S|}
  \cdot \frac{1}{2}
  \sum_{\iota\in\mathcal I_G(P,S)}\partial_\iota
  \le
  \frac{\Delta_G}{2}\frac{\delta}{\gamma}+o(1).
\]
In particular, for a fixed observer \(|P|=O(1)\), the averaged gauge contribution
is \(o(1)\).
\end{corollary}

\begin{proof}
Divide the high-probability gauge derivative budget by \(2|S|\).  The program
length term is bounded by
\[
  \frac{\Delta_G}{2}\frac{|P|}{|S|}
  \le
  \frac{\Delta_G}{2}\frac{\delta t}{\gamma t}
  =
  \frac{\Delta_G}{2}\frac{\delta}{\gamma}.
\]
The terms \(\varepsilon_{\le Q}\), \(A\log_2m\), and \(O(1)\), divided by
\(|S|=\Theta(m)\), are \(o(1)\) under the stated assumptions.  If \(|P|=O(1)\),
the program-length term is also \(o(1)\).
\end{proof}

\subsection{Verification of bounded incidence in the locked construction}
\label{subsec:appF-verifying-incidence}

This subsection records the finite design condition that Appendix~\ref{app:hidden-gauge-buffered-locked-construction}
must verify.  It is not a new probabilistic estimate; it is a bounded-degree
property of the readout/gauge support hypergraph.

\begin{definition}[Readout support hypergraph]
\label{def:appF-readout-support-hypergraph}
The readout support hypergraph has left vertices the switched target coordinates
\(S\), right vertices the gauge coordinates \(V\), and an edge \((j,v)\) whenever
coordinate \(v\) lies in the normalized raw gauge support of some witness or
quotient fact from which the target bit \(B_j\) can be resolved.
\end{definition}

\begin{hypothesis}[Bounded readout overlap]
\label{hyp:appF-bounded-readout-overlap}
There is a constant \(\Delta_G\) such that every gauge coordinate
\(v\in V\) belongs to the readout support of at most \(\Delta_G\) switched target
coordinates.
\end{hypothesis}

\begin{lemma}[Bounded readout overlap implies bounded gauge incidence]
\label{lem:appF-readout-overlap-implies-incidence}
Hypothesis~\ref{hyp:appF-bounded-readout-overlap} implies
Hypothesis~\ref{hyp:appF-bounded-gauge-incidence}.
\end{lemma}

\begin{proof}
By gauge faithfulness, any normalized target-relevant gauge leaf for target
coordinate \(j\) must use raw gauge support lying in the readout support of
\(B_j\).  Therefore, if \(\Inc(j,v)\) holds, then \((j,v)\) is an edge of the
readout support hypergraph.  The bounded right degree of that hypergraph is
exactly the desired bounded incidence condition.
\end{proof}

\begin{remark}[How to enforce bounded overlap]
\label{rem:appF-enforcing-bounded-overlap}
The construction can enforce Hypothesis~\ref{hyp:appF-bounded-readout-overlap}
by allocating protected gauge pads to readout coordinates with bounded reuse, or
by using a bounded-degree expander-style incidence design whose right degree is
constant.  The construction must not use a dense global parity of all gauge bits
as a single target coordinate unless it also changes the incidence theorem, because
such a dense parity would make many targets incident to many of the same gauge
coordinates.
\end{remark}

\subsection{Interface exported to Section 7}
\label{subsec:appF-exported-interface}

Appendix~\ref{app:gauge-rank-incidence} exports the following finite theorem
package.

\begin{enumerate}[label=(\roman*)]
\item \textbf{Fresh-support principle.}  A normalized gauge leaf can separate a
      surviving message-opposite pair only through fresh hidden gauge support.
      Repeated gauge mentions on the same branch have zero derivative.

\item \textbf{Bounded incidence.}  If the target-to-gauge incidence graph has
      right degree at most \(\Delta_G\), then
      \[
        \sum_{\iota\in\mathcal I_G}\partial_\iota
        \le
        \Delta_G\rankG(h^\Delta).
      \]

\item \textbf{Safe leakage composition.}  One-step max-divergence leakages
      \(\varepsilon_r\) compose additively into
      \[
        \varepsilon_{\le Q}=\sum_r\varepsilon_r.
      \]

\item \textbf{Gauge-rank entropy.}  For all programs of length at most \(L\),
      \[
        \Prb[\exists P:\rankG(h^\Delta_P)\ge R]
        \le
        2^{L+O(1)}e^{\varepsilon_{\le Q}}2^{-R}.
      \]
      Hence, with probability at least \(1-m^{-A}\),
      \[
        \rankG(h^\Delta_P)
        \le
        L+\frac{\varepsilon_{\le Q}}{\ln 2}+A\log_2m+O(1)
      \]
      simultaneously for all such programs.

\item \textbf{Gauge derivative budget.}  Combining (ii) and (iv), the total
      gauge derivative contribution is at most
      \[
        \Delta_G
        \left(
          L+\frac{\varepsilon_{\le Q}}{\ln 2}+A\log_2m+O(1)
        \right)
      \]
      with probability at least \(1-m^{-A}\).
\end{enumerate}

Together with neutral-zero derivative and the safe-buffer derivative estimate
from Appendix~\ref{app:max-qssm-dobrushin}, this is exactly the non-neutral
part of the Atomic Evidence Budget.

\subsection{Lean-facing finite skeleton}
\label{subsec:appF-lean-facing}

A Lean formalization can split this appendix into the following modules.

\begin{enumerate}[label=(\arabic*)]
\item \textbf{GaugeClaims.lean.}  Finite gauge coordinate type, partial
      assignments, compatibility, union, truth in a hidden vector, and rank.
\item \textbf{FreshGaugeSupport.lean.}  Gauge prefix claims, fresh support, old
      support zero-derivative, and the fresh-coordinate separation lemma.
\item \textbf{GaugeIncidence.lean.}  Finite target-to-gauge incidence relation,
      bounded right degree, coordinate allocation of separated pairs,
      first-exposure disjointness, and aggregate derivative charge.
\item \textbf{SafeLeakagePosterior.lean.}  Max-divergence transcript leakage,
      adaptive composition, Bayes posterior amplification, and
      transcript-dependent claim bounds.
\item \textbf{GaugeRankEntropy.lean.}  One-program rank tail, prefix-code
      counting, family union bound, and high-probability rank theorem.
\item \textbf{GaugeBudgetInterface.lean.}  Combination of bounded incidence and
      rank entropy, averaged phase-gap corollaries, and the exact interface
      imported by the Atomic Evidence Budget.
\end{enumerate}

Every theorem in these modules is a finite-sum, finite-probability, or finite
counting statement.  The only analytic input is the max-qSSM leakage bound
imported from Appendix~\ref{app:max-qssm-dobrushin}.

\section{Boundary-Law Mixing}
\label{app:boundary-law-mixing}
\providecommand{\bits}{\{0,1\}}
\providecommand{\Prb}{\mathbb P}
\providecommand{\E}{\mathbb E}
\providecommand{\Law}{\mathcal L}
\providecommand{\TV}{\operatorname{TV}}
\providecommand{\Dinf}{D_\infty}
\providecommand{\Range}{\operatorname{Range}}
\providecommand{\Good}{\mathsf{Good}}
\providecommand{\Bad}{\mathsf{Bad}}
\providecommand{\supp}{\operatorname{supp}}

This appendix proves the boundary-law mixing interface used in
Section~\ref{sec:mixing} and exported to the product argument in
Section~\ref{sec:product-small-success}.  Boundary-law mixing is independent of
CD trace capture.  It is the visible baseline theorem: after the target phase is
separated from the pivot by a depth-
\(L\) locally mixing buffer, any predictor using only the pivot-visible summary
has success at most
\[
  \frac{1}{2}+\varepsilon_{\mathrm{mix}}(m,L),
  \qquad
  \varepsilon_{\mathrm{mix}}(m,L)=m^{-\Omega(1)}
\]
for \(L=\Theta(\log m)\).

The proof is finite throughout.  Sigma-fields are finite partitions, conditional
expectations are finite sums over atoms, and Markov kernels are finite
stochastic matrices.  This is the intended Lean-facing interpretation.

\subsection{Finite Bayes testing}
\label{subsec:appG-finite-bayes-testing}

\begin{definition}[Posterior of a finite feature]
\label{def:appG-posterior-feature}
Let \((\Omega,\mu)\) be a finite probability space, let
\(B:\Omega\to\bits\), and let \(Z:\Omega\to\mathcal Z\) be a finite feature.
For \(z\in\mathcal Z\) with \(\Prb[Z=z]>0\), define
\[
  p_Z(z):=\Prb[B=1\mid Z=z].
\]
For zero-mass values of \(z\), set \(p_Z(z)=1/2\).  This convention does not
change any expectation.
\end{definition}

\begin{lemma}[Bayes success for a balanced bit]
\label{lem:appG-bayes-success-balanced}
Assume \(\Prb[B=0]=\Prb[B=1]=1/2\).  Then
\[
  \sup_{h:\mathcal Z\to\bits}\Prb[h(Z)=B]
  =
  \frac{1}{2}+
  \E\left[\left|p_Z(Z)-\frac{1}{2}\right|\right].
\]
The supremum is attained by
\[
  h^*(z)=1
  \quad\Longleftrightarrow\quad
  p_Z(z)\ge \frac{1}{2},
\]
with arbitrary tie-breaking.
\end{lemma}

\begin{proof}
Condition on \(Z=z\).  If \(h(z)=1\), the conditional success probability is
\(p_Z(z)\).  If \(h(z)=0\), it is \(1-p_Z(z)\).  The best conditional success at
\(z\) is therefore
\[
  \max\{p_Z(z),1-p_Z(z)\}
  =
  \frac{1}{2}+\left|p_Z(z)-\frac{1}{2}\right|.
\]
Averaging over \(z\) gives the identity and the maximizing rule.
\end{proof}

\begin{lemma}[Bayes success and total variation]
\label{lem:appG-bayes-tv}
Assume \(\Prb[B=0]=\Prb[B=1]=1/2\).  Let
\[
  \nu_b:=\Law(Z\mid B=b),
  \qquad b\in\bits.
\]
Then
\[
  \sup_{h:\mathcal Z\to\bits}\Prb[h(Z)=B]
  =
  \frac{1}{2}+\frac{1}{2}\TV(\nu_1,\nu_0),
\]
where
\[
  \TV(\nu_1,\nu_0)
  :=\frac{1}{2}\sum_{z\in\mathcal Z}|\nu_1(z)-\nu_0(z)|.
\]
Consequently,
\[
  \E\left[\left|p_Z(Z)-\frac{1}{2}\right|\right]
  =\frac{1}{2}\TV(\nu_1,\nu_0).
\]
\end{lemma}

\begin{proof}
For a predictor \(h\), let \(U_h=\{z:h(z)=1\}\).  Since the priors are balanced,
\[
\begin{aligned}
  \Prb[h(Z)=B]
  &= \frac{1}{2}\nu_1(U_h)+\frac{1}{2}\nu_0(U_h^c) \\
  &= \frac{1}{2}+\frac{1}{2}(\nu_1(U_h)-\nu_0(U_h)).
\end{aligned}
\]
The supremum over subsets \(U_h\subseteq\mathcal Z\) is obtained by taking the
positive part of \(\nu_1-\nu_0\), and equals
\(1/2+(1/2)\TV(\nu_1,\nu_0)\).  Comparing with
Lemma~\ref{lem:appG-bayes-success-balanced} gives the posterior-deviation
identity.
\end{proof}

\begin{lemma}[Conditional Bayes testing]
\label{lem:appG-conditional-bayes}
Let \(\mathcal C\) be a finite sigma-field.  Suppose
\[
  \Prb[B=0\mid\mathcal C]=\Prb[B=1\mid\mathcal C]=\frac{1}{2}
\]
almost surely.  Then, almost surely on \(\mathcal C\),
\[
  \sup_{h:\mathcal Z\to\bits}
  \Prb[h(Z)=B\mid\mathcal C]
  =
  \frac{1}{2}+
  \E\left[
    \left|
      \Prb[B=1\mid Z,\mathcal C]-\frac{1}{2}
    \right|
    \middle|\mathcal C
  \right]
\]
and equivalently
\[
  \sup_h\Prb[h(Z)=B\mid\mathcal C]
  =
  \frac{1}{2}+\frac{1}{2}
  \TV(\Law(Z\mid B=1,\mathcal C),
      \Law(Z\mid B=0,\mathcal C)).
\]
\end{lemma}

\begin{proof}
Apply Lemmas~\ref{lem:appG-bayes-success-balanced} and
\ref{lem:appG-bayes-tv} inside each positive-probability atom of
\(\mathcal C\).  The displayed identities are atomwise identities, so they hold
as conditional identities almost surely.
\end{proof}

\subsection{Data processing and dithered quantization}
\label{subsec:appG-data-processing}

\begin{definition}[Finite Markov postprocessing]
\label{def:appG-finite-markov-postprocessing}
Let \(R\) take values in a finite set \(\mathcal R\).  A finite Markov
postprocessing of \(R\) is a stochastic matrix
\[
  K:\mathcal R\leadsto\mathcal Z,
  \qquad
  K(z\mid r)\ge 0,
  \qquad
  \sum_{z\in\mathcal Z}K(z\mid r)=1.
\]
The feature \(Z\) is generated by first sampling \(R\), then sampling
\(Z\) from \(K(\cdot\mid R)\).  The kernel randomness is independent of all
other variables conditional on \(R\).
\end{definition}

\begin{lemma}[Total-variation data processing]
\label{lem:appG-tv-data-processing}
Let \(\alpha\) and \(\beta\) be probability laws on a finite set
\(\mathcal R\), and let \(K:\mathcal R\leadsto\mathcal Z\) be a finite Markov
kernel.  Then
\[
  \TV(\alpha K,\beta K)\le \TV(\alpha,\beta).
\]
\end{lemma}

\begin{proof}
For each \(z\),
\[
  (\alpha K)(z)-(\beta K)(z)
  =
  \sum_{r\in\mathcal R}(\alpha(r)-\beta(r))K(z\mid r).
\]
Thus
\[
\begin{aligned}
  \TV(\alpha K,\beta K)
  &= \frac{1}{2}\sum_z
     \left|\sum_r(\alpha(r)-\beta(r))K(z\mid r)\right| \\
  &\le \frac{1}{2}\sum_z\sum_r
      |\alpha(r)-\beta(r)|K(z\mid r) \\
  &= \frac{1}{2}\sum_r|\alpha(r)-\beta(r)|
      \sum_zK(z\mid r) \\
  &= \TV(\alpha,\beta).
\end{aligned}
\]
\end{proof}

\begin{lemma}[Prediction data processing]
\label{lem:appG-prediction-data-processing}
Let \(B\) be balanced, let \(R\) be a finite feature, and let \(Z\) be a finite
Markov postprocessing of \(R\).  Then
\[
  \sup_h\Prb[h(Z)=B]
  \le
  \sup_g\Prb[g(R)=B].
\]
The same statement holds conditionally on any finite sigma-field on whose atoms
\(B\) is balanced.
\end{lemma}

\begin{proof}
The phase-conditioned laws of \(Z\) are obtained by applying the same Markov
kernel to the phase-conditioned laws of \(R\).  Lemma~\ref{lem:appG-tv-data-processing}
therefore gives
\[
  \TV(\Law(Z\mid B=1),\Law(Z\mid B=0))
  \le
  \TV(\Law(R\mid B=1),\Law(R\mid B=0)).
\]
Now apply Lemma~\ref{lem:appG-bayes-tv}.  The conditional form is the same
argument on each atom of the conditioning field.
\end{proof}

\begin{definition}[Dithered quantized summary]
\label{def:appG-dithered-summary}
Let \(R_{L,j}\) be an exact finite pivot boundary-law summary.  A dithered
quantized pivot summary is any finite Markov postprocessing
\[
  Q_{\delta_m}:\Range(R_{L,j})\leadsto\mathcal Z_{L,m}
\]
obtained by adjoining an independent finite dither seed and recording the grid
cell, of mesh \(\delta_m\), containing \(R_{L,j}\).  We write
\[
  Z_{L,j}:=Q_{\delta_m}(R_{L,j}).
\]
For the proof below, only the Markov-postprocessing property is used.
\end{definition}

\begin{corollary}[Quantization cannot increase phase advantage]
\label{cor:appG-quantization-data-processing}
Let \(\mathcal C\) be a finite field over which \(B_j\) is balanced.  If
\(Z_{L,j}\) is a dithered quantized summary of \(R_{L,j}\), then
\[
  \sup_h\Prb[h(Z_{L,j})=B_j\mid\mathcal C]
  \le
  \sup_g\Prb[g(R_{L,j})=B_j\mid\mathcal C].
\]
\end{corollary}

\begin{proof}
Apply Lemma~\ref{lem:appG-prediction-data-processing} conditionally on
\(\mathcal C\).
\end{proof}

\subsection{Pivot boundary-law variables}
\label{subsec:appG-pivot-variables}

The raw exterior boundary condition at a cut may itself carry phase information;
it is not the visible surface.  The visible surface is obtained only after the
raw condition has been propagated through a locally mixing depth-\(L\) core.
This distinction is important for the correctness of the mixing statement.

\begin{definition}[Target bit and pivot ball]
\label{def:appG-target-bit-pivot-ball}
For a switched coordinate \(j\), write
\[
  B_j:=\ell_j(M(Y))\in\bits.
\]
The ensemble provides a designated pivot region and a depth-\(L\) pivot ball
\(\mathbb B_{L,j}\) in the buffer/core geometry.  Its outer cut is denoted
\(\partial\mathbb B_{L,j}\).
\end{definition}

\begin{definition}[Raw cut state and exact boundary-law summary]
\label{def:appG-raw-cut-exact-summary}
The raw exterior state on the cut is a hidden finite object
\[
  \Xi_{L,j}.
\]
The exact pivot boundary-law summary is a finite object
\[
  R_{L,j}=\mathsf{BLaw}_{L,j}(Y_{\mathrm{neu}},\Xi_{L,j})
\]
computed from the neutral public geometry and the raw cut state by the fixed
finite message recursion through the depth-\(L\) buffer/core.  Examples of
\(R_{L,j}\) include a root belief vector, a tuple of root incoming messages, or a
finite exact pivot law before dithered quantization.
\end{definition}

\begin{remark}[The raw cut state is not exposed]
\label{rem:appG-raw-cut-not-exposed}
The variable \(\Xi_{L,j}\) is not an allowed pivot-visible summary.  It may be
strongly correlated with the hidden phase.  Boundary-law mixing asserts that the
propagated object \(R_{L,j}\), and hence any postprocessing \(Z_{L,j}\) of it,
is nearly phase-blind after depth \(L\) contraction.
\end{remark}

\begin{definition}[Admissible conditioning]
\label{def:appG-admissible-conditioning}
A finite sigma-field \(\mathcal C\) is admissible for coordinate \(j\) if:
\begin{enumerate}[label=(\roman*)]
\item \(B_j\) is conditionally balanced over \(\mathcal C\):
\[
  \Prb[B_j=0\mid\mathcal C]
  =
  \Prb[B_j=1\mid\mathcal C]
  =\frac{1}{2};
\]
\item \(\mathcal C\) is generated by neutral public data, previous switched
      blocks, fixed observer-side randomness, and bookkeeping outside the
      current protected pivot core;
\item \(\mathcal C\) contains no charged non-neutral evidence from the current
      coordinate \(j\).
\end{enumerate}
\end{definition}

\subsection{Boundary-law contraction}
\label{subsec:appG-boundary-law-contraction}

\begin{definition}[Boundary-law contraction]
\label{def:appG-boundary-law-contraction}
Boundary-law contraction at depth \(L\) holds with error
\(\varepsilon_{\mathrm{BL}}(m,L)\) if, for every switched coordinate \(j\) and
every admissible conditioning field \(\mathcal C\),
\[
  \TV\left(
    \Law(R_{L,j}\mid B_j=1,\mathcal C),
    \Law(R_{L,j}\mid B_j=0,\mathcal C)
  \right)
  \le
  2\varepsilon_{\mathrm{BL}}(m,L)
\]
almost surely.
\end{definition}

\begin{lemma}[TV contraction equals posterior contraction]
\label{lem:appG-tv-posterior-equivalence}
Assume \(\mathcal C\) is admissible for \(j\).  Then boundary-law contraction
with error \(\varepsilon\) is equivalent to
\[
  \E\left[
    \left|
      \Prb[B_j=1\mid R_{L,j},\mathcal C]-\frac{1}{2}
    \right|
    \middle|\mathcal C
  \right]
  \le
  \varepsilon.
\]
\end{lemma}

\begin{proof}
By admissibility, \(B_j\) is balanced on every atom of \(\mathcal C\).  Apply
Lemma~\ref{lem:appG-conditional-bayes} with \(Z=R_{L,j}\).  The expected
posterior deviation is one half of the total variation between the two
phase-conditioned laws.
\end{proof}

\begin{definition}[Good geometry]
\label{def:appG-good-geometry}
Let \(\Good_{L,j}\) be the event that the depth-\(L\) pivot region has all
regularity properties required by the local mixing theorem: bounded degree,
legal separation from locked and gauge regions, no forbidden short collision,
and valid buffer boundary conditions.  Write
\[
  \tau_{\mathrm{geo}}(m,L)
  :=
  \sup_{j,b,\mathcal C}
  \Prb[\neg\Good_{L,j}\mid B_j=b,\mathcal C],
\]
where the supremum ranges over switched coordinates, phases, and admissible
conditioning fields for which the conditional probability is defined.
\end{definition}

\begin{definition}[Root-kernel contraction on the good event]
\label{def:appG-root-kernel-contraction}
The locally mixing core has root-kernel contraction if there are constants
\[
  C_{\mathrm{root}}<\infty,
  \qquad
  0<\rho_{\mathrm{root}}<1,
\]
such that, on \(\Good_{L,j}\), for every atom of every admissible
\(\mathcal C\),
\[
  \TV\left(
    \Law(R_{L,j}\mid B_j=1,\mathcal C,\Good_{L,j}),
    \Law(R_{L,j}\mid B_j=0,\mathcal C,\Good_{L,j})
  \right)
  \le
  2C_{\mathrm{root}}\rho_{\mathrm{root}}^L.
\]
\end{definition}

\begin{proposition}[Good-event contraction plus bad-event error]
\label{prop:appG-good-plus-bad}
Assume root-kernel contraction.  Then boundary-law contraction holds with
\[
  \varepsilon_{\mathrm{BL}}(m,L)
  :=
  C_{\mathrm{root}}\rho_{\mathrm{root}}^L+\tau_{\mathrm{geo}}(m,L).
\]
\end{proposition}

\begin{proof}
Fix a switched coordinate \(j\) and an admissible atom of \(\mathcal C\).  Let
\[
  P_b:=\Law(R_{L,j}\mid B_j=b,\mathcal C),
  \qquad
  P_b^G:=\Law(R_{L,j}\mid B_j=b,\mathcal C,\Good_{L,j}).
\]
Changing a law by conditioning on an event of failure probability \(\eta_b\)
changes it in total variation by at most \(\eta_b\), where
\[
  \eta_b:=\Prb[\neg\Good_{L,j}\mid B_j=b,\mathcal C].
\]
Thus
\[
  \TV(P_b,P_b^G)\le \eta_b.
\]
The triangle inequality and root-kernel contraction give
\[
\begin{aligned}
  \TV(P_1,P_0)
  &\le \TV(P_1,P_1^G)+\TV(P_1^G,P_0^G)+\TV(P_0^G,P_0) \\
  &\le \eta_1+2C_{\mathrm{root}}\rho_{\mathrm{root}}^L+
        \eta_0 \\
  &\le 2C_{\mathrm{root}}\rho_{\mathrm{root}}^L+2\tau_{\mathrm{geo}}(m,L) \\
  &= 2\varepsilon_{\mathrm{BL}}(m,L).
\end{aligned}
\]
This is exactly Definition~\ref{def:appG-boundary-law-contraction}.
\end{proof}

\begin{definition}[Log-likelihood contraction, optional form]
\label{def:appG-log-likelihood-contraction}
A stronger sufficient condition is the following.  On \(\Good_{L,j}\), for
each admissible atom of \(\mathcal C\), there is a reference law
\(\lambda_{L,j,\mathcal C}\) on \(\Range(R_{L,j})\) such that, for both
\(b\in\bits\),
\[
  \Dinf\left(
    \Law(R_{L,j}\mid B_j=b,\mathcal C,\Good_{L,j})
    \middle\|
    \lambda_{L,j,\mathcal C}
  \right)
  \le
  \kappa_L,
  \qquad
  \kappa_L=C_{\log}\rho_{\log}^L.
\]
\end{definition}

\begin{lemma}[Log-likelihood contraction implies root-kernel contraction]
\label{lem:appG-loglik-implies-root}
Assume the optional log-likelihood contraction condition with \(\kappa_L\le1\).
Then root-kernel contraction holds, after changing constants, with the same
exponential rate \(\rho_{\log}\).
\end{lemma}

\begin{proof}
Let \(P_b=\Law(R_{L,j}\mid B_j=b,\mathcal C,\Good_{L,j})\) and let
\(\lambda=\lambda_{L,j,\mathcal C}\).  The max-divergence bound gives
\[
  P_b(r)
  \le
  e^{\kappa_L}\lambda(r)
\]
for every \(r\).  Therefore
\[
  \TV(P_b,\lambda)
  =\frac{1}{2}\sum_r|P_b(r)-\lambda(r)|
  \le e^{\kappa_L}-1
  \le 2\kappa_L.
\]
By the triangle inequality,
\[
  \TV(P_1,P_0)
  \le \TV(P_1,\lambda)+\TV(P_0,\lambda)
  \le 4\kappa_L.
\]
Since \(\kappa_L=C_{\log}\rho_{\log}^L\), this is root-kernel contraction after
absorbing the fixed factor into \(C_{\mathrm{root}}\).
\end{proof}

\subsection{Boundary-law mixing theorem}
\label{subsec:appG-mixing-theorem}

\begin{theorem}[Boundary-law mixing]
\label{thm:appG-boundary-law-mixing}
Assume boundary-law contraction with error
\(\varepsilon_{\mathrm{BL}}(m,L)\).  Let
\[
  Z_{L,j}=Q_{\delta_m}(R_{L,j})
\]
be any finite Markov postprocessing of the exact boundary-law summary.  Then
for every admissible conditioning field \(\mathcal C\) and every predictor
\(h:\mathcal Z_{L,m}\to\bits\),
\[
  \Prb[h(Z_{L,j})=B_j\mid\mathcal C]
  \le
  \frac{1}{2}+\varepsilon_{\mathrm{BL}}(m,L)
\]
almost surely.  Equivalently,
\[
  \TV\left(
    \Law(Z_{L,j}\mid B_j=1,\mathcal C),
    \Law(Z_{L,j}\mid B_j=0,\mathcal C)
  \right)
  \le
  2\varepsilon_{\mathrm{BL}}(m,L).
\]
\end{theorem}

\begin{proof}
Condition on an atom of \(\mathcal C\).  By admissibility, \(B_j\) is balanced
on that atom.  Since \(Z_{L,j}\) is a finite Markov postprocessing of
\(R_{L,j}\), Lemma~\ref{lem:appG-tv-data-processing} gives
\[
  \TV(\Law(Z_{L,j}\mid B_j=1,\mathcal C),
       \Law(Z_{L,j}\mid B_j=0,\mathcal C))
  \le
  \TV(\Law(R_{L,j}\mid B_j=1,\mathcal C),
       \Law(R_{L,j}\mid B_j=0,\mathcal C)).
\]
Boundary-law contraction bounds the right-hand side by
\(2\varepsilon_{\mathrm{BL}}(m,L)\).  Lemma~\ref{lem:appG-conditional-bayes}
then gives the predictor-success bound.
\end{proof}

\begin{corollary}[Mixing from root contraction]
\label{cor:appG-mixing-from-root-contraction}
Assume root-kernel contraction and the good-geometry error bound.  Then
Theorem~\ref{thm:appG-boundary-law-mixing} holds with
\[
  \varepsilon_{\mathrm{mix}}(m,L)
  :=
  C_{\mathrm{root}}\rho_{\mathrm{root}}^L+\tau_{\mathrm{geo}}(m,L).
\]
\end{corollary}

\begin{proof}
This is Proposition~\ref{prop:appG-good-plus-bad} followed by
Theorem~\ref{thm:appG-boundary-law-mixing}.
\end{proof}

\begin{corollary}[Logarithmic depth gives polynomial decay]
\label{cor:appG-log-depth}
Let
\[
  L=c_L\log m.
\]
If
\[
  \tau_{\mathrm{geo}}(m,L)\le C_{\mathrm{geo}}m^{-a_{\mathrm{geo}}}
\]
for some \(a_{\mathrm{geo}}>0\), then
\[
  \varepsilon_{\mathrm{mix}}(m,L)
  \le
  C_{\mathrm{root}}m^{-c_L|\log\rho_{\mathrm{root}}|}
  +C_{\mathrm{geo}}m^{-a_{\mathrm{geo}}}
  =m^{-\Omega(1)}.
\]
For any fixed \(A<a_{\mathrm{geo}}\), choosing \(c_L\) large enough gives
\(\varepsilon_{\mathrm{mix}}(m,L)\le m^{-A}\) for all sufficiently large
\(m\), up to constants.
\end{corollary}

\begin{proof}
Since \(0<\rho_{\mathrm{root}}<1\),
\[
  \rho_{\mathrm{root}}^{c_L\log m}
  =
  m^{-c_L|\log\rho_{\mathrm{root}}|}
\]
with the fixed convention for the logarithm base.  The displayed estimate
follows immediately.
\end{proof}

\subsection{Conditional switched-sequence form}
\label{subsec:appG-switched-sequence-form}

\begin{definition}[Ordered switched history]
\label{def:appG-ordered-switched-history}
Let
\[
  S=(j_1,\ldots,j_s)
\]
be an ordered switched set.  Let \(\mathcal F_{\ell-1}\) be the finite
sigma-field generated by neutral public data, fixed observer-side bookkeeping,
and all previous switched target bits, summaries, and observer outputs for
\(j_1,\ldots,j_{\ell-1}\).
\end{definition}

\begin{hypothesis}[Admissible switched histories]
\label{hyp:appG-admissible-histories}
For every \(\ell\), the history \(\mathcal F_{\ell-1}\) is admissible for the
current coordinate \(j_\ell\).
\end{hypothesis}

\begin{theorem}[Conditional boundary-law mixing]
\label{thm:appG-conditional-boundary-law-mixing}
Assume Hypothesis~\ref{hyp:appG-admissible-histories} and boundary-law
contraction.  Then, for every \(\ell\) and every pivot-visible predictor
\(h:\mathcal Z_{L,m}\to\bits\),
\[
  \Prb[h(Z_{L,j_\ell})=B_{j_\ell}\mid\mathcal F_{\ell-1}]
  \le
  \frac{1}{2}+\varepsilon_{\mathrm{mix}}(m,L)
\]
almost surely.
\end{theorem}

\begin{proof}
Apply Theorem~\ref{thm:appG-boundary-law-mixing} with
\(\mathcal C=\mathcal F_{\ell-1}\).  The admissibility hypothesis gives the
conditional balance of \(B_{j_\ell}\).
\end{proof}

\begin{corollary}[Visible baseline exported to the product theorem]
\label{cor:appG-visible-baseline-export}
Under the same hypotheses,
\[
  \sup_{h:\mathcal Z_{L,m}\to\bits}
  \Prb[h(Z_{L,j_\ell})=B_{j_\ell}\mid\mathcal F_{\ell-1}]
  \le
  \frac{1}{2}+\varepsilon_{\mathrm{mix}}(m,L)
\]
for every \(\ell\).  For \(L=c_L\log m\) with parameters chosen as in
Corollary~\ref{cor:appG-log-depth}, this is
\[
  \sup_h
  \Prb[h(Z_{L,j_\ell})=B_{j_\ell}\mid\mathcal F_{\ell-1}]
  \le
  \frac{1}{2}+m^{-\Omega(1)}.
\]
\end{corollary}

\begin{proof}
Take the supremum in Theorem~\ref{thm:appG-conditional-boundary-law-mixing}
and apply Corollary~\ref{cor:appG-log-depth}.
\end{proof}

\begin{remark}[Use in Section 9]
\label{rem:appG-use-in-section9}
The Atomic Evidence Budget controls the excess of the actual observer over the
best pivot-visible predictor.  Appendix~\ref{app:boundary-law-mixing} controls
the best pivot-visible predictor itself.  Combining the two estimates gives a
conditional per-coordinate bound of the form
\[
  \Prb[A_{j_\ell}=B_{j_\ell}\mid\mathcal F_{\ell-1}]
  \le
  \frac{1}{2}
  +\varepsilon_{\mathrm{mix}}(m,L)
  +\varepsilon_{\mathrm{AEB},\ell},
\]
where the average of \(\varepsilon_{\mathrm{AEB},\ell}\) is bounded in
Section~\ref{sec:aeb}.  The tower-product argument in
Section~\ref{sec:product-small-success} then gives the exponential small-success
estimate.
\end{remark}

\subsection{Lean-facing package}
\label{subsec:appG-lean-facing-package}

A Lean formalization can split this appendix into five finite modules.

\begin{enumerate}[label=(\roman*)]
\item \texttt{FiniteBayes.lean}: posterior, Bayes success, and the balanced-bit
      TV identity.
\item \texttt{FiniteKernelDataProcessing.lean}: TV contraction under finite
      Markov kernels and the prediction data-processing corollary.
\item \texttt{BoundaryLawObjects.lean}: target bit, pivot ball, raw cut state,
      exact boundary-law summary, dithered summary, and admissible conditioning.
\item \texttt{BoundaryLawContraction.lean}: root-kernel contraction, good-geometry
      error, and the TV boundary-contraction estimate.
\item \texttt{BoundaryLawMixing.lean}: the final mixing theorem and the
      conditional switched-sequence export.
\end{enumerate}

The first two modules contain only finite probability.  The ensemble-specific
obligations are isolated in root-kernel contraction and admissibility of the
switched histories.

\section{Product Small-Success and Compression-from-Success}
\label{app:product-compression}
\providecommand{\bits}{\{0,1\}}
\providecommand{\Prb}{\mathbb P}
\providecommand{\E}{\mathbb E}
\providecommand{\Kp}{K_{\mathrm{poly}}}
\providecommand{\poly}{\operatorname{poly}}
\providecommand{\Gap}{\operatorname{Gap}}
\providecommand{\Base}{\operatorname{Base}}
\providecommand{\Excess}{\operatorname{Excess}}
\providecommand{\Succ}{\mathsf{Succ}}
\providecommand{\Bad}{\mathsf{Bad}}
\providecommand{\Good}{\mathsf{Good}}
\providecommand{\Range}{\operatorname{Range}}
\providecommand{\rankG}{\operatorname{rank}_G}

This appendix expands the final probabilistic and coding steps used in
Sections~\ref{sec:product-small-success} and~\ref{sec:message-incompressibility}.
The inputs are the two middle estimates proved earlier:

\begin{enumerate}[label=(\roman*)]
\item \textbf{ACCEI/PNLD.}  After conditioning on the pivot-visible summary, the
      residual exterior computation has only the phase gap allowed by the CD
      Atomic Evidence Budget.

\item \textbf{Boundary-law mixing.}  The pivot-visible summary itself predicts
      each switched target bit only with success
      \(1/2+\varepsilon_{\mathrm{mix}}(m,L)\).
\end{enumerate}

The output is an exact-success estimate
\[
  \Prb[A_S=B_S]\le 2^{-\Omega(t)}
\]
for every short polynomial-time observer, followed by the standard conversion
from exponentially small exact success to a high-probability lower bound on
clocked polytime-capped conditional description length.  The appendix is finite
throughout: histories are finite partitions, conditional probabilities are finite
sums over atoms, and programs are counted as finite prefix strings.

\subsection{Finite histories and one-step success}
\label{subsec:appH-histories-one-step}

Let
\[
  S=(j_1,\ldots,j_s)
\]
be an ordered switched set with
\[
  s=|S|\ge \gamma t,
  \qquad
  t=\Theta(m).
\]
Write
\[
  B_\ell:=B_{j_\ell},
  \qquad
  A_\ell:=A_{j_\ell},
  \qquad
  Z_\ell:=Z_{L,j_\ell}.
\]
Here \(B_\ell\) is the target bit, \(A_\ell\) is the observer's output bit, and
\(Z_\ell\) is the pivot-visible boundary summary.

\begin{definition}[Switched histories]
\label{def:appH-switched-histories}
For each \(\ell\), let \(\mathcal F_{\ell-1}\) be a finite conditioning field
containing the public bookkeeping and all switched coordinates already processed
before \(j_\ell\).  Let
\[
  \mathcal G_\ell:=\mathcal F_{\ell-1}\vee \sigma(Z_\ell)
\]
be the same history augmented by the pivot-visible summary.
\end{definition}

\begin{hypothesis}[Admissible ordered histories]
\label{hyp:appH-admissible-histories}
For every \(\ell\), the field \(\mathcal F_{\ell-1}\) is admissible for the
coordinate \(j_\ell\) in the sense of
Definition~\ref{def:admissible-conditioning-section8}.  In particular,
\[
  \Prb[B_\ell=0\mid\mathcal F_{\ell-1}]
  =
  \Prb[B_\ell=1\mid\mathcal F_{\ell-1}]
  =\frac12
\]
on every positive-probability atom.  The augmented field
\(\mathcal G_\ell\) is admissible for the residual exterior trace.
\end{hypothesis}

\begin{definition}[Pivot-visible Bayes baseline]
\label{def:appH-visible-bayes-baseline}
The best conditional success available from the pivot-visible summary alone is
\[
  \Base_\ell
  :=
  \sup_{h:\Range(Z_\ell)\to\bits}
  \Prb[h(Z_\ell)=B_\ell\mid\mathcal F_{\ell-1}].
\]
Equivalently,
\[
  \Base_\ell
  =
  \E\left[
    \max\{\Prb[B_\ell=0\mid\mathcal G_\ell],
          \Prb[B_\ell=1\mid\mathcal G_\ell]\}
    \middle|\mathcal F_{\ell-1}
  \right].
\]
\end{definition}

\begin{definition}[Pivot-fiber phase gap]
\label{def:appH-pivot-fiber-gap}
For a Boolean output \(A_\ell\), define
\[
  \Gap_\ell^Z(A_\ell)
  :=
  \frac12
  \left|
    \Prb[A_\ell=1\mid B_\ell=1,\mathcal G_\ell]
    -
    \Prb[A_\ell=1\mid B_\ell=0,\mathcal G_\ell]
  \right|.
\]
This is a \(\mathcal G_\ell\)-measurable random variable.  On zero-phase atoms,
choose an arbitrary value; those atoms do not contribute to the finite sums.
\end{definition}

\begin{lemma}[Bayes-vs-test inequality on an atom]
\label{lem:appH-bayes-vs-test-atom}
Let \(A,B\in\bits\), and condition on one atom \(G\) of a finite field.  Put
\[
  p:=\Prb[B=1\mid G],
  \qquad
  a:=\Prb[A=1\mid B=1,G],
  \qquad
  b:=\Prb[A=1\mid B=0,G].
\]
Then
\[
  \Prb[A=B\mid G]
  \le
  \max\{p,1-p\}+\frac12|a-b|.
\]
\end{lemma}

\begin{proof}
On the atom \(G\),
\[
  \Prb[A=B\mid G]=pa+(1-p)(1-b).
\]
Assume first that \(p\ge 1/2\).  If \(a\le b\), then
\[
  pa+(1-p)(1-b)
  \le
  pb+(1-p)(1-b)
  \le p,
\]
because \(p\ge 1-p\).  If \(a>b\), write \(d=a-b\).  Then
\[
\begin{aligned}
  pa+(1-p)(1-b)-p
  &= (1-p)(1-b)-p(1-a)  \\
  &\le (1-p)d
  \le \frac d2.
\end{aligned}
\]
This proves the claim for \(p\ge 1/2\).  The case \(p\le 1/2\) is symmetric,
with the two target phases exchanged.
\end{proof}

\begin{proposition}[PNLD from conditional phase gap]
\label{prop:appH-pnld-from-gap}
For each switched coordinate,
\[
  \Prb[A_\ell=B_\ell\mid\mathcal F_{\ell-1}]
  \le
  \Base_\ell+
  \E[\Gap_\ell^Z(A_\ell)\mid\mathcal F_{\ell-1}].
\]
\end{proposition}

\begin{proof}
Apply Lemma~\ref{lem:appH-bayes-vs-test-atom} on every atom of
\(\mathcal G_\ell\), then average conditional on \(\mathcal F_{\ell-1}\).  The
Bayes part averages to \(\Base_\ell\), and the half-difference part is precisely
\(\Gap_\ell^Z(A_\ell)\).
\end{proof}

\subsection{ACCEI envelopes and boundary-law baselines}
\label{subsec:appH-accei-baseline}

The Atomic Evidence Budget is used through a deterministic envelope.  This is
convenient for formalization because the later tower-product step needs a
pointwise one-step bound, not merely an unconditional average.

\begin{definition}[ACCEI envelope]
\label{def:appH-accei-envelope}
A sequence of nonnegative deterministic numbers
\(\eta_1,\ldots,\eta_s\) is an ACCEI envelope for an observer \(A\) on the
ordered switched set if, almost surely, for every \(\ell\),
\[
  \E[\Gap_\ell^Z(A_\ell)\mid\mathcal F_{\ell-1}]
  \le
  \eta_\ell.
\]
Its average is
\[
  \overline\eta_S:=\frac1s\sum_{\ell=1}^s\eta_\ell.
\]
\end{definition}

\begin{hypothesis}[Conditional boundary-law baseline]
\label{hyp:appH-boundary-baseline}
For every admissible history,
\[
  \Base_\ell
  \le
  \frac12+\varepsilon_{\mathrm{mix}}(m,L).
\]
This is the exported form of Appendix~\ref{app:boundary-law-mixing}.
\end{hypothesis}

\begin{lemma}[One-step PNLD bound from an envelope]
\label{lem:appH-one-step-pnld-envelope}
If \((\eta_\ell)\) is an ACCEI envelope for \(A\), then for every \(\ell\),
\[
  \Prb[A_\ell=B_\ell\mid\mathcal F_{\ell-1}]
  \le
  \frac12+\varepsilon_{\mathrm{mix}}(m,L)+\eta_\ell.
\]
\end{lemma}

\begin{proof}
Combine Proposition~\ref{prop:appH-pnld-from-gap} with
Hypothesis~\ref{hyp:appH-boundary-baseline} and the definition of an ACCEI
envelope.
\end{proof}

\subsection{Stopping and deterministic rank envelopes}
\label{subsec:appH-stopping}

The gauge-rank entropy theorem gives a tail estimate.  To get a clean product
bound, we stop the observer above a chosen rank threshold and then add the tail
probability at the end.

\begin{definition}[Stopped observer]
\label{def:appH-stopped-observer}
Fix \(\lambda\ge0\).  Let
\[
  R_\lambda
  :=
  |P|+\frac{\varepsilon_{\le Q}}{\ln 2}+\lambda+C_0,
\]
where \(C_0\) is the universal prefix-counting constant in the gauge-rank
entropy estimate.  The stopped observer \(A^{\le\lambda}\) follows the original
observer until the next normalized gauge-evidence leaf would make the exposed
hidden gauge rank exceed \(R_\lambda\).  From then on it outputs a fixed default
bit on every remaining switched coordinate.  Let
\[
  \Bad_\lambda:=\{A^{\le\lambda}_S\ne A_S\}
\]
be the event that stopping changes at least one switched output.
\end{definition}

\begin{hypothesis}[Gauge-rank stopping tail]
\label{hyp:appH-stopping-tail}
For every \(\lambda\ge0\),
\[
  \Prb[\Bad_\lambda]
  \le
  2^{-\lambda+C_{\mathrm{tail}}}
\]
for a constant \(C_{\mathrm{tail}}\) independent of \(m,t,P,S\).
\end{hypothesis}

\begin{definition}[CD budget for the stopped observer]
\label{def:appH-cd-budget-stopped}
For the stopped observer define
\[
  \varepsilon_{\mathrm{cd}}(m;P,S,\lambda)
  :=
  \frac{1}{2s}
  \left[
    Q_{\mathrm{tot}}\varepsilon_{\mathrm{step}}(m)
    +
    \Delta_G
    \left(
      |P|+\frac{\varepsilon_{\le Q}}{\ln2}+\lambda+C_1
    \right)
  \right],
\]
where \(C_1\) is an absolute constant absorbing the fixed normal-form and
prefix-coding overheads.
\end{definition}

\begin{hypothesis}[Averaged Atomic Evidence Budget]
\label{hyp:appH-averaged-aeb}
For the stopped observer \(A^{\le\lambda}\), there is an ACCEI envelope
\((\eta_\ell)_{\ell=1}^s\) satisfying
\[
  \overline\eta_S
  \le
  \varepsilon_{\mathrm{cd}}(m;P,S,\lambda).
\]
This is the exported conditional form of the Atomic Evidence Budget from
Section~\ref{sec:aeb} and Appendix~\ref{app:gauge-rank-entropy-incidence}.
\end{hypothesis}

\subsection{From averaged envelope to product decay}
\label{subsec:appH-product-decay}

\begin{definition}[Pruned set]
\label{def:appH-pruned-set}
Given an envelope \((\eta_\ell)\) and a threshold \(\theta>0\), define
\[
  S_\theta:=\{j_\ell\in S:\eta_\ell\le\theta\},
  \qquad
  s_\theta:=|S_\theta|.
\]
The order on \(S_\theta\) is inherited from \(S\).
\end{definition}

\begin{lemma}[Markov pruning]
\label{lem:appH-markov-pruning}
For every \(\theta>0\),
\[
  s_\theta
  \ge
  \left(1-\frac{\overline\eta_S}{\theta}\right)s.
\]
\end{lemma}

\begin{proof}
At most \(s\overline\eta_S/\theta\) indices can have \(\eta_\ell>\theta\), since
otherwise their contribution to \(\sum_\ell\eta_\ell\) would exceed
\(s\overline\eta_S\).
\end{proof}

\begin{lemma}[Sequential tower product]
\label{lem:appH-sequential-tower-product}
Let \(F_1,\ldots,F_r\) be events, and let
\(\mathcal H_0\subseteq\cdots\subseteq\mathcal H_{r-1}\) be finite histories
with \(F_1\cap\cdots\cap F_{a-1}\in\mathcal H_{a-1}\).  If
\[
  \Prb[F_a\mid\mathcal H_{a-1}]\le q
\]
almost surely for every \(a\), then
\[
  \Prb\left[\bigcap_{a=1}^rF_a\right]
  \le q^r.
\]
\end{lemma}

\begin{proof}
Let \(G_a:=F_1\cap\cdots\cap F_a\), with \(G_0=\Omega\).  Since
\(G_{a-1}\) is \(\mathcal H_{a-1}\)-measurable,
\[
\begin{aligned}
  \Prb[G_a]
  &=\E[\mathbf 1_{G_{a-1}}\mathbf 1_{F_a}] \\
  &=\E[\mathbf 1_{G_{a-1}}\Prb[F_a\mid\mathcal H_{a-1}]] \\
  &\le q\Prb[G_{a-1}].
\end{aligned}
\]
Induction gives \(\Prb[G_r]\le q^r\).
\end{proof}

\begin{theorem}[Product bound with stopping]
\label{thm:appH-product-with-stopping}
Let \((\eta_\ell)\) be an ACCEI envelope for \(A^{\le\lambda}\).  For every
\(\theta>0\),
\[
  \Prb[A_S=B_S]
  \le
  \left(\frac12+\varepsilon_{\mathrm{mix}}(m,L)+\theta\right)^{s_\theta}
  +
  2^{-\lambda+C_{\mathrm{tail}}}.
\]
\end{theorem}

\begin{proof}
On the complement of \(\Bad_\lambda\), the stopped and unstopped observers have
the same switched tuple.  Hence
\[
  \Prb[A_S=B_S]
  \le
  \Prb[A^{\le\lambda}_{S_\theta}=B_{S_\theta}]
  +
  \Prb[\Bad_\lambda].
\]
For every retained coordinate, \(\eta_\ell\le\theta\), so
Lemma~\ref{lem:appH-one-step-pnld-envelope} gives
\[
  \Prb[A^{\le\lambda}_\ell=B_\ell\mid\mathcal F_{\ell-1}]
  \le
  \frac12+\varepsilon_{\mathrm{mix}}(m,L)+\theta.
\]
Apply Lemma~\ref{lem:appH-sequential-tower-product} to the retained ordered
coordinates and then use Hypothesis~\ref{hyp:appH-stopping-tail}.
\end{proof}

\begin{proposition}[Linear-envelope product lemma]
\label{prop:appH-linear-envelope-product}
Suppose that \(s\ge\gamma t\), \(\varepsilon_{\mathrm{mix}}(m,L)=o(1)\), and for
a sequence of observers the stopped-envelope average satisfies
\[
  \overline\eta_S\le \eta_0+o(1)
\]
for some \(\eta_0>0\).  If \(\theta:=\sqrt{2\eta_0}<1/4\), then for all large
\(m\),
\[
  \Prb[A_S=B_S]
  \le
  \left(\frac12+2\theta\right)^{(1-\theta)\gamma t}
  +
  2^{-\lambda+C_{\mathrm{tail}}}.
\]
In particular, if \(\lambda=\alpha t\) with \(\alpha>0\), then
\[
  \Prb[A_S=B_S]\le 2^{-c t}
\]
for some \(c>0\), provided \(\eta_0\) and \(\alpha\) are fixed small enough.
\end{proposition}

\begin{proof}
For all large \(m\), \(\overline\eta_S\le 2\eta_0\).  With
\(\theta=\sqrt{2\eta_0}\), Lemma~\ref{lem:appH-markov-pruning} gives
\[
  s_\theta\ge(1-\theta)s\ge(1-\theta)\gamma t.
\]
For large \(m\), \(\varepsilon_{\mathrm{mix}}(m,L)\le\theta\), so the product
term in Theorem~\ref{thm:appH-product-with-stopping} is at most
\[
  \left(\frac12+2\theta\right)^{(1-\theta)\gamma t}.
\]
If \(\theta<1/4\), then \(1/2+2\theta<1\), so this term is
\(2^{-\Omega(t)}\).  The tail term is \(2^{-\alpha t+O(1)}\), also
\(2^{-\Omega(t)}\).
\end{proof}

\subsection{Uniform product small-success for short observers}
\label{subsec:appH-uniform-small-success}

The preceding proposition becomes the product small-success theorem after
substituting the explicit CD budget.

\begin{theorem}[Fixed-observer small-success]
\label{thm:appH-fixed-observer-small-success}
Let \(A\) be a fixed deterministic polynomial-time observer.  Choose
\[
  L=c_L\log m,
  \qquad
  R_{\mathrm{safe}}=c_R\log m
\]
so that
\[
  Q_{\mathrm{tot}}\varepsilon_{\mathrm{step}}(m)=o(t),
  \qquad
  \varepsilon_{\le Q}=o(t),
  \qquad
  \varepsilon_{\mathrm{mix}}(m,L)=o(1).
\]
Then there is a constant \(c_A>0\) such that, for all sufficiently large \(m\),
\[
  \Prb[A_S=B_S]
  \le
  2^{-c_A t}.
\]
\end{theorem}

\begin{proof}
Let \(P\) be the fixed program representing \(A\), and take
\(\lambda=\alpha t\).  Since \(|P|=O(1)\) and \(s\ge\gamma t\), the CD budget
satisfies
\[
  \overline\eta_S
  \le
  \frac{\Delta_G}{2\gamma}\alpha+o(1).
\]
Choose \(\alpha>0\) so small that
\[
  \eta_0:=\frac{\Delta_G}{2\gamma}\alpha<\frac1{1000}.
\]
The conclusion follows from Proposition~\ref{prop:appH-linear-envelope-product}.
\end{proof}

\begin{theorem}[Linear-length observer small-success]
\label{thm:appH-linear-length-observer-small-success}
There are constants \(\delta_0,c_0>0\) such that the following holds.  For every
deterministic observer represented by a program \(P\) with
\[
  |P|\le\delta_0 t,
\]
after the same parameter choices,
\[
  \Prb[A_S=B_S]
  \le
  2^{-c_0 t}
\]
for all sufficiently large \(m\).  The constants \(\delta_0,c_0\) are independent
of the particular program \(P\).
\end{theorem}

\begin{proof}
Take \(\lambda=\alpha t\).  The CD budget gives
\[
  \overline\eta_S
  \le
  \frac{\Delta_G}{2\gamma}(\delta_0+\alpha)+o(1).
\]
Choose \(\alpha>0\) and then \(\delta_0>0\) so that
\[
  \eta_0:=\frac{\Delta_G}{2\gamma}(\delta_0+\alpha)<\frac1{1000}.
\]
Proposition~\ref{prop:appH-linear-envelope-product} then gives a bound
\(2^{-c_0t}\), with \(c_0>0\) depending only on
\(\gamma,\Delta_G,\alpha,\delta_0\), and the asymptotic constants in the mixing
and safe-leakage estimates.
\end{proof}

\begin{remark}[Randomized observers]
\label{rem:appH-randomized-observers}
Randomized observers are reduced to deterministic observers by fixing their
random coins.  If a randomized observer had success larger than the stated
bound, then at least one deterministic coin string would have at least that
success.
\end{remark}

\subsection{Coding from partial success}
\label{subsec:appH-coding-from-success}

This subsection records the coding lemma behind the name
Compression-from-Success.  The exact-success union bound in the next subsection
is the form used for the final incompressibility theorem, but the patching code is
useful for checking that the information-theoretic direction is the expected one.

\begin{lemma}[Subset-rank code]
\label{lem:appH-subset-rank-code}
For every \(0\le r\le s\), every subset \(R\subseteq[s]\) of size \(r\) can be
encoded by
\[
  \left\lceil \log_2 {s\choose r}\right\rceil+O(1)
\]
bits, given \(s\) and \(r\), using the rank of \(R\) in a fixed lexicographic
enumeration.
\end{lemma}

\begin{proof}
There are \({s\choose r}\) such subsets.  A fixed lexicographic enumeration and
its inverse are computable in time polynomial in \(s\) and the bitlength of the
rank.  The self-delimiting overhead is constant when \(s,r\) are already known,
and \(O(\log s)\) when they must be encoded.
\end{proof}

\begin{theorem}[Compression-from-Success, selected bit tuple]
\label{thm:appH-compression-from-success-bits}
Let \(Y\) be public side information, let \(B\in\bits^s\), and let \(P\) be a
program of length \(L\) outputting \(\widehat B=P(Y)\in\bits^s\).  Suppose
\(T\subseteq[s]\) is a trusted success set:
\[
  \widehat B_i=B_i
  \qquad(i\in T).
\]
Then
\[
  \Kp(B\mid Y)
  \le
  L+
  \left\lceil \log_2 {s\choose |T|}\right\rceil
  +(s-|T|)+O(\log s).
\]
\end{theorem}

\begin{proof}
The decoder receives the program \(P\), the set \(T\) encoded by its rank, and
the verbatim patch bits \((B_i)_{i\notin T}\).  On input \(Y\), it runs \(P\) to
obtain \(\widehat B\).  It outputs \(\widehat B_i\) on \(T\), and the patch bit
on \([s]\setminus T\).  The length is the displayed quantity, including the
self-delimiting encoding of the necessary lengths.  The runtime is polynomial in
\(|Y|+s\) plus the runtime of \(P\).
\end{proof}

\begin{corollary}[Exact-error enumerative code]
\label{cor:appH-exact-error-code}
Let
\[
  E:=\{i:\widehat B_i\ne B_i\}
\]
be the exact error set.  Then
\[
  \Kp(B\mid Y)
  \le
  L+
  \left\lceil\log_2{s\choose |E|}\right\rceil+O(\log s).
\]
In particular,
\[
  \Kp(B\mid Y)
  \le
  L+sH_2(|E|/s)+O(\log s),
\]
where \(H_2\) is binary entropy.
\end{corollary}

\begin{proof}
Encode the error set \(E\) and flip exactly those predicted bits.  The entropy
bound is the standard estimate
\(\log_2{s\choose e}\le sH_2(e/s)\).
\end{proof}

\subsection{Small exact success implies incompressibility}
\label{subsec:appH-small-success-incompressibility}

The final lower bound uses exact-success rarity rather than a high-Hamming-error
statement.  This is the cleanest form for the upper-lower clash: a short
description of the target tuple is itself a predictor that succeeds exactly.

\begin{definition}[Clocked exact success]
\label{def:appH-clocked-exact-success}
Fix a universal machine \(U\), a clock exponent \(D\), and a random pair
\((Y,B)\) with \(B\in\bits^s\).  For a program \(p\), define
\[
  \Succ_{p}^{(D)}
  :=
  \{Y:U(p,Y)=B(Y)\text{ within }(|Y|+2)^D\text{ steps}\}.
\]
If \(U(p,Y)\) does not halt in the clock or does not output an \(s\)-bit string,
then \(Y\notin\Succ_p^{(D)}\).
\end{definition}

\begin{lemma}[Prefix-program count]
\label{lem:appH-prefix-program-count}
There is an absolute constant \(C_{\mathrm{cnt}}\) such that, for every \(a>0\),
\[
  |\{p:|p|\le at\}|
  \le
  2^{at+C_{\mathrm{cnt}}}.
\]
\end{lemma}

\begin{proof}
For prefix programs, the number of codewords of length at most \(n\) is at most
\(2^{n+1}\).  Absorb the harmless additive constant into
\(C_{\mathrm{cnt}}\).
\end{proof}

\begin{theorem}[Small exact success implies clocked incompressibility]
\label{thm:appH-small-success-to-kpoly}
Let \((Y,B(Y))\) be a finite random public-target pair with \(B(Y)\in\bits^s\).
Assume that, for some \(c>0\), every program \(p\) of length at most
\(a t\) satisfies
\[
  \Prb[\Succ_p^{(D)}]\le 2^{-ct}.
\]
Then
\[
  \Prb[\Kp^{(D)}(B(Y)\mid Y)
       \le at]
  \le
  2^{-(c-a)t+C_{\mathrm{cnt}}}.
\]
In particular, if \(a<c\), then
\[
  \Prb[\Kp^{(D)}(B(Y)\mid Y)>at]
  \ge
  1-2^{-\Omega(t)}.
\]
\end{theorem}

\begin{proof}
If \(\Kp^{(D)}(B(Y)\mid Y)\le at\), then some program \(p\) of length at most
\(at\) outputs \(B(Y)\) from \(Y\) within the \(D\)-clock.  Therefore the bad
compression event is contained in
\[
  \bigcup_{|p|\le at}\Succ_p^{(D)}.
\]
Using Lemma~\ref{lem:appH-prefix-program-count} and the assumed success bound,
\[
  \Prb[\Kp^{(D)}(B(Y)\mid Y)\le at]
  \le
  2^{at+C_{\mathrm{cnt}}}2^{-ct}
  =
  2^{-(c-a)t+C_{\mathrm{cnt}}}.
\]
\end{proof}

\subsection{Applying the union bound to switched message coordinates}
\label{subsec:appH-switched-coordinate-incompressibility}

Now specialize \(B(Y)\) to the switched coordinate tuple of the global message.
Let
\[
  B_S(Y)=\bigl(B_{j_1}(Y),\ldots,B_{j_s}(Y)\bigr)
  =\pi_S(M(Y)).
\]

\begin{lemma}[A decompressor induces an observer]
\label{lem:appH-decompressor-induces-observer}
Fix a clock exponent \(D\).  There are constants \(C_{\mathrm{obs}}\) and
\(D_{\mathrm{obs}}\), depending only on the universal machine and tuple
conventions, such that every program \(p\) induces a deterministic observer
\(A^p\) satisfying
\[
  |A^p|\le |p|+C_{\mathrm{obs}},
\]
and
\[
  \Succ_p^{(D)}
  \subseteq
  \{Y:A^p_S(Y)=B_S(Y)\}.
\]
\end{lemma}

\begin{proof}
The observer contains \(p\), simulates it for \((|Y|+2)^D\) steps, checks that
the output is an \(s\)-bit string, and distributes that string to the ordered
switched coordinates.  If the simulation fails, it outputs a fixed default tuple.
This adds only constant source length and a fixed polynomial overhead.  Whenever
\(p\) succeeds as a decompressor for \(B_S(Y)\), the induced observer predicts
\(B_S(Y)\) exactly.
\end{proof}

\begin{theorem}[Switched-tuple incompressibility]
\label{thm:appH-switched-tuple-incompressibility}
Fix \(D\).  Choose the ensemble parameters for the corresponding observer
runtime exponent.  Let \(\delta_0,c_0\) be the constants from
Theorem~\ref{thm:appH-linear-length-observer-small-success}.  For every
\[
  0<a<\min\{\delta_0,c_0\},
\]
there is \(\kappa>0\) such that
\[
  \Prb[\Kp^{(D)}(B_S(Y)\mid Y)
       \le at]
  \le
  2^{-\kappa t}
\]
for all sufficiently large \(t\).
\end{theorem}

\begin{proof}
Let \(p\) have length at most \(at\).  By
Lemma~\ref{lem:appH-decompressor-induces-observer}, \(p\) induces an observer
of length at most \(at+C_{\mathrm{obs}}\).  Since \(a<\delta_0\), this is at
most \(\delta_0t\) for all large \(t\).  Therefore
Theorem~\ref{thm:appH-linear-length-observer-small-success} gives
\[
  \Prb[\Succ_p^{(D)}]
  \le
  2^{-c_0t}.
\]
Apply Theorem~\ref{thm:appH-small-success-to-kpoly} with \(c=c_0\).  Any
\(\kappa<c_0-a\) works for all large \(t\).
\end{proof}

\subsection{Projection transfer to the full message}
\label{subsec:appH-projection-transfer}

The tuple \(B_S(Y)\) is a fixed polynomial-time projection of the whole message.
Therefore incompressibility of \(B_S(Y)\) transfers to incompressibility of
\(M(Y)\).

\begin{lemma}[Clocked projection transfer]
\label{lem:appH-clocked-projection-transfer}
For every clock exponent \(D_M\) there are a clock exponent \(D_S\) and a
constant \(C_\pi\), depending only on \(D_M\) and on the fixed projection
\(\pi_S\), such that
\[
  \Kp^{(D_S)}(B_S(Y)\mid Y)
  \le
  \Kp^{(D_M)}(M(Y)\mid Y)+C_\pi
\]
for every public instance \(Y\).  Consequently,
\[
  \Kp^{(D_S)}(B_S(Y)\mid Y)>at
  \quad\Longrightarrow\quad
  \Kp^{(D_M)}(M(Y)\mid Y)>at-C_\pi.
\]
\end{lemma}

\begin{proof}
A constant-size wrapper first runs a \(D_M\)-clocked program producing
\(M(Y)\) from \(Y\), and then applies the fixed polynomial-time projection
\(\pi_S\).  This adds a constant number of bits and increases the clock exponent
to some fixed \(D_S\).  The implication is the contrapositive of the displayed
inequality.
\end{proof}

\begin{theorem}[Global message incompressibility]
\label{thm:appH-global-message-incompressibility}
For every fixed clock exponent \(D_M\), after choosing the ensemble parameters
for the corresponding projection exponent, there are constants \(\eta,\kappa>0\)
such that
\[
  \Prb\left[
    \Kp^{(D_M)}(M(Y)\mid Y)
    \ge
    \eta t
  \right]
  \ge
  1-2^{-\kappa t}.
\]
\end{theorem}

\begin{proof}
Let \(D_S\) and \(C_\pi\) be supplied by
Lemma~\ref{lem:appH-clocked-projection-transfer}.  Apply
Theorem~\ref{thm:appH-switched-tuple-incompressibility} at exponent \(D_S\),
choosing \(a>0\) smaller than both \(\delta_0\) and \(c_0\).  Except on an event
of probability \(2^{-\Omega(t)}\),
\[
  \Kp^{(D_S)}(B_S(Y)\mid Y)>at.
\]
By projection transfer,
\[
  \Kp^{(D_M)}(M(Y)\mid Y)>at-C_\pi.
\]
For all sufficiently large \(t\), the right side is at least \((a/2)t\).  Set
\(\eta=a/2\), and take \(\kappa\) to be any positive exponent below the union-bound
exponent obtained for the switched tuple.
\end{proof}

\begin{corollary}[Positive-probability hard support]
\label{cor:appH-positive-hard-support}
For every fixed clock exponent \(D_M\) and all sufficiently large \(t\), there is
at least one public instance \(Y\) in the support of the ensemble such that
\[
  \Kp^{(D_M)}(M(Y)\mid Y)\ge \eta t.
\]
\end{corollary}

\begin{proof}
The lower-bound event in
Theorem~\ref{thm:appH-global-message-incompressibility} has probability at
least \(1-2^{-\kappa t}\), which is positive for all large \(t\).  Hence the event
contains at least one point of the finite support.
\end{proof}

\subsection{Exported final-lower-bound interface}
\label{subsec:appH-exported-interface}

Appendix~H exports the following theorem to the final upper-bound appendix.

\begin{theorem}[Appendix H lower-bound interface]
\label{thm:appH-lower-bound-interface}
For every fixed polynomial clock exponent \(D\), after choosing the ensemble
parameters for that exponent, there exist constants \(\eta_D,\kappa_D>0\) such
that
\[
  \Prb\left[
    \Kp^{(D)}(M(Y)\mid Y)
    \ge
    \eta_D t
  \right]
  \ge
  1-2^{-\kappa_D t}.
\]
In particular, for all sufficiently large \(t\), the support contains at least one
public instance satisfying this lower bound.
\end{theorem}

\begin{proof}
This is Theorem~\ref{thm:appH-global-message-incompressibility}, followed by
Corollary~\ref{cor:appH-positive-hard-support}.
\end{proof}

\begin{remark}[Relation to the final contradiction]
\label{rem:appH-relation-final-contradiction}
This appendix proves only the lower-bound half of the final clash.  The remaining
upper-bound half is purely SAT-theoretic: under \(P=NP\), a fixed SAT
self-reduction finds a satisfying witness of the realized CNF \(F_Y\), and the
single-message promise lets a fixed projection read \(M(Y)\) from that witness.
That step is the subject of the SAT-realization and self-reduction appendix.
\end{remark}

\begin{remark}[Lean-facing decomposition]
\label{rem:appH-lean-facing}
A Lean formalization can split Appendix~H into the following modules.
\begin{enumerate}[label=(\roman*)]
\item Finite conditional Bayes-vs-test inequality.
\item ACCEI envelope plus boundary-law baseline gives one-step PNLD.
\item Stopped-observer tail and deterministic rank envelope.
\item Markov pruning and sequential tower-product bound.
\item Uniform small-success for observers of length \(\le\delta_0t\).
\item Prefix-program counting and exact-success union bound.
\item Decompressor-to-observer wrapper.
\item Projection transfer from \(B_S(Y)\) to \(M(Y)\).
\end{enumerate}
Only the first four modules use probability.  The last four are finite program
counting and clocked description-length bookkeeping.
\end{remark}


\section{SAT Realization and the \boldmath{$P=NP$} Self-Reduction}
\label{app:sat-realization-self-reduction}
\providecommand{\Prb}{\mathbb P}
\providecommand{\bits}{\{0,1\}}
\providecommand{\Kp}{K_{\mathrm{poly}}}
\providecommand{\poly}{\operatorname{poly}}
\providecommand{\SAT}{\operatorname{SAT}}
\providecommand{\Var}{\operatorname{Var}}
\providecommand{\Supp}{\operatorname{Supp}}
\providecommand{\CNF}{\operatorname{CNF}}

This appendix supplies the purely SAT-theoretic end of the proof.  The lower
bound is distributional and evidence-theoretic: it says that, for the public
instances sampled by the gauge-buffered locked ensemble, no short polynomial-time
observer can recover the locked message.  The upper bound under the assumption
\(P=NP\) is different in kind.  It uses only the fact that the same public
instance defines a satisfiable SAT formula whose satisfying assignments all carry
the same message.

The appendix proves three interfaces.

\begin{enumerate}[label=(\roman*)]
\item The witness relation \(\mathcal R(Y,W)=1\) has a uniform polynomial-size
      CNF realization \(F_Y\).  A satisfying assignment of \(F_Y\) projects to a
      valid witness, and every valid witness extends to a satisfying assignment.

\item The message projection is fixed and witness-independent on the support:
      every satisfying assignment of \(F_Y\) has the same projected message
      \(M(Y)\).

\item If \(P=NP\), the standard bit-fixing self-reduction produces a satisfying
      assignment of \(F_Y\) in polynomial time.  Composing this SAT search with
      the fixed message projection gives a constant-length clocked program for
      \(M(Y)\) from \(Y\), contradicting the lower bound.
\end{enumerate}

The upper-bound proof does not use uniqueness of the satisfying witness.  It
uses only uniqueness of the projected message.  It also does not classify public
CNF syntax as safe lower-bound evidence.  Public clauses describe a verifier;
they do not themselves assign values to hidden witness or message variables.

\subsection{Boolean encodings and local CNF tables}
\label{subsec:appI-boolean-encodings}

We fix effective encodings of variables, literals, clauses, CNF formulas, finite
tuples, and finite assignments.  All encoding and decoding routines run in time
polynomial in the encoded length.  At each fixed parameter value, every formula
uses only a finite ordered set of variables.

\begin{definition}[CNF syntax]
\label{def:appI-cnf-syntax}
A literal over a finite variable set \(U\) is either \(u\) or \(\neg u\), where
\(u\in U\).  A clause is a finite disjunction of literals, and a CNF formula is a
finite conjunction of clauses.  An assignment \(\alpha:U\to\bits\) satisfies a
positive literal \(u\) when \(\alpha(u)=1\), and satisfies \(\neg u\) when
\(\alpha(u)=0\).  It satisfies a clause when it satisfies at least one literal,
and satisfies a CNF formula when it satisfies every clause.
\end{definition}

\begin{definition}[Forbidden-pattern clause]
\label{def:appI-forbidden-pattern-clause}
Let \(u_1,\ldots,u_k\) be distinct Boolean variables and let
\(a=(a_1,\ldots,a_k)\in\bits^k\).  Define
\[
  C_a(u_1,\ldots,u_k)
  :=
  \bigvee_{i:a_i=0} u_i
  \vee
  \bigvee_{i:a_i=1} \neg u_i .
\]
If one of the two displayed disjunctions is empty, it is omitted.
\end{definition}

\begin{lemma}[Forbidden-pattern correctness]
\label{lem:appI-forbidden-pattern-correct}
For every \(b\in\bits^k\),
\[
  b\models C_a
  \quad\Longleftrightarrow\quad
  b\ne a.
\]
\end{lemma}

\begin{proof}
If \(b=a\), then for each coordinate with \(a_i=0\), the literal \(u_i\) is
false, and for each coordinate with \(a_i=1\), the literal \(\neg u_i\) is
false.  Hence \(C_a\) is false.  Conversely, if \(b\ne a\), choose an index
\(i\) with \(b_i\ne a_i\).  If \(a_i=0\), then \(b_i=1\), so \(u_i\) is true.
If \(a_i=1\), then \(b_i=0\), so \(\neg u_i\) is true.  Therefore the clause is
true.
\end{proof}

\begin{definition}[Truth-table CNF]
\label{def:appI-truth-table-cnf}
Let \(R\subseteq\bits^k\) be a Boolean relation.  Its truth-table CNF is
\[
  \CNF(R)(u_1,\ldots,u_k)
  :=
  \bigwedge_{a\in\bits^k\setminus R} C_a(u_1,\ldots,u_k).
\]
If \(R=\bits^k\), the empty conjunction is interpreted as true.
\end{definition}

\begin{lemma}[Truth-table CNF correctness]
\label{lem:appI-truth-table-cnf-correct}
For every \(b\in\bits^k\),
\[
  b\models \CNF(R)
  \quad\Longleftrightarrow\quad
  b\in R.
\]
\end{lemma}

\begin{proof}
The input \(b\) satisfies \(\CNF(R)\) exactly when it satisfies every clause
\(C_a\) with \(a\notin R\).  By
Lemma~\ref{lem:appI-forbidden-pattern-correct}, this is exactly the condition
that \(b\ne a\) for every forbidden pattern \(a\notin R\).  Since the domain is
\(\bits^k\), that condition is equivalent to \(b\in R\).
\end{proof}

\begin{lemma}[Constant size for bounded arity]
\label{lem:appI-bounded-arity-size}
If \(k\le k_0\), where \(k_0\) is a fixed constant, then \(\CNF(R)\) has at
most \(2^{k_0}\) clauses, each of width at most \(k_0\).  Thus its size is
bounded by a constant depending only on \(k_0\).
\end{lemma}

\begin{proof}
There is one clause for each forbidden pattern in \(\bits^k\setminus R\), and
there are at most \(2^k\le 2^{k_0}\) such patterns.  Each clause mentions at
most one literal for each of the \(k\le k_0\) variables.
\end{proof}

\begin{remark}[The XOR equation]
\label{rem:appI-xor-cnf}
The gauge equation
\[
  z=x\oplus g
\]
is the arity-three relation
\[
  R_{\oplus}:=\{(x,g,z)\in\bits^3:z=x\oplus g\}.
\]
The truth-table CNF is constant size.  Explicitly, one equivalent CNF is
\[
  (x\vee g\vee \neg z)
  \wedge
  (x\vee \neg g\vee z)
  \wedge
  (\neg x\vee g\vee z)
  \wedge
  (\neg x\vee \neg g\vee \neg z).
\]
\end{remark}

\begin{remark}[Arbitrary CNF is enough]
\label{rem:appI-arbitrary-cnf-enough}
The final upper-bound argument uses a polynomial-time decision procedure for
CNF satisfiability.  Therefore constant-width CNF is already sufficient.  If one
wants a 3-CNF presentation, each bounded-width clause can be converted to an
equisatisfiable 3-CNF gadget with only a constant number of auxiliary variables.
The projection from any satisfying assignment of the refined formula to the
original variables preserves the witness and message.  Parsimony of this optional
3-CNF refinement is not needed for the self-reduction.
\end{remark}

\subsection{Local presentations of the witness relation}
\label{subsec:appI-local-presentations}

The ensemble verifier is already a bounded-arity local constraint system.  We
record a finite compiler for such systems.

\begin{definition}[Constraint record]
\label{def:appI-constraint-record}
A constraint record over a finite variable set \(U\) is a tuple
\[
  \chi=(\mathrm{kind},(u_1,\ldots,u_k),\theta),
\]
where \(u_i\in U\), the arity \(k\) is bounded by a fixed constant
\(k_{\max}\), \(\mathrm{kind}\) belongs to a fixed finite list of constraint
kinds, and \(\theta\) is a public finite parameter.  The record determines a
relation
\[
  R_{\chi}\subseteq\bits^k.
\]
The table of \(R_{\chi}\) is computable from \((\mathrm{kind},\theta)\) in time
polynomial in the length of \(\theta\).  In the ensemble, the total public
parameter length is polynomial in \(|Y|\).
\end{definition}

\begin{definition}[Local presentation]
\label{def:appI-local-presentation}
A local presentation at public input \(Y\) consists of:
\begin{enumerate}[label=(\roman*)]
\item a finite variable set \(U_Y\);
\item a distinguished tuple of witness variables inside \(U_Y\);
\item a finite list of constraint records
      \[
        \mathsf{Loc}(Y)=(\chi_1,\ldots,
        \chi_N)
      \]
      over \(U_Y\).
\end{enumerate}
An assignment \(\alpha:U_Y\to\bits\) is locally accepting if, for every record
\(\chi_s\), the tuple of values of \(\alpha\) on the scope of \(\chi_s\) belongs
to \(R_{\chi_s}\).
\end{definition}

\begin{definition}[CNF compilation of a local presentation]
\label{def:appI-local-cnf-compilation}
Given \(\mathsf{Loc}(Y)=(\chi_1,\ldots,
\chi_N)\), define
\[
  F_Y
  :=
  \bigwedge_{s=1}^N
  \CNF(R_{\chi_s})(\mathrm{scope}(\chi_s)).
\]
The notation means that the scoped variables of \(\chi_s\) are substituted into
the truth-table CNF for \(R_{\chi_s}\).
\end{definition}

\begin{lemma}[Correctness of local CNF compilation]
\label{lem:appI-local-cnf-correct}
For every assignment \(\alpha:U_Y\to\bits\),
\[
  \alpha\models F_Y
  \quad\Longleftrightarrow\quad
  \alpha\text{ is locally accepting for }\mathsf{Loc}(Y).
\]
\end{lemma}

\begin{proof}
The formula \(F_Y\) is the conjunction of the truth-table CNFs for the local
relations.  By Lemma~\ref{lem:appI-truth-table-cnf-correct}, \(\alpha\)
satisfies the \(s\)-th conjunct exactly when the values on the scope of
\(\chi_s\) belong to \(R_{\chi_s}\).  Satisfying all conjuncts is therefore exactly
local acceptance.
\end{proof}

\begin{lemma}[Polynomial size and uniformity]
\label{lem:appI-local-cnf-size-uniform}
Assume \(\mathsf{Loc}(Y)\) contains at most \(|Y|^c\) constraint records, all of
arity at most \(k_{\max}\), and that the list \(\mathsf{Loc}(Y)\) is computable
from \(Y\) in time \(|Y|^{c'}\).  Then the compiled formula \(F_Y\) has size
\(\poly(|Y|)\) and is computable from \(Y\) in time \(\poly(|Y|)\).
\end{lemma}

\begin{proof}
Each constraint contributes constant-size CNF by
Lemma~\ref{lem:appI-bounded-arity-size}.  Since there are polynomially many
constraints, the total number of clauses and literals is polynomial in \(|Y|\).
The compiler first constructs the local list and then replaces each record by its
constant-size truth-table CNF, so the construction time is polynomial.
\end{proof}

\begin{remark}[Relation to Cook--Levin]
\label{rem:appI-cook-levin}
The local compiler is enough here because the verifier is already presented as a
bounded-arity constraint system.  Starting from an arbitrary polynomial-time
verifier, one would first use the standard Cook--Levin tableau or circuit
construction and then apply bounded-arity CNF conversion gate by gate.  The
present construction is a direct local instance of that standard mechanism.
\end{remark}

\subsection{The gauge-buffered locked relation as a local presentation}
\label{subsec:appI-gauge-buffered-local-presentation}

We now instantiate the local compiler with the witness relation of the
construction.  The witness has the form
\[
  W=(x,g,z,\xi,M),
  \qquad
  x,g,z\in\bits^V,
  \qquad
  M\in\bits^{r_t}.
\]
The public instance \(Y\) fixes the geometry, the locked-layer local constraints,
the buffer local constraints, and the readout local constraints.

\begin{definition}[Variable blocks]
\label{def:appI-variable-blocks}
For each public instance \(Y\), let the CNF variable set be partitioned as
\[
  U_Y
  =
  U_x\sqcup U_g\sqcup U_z\sqcup U_{\xi}\sqcup U_M\sqcup U_{\mathrm{aux}}.
\]
The first five blocks encode \((x,g,z,\xi,M)\).  The block
\(U_{\mathrm{aux}}\) contains Tseitin and local-gadget auxiliary variables.  The
message projection is the fixed coordinate projection
\[
  \pi_M:\bits^{U_Y}\to\bits^{r_t}
\]
which reads \(U_M\) in canonical order.
\end{definition}

\begin{definition}[Gauge-buffered locked local list]
\label{def:appI-gbl-local-list}
The local list \(\mathsf{Loc}_{\mathrm{GBL}}(Y)\) is the concatenation of these
bounded-arity constraint families.
\begin{enumerate}[label=(\roman*)]
\item For every \(v\in V\), include the gauge equation
      \[
        z_v=x_v\oplus g_v.
      \]
\item Include the local constraints defining
      \(\mathrm{Lock}_Y(z,\xi_{\mathrm{lock}},M)=1\).
\item Include the local constraints defining
      \(\mathrm{Buf}_Y(z,\xi_{\mathrm{buf}})=1\).
\item Include the local constraints defining
      \(\mathrm{Read}_Y(z,\xi_{\mathrm{lock}},M)=1\).
\item Include the Tseitin and local-gadget consistency constraints tying the
      auxiliary variables to the intended local predicates.
\end{enumerate}
All arities are bounded by a fixed constant, and the number of records is
\(\poly(|Y|)\).
\end{definition}

\begin{definition}[Gauge-buffered locked CNF]
\label{def:appI-gbl-cnf}
Let
\[
  F_Y:=\CNF(\mathsf{Loc}_{\mathrm{GBL}}(Y))
\]
be the CNF obtained by Definition~\ref{def:appI-local-cnf-compilation}.
\end{definition}

\begin{definition}[Witness extraction]
\label{def:appI-witness-extraction}
Given an assignment \(\alpha:U_Y\to\bits\), define
\[
  \mathsf{Wit}_Y(\alpha)
  :=
  (x^{\alpha},g^{\alpha},z^{\alpha},\xi^{\alpha},M^{\alpha})
\]
by reading the blocks \(U_x,U_g,U_z,U_{\xi},U_M\) and ignoring
\(U_{\mathrm{aux}}\).  Thus
\[
  M^{\alpha}=\pi_M(\alpha).
\]
\end{definition}

\begin{proposition}[CNF realization of the witness relation]
\label{prop:appI-cnf-realizes-witness-relation}
For every public instance \(Y\) and every assignment \(\alpha:U_Y\to\bits\),
\[
  \alpha\models F_Y
  \quad\Longrightarrow\quad
  \mathcal R(Y,\mathsf{Wit}_Y(\alpha))=1.
\]
Conversely, if \(W=(x,g,z,\xi,M)\) satisfies \(\mathcal R(Y,W)=1\), then
there exists an extension \(\widehat\alpha:U_Y\to\bits\), agreeing with \(W\) on
the witness blocks, such that
\[
  \widehat\alpha\models F_Y.
\]
\end{proposition}

\begin{proof}
If \(\alpha\models F_Y\), Lemma~\ref{lem:appI-local-cnf-correct} implies that
all records in \(\mathsf{Loc}_{\mathrm{GBL}}(Y)\) are locally satisfied.  The gauge
records give \(z_v=x_v\oplus g_v\) for all \(v\).  The locked, buffer, readout,
and auxiliary records give the other clauses of the witness relation.  Hence the
extracted witness satisfies \(\mathcal R\).

Conversely, suppose \(\mathcal R(Y,W)=1\).  The witness values satisfy every
semantic local predicate in the gauge, locked, buffer, and readout layers.  For
each auxiliary variable, choose the value dictated by the corresponding Tseitin
or local-gadget definition.  With these auxiliary values added, every constraint
record in \(\mathsf{Loc}_{\mathrm{GBL}}(Y)\) is locally accepting.  Therefore
Lemma~\ref{lem:appI-local-cnf-correct} gives \(\widehat\alpha\models F_Y\).
\end{proof}

\begin{corollary}[Polynomial-size uniform SAT realization]
\label{cor:appI-polysize-uniform-sat-realization}
The map \(Y\mapsto F_Y\) is computable in polynomial time and
\[
  |F_Y|\le \poly(|Y|).
\]
Moreover,
\[
  F_Y\text{ is satisfiable}
  \quad\Longleftrightarrow\quad
  \exists W\;\mathcal R(Y,W)=1.
\]
\end{corollary}

\begin{proof}
The size and uniformity are Lemma~\ref{lem:appI-local-cnf-size-uniform} applied
to \(\mathsf{Loc}_{\mathrm{GBL}}(Y)\).  The satisfiability equivalence is
Proposition~\ref{prop:appI-cnf-realizes-witness-relation}: satisfying CNF
assignments project to satisfying witnesses, and satisfying witnesses extend to
satisfying CNF assignments.
\end{proof}

\subsection{Single-message readout at the CNF level}
\label{subsec:appI-single-message-cnf-level}

The CNF formula may have many satisfying assignments.  The proof needs only
that all of them agree on the projected message block.

\begin{definition}[Public support]
\label{def:appI-public-support}
For parameter \(m\), let \(\mathcal D_m\) be the public marginal of the
witnessed sampler, and let
\[
  \Supp(\mathcal D_m)
\]
be its finite support.  Thus \(Y\in\Supp(\mathcal D_m)\) means that the sampler
outputs \(Y\) with positive probability.
\end{definition}

\begin{lemma}[Satisfiability on public support]
\label{lem:appI-satisfiable-on-support}
For every \(Y\in\Supp(\mathcal D_m)\), the formula \(F_Y\) is satisfiable.
\end{lemma}

\begin{proof}
By definition of the witnessed sampler, every sampled public instance is output
with at least one witness \(W\) satisfying \(\mathcal R(Y,W)=1\).  Thus such a
witness exists for every support point.  The implication to satisfiability is
Corollary~\ref{cor:appI-polysize-uniform-sat-realization}.
\end{proof}

\begin{lemma}[CNF-level single-message property]
\label{lem:appI-cnf-level-single-message}
Assume the single-message promise:
\[
  \mathcal R(Y,W)=1
  \wedge
  \mathcal R(Y,W')=1
  \quad\Longrightarrow\quad
  M(W)=M(W').
\]
If \(\alpha\models F_Y\) and \(\alpha'\models F_Y\), then
\[
  \pi_M(\alpha)=\pi_M(\alpha').
\]
\end{lemma}

\begin{proof}
By Proposition~\ref{prop:appI-cnf-realizes-witness-relation}, the extracted
witnesses
\[
  W=\mathsf{Wit}_Y(\alpha),
  \qquad
  W'=\mathsf{Wit}_Y(\alpha')
\]
satisfy \(\mathcal R(Y,W)=1\) and \(\mathcal R(Y,W')=1\).  The single-message
promise gives \(M(W)=M(W')\).  By the definition of witness extraction,
\(M(W)=\pi_M(\alpha)\) and \(M(W')=\pi_M(\alpha')\).
\end{proof}

\begin{definition}[Message of a public instance]
\label{def:appI-message-of-public-instance}
For \(Y\in\Supp(\mathcal D_m)\), define
\[
  M(Y):=\pi_M(\alpha)
\]
for any satisfying assignment \(\alpha\models F_Y\).  This is well-defined by
Lemma~\ref{lem:appI-satisfiable-on-support} and
Lemma~\ref{lem:appI-cnf-level-single-message}.
\end{definition}

\begin{proposition}[Fixed projection readout]
\label{prop:appI-fixed-projection-readout}
For every \(Y\in\Supp(\mathcal D_m)\) and every satisfying assignment
\(\alpha\models F_Y\),
\[
  \pi_M(\alpha)=M(Y).
\]
The projection \(\pi_M\) is computable in time polynomial in \(|\alpha|\), and
its program description is independent of \(m,t,Y\).
\end{proposition}

\begin{proof}
The equality is Definition~\ref{def:appI-message-of-public-instance}.  The
projection reads a canonically specified block of message variables from the
assignment encoding.  The ordering convention is fixed once and for all, so the
projection routine is a fixed polynomial-time program.
\end{proof}

\begin{theorem}[Uniform SAT realization and readout]
\label{thm:appI-uniform-sat-realization}
For every \(Y\in\Supp(\mathcal D_m)\):
\begin{enumerate}[label=(\roman*)]
\item \(F_Y\) is satisfiable;
\item every satisfying assignment \(\alpha\models F_Y\) projects to a witness
      satisfying \(\mathcal R(Y,\mathsf{Wit}_Y(\alpha))=1\);
\item every satisfying assignment \(\alpha\models F_Y\) has message projection
      \(\pi_M(\alpha)=M(Y)\);
\item the map \(Y\mapsto F_Y\) is uniform polynomial-time and
      \(|F_Y|\le\poly(|Y|)\).
\end{enumerate}
\end{theorem}

\begin{proof}
Item (i) is Lemma~\ref{lem:appI-satisfiable-on-support}.  Item (ii) is
Proposition~\ref{prop:appI-cnf-realizes-witness-relation}.  Item (iii) is
Proposition~\ref{prop:appI-fixed-projection-readout}.  Item (iv) is
Corollary~\ref{cor:appI-polysize-uniform-sat-realization}.
\end{proof}

\begin{remark}[No uniqueness-of-witness requirement]
\label{rem:appI-no-unique-witness}
The formula \(F_Y\) may have many satisfying assignments.  The upper bound does
not need a unique assignment.  A SAT search procedure may return any satisfying
assignment, because all satisfying assignments have the same projected message.
\end{remark}

\subsection{SAT decision gives SAT search}
\label{subsec:appI-sat-decision-to-search}

We now record the standard bit-fixing self-reduction.  It is stated in a clocked
form because the final clash uses clocked polytime-capped description length.

\begin{definition}[Variable order and restriction]
\label{def:appI-variable-order-restriction}
For a CNF formula \(F\), let
\[
  \Var(F)=(u_1,\ldots,u_N)
\]
be the canonical lexicographic ordering of its variables.  For a prefix
\(a=(a_1,\ldots,a_i)\in\bits^i\), define
\[
  F[a]
\]
to be the CNF obtained from \(F\) by setting
\(u_1=a_1,\ldots,u_i=a_i\) and simplifying the result.
\end{definition}

\begin{definition}[Bit-fixing search]
\label{def:appI-bit-fixing-search}
Let \(D_{\SAT}\) be a deterministic SAT decision procedure.  Define
\(\mathsf{FindSat}_{D_{\SAT}}(F)\), for satisfiable CNF formulas \(F\), as
follows.  Starting with the empty prefix, process variables in the order
\(u_1,\ldots,u_N\).  At step \(i\), query \(D_{\SAT}\) on
\[
  F[(\alpha_1,\ldots,\alpha_{i-1},0)].
\]
If the restricted formula is satisfiable, set \(\alpha_i=0\); otherwise set
\(\alpha_i=1\).  After all variables are assigned, output
\(\alpha=(\alpha_1,\ldots,
\alpha_N)\).
\end{definition}

\begin{lemma}[Bit-fixing invariant]
\label{lem:appI-bit-fixing-invariant}
Assume \(D_{\SAT}\) decides satisfiability correctly.  If \(F\) is satisfiable,
then after the first \(i\) stages of \(\mathsf{FindSat}_{D_{\SAT}}\), the
restricted formula
\[
  F[(\alpha_1,\ldots,
  \alpha_i)]
\]
is satisfiable.
\end{lemma}

\begin{proof}
Induct on \(i\).  The case \(i=0\) is the assumption that \(F\) is satisfiable.
Assume the invariant after \(i-1\) stages.  If the restriction with
\(u_i=0\) is satisfiable, the procedure sets \(\alpha_i=0\), and the invariant
continues to hold.  If the restriction with \(u_i=0\) is not satisfiable, then
because the prefix restriction before assigning \(u_i\) is satisfiable, some
satisfying completion must set \(u_i=1\).  Hence the restriction with
\(u_i=1\) is satisfiable, and the invariant again holds.
\end{proof}

\begin{proposition}[Correctness of bit-fixing]
\label{prop:appI-bit-fixing-correct}
If \(D_{\SAT}\) decides satisfiability correctly and \(F\) is satisfiable, then
\(\mathsf{FindSat}_{D_{\SAT}}(F)\) outputs a satisfying assignment of \(F\).
\end{proposition}

\begin{proof}
Apply Lemma~\ref{lem:appI-bit-fixing-invariant} at \(i=N\).  A complete
assignment leaves a satisfiable restricted formula if and only if the complete
assignment satisfies the original formula.
\end{proof}

\begin{lemma}[Polynomial runtime of bit-fixing]
\label{lem:appI-bit-fixing-runtime}
If \(D_{\SAT}\) runs in time at most \((|F|+2)^d\), then there are constants
\(d'\) and \(C\), depending only on \(d\) and on the encoding conventions, such
that \(\mathsf{FindSat}_{D_{\SAT}}(F)\) runs in time at most
\[
  C(|F|+2)^{d'}.
\]
\end{lemma}

\begin{proof}
There are at most \(|F|\) variables.  At each stage the restricted formula has
size polynomially bounded by \(|F|\), and computing it costs polynomial time.
The SAT decision call costs a fixed polynomial in the restricted formula size.
A polynomial number of polynomial-time stages is polynomial-time.
\end{proof}

\begin{proposition}[SAT search from \(P=NP\)]
\label{prop:appI-sat-search-from-pnp}
Assume \(P=NP\).  Then there is a fixed deterministic polynomial-time program
\[
  \mathsf{FindSat}_*
\]
which, on every satisfiable CNF formula \(F\), outputs a satisfying assignment
of \(F\).
\end{proposition}

\begin{proof}
Since \(\SAT\in NP\), the assumption \(P=NP\) gives a deterministic
polynomial-time decider \(D_{\SAT,*}\) for CNF satisfiability.  Define
\(\mathsf{FindSat}_*:=\mathsf{FindSat}_{D_{\SAT,*}}\).  Correctness is
Proposition~\ref{prop:appI-bit-fixing-correct}, and polynomial runtime is
Lemma~\ref{lem:appI-bit-fixing-runtime}.
\end{proof}

\subsection{The constant program computing the message under \boldmath{$P=NP$}}
\label{subsec:appI-pnp-message-decoder}

\begin{definition}[The \(P=NP\) message decoder]
\label{def:appI-pnp-message-decoder}
Assume \(P=NP\), and let \(\mathsf{FindSat}_*\) be the fixed search procedure
from Proposition~\ref{prop:appI-sat-search-from-pnp}.  Define
\(\mathsf{MsgDec}_*\) on input \(Y\) by:
\begin{enumerate}[label=(\roman*)]
\item construct the CNF formula \(F_Y\);
\item run \(\mathsf{FindSat}_*(F_Y)\), obtaining a satisfying assignment
      \(\alpha\);
\item output \(\pi_M(\alpha)\).
\end{enumerate}
On inputs outside the public support, the procedure may behave arbitrarily;
the upper-bound claim is support-wise.
\end{definition}

\begin{lemma}[Correctness of the message decoder]
\label{lem:appI-message-decoder-correct}
Assume \(P=NP\).  For every \(Y\in\Supp(\mathcal D_m)\),
\[
  \mathsf{MsgDec}_*(Y)=M(Y).
\]
\end{lemma}

\begin{proof}
For \(Y\in\Supp(\mathcal D_m)\), the formula \(F_Y\) is satisfiable by
Theorem~\ref{thm:appI-uniform-sat-realization}.  The procedure
\(\mathsf{FindSat}_*\) therefore returns a satisfying assignment \(\alpha\).  By
Theorem~\ref{thm:appI-uniform-sat-realization} again,
\(\pi_M(\alpha)=M(Y)\).  That is the output of \(\mathsf{MsgDec}_*\).
\end{proof}

\begin{lemma}[Clock exponent of the message decoder]
\label{lem:appI-message-decoder-clock}
Assume \(P=NP\).  There is an exponent \(D_*\ge 1\), depending only on the
fixed SAT decider supplied by \(P=NP\), the fixed CNF compiler, and the fixed
message projection, such that \(\mathsf{MsgDec}_*(Y)\) runs within
\[
  (|Y|+2)^{D_*}
\]
steps for every \(Y\in\Supp(\mathcal D_m)\).
\end{lemma}

\begin{proof}
The map \(Y\mapsto F_Y\) runs in polynomial time and produces a formula of
size polynomial in \(|Y|\).  The fixed search procedure \(\mathsf{FindSat}_*\)
runs in polynomial time in \(|F_Y|\).  The projection \(\pi_M\) is fixed and
polynomial-time.  Also \(|M(Y)|=r_t\) is polynomial in the ensemble parameter
and hence polynomial in \(|Y|\).  Composing these polynomial-time routines gives
a fixed polynomial-time clock in \(|Y|\).  Let \(D_*\) be any sufficiently large
exponent.
\end{proof}

\begin{theorem}[Clocked \(P=NP\) upper bound]
\label{thm:appI-clocked-pnp-upper-bound}
Assume \(P=NP\), and let \(D_*\) be the exponent from
Lemma~\ref{lem:appI-message-decoder-clock}.  There is a constant
\(c_{\mathrm{up}}\), independent of \(m,t\), and \(Y\), such that for every
\(Y\in\Supp(\mathcal D_m)\),
\[
  \Kp^{(D_*)}(M(Y)\mid Y)
  \le
  c_{\mathrm{up}}.
\]
\end{theorem}

\begin{proof}
Use the fixed program \(\mathsf{MsgDec}_*\) as the conditional description.  Its
code contains only the fixed SAT search routine, the fixed CNF compiler, and the
fixed message projection.  Thus its length is a constant \(c_{\mathrm{up}}\),
independent of the parameters and of \(Y\).  By
Lemma~\ref{lem:appI-message-decoder-correct}, it outputs \(M(Y)\) on every
\(Y\in\Supp(\mathcal D_m)\).  By Lemma~\ref{lem:appI-message-decoder-clock},
it runs within the fixed \(D_*\)-clock.  This is exactly the displayed
\(\Kp^{(D_*)}\) bound.
\end{proof}

\begin{corollary}[Unclocked form]
\label{cor:appI-unclocked-upper-bound}
Assume \(P=NP\).  Then, in the unclocked convention of
Definition~\ref{def:kpoly},
\[
  \Kp(M(Y)\mid Y)=O(1)
\]
for all \(Y\in\Supp(\mathcal D_m)\).
\end{corollary}

\begin{proof}
The unclocked description may include the fixed exponent \(D_*\) at constant
additional cost.  The clocked upper bound from
Theorem~\ref{thm:appI-clocked-pnp-upper-bound} therefore gives the unclocked
bound.
\end{proof}

\begin{remark}[Support-wise correctness]
\label{rem:appI-support-wise-correctness}
The upper bound is asserted for \(Y\) in the support of the public ensemble.
Outside the support, \(F_Y\) may be unsatisfiable or may fail the sampler's
promise conditions.  This is exactly what the final clash needs: the lower bound
also concerns the ensemble distribution, and the contradiction selects a public
instance from a positive-probability lower-bound event inside the support.
\end{remark}

\subsection{Compatibility with the lower-bound proof}
\label{subsec:appI-compatibility-lower-bound}

The existence of a SAT realization does not conflict with the lower-bound
claim that the public instance does not cheaply reveal the message.  The lower
bound is resource-bounded and evidence-accounted; the upper bound assumes
\(P=NP\) and obtains a satisfying assignment by global SAT search.

\begin{proposition}[SAT realization is neutral to the evidence budget]
\label{prop:appI-sat-realization-neutral-to-budget}
The construction of \(F_Y\) in Theorem~\ref{thm:appI-uniform-sat-realization}
adds no new lower-bound primitive of the form
\[
  M_a=b,
  \qquad
  g_v=\gamma,
  \qquad
  x_v=\gamma
\]
as public target evidence.  All target-relevant use of witness variables in a
normalized evidence trace remains governed by the CD-ENF atom-completeness
interface: public template syntax is neutral, legal safe-buffer reads are safe
atoms, and quotient or witness-value use exposing raw support is charged as
hidden-gauge evidence.
\end{proposition}

\begin{proof}
The CNF construction is a uniform syntactic presentation of the same verifier
relation \(\mathcal R(Y,W)=1\).  It introduces public variable names, public
clauses describing local predicates, and auxiliary Tseitin variables.  These are
public template facts.  A clause mentioning a message variable is not a public
assignment to that variable; it is a constraint that a satisfying witness must
obey.  To turn such constraints into an actual message value, an observer must
produce enough evidence about a satisfying assignment or otherwise close an
opposite-message pair.  In the CD-ENF interface, that target-relevant use
normalizes to safe-buffer or gauge-support evidence and is charged by the
Atomic Evidence Budget.  Thus the SAT presentation changes syntax, not the
allowed evidence classes.
\end{proof}

\begin{remark}[No claim that all CNF cells are qSSM-small]
\label{rem:appI-no-all-cells-qssm}
The lower-bound proof does not require every public CNF clause or every raw
syntax cell to be a safe local buffer observation.  Public syntax may define a
global constraint system whose satisfying assignments have a unique message.
The proof only requires that target advantage acquired by an actual polynomial-time
observer be captured by the CD trace and charged through neutral, safe-buffer,
and hidden-gauge atoms.  The \(P=NP\) upper bound bypasses this budget only
under the contradictory assumption that SAT search itself is polynomial-time.
\end{remark}

\subsection{The final upper-lower clash}
\label{subsec:appI-final-clash}

Appendix~\ref{app:product-compression} exports the lower-bound interface: for
every fixed clock exponent \(D\), after choosing the ensemble parameters for
that exponent, there are constants \(\eta_D,\kappa_D>0\) such that
\[
  \Prb\left[
    \Kp^{(D)}(M(Y)\mid Y)
    \ge
    \eta_D t
  \right]
  \ge
  1-2^{-\kappa_D t}.
\]
We instantiate this lower bound at the clock exponent supplied by the
hypothetical \(P=NP\) SAT solver.

\begin{theorem}[Final contradiction]
\label{thm:appI-final-contradiction}
The assumption \(P=NP\) is false.
\end{theorem}

\begin{proof}
Assume \(P=NP\).  By Theorem~\ref{thm:appI-clocked-pnp-upper-bound}, there are
an exponent \(D_*\) and a constant \(c_{\mathrm{up}}\) such that
\[
  \Kp^{(D_*)}(M(Y)\mid Y)
  \le
  c_{\mathrm{up}}
\]
for every \(Y\in\Supp(\mathcal D_m)\).

Now choose the ensemble parameters for this fixed exponent \(D_*\), as allowed
by the lower-bound theorem.  Then there are constants
\(\eta_{D_*},\kappa_{D_*}>0\) such that
\[
  \Prb\left[
    \Kp^{(D_*)}(M(Y)\mid Y)
    \ge
    \eta_{D_*}t
  \right]
  \ge
  1-2^{-\kappa_{D_*}t}.
\]
Choose \(t\) large enough that
\[
  \eta_{D_*}t>c_{\mathrm{up}}
  \qquad\text{and}\qquad
  1-2^{-\kappa_{D_*}t}>0.
\]
The lower-bound event then has positive probability, so it contains at least one
support point \(Y\).  For that \(Y\),
\[
  \Kp^{(D_*)}(M(Y)\mid Y)
  \ge
  \eta_{D_*}t
  >
  c_{\mathrm{up}},
\]
contradicting the pointwise upper bound on the support.  Hence \(P=NP\) cannot
hold.
\end{proof}

\begin{corollary}[Main theorem]
\label{cor:appI-main-theorem}
\[
  P\ne NP.
\]
\end{corollary}

\begin{proof}
This is Theorem~\ref{thm:appI-final-contradiction}.
\end{proof}

\subsection{Clocked final-clash interface}
\label{subsec:appI-clocked-final-clash-interface}

The main text uses the following exported interface.

\begin{theorem}[SAT realization and self-reduction interface]
\label{thm:appI-sat-selfreduction-interface}
For the gauge-buffered locked ensemble:
\begin{enumerate}[label=(\roman*)]
\item the verifier relation \(\mathcal R(Y,W)=1\) has a uniformly computable
      polynomial-size CNF realization \(F_Y\);
\item if \(Y\in\Supp(\mathcal D_m)\), then \(F_Y\) is satisfiable;
\item every satisfying assignment of \(F_Y\) projects to a valid satisfying
      witness;
\item every satisfying assignment of \(F_Y\) has the same message projection
      \(M(Y)\);
\item if \(P=NP\), then for some fixed polynomial clock exponent \(D_*\),
      independent of \(m,t,Y\),
      \[
        \Kp^{(D_*)}(M(Y)\mid Y)=O(1)
      \]
      for all \(Y\in\Supp(\mathcal D_m)\).
\end{enumerate}
\end{theorem}

\begin{proof}
Items (i)--(iv) are Theorem~\ref{thm:appI-uniform-sat-realization}.  Item (v)
is Theorem~\ref{thm:appI-clocked-pnp-upper-bound}.
\end{proof}

\begin{corollary}[Clocked contradiction interface]
\label{cor:appI-clocked-contradiction-interface}
Suppose the lower-bound theorem of Appendix~\ref{app:product-compression} is
available for every fixed polynomial clock exponent.  If \(P=NP\), let \(D_*\)
be the exponent from Theorem~\ref{thm:appI-clocked-pnp-upper-bound} and
instantiate the lower-bound parameters at \(D_*\).  Then the lower bound gives
\[
  \Kp^{(D_*)}(M(Y)\mid Y)
  \ge
  \eta t
\]
with positive probability for all sufficiently large \(t\), while
Theorem~\ref{thm:appI-clocked-pnp-upper-bound} gives
\[
  \Kp^{(D_*)}(M(Y)\mid Y)
  \le
  c_{\mathrm{up}}
\]
pointwise on the support.  For \(t>c_{\mathrm{up}}/\eta\), these two statements
are incompatible.
\end{corollary}

\begin{proof}
The lower-bound event has probability at least \(1-2^{-\Omega(t)}\), hence
positive probability for all sufficiently large \(t\).  Choose a public instance
in that event.  The lower-bound inequality and the pointwise upper-bound
inequality both apply to that same instance and contradict each other once
\(\eta t>c_{\mathrm{up}}\).
\end{proof}

\subsection{Lean-facing theorem package}
\label{subsec:appI-lean-facing}

A Lean formalization can split this appendix into the following modules.

\begin{enumerate}[label=(\roman*)]
\item \textbf{Finite CNF syntax.}  Variables, literals, clauses, CNF formulas,
      finite assignments, satisfaction, and restrictions by partial assignments.

\item \textbf{Truth-table CNF.}  For every bounded-arity relation
      \(R\subseteq\bits^k\), construct \(\CNF(R)\) and prove
      Lemma~\ref{lem:appI-truth-table-cnf-correct}.

\item \textbf{Local compiler.}  Define constraint records, local presentations,
      compiled CNFs, and prove Lemma~\ref{lem:appI-local-cnf-correct} plus
      polynomial-size bookkeeping.

\item \textbf{Gauge-buffered realization.}  Instantiate the local compiler with
      the gauge, locked, buffer, readout, and auxiliary constraint lists, proving
      Proposition~\ref{prop:appI-cnf-realizes-witness-relation}.

\item \textbf{Single-message projection.}  Prove the CNF-level single-message
      property, Lemma~\ref{lem:appI-cnf-level-single-message}, and the fixed
      projection readout, Proposition~\ref{prop:appI-fixed-projection-readout}.

\item \textbf{SAT search.}  Formalize the bit-fixing self-reduction and prove
      Proposition~\ref{prop:appI-bit-fixing-correct}.

\item \textbf{Clocked upper bound.}  Package the fixed \(P=NP\) SAT search,
      the uniform CNF compiler, and the message projection into one
      constant-length program, proving
      Theorem~\ref{thm:appI-clocked-pnp-upper-bound}.

\item \textbf{Final clash interface.}  Import the lower-bound theorem at the
      exponent \(D_*\), select a positive-probability hard instance, and derive
      the contradiction.
\end{enumerate}

\end{document}